\documentclass[aps,prd,reprint,showpacs,showkeys,amsmath,amssymb,eqsecnum,nofootinbib,notitlepage,superscriptaddress,floatfix]{revtex4-1}

\usepackage{hyperref}
\usepackage{dcolumn}	% alignment of numbers on decimal points
\usepackage{graphicx}
\usepackage{url}
\usepackage{bm}		% bold math symbols
\usepackage{color}
\usepackage[usenames,dvipsnames,svgnames,table]{xcolor}

\newcommand{\ijcoord}[1]{\textsf{#1}}

\newcommand{\SubroutineLibrary}[1]{\textsf{#1}}
\newcommand{\subroutine}[1]{\textsf{#1}}
\newcommand{\variablename}[1]{\texttt{#1}}
\newcommand{\allbf}[1]{\textbf{\mathversion{bold}{#1}}}

\newcommand{\cf}{cf.\hbox{}}
\newcommand{\etal}{{\it et~al.\hbox{}\/}}

\newcommand{\Cplusplus}{\hbox{C\raise.2ex\hbox{\footnotesize ++}}}
\renewcommand{\P}[1]{\phantom{#1}}
\newcommand{\Z}{\phantom{0}}		% useful for aligning numeric tables

\newcommand{\Coeff}[1]{C[#1]}
\newcommand{\Smooth}[1]{\mathcal{S} \left[ #1 \right]}
\newcommand{\noise}{\text{noise}}
\newcommand{\Noise}[1]{\text{noise} \left[ #1 \right]}
\newcommand{\RelativeNoise}[1]{\text{relative\_noise} \left[ #1 \right]}

\newcommand{\half}{\frac{1}{2}}
\newcommand{\thalf}{\tfrac{1}{2}}
\newcommand{\tthird}{\tfrac{1}{3}}

\newcommand{\apoastron}{\text{apoastron}}
\newcommand{\cons}{\text{cons}}
\newcommand{\diss}{\text{diss}}
\newcommand{\effective}{\text{effective}}

\newcommand{\erf}{\mathop{\text{erf}}}
\newcommand{\erfc}{\mathop{\text{erfc}}}
\newcommand{\equator}{\text{equator}}
\newcommand{\even}{\text{even}}
\newcommand{\fit}{\mathop{\text{fit}}}

\newcommand{\main}{\text{main}}
\newcommand{\maxmove}{\text{max-move}}
\newcommand{\move}{\text{move}}
\newcommand{\numerical}{\text{num}}

\newcommand{\odd}{\text{odd}}
\newcommand{\particle}{\text{particle}}
\newcommand{\preliminary}{\text{preliminary}}
\newcommand{\puncture}{\text{puncture}}
\newcommand{\relative}{\text{relative}}
\newcommand{\residual}{\text{residual}}
\newcommand{\regular}{\text{regular}}
\newcommand{\sample}{\text{sample}}

\newcommand{\singular}{\text{singular}}
\newcommand{\temp}{\text{temp}}
\newcommand{\myinput}{\text{(input)}}
\newcommand{\initial}{\text{initial}}

\newcommand{\startup}{\text{startup}}
\newcommand{\tailset}{\text{tail-set}}
\newcommand{\WTcenter}{\variablename{WT\_center}}
\newcommand{\FF}{\mathsf{F}}
\newcommand{\GG}{\mathsf{G}}

\newcommand{\kk}{\mathsf{k}}
\renewcommand{\O}{\mathcal{O}}
\newcommand{\uu}{\mathsf{u}}
\newcommand{\boxop}{\Box}
\newcommand{\boundary}[1]{{\partial #1}}
\newcommand{\del}{\nabla}
\newcommand{\Diss}{\mathsf{D}}
\newcommand{\conjugate}[1]{\mathop{\text{conj}}\left[#1\right]}
\newcommand{\realpart}[1]{\mathop{\text{Re}}\left[#1\right]}
\newcommand{\imagpart}[1]{\mathop{\text{Im}}\left[#1\right]}
\newcommand{\sun}{\odot}

\newcommand{\ltsim}{\lesssim}
\newcommand{\gtsim}{\gtrsim}

\newcommand{\RMS}{\mathop{\text{RMS}}}

\newcommand{\Hbar}{{\,\overline{\!{\vrule height 1.8ex depth 0ex width 0em}H\!}\,}}

\newcommand{\Scri}{\mathcal{J}}
\newcommand{\Scriplus}{{\Scri^+}}

\newcommand{\CenterWithSizeOf}[2]{{\setbox0=\hbox{#2}\hbox to\wd0{\hss{#1}\hss}}}

\begin{document}
\title{Scalar self-force for highly eccentric equatorial orbits in Kerr spacetime}

\author{Jonathan Thornburg}
\email{jthorn@astro.indiana.edu}
\affiliation{Department of Astronomy and Center for Spacetime Symmetries, Indiana University, Bloomington, Indiana 47405, USA}

\author{Barry Wardell}
\email{barry.wardell@gmail.com}
\affiliation{School of Mathematics and Statistics and Institute for Discovery, University College Dublin, Belfield, Dublin 4, Ireland}
\affiliation{Department of Astronomy, Cornell University, Ithaca, NY 14853, USA}

%%%%%%%%%%%%%%%%%%%%%%%%%%%%%%%%%%%%%%%%

\begin{abstract}
If a small ``particle'' of mass $\mu M$ (with $\mu \ll 1$) orbits
a black hole of mass $M$, the leading-order radiation-reaction effect
is an $\O(\mu^2)$ ``self-force'' acting on the particle, with a
corresponding $\O(\mu)$ ``self-acceleration'' of the particle away
from a geodesic.
Such ``extreme--mass-ratio inspiral'' systems are likely to be
important gravitational-wave sources for future space-based
gravitational-wave detectors.
Here we consider the ``toy model'' problem of computing the self-force
for a scalar-field particle on a bound eccentric orbit in Kerr spacetime.
We use the Barack-Golbourn-Vega-Detweiler effective-source regularization
with a 4th~order puncture field,
followed by an $e^{im\phi}$ (``m-mode'') Fourier decomposition and a
separate time-domain numerical evolution in $2{+}1$~dimensions for each~$m$.
We introduce a finite worldtube that surrounds the particle worldline
and define our evolution equations in a piecewise manner so that the
effective source is only used within the worldtube.
Viewed as a spatial region, the worldtube moves to follow the
particle's orbital motion.
We use slices of constant Boyer-Lindquist time in the region of the
particle's motion, deformed to be asymptotically hyperboloidal and
compactified near the horizon and $\Scri^+$.
Our numerical evolution uses Berger-Oliger mesh refinement with
4th~order finite differencing in space and time.
Our computational scheme allows computation for highly eccentric orbits
and should be generalizable to orbital evolution in the future.
Our present implementation is restricted to equatorial geodesic orbits,
but this restriction is not fundamental.
We present numerical results for a number of test cases
with orbital eccentricities as high as~$0.98$.
In some cases we find large oscillations (``wiggles'') in the self-force
on the outgoing leg of the orbit shortly after periastron passage; these
appear to be caused by the passage of the orbit through the strong-field
region close to the background Kerr black hole.
\end{abstract}

%%%%%%%%%%%%%%%%%%%%%%%%%%%%%%%%%%%%%%%%

% PACS numbers from 2003 edition

\maketitle

%%%%%%%%%%%%%%%%%%%%%%%%%%%%%%%%%%%%%%%%%%%%%%%%%%%%%%%%%%%%%%%%%%%%%%%%%%%%%%%%

\begin{quote}
\textit{This paper is dedicated to the memory of
our late friends and colleagues Thomas Radke and Steven Detweiler.}
\end{quote}

%%%%%%%%%%%%%%%%%%%%%%%%%%%%%%%%%%%%%%%%%%%%%%%%%%%%%%%%%%%%%%%%%%%%%%%%%%%%%%%%

\section{Introduction}
\label{sect:introduction}

%%%%%%%%%%%%%%%%%%%%%%%%%%%%%%%%%%%%%%%%%%%%%%%%%%%%%%%%%%%%%%%%%%%%%%%%%%%%%%%%

Consider a small (compact) body of mass $\mu M$ (with $0 < \mu \ll 1$)
moving freely in an asymptotically flat background spacetime
(e.g., Kerr spacetime) of mass~$M$.
This system emits gravitational radiation, and there is a corresponding
radiation-reaction influence on the small body's motion.  Self-consistently
calculating this motion and the emitted gravitational radiation
(and in general, the perturbed spacetime) is a long-standing research
question in general relativity.

There is also an astrophysical motivation for this calculation:
If a neutron star or stellar-mass black hole of mass~${\sim}\, 1$--$100 M_\sun$
orbits a massive black hole of mass~${\sim}\, 10^5$--$10^7 M_\sun$,%%%
\footnote{%%%
	 $M_\sun$ denotes the solar mass.
	 }%%%
{} the resulting ``extreme--mass-ratio inspiral'' (EMRI) system is
expected to be a strong astrophysical gravitational-wave (GW) source
detectable by the planned Laser Interferometer Space Array (LISA)
space-based gravitational-wave detector.%%%
\footnote{%%%
	 The LISA proposal has had various design and name
	 changes during its lifetime.  For a time it was known
	 as the New Gravitational-Wave Observatory (NGO) or
	 evolved LISA (eLISA), but recently it has returned to
	 the original name, LISA.%%%
	 }%%%
{}  LISA is expected to observe many such systems, some of
them at quite high signal/noise ratios
(\cite{Gair-etal-2004:LISA-EMRI-event-rates,Barack-Cutler-2004,%%%
Amaro-Seoane-etal-2007:LISA-IMRI-and-EMRI-review,%%%
Gair-2009:LISA-EMRI-event-rates}).
The data analysis for, and indeed the detection of, such systems
will generally require matched-filtering the detector data stream
against appropriate precomputed GW templates.  The problem of
computing such templates provides the astrophysical motivation for
our calculation.

We are particularly concerned with the case where the small body's
orbit is highly relativistic, so post-Newtonian methods (see, for
example,~\cite[section~6.10]{Damour-in-Hawking-Israel-1987};
\cite{Blanchet-2014-living-review,Futamase-Itoh-2007:PN-review,%%%
Blanchet-2009:PN-review,Schaefer-2009:PN-review} and references therein)
%% note Blanchet-2009:PN-review is article in Mass and Motion volume
are not reliably accurate.  Since the timescale for radiation reaction
to shrink the orbit is very long ($\sim \mu^{-1} M$) while the required
resolution near the small body is very high ($\sim \mu M$), a direct
``numerical relativity'' integration of the Einstein equations
(see, for example,~\cite{Pretorius-2007:2BH-review,%%%
Hannam-etal-2009:Samurai-project,Hannam-2009:2BH-review,%%%
Hannam-Hawke-2010:2BH-in-era-of-Einstein-telescope-review,%%%
Campanelli-etal-2010:2BH-numrel-review} and references therein)
would be prohibitively expensive (and probably insufficiently accurate)
for this problem.%%%
\footnote{%%%
	 A number of researchers have attempted direct
	 numerical-relativity binary black hole simulations
	 for systems with ``intermediate'' mass ratios up to
	 $100\,{:}\,1$ ($\mu = 0.01$), (see, for example,
\protect\cite{Bishop-etal-2003,Bishop-etal-2005,%%%
Sopuerta-etal-2006,Sopuerta-Laguna-2006,%%%
Lousto-etal-2010:intermediate-mass-2BH-numrel-Lazarus,%%%
Lousto-Zlochower-2011:100-to-1-mass-ratio-2BH}).
	 However, it has not (yet) been possible to extend
	 these results to the extreme-mass-ratio case
	 nor to accurately evolve even the~$100\,{:}\,1$
	 case for a radiation-reaction time scale.%%%
	 }%%%

Instead, we use black hole perturbation theory, treating the small body
as an $\O(\mu)$~perturbation on the background spacetime.  For this
work we attempt to calculate leading-order radiation-reaction effects,
i.e., $\O(\mu)$~field perturbations and $\O(\mu^2)$~radiation-reaction
``self-forces'' acting on the small body.  Because of the technical
difficulty of controlling gauge effects in gravitational perturbations,
in this work we use a scalar-field ``toy model'' system with the expectation
that the techniques developed and discoveries made in the scalar case
will carry over to the gravitational case.

The obvious way to model the small body is as a small black hole.
While conceptually elegant, this approach is technically somewhat
complicated~\cite{Poisson-Pound-Vega-2011:living-review}.  Instead,
we model the small body as a point particle.  Although one may be
concerned about potential foundational issues with this approach,%%%
\footnote{%%%
	 Geroch and Traschen~\protect\cite{Geroch-Traschen-1987}
	 have shown that point particles in general relativity can
	 \emph{not} consistently be described by metrics with
	 $\delta$-function stress-energy tensors.  More general
	 Colombeau-algebra methods may be able to resolve this
	 problem~\protect\cite{Steinbauer-Vickers-2006}, but the
	 precise meaning of the phrase ``point particle'' in
	 general relativity remains a delicate question.%%%
	 }%%%
{} in practice it works well and, importantly, it agrees with
rigorous derivations that do not rely on the use of point particles.

The $\O(\mu)$ ``MiSaTaQuWa'' equations of motion for a gravitational
point particle in a (strong-field) curved spacetime were first derived
by Mino, Sasaki, and Tanaka~\cite{Mino-Sasaki-Tanaka-1997} and
Quinn and Wald~\cite{Quinn-Wald-1997} (also see Detweiler's
analysis~\cite{Detweiler-2001:radiation-reaction-and-self-force})
and have recently been rederived in a more rigorous manner by
Gralla and Wald~\cite{Gralla-Wald-2008}.%%%
\footnote{%%%
	 Gralla, Harte, and Wald~\protect\cite{Gralla-Harte-Wald-2009}
	 have also recently obtained a rigorous derivation of
	 the electromagnetic self-force in a curved spacetime.%%%
	 }%%%
{}  See \cite{Poisson-Pound-Vega-2011:living-review,Detweiler-2005,%%%
Barack-2009:self-force-review,Barack-2009:self-force-review2,%%%
Burko-2009:self-force-review,Detweiler-2009:self-force-review,%%%
Poisson-2009:self-force-review,Wald-2009:self-force-review}
for general reviews of gravitational radiation-reaction dynamics.

The particle's motion may be modelled as either (i) non-geodesic motion
in the background Schwarzschild/Kerr spacetime under the influence of
a radiation-reaction ``self-force'', or (ii) geodesic motion in a
perturbed spacetime.  These two perspectives (which are in some ways
analogous to Eulerian versus Lagrangian formulations of fluid dynamics)
are equivalent~\cite{Sago-Barack-Detweiler-2008}; in this work we use
the formulation~(i).
The MiSaTaQuWa equations then give the self-force in terms of (the
gradient of) the metric perturbation due to the particle, which must
be computed using black-hole perturbation theory.

The computation of the field perturbation due to a point particle
is particularly difficult because the ``perturbation'' is formally
infinite at the particle and thus must be regularized.  There are
several different, but equivalent, regularization schemes known for this
problem, notably the ``mode-sum'' or ``$\ell$-mode'' scheme developed by
Barack and Ori~\cite{Barack-Ori-2000,Barack-2000,Barack-etal-2002,%%%
Barack-Ori-2002,Barack-Ori-2003}, Detweiler, Messaritaki, and
Whiting~\cite{Detweiler-Whiting-2003,Detweiler-Messaritaki-Whiting-2003},
and Haas and Poisson~\cite{Haas-Poisson-2006};
the Green-function approach~\cite{Anderson:2005gb,Casals:2009zh,%%%
Casals:2013mpa,Wardell-etal-2014:self-force-via-Green-fn}; and
the ``effective-source'' scheme of
Barack and Golbourn~\cite{Barack-Golbourn-2007} and
Vega and Detweiler~\cite{Vega-Detweiler-2008:self-force-regularization}.

For a detailed presentation of the different regularization/computation
schemes and their advantages and disadvantages, see~\cite{Wardell:2015kea}.
In the present context we observe that for a Kerr background the
traditional mode-sum scheme becomes less desirable because the mode
equations don't separate: all the (infinite set of) modes remain coupled.
While the coupled modes can still be treated numerically
(see, e.g.,~\cite{Warburton-Barack-2011}), here we adopt a
different approach, the effective-source regularization scheme.

As discussed in detail in Sec.~\ref{sect:theory/effective-src}, the
effective-source scheme's basic concept is to analytically compute a
``puncture field'' which approximates the particle's Detweiler-Whiting
singular field~\cite{Detweiler-Whiting-2003}, then numerically
solve for the difference between the actual field perturbation and
the puncture field.  We have previously described many of the details
of the computation of the puncture field~\cite{Wardell:2011gb};
in this work we focus on the application of this scheme to a
particular class of self-force computations.

Depending on how the partial differential equations (PDEs) are solved,
there are two broad classes of
self-force computations: frequency-domain and time-domain.
Frequency-domain computations involve a Fourier transform of the
PDEs in time, reducing the numerical computation to the solution
of a set of ordinary differential equations (ODEs) (see, for example,
\cite{Detweiler-Messaritaki-Whiting-2003}).  The resulting computations
are typically very efficient and accurate for circular or near-circular
particle orbits,%%%
\footnote{%%%
	 As notable examples of this accuracy,
Blanchet~\etal{}~\protect\cite{Blanchet-etal-2010:cmp-3PN-with-self-force}
	 and Shah~\etal{}~\protect\cite{Shah-etal-2011} have both
	 recently computed the gravitational self-force
	 for circular geodesic orbits in Schwarzschild
	 spacetime to a relative accuracy of approximately
	 one part in $10^{13}$, and
	 Heffernan, Ottewill, and Wardell~\protect\cite{Heffernan:2012su}
	 (building on earlier work by
Detweiler, Messaritaki, and Whiting~\protect\cite{Detweiler-Messaritaki-Whiting-2003})
	 have extended this to a few parts in $10^{17}$.
Johnson-McDaniel, Shah, and Whiting~\protect\cite{Johnson-McDaniel-Shah-Whiting-2015}
	 describe an ``experimental mathematics'' approach
	 to computing post-Newtonian expansions of various
	 invariants (again for circular geodesic orbits in
	 Schwarzschild spacetime) by applying an integer-relation
	 algorithm to numerical results calculated using up to
	 $5000$~decimal digits of precision.
 	 }%%%
{} but degrade rapidly in efficiency with increasing eccentricity
of the particle's orbit, becoming impractical for highly eccentric
orbits~\cite{Glampedakis-Kennefick-2002,Barack-Lousto-2005}.%%%
\footnote{%%%
	 Barack, Ori, and Sago~\protect\cite{Barack-Ori-Sago-2008}
	 have found an elegant solution for some other
	 limitations which had previously affected
	 frequency-domain calculations.%%%
	 }%%%
{}  In contrast, time-domain computations involve a direct numerical
time-integration of the PDEs and are generally less efficient and accurate
than frequency-domain computations.  However, time-domain computations
can accommodate arbitrary particle orbits with only modest penalties
in performance and accuracy
\cite{Barton-etal:cmp-EMRI-frequency-vs-time-domain-methods},
with some complications in the numerical schemes
(see, for example, \cite{Haas-2007,Barack-Sago-2010}).

In this work our goal is to consider highly eccentric orbits,%%%
\footnote{%%%
	 Hopman and Alexander~\protect\cite{Hopman-Alexander-2005}
	 find that LISA EMRIs are likely to have
	 eccentricities up to $e \sim 0.8$.
	 \emph{Intermediate}--mass-ratio-inspirals
	 (where the small body has a mass
	 $100 M_\sun \ltsim \mu M \ltsim 10^4 M_\sun$)
	 are likely to have very high eccentricities
	 $0.995 \ltsim e \ltsim 0.998$; these systems
	 are likely much rarer than EMRIs, but are also
	 much stronger GW sources.
	 }%%%
{}, so we follow the time-domain approach.
We use standard Berger-Oliger mesh refinement techniques and
compactified hyperboloidal slices for improved accuracy and efficiency.

The remainder of this paper is organized as follows:

Section~\ref{sect:introduction/notation} summarizes our notation.

Section~\ref{sect:theory} gives a detailed description of our
theoretical and computational formalism for self-force computations,
with subsections on
the effective source regularization (\ref{sect:theory/effective-src}),
the $m$-mode Fourier decomposition (\ref{sect:theory/m-mode}),
the worldtube (\ref{sect:theory/worldtube}),
moving the worldtube (\ref{sect:theory/moving-worldtube}),
hyperboloidal slices and compactification (\ref{sect:theory/compactification}),
our reduction to a 1st-order-in-time system of evolution equations
   (\ref{sect:theory/1st-order-in-time}),
the computation of the puncture field and effective source
   (\ref{sect:theory/computing-puncture-fn-and-effective-src}),
the computation of the effective source close to the particle
   (\ref{sect:theory/esrc-close-to-particle}),
boundary conditions (\ref{sect:theory/BCs}),
initial data (\ref{sect:theory/initial-data}),
how the self-force is computed from our evolved field variables
  (\ref{sect:theory/computing-self-force-from-evolved-fields}),
the large-$m$ ``tail series'' (\ref{sect:theory/tail-series}),
selecting the time interval for analysis within an evolution
  (\ref{sect:theory/selecting-time-interval-for-analysis}),
selecting a ``low-noise'' subset of times within an evolution
  (\ref{sect:theory/selecting-low-noise-times}),
how we split the self-force into dissipative and conservative parts
  (\ref{sect:theory/diss-cons}),
and a summary of our computation and data analysis
  (\ref{sect:theory/summary-of-cmpt-and-data-analysis}).

Section~\ref{sect:results} presents our numerical results
and compares them to values obtained by other authors,
with subsections on
our test configurations and parameters (\ref{sect:results/configs-and-pars}),
an example of our data analysis (\ref{sect:results/data-analysis-eg}),
the convergence of our results with numerical resolution
  (\ref{sect:results/convergence}),
a numerical verification that our results are independent
of the choice of worldtube and other numerical parameters
  (\ref{sect:results/verify-results-ne-fn(worldtube-etal)}),
comparison of our results with those of other researchers
  (\ref{sect:results/cmp-with-other-researchers}),
an overview of our computed self-force for each configuration
  (\ref{sect:results/overview-of-self-forces}),
our results for highly eccentric orbits
  (\ref{sect:results/highly-eccentric-orbits}),
our results for zoom-whirl orbits (\ref{sect:results/zoom-whirl-orbits}),
and strong oscillations (``wiggles'') in the self-force shortly
after periastron (\ref{sect:results/wiggles}).

Section~\ref{sect:discussion} presents a general discussion
of this work, the conclusions to be drawn from it, and some directions
for future research.

Appendix~\ref{app:phi-tilde-derivs} describes the transformation
between $\tilde{\phi}$ and $\phi$~derivatives, where $\tilde{\phi}$~is
the ``untwisted'' azimuthal coordinate defined by~\eqref{eqn:phi-tilde-defn}.

Appendix~\ref{app:details} describes our computational scheme
in more detail, with subsections on
the numerical computation of $r(r_*)$ (\ref{app:details/computing-r(rstar)}),
the numerical integration of equatorial eccentric Kerr geodesics
  (\ref{app:details/integrating-Kerr-geodesics}),
gradual turnon of the effective source
  (\ref{app:details/gradual-turnon-of-effective-src}),
our algorithm for moving the worldtube
  (\ref{app:details/moving-worldtube}),
constraints on moving the worldtube early in the time evolution
  (\ref{app:details/constraints-on-moving-WT-early-in-evolution}),
finite differencing across the worldtube boundary
  (\ref{app:details/FD-across-worldtube-boundary}),
computing the set of grid points where adjusted finite differencing
is needed
  (\ref{app:details/computing-where-adjusted-FD-needed}),
computing the set of grid points where the puncture field is needed
  (\ref{app:details/computing-where-pfn-needed}),
the numerical time-evolution using Berger-Oliger mesh refinement
  (\ref{app:details/numerical-time-evolution}),
finite differencing near the particle (\ref{app:details/FD-near-particle}),
and implicit-explicit (IMEX) evolution schemes
  (\ref{app:details/IMEX-evolution-schemes}).

%%%%%%%%%%%%%%%%%%%%%%%%%%%%%%%%%%%%%%%%

\subsection{Notation}
\label{sect:introduction/notation}

We generally follow the sign and notation conventions of Wald~\cite{Wald-1984},
with $G = c = 1$ units and a $(-,+,+,+)$ metric signature.  We use the
Penrose abstract-index notation, with indices $abcd$ running over
spacetime coordinates, $ijk$ running over the spatial coordinates,
$\ell$ running over only the $m$-mode coordinates $(t,r,\theta)$,
and $s$ running over only the spatial $m$-mode coordinates $(r,\theta)$
(in both of the latter cases, the coordinates are
defined by~\eqref{eqn:Kerr-Boyer-Lindquist-coords} below).
$\del_a$ is the (spacetime) covariant derivative operator.
$X := Y$ means that $X$ is defined to be $Y$.
$\boxop := \del_a \del^a$ is the 4-dimensional (scalar) wave
operator~\cite{Brill-etal-1972,Teukolsky73}.
%%$\| \cdot \|_\rms$ is the root-mean-square norm on $\Re^n$,
%%$\big\| \{x_k\} \big\|_\rms := \sqrt{ \left( \sum_k x_k^2 \right)/n}$.
%%$\| \cdot \|$ is the magnitude of a complex number.
$\conjugate{z}$ is the complex conjugate of the complex number~$z$.
$\boundary{S}$ is the boundary of the set~$S$.
$(a)_n$ denotes the Pochhammer symbol $\displaystyle \Pi_{k=a}^{a+n-1} k$.

We use Boyer-Lindquist coordinates $(t,r,\theta,\phi)$ on Kerr spacetime,
defined by the line element
\begin{align}
ds^2	= {} &
		- \left( 1 - \frac{2Mr}{\Sigma} \right) \, dt^2
		- 4M^2 \tilde{a} \frac{r \sin^2\theta}{\Sigma} \, dt \, d\phi
								\nonumber\\*
	&
		+ \frac{\Sigma}{\Delta} \, dr^2
		+ \Sigma \, d\theta^2
								\nonumber\\*
	&
		+ \left(
		  r^2 + M^2 \tilde{a}^2 + 2M^3 \tilde{a}^2 \frac{r \sin^2\theta}{\Sigma}
		  \right) \sin^2\theta \, d\phi^2,
					 \label{eqn:Kerr-Boyer-Lindquist-coords}
\end{align}
where $M$~is the spacetime mass,
$\tilde{a} = J/M^2$ is the dimensionless spin of the black hole
(limited to $|\tilde{a}| < 1$),
$\Sigma = r^2 + M^2 \tilde{a}^2 \cos^2\theta$,
and
$\Delta = r^2 - 2Mr + M^2 \tilde{a}^2$.
In Boyer-Lindquist coordinates the event horizon is the coordinate
sphere $r = r_h = r_+ = M \left(1 + \sqrt{1 - \tilde{a}^2}\right)$
and the inner horizon is the coordinate sphere
$r = r_- = M \left(1 - \sqrt{1 - \tilde{a}^2}\right)$.

%% it would be nice to introduce this paragraph,
%% but that's too disruptive for the PRD page-proofs process :(
%% We also use the usual tortise coordinate $r_*$
%% (defined by~\eqref{eqn:rstar(r)}),
%% the compactified tortise coordinate $R_*$
%% (defined by~\eqref{eqn:r_*(R_*,Omega)} and~\eqref{eqn:Omega(R_*)}),
%% and the compactified time coordinate $T$ (defined by~\eqref{eqn:T(t)}.

We take the particle to orbit in the equatorial plane in the
$d\phi/dt > 0$ direction, with $\tilde{a} > 0$ for prograde orbits
and $\tilde{a} < 0$ for retrograde orbits.
We parameterize the particle's (bound equatorial geodesic) orbit
by the usual dimensionless semi-latus rectum $p$ and eccentricity $e$; these are
defined in detail in Appendix~\ref{app:details/integrating-Kerr-geodesics}.
We refer to the combination of a spacetime and a particle orbit as a
``configuration'', and parameterize it with the triplet $(\tilde{a},p,e)$.
We define $T_r$ to be the coordinate-time period of the particle's
radial motion; we usually refer to $T_r$ as the particle's ``orbital period''.
We define the ``modulo time'' to be the coordinate time modulo $T_r$.

To aid in assessing the accuracy of our computed self-forces, we define
a positive-definite pointwise norm on covariant or contravariant 4-vectors,
\begin{subequations}
\begin{align}
\| v_a \|_+
	& :=	\left( |v_t v^t| + |v_i v^i| \right)^{1/2}		\\
\| v^a \|_+
	& :=	\left( |v_t v^t| + |v_i v^i| \right)^{1/2},		%%%\\
\end{align}
\end{subequations}
where all indices are raised and lowered with the Boyer-Lindquist 4-metric.

We use $x^a_\particle(t)$ to denote the particle's worldline, which we consider
to be known in advance, i.e., we do \emph{not} consider changes to
the particle's worldline induced by the self-force.
$\mathcal{E}$ and $\mathcal{L}$ are the particle's specific energy
and specific angular momentum (i.e., the particle's energy and angular
momentum per unit mass).

\begingroup
% macros for typesetting algorithms
\newcommand{\assign}{\leftarrow}	% assignment operator
\newcommand{\var}[1]{\texttt{#1}}	% variable name
\newcommand{\kw}[1]{\textbf{#1}}	% keyword

When referring to finite difference molecules (stencils) we use $\var{i}$
and $\var{j}$ as generic integer grid coordinates in the radial~($R_*$) and
angular~($\theta$) directions, respectively
(where $R_*$ is the compactified tortise coordinate defined
by~\eqref{eqn:rstar(r)}, \eqref{eqn:r_*(R_*,Omega)},
and~\eqref{eqn:Omega(R_*)}).
Considering a finite difference molecule
evaluated at the grid point $(\var{i}, \var{j})$, we define the
molecule's ``radius'' in a given direction
($\var{i}+$, $\var{i}-$, $\var{j}+$, or $\var{j}-$)
as the maximum integer $\delta \ge 0$ such that the molecule has a nonzero
coefficient at $\var{i} \pm \delta$ or $\var{j} \pm \delta$, respectively,
and we refer to these as $R_{\var{i}+}$, $R_{\var{i}-}$,
$R_{\var{j}+}$, and $R_{\var{j}-}$ respectively.
For example, the usual 3-point centered 2nd-order molecule approximating
the radial partial derivative $\partial_{R_*}$ has
$R_{\var{i}+} = R_{\var{i}-} = 1$ and $R_{\var{j}+} = R_{\var{j}-} = 0$.

We use a pseudocode notation to describe algorithms:
%%This notation borrows from many common programming languages
%%including Fortran, C, \Cplusplus{}, PL/I, and BASIC:
Lines are numbered for reference,
but the line numbers are not used in the algorithm itself.
\# marks comment lines, while
keywords are typeset in \kw{bold font}.
Procedures are marked with the keyword \kw{procedure}
and have bodies delimited by ``$\{$'' and ``$\}$''.
Code layout and indentation are solely for clarity and
(unlike Python) do not have any explicit semantics.
Procedure names are typeset in \var{typewriter font}.
Value-returning procedures (functions) have an explicitly-declared
return type (e.g., ``\kw{boolean} \kw{procedure}'') and return
a value with a \kw{return} statement.
When referring to a procedure as a noun in a figure caption or in the
main text of this paper, the procedure name is suffixed with ``\var{()}'',
as in ``\var{foo()}''.

Variable names are either mathematical expressions, such as ``$R_{i+}$'',
or are typeset in \var{typewriter font}.
\hbox{``$\var{var} \assign X$''} means that the variable $\var{var}$
is assigned the value of the expression $X$.
Variables are always declared before use.
The declaration of a variable explicitly states the variable's type
(\kw{integer}, \kw{floating\_point}, \kw{interval}, or \kw{region},
the last of these being a rectangular region in the integer plane
$\mathbb{Z} \,{\times}\, \mathbb{Z}$)
and may also be combined with the assignment of an initial value, as in
\hbox{``\kw{region} $W \assign \text{worldtube region}$''}.
Conditional expressions have C-style syntax and semantics,
\hbox{\textit{condition} ? \textit{expression-if-true} : \textit{expression-if-false}},
while conditional statements have explicit \kw{if}, \kw{then}, and
\kw{else} keywords.
\endgroup

In Appendix~\ref{app:details/IMEX-evolution-schemes} we use lower-case
sans-serif letters $\uu$, $\kk$, and $\tilde{\kk}$ for state vectors,
and upper-case sans-serif letters $\FF$ and $\GG$ for state-vector-valued
functions.

%%%%%%%%%%%%%%%%%%%%%%%%%%%%%%%%%%%%%%%%%%%%%%%%%%%%%%%%%%%%%%%%%%%%%%%%%%%%%%%%

\section{Theoretical formalism}
\label{sect:theory}

Ignoring questions of divergence and regularization near the particle,
in general the (4-vector) radiation-reaction self-force on a scalar particle
moving in an arbitrary (specified) background spacetime is given by
\begin{equation}
F_a = q \, \bigl. (\del_a \Phi) \bigr|_\particle,
							   \label{eqn:F=del-Phi}
\end{equation}
where the particle's scalar charge is $q$ (which may vary along the
particle's worldline), and the (real) scalar field $\Phi$ satisfies
the wave equation
\begin{equation}
\boxop \Phi = q \, \delta\bigl(x^a - x^a_\particle(t) \bigr),
						    \label{eqn:box-Phi=delta-fn}
\end{equation}
where $\boxop$ is the curved-space wave operator in the background
spacetime~\cite{Brill-etal-1972}.

Because of the $\delta$-function source in~\eqref{eqn:box-Phi=delta-fn},
$\Phi$ diverges on the particle's worldline, so that some type of
regularization is essential in order to obtain a finite self-force.

%%%%%%%%%%%%%%%%%%%%%%%%%%%%%%%%%%%%%%%%

\subsection{Effective source regularization}
\label{sect:theory/effective-src}

We use the ``effective-source'' or ``puncture-field'' regularization
scheme introduced by Barack and Golbourn~\cite{Barack-Golbourn-2007} and
Vega and Detweiler~\cite{Vega-Detweiler-2008:self-force-regularization}
(see~\cite{Vega-Wardell-Diener-2011:effective-source-for-self-force}
for a recent review).
This regularization is based on the Detweiler-Whiting
decomposition~\cite{Detweiler-Whiting-2003} of $\Phi$
into the sum of a ``singular'' and a ``regular'' field,
$\Phi = \Phi_\singular + \Phi_\regular$,
with the following properties:
\begin{itemize}
\item	The singular field is divergent on the particle's worldline
	but is (in a suitable sense) spherically symmetric at the
	particle and hence exerts no self-force.
\item	The regular field is finite -- in fact $C^\infty$ -- at the
	particle and exerts the entire self-force.  That is, the
	correct self-force may be obtained by applying~\eqref{eqn:F=del-Phi}
	to the regular field,
	\begin{equation}
	F_a = q \, \bigl. (\del_a \Phi_\regular) \bigr|_\particle.
						   \label{eqn:F=del-Phi-regular}
	\end{equation}
\end{itemize}

Unfortunately, it is very difficult to compute the exact Detweiler-Whiting
singular or regular fields in Schwarzschild or Kerr spacetime.
The basic concept of the effective-source regularization
is to instead compute a ``puncture field'' approximation
$\Phi_\puncture \approx \Phi_\singular$, chosen
(in a manner to be described in detail below) so that the
``residual field'' $\Phi_\residual := \Phi - \Phi_\puncture$
is finite and ``somewhat differentiable'' (in our case $C^2$)
in a neighborhood of the particle.  We then have
\begin{widetext}
\begin{align}
\boxop \Phi_\residual
	& =	\boxop \Phi - \boxop \Phi_\puncture
								\nonumber\\*
	& =	q \, \delta\bigl( x - x_\particle(t) \bigr)
		- \boxop \Phi_\puncture
								\nonumber\\*
	& =	\begin{cases}
		0	& \begin{tabular}[t]{@{}l@{}}
			  on the particle worldline	\\[-0.5ex]
			  (our choice of $\Phi_\puncture$ will ensure this)
							%%%\\
			  \end{tabular}
								\\
		- \boxop \Phi_\puncture	& \text{elsewhere}	%%%\\*
		\end{cases}
			\label{eqn:box-Phi-residual=what-we'll-call-S-effective}
									\\*
	& :=	S_\effective,
					\label{eqn:box-Phi-residual=S-effective}
									%%%\\*
\end{align}
\end{widetext}
where we define the ``effective source'' $S_\effective$ to be the right
hand side of~\eqref{eqn:box-Phi-residual=what-we'll-call-S-effective}.

In more detail, we choose $\Phi_\puncture$ so that for some chosen
integer~$n \ge 3$,
\begin{equation}
\Phi_\puncture - \Phi_\singular
	= \O\bigl( \| x \,{-}\,x_\particle(t) \|^{n-1} \bigr)
				    \label{eqn:Phi-puncture-approx-Phi-singular}
\end{equation}
in a neighborhood of the particle.  (This is equivalent to choosing
$\Phi_\puncture$ so that its Laurent series about the particle position
matches the first $n$~terms of $\Phi_\singular$'s Laurent series;
both series begin with $\bigl\| x \,{-}\,x_\particle(t) \bigr\|^{-1}$
terms.)  Since $\Phi_\regular$ is~$C^\infty$ at the particle and
$\Phi_\residual
	= \Phi_\regular + (\Phi_\singular - \Phi_\puncture)
	= \Phi_\regular
	  + \O\bigl(\| x \,{-}\,x_\particle(t) \|^{n-1}\bigr)$
in a neighborhood of the particle, we have
$\bigl. (\del \Phi_\residual) \bigr|_\particle
	= \bigl. (\del \Phi_\regular) \bigr|_\particle$.
By virtue of~\eqref{eqn:F=del-Phi-regular} the radiation-reaction
self-force is thus given by
\begin{equation}
F_a = q \, \bigl. (\del_a \Phi_\residual) \bigr|_\particle.
						  \label{eqn:F=del-Phi-residual}
\end{equation}

In this work we choose $n = 4$, so that $\Phi_\residual$ is~$C^2$ at the particle
and $S_\effective$ is $C^0$ at the particle.
Note, however, that the criterion~\eqref{eqn:Phi-puncture-approx-Phi-singular} still leaves
considerable freedom in the choice (definition) of~$\Phi_\puncture$.
We describe our choice in detail in
section~\ref{sect:theory/computing-puncture-fn-and-effective-src}.

%%%%%%%%%%%%%%%%%%%%%%%%%%%%%%%%%%%%%%%%

\subsection{\texorpdfstring{$m$}{m}-mode Fourier decomposition}
\label{sect:theory/m-mode}

Given the basic effective-source formalism,
some authors (e.g.,~\cite{Vega-Detweiler-2008:self-force-regularization,%%%
Vega-etal-2009:self-force-3+1-primer,%%%
Vega-Wardell-Diener-2011:effective-source-for-self-force,%%%
Diener-etal-2012:self-consistent-Schw-orbital-evolution,%%%
Vega-etal-2013:Schwarzschild-scalar-self-force-via-effective-src})
choose to solve~\eqref{eqn:box-Phi-residual=S-effective} via a direct
numerical integration in $3{+}1$~dimensions.  However,
following~\cite{Barack-Golbourn-2007,Barack-Golbourn-Sago-2007,%%%
Dolan-Barack-2011,Dolan-Barack-Wardell-2011,Dolan-Barack-2013},
we prefer instead to exploit the axisymmetry of the background (Kerr)
spacetime and introduce an $m$-mode (Fourier) decomposition.

To avoid infinite twisting of the Boyer-Lindquist $\phi$~coordinate
at the event horizon, we
follow~\cite{Krivan-Laguna-Papadopuloos-1996:scalar-field-on-Kerr-background}
by introducing an ``untwisted'' azimuthal coordinate
\begin{equation}
\tilde{\phi} = \phi + f(r)
						      \label{eqn:phi-tilde-defn}
\end{equation}
with the function~$f$ chosen such that
\begin{equation}
d\tilde{\phi} = d\phi + \frac{M \tilde{a}}{\Delta} dr.
						     \label{eqn:dphi-tilde-defn}
\end{equation}
It is straightforward to integrate this to give
\begin{equation}
f(r) = \frac{\tilde{a}}{2\sqrt{1-\tilde{a}^2}}
       \ln \left| \frac{r-r_+}{r-r_-} \right|
       + \text{constant}.
\end{equation}

Using the $\tilde{\phi}$-derivative transformations derived in
Appendix~\ref{app:phi-tilde-derivs}, $\boxop \Phi$ can be written in
$(t,r,\theta,\tilde{\phi})$ coordinates
\footnote{%%%
	 In an early version of our theoretical formalism we
	 wrote the equations using $\eta = \cos\theta$ as an angular
	 variable.  Provided that $\Phi$ is a nonsingular function
	 of $\eta$ near the $z$~axis, this automatically enforces
	 the boundary condition $\partial_\theta \Phi = 0$ there
	 (\cf{}~section~\protect\ref{sect:theory/BCs}).
	 However,
	 $\partial_{\theta\theta} \Phi
		= \sin^2 \theta \, \partial_{\eta\eta} \Phi
		  - \cos \theta \, \partial_\eta \Phi$,
	 so that on the $z$~axis
	 $\partial_{\theta\theta} \Phi = - \partial_\eta \Phi$.
	 This means that specifying $\partial_\eta \Phi$ on the
	 $z$~axis (which should \textit{a priori} be a reasonable
	 boundary condition) would implicitly also specify
	 $\partial_{\theta\theta} \Phi$ there, which should
	 actually be determined by the field (evolution) equations.
	 In other words, such a ``boundary condition'' would in
	 fact over-constrain the evolution system.  To avoid the
	 possibility of such an over-constraint, we abandoned
	 the $\eta = \cos\theta$ scheme.%%%
	 }%%%
{} as
\begin{align}
\Sigma \boxop \Phi
	= {} &	- \left[
		  \frac{(r^2+M^2 \tilde{a}^2)^2}{\Delta}
		  -
		  M^2 \tilde{a}^2 \sin^2 \theta
		  \right]
		  \partial_{tt} \Phi
								\nonumber\\*
	&
		- \frac{4M^2 \tilde{a}r}{\Delta} \partial_{t \tilde{\phi}} \Phi
		+ \partial_r \Bigl( \Delta \partial_r \Phi \Bigr)
		+ 2M\tilde{a} \partial_{r\tilde{\phi}} \Phi
								\nonumber\\*
	&
		+ \partial_{\theta\theta} \Phi
		+ \cot\theta \, \partial_\theta \Phi
		+ \frac{1}{\sin^2 \theta} \partial_{\tilde{\phi}\tilde{\phi}} \Phi.
					       \label{eqn:4D-box-Phi(phi-tilde)}
\end{align}

We Fourier-decompose the field in $e^{im\tilde{\phi}}$ modes, writing
\begin{equation}
\Phi(t,r,\theta,\phi)
	= \sum_{m=-\infty}^\infty
	  e^{im\tilde{\phi}} \Psi_m(t,r,\theta)
						       \label{eqn:m-mode-decomp}
\end{equation}
and analogously for the other fields $\Phi_\puncture$, $\Phi_\residual$,
and $S_\effective$.
For each integer~$m$, the (complex) $m$-mode fields are given by
\begin{equation}
\Psi_m(t,r,\theta)
	 = \frac{1}{2\pi}
	   \int_{-\pi}^\pi
	   \Phi(t,r,\theta,\phi) e^{-im\tilde{\phi}}
	   \, d\tilde{\phi}
						\label{eqn:m-mode-decomp-invert}
\end{equation}
and analogously for the other fields $\Psi_{\puncture,m}$,
$\Psi_{\residual,m}$, and $S_{\effective,m}$.
We then introduce the (complex) radial-factored field
\begin{equation}
\varphi_m = r \Psi_m
							 \label{eqn:varphi-defn}
\end{equation}
(and analogously for $\varphi_{\puncture,m}$ and $\varphi_{\residual,m}$)
so that the far-field falloffs around an asymptotically-flat system
are $\varphi_m = \O(1)$ when $\Psi_m = \O(1/r)$.

Following~\cite{Sundararajan-Khanna-Hughes-2007},
we introduce the tortise coordinate~$r_*$
defined (up to an arbitrary additive constant) by
\begin{equation}
\frac{dr_*}{dr} = \frac{r^2+M^2 \tilde{a}^2}{\Delta}.
							  \label{eqn:rstar-defn}
\end{equation}
Again following~\cite{Sundararajan-Khanna-Hughes-2007},
we fix the additive constant by choosing
\begin{align}
r_*	= {}&	r
		+ 2M\frac{r_+}{r_+ - r_-} \ln\left( \frac{r - r_+}{2M} \right)
								\nonumber\\
	&	\phantom{r}
		- 2M\frac{r_-}{r_+ - r_-} \ln\left( \frac{r - r_-}{2M} \right).
							    \label{eqn:rstar(r)}
\end{align}
We describe the numerical computation of $r(r_*)$ in
Appendix~\ref{app:details/computing-r(rstar)}.
For any scalar quantity~$Q$ we have (using the chain rule)
\begin{equation}
\frac{\partial Q}{\partial r}
	= \frac{r^2+M^2 \tilde{a}^2}{\Delta} \frac{\partial Q}{\partial r_*}.
\end{equation}

The scalar wave operator $\boxop \Phi$ then becomes
\begin{equation}
\boxop \Phi = \sum_{m=-\infty}^\infty
	      \frac{e^{im\tilde{\phi}}}{r}
	      \, \boxop_m \varphi_m
						     \label{eqn:4D-box-varphi-m}
\end{equation}
and each $m$-mode of the residual field satisfies
\begin{equation}
\boxop_m \varphi_{\residual,m} = S_{\effective,m} \,,
					  \label{eqn:box-m-varphi=S-effective-m}
\end{equation}
where
\begin{widetext}
\begin{align}
\boxop_m \varphi_
	={} &	- \frac{1}{r\Sigma}
		  \left[
		  \frac{(r^2+M^2 \tilde{a}^2)^2}{\Delta} - M^2 \tilde{a}^2 \sin^2 \theta
		  \right] \partial_{tt} \varphi
		- 4im \frac{M^2 \tilde{a}}{\Delta\Sigma} \partial_t \varphi
								\nonumber\\*
	&	+ \frac{(r^2+M^2 \tilde{a}^2)^2}{r\Delta\Sigma}
		  \partial_{r_*r_*} \varphi
		+ \left[
		  - 2 \frac{M^2 \tilde{a}^2}{r^2\Sigma}
		  + 2imM\tilde{a} \frac{r^2+M^2 \tilde{a}^2}{r\Delta\Sigma}
		  \right] \partial_{r_*} \varphi
								\nonumber\\*
	&	+ \frac{1}{r\Sigma} \partial_{\theta\theta} \varphi
		+ \frac{\cot\theta}{r\Sigma} \partial_\theta \varphi
		- \left[
		  \frac{2}{r^2\Sigma} \left(
				      M - \frac{M^2 \tilde{a}^2}{r}
				      \right)
		  + \frac{m^2}{r\Sigma \sin^2 \theta}
		  + 2im\frac{M \tilde{a}}{r^2\Sigma}
		  \right] \varphi.
					       \label{eqn:box-m-coordinate-form}
									%%%\\
\end{align}
\end{widetext}

%%%%%%%%%%%%%%%%%%%%%%%%%%%%%%%%%%%%%%%%

\subsection{The worldtube}
\label{sect:theory/worldtube}

Our construction of the puncture field and effective source
(\cite{Wardell:2011gb} and
Sec.~\ref{sect:theory/computing-puncture-fn-and-effective-src})
is only valid in an finite $(r,\theta)$~neighborhood of the particle.
Moreover, it is not clear what far-field boundary conditions the
residual field should satisfy.  Therefore, rather than
solving~\eqref{eqn:box-m-varphi=S-effective-m} directly, for each~$m$
we introduce a finite worldtube $W_m$ chosen so that its interior
contains the particle worldline, and the puncture field and
effective source are defined everywhere in the worldtube.  (Notice
that $W_m$ logically ``lives'' in the $m$-mode $(t,r,\theta)$ space,
\emph{not} in spacetime.)

For each~$m$ we define the piecewise ``numerical field''
\begin{equation}
\varphi_{\numerical,m}
	= \begin{cases}
	  \varphi_{\residual,m}	& \text{in the worldtube}	\\*
	  \varphi_m		& \text{outside the worldtube}	%%%\\*
	  \end{cases}.
\end{equation}
This field has a jump discontinuity across the worldtube boundary,
% note we don't want \left[ ... \right] in this next equation
% because that doesn't look good: it puts the top of the [ and ] very high up
\begin{equation}
\lim_{\substack{x^\ell \to b^\ell \\ x^\ell \in W_m}}
  \!\! \varphi_{\numerical,m}(x^\ell)
	= \biggl[
	  \lim_{\substack{x^\ell \to b^\ell \\ x^\ell \not\in W_m}}
	    \!\! \varphi_{\numerical,m}(x^\ell)
	  \biggr]
	  - \varphi_{\puncture,m}(b^\ell)
					 \label{eqn:varphi-jump-across-WT-bndry}
\end{equation}
for any worldtube-boundary point $b^\ell \in \boundary{W_m}$, and it also satisfies
\begin{equation}
\boxop_m \varphi_{\numerical,m}
	= \begin{cases}
	  S_{\effective,m}	& \text{inside the worldtube}	\\*
	  0			& \text{outside the worldtube}.	%%%\\*
	  \end{cases}
				    \label{eqn:box-varphi=S-effective-piecewise}
\end{equation}

We numerically solve~\eqref{eqn:box-varphi=S-effective-piecewise}
via a separate Cauchy time-evolution for each~$m$.  The form
of~\eqref{eqn:box-varphi=S-effective-piecewise} ensures that the
effective source only needs to be computed inside the worldtube, and
(as discussed in detail in
Sec.~\ref{sect:theory/moving-worldtube} and
Appendices~\ref{app:details/FD-across-worldtube-boundary}
and~\ref{app:details/computing-where-pfn-needed})
the puncture field only needs to be computed within a small
neighborhood of the worldtube boundary.

The precise choice of the worldtube may be made for computational
convenience; by construction, the computed self-force is independent
of this choice
(see Sec.~\ref{sect:results/verify-results-ne-fn(worldtube-etal)}
for a numerical verification of this independence).
The worldtube's size should reflect a tradeoff between
numerical cost and accuracy:
\begin{itemize}
\item	A larger worldtube requires computing $S_{\effective,m}$
	(which is expensive) at a larger set of events.
\item	A smaller worldtube (more precisely, one whose
	complement includes points closer to the particle)
	requires numerically computing -- and hence finite differencing --
	$\varphi_m$ closer to its singularity at the particle,
	leading to larger numerical errors.
\end{itemize}
For a given worldtube shape and size, the best accuracy is generally
obtained by choosing the worldtube to be approximately centered on the
particle.

In practice we typically choose a worldtube which is a rectangle in
$(r_*,\theta)$ of half-width~$5\,M$ in $r_*$ and approximately $\pi/8$
in~$\theta$.

Since we use Berger-Oliger mesh refinement
(Appendix~\ref{app:details/numerical-time-evolution}), the question
arises of how the worldtube should interact with the mesh refinement.
In particular, should the worldtube differ from one refinement level
to another?  For simplicity we have chosen a computational scheme
where this is \emph{not} the case -- in our scheme the worldtube
is the same at all refinement levels.  This means that the
Berger-Oliger mesh-refinement algorithm does \emph{not} need to
make the adjustment~\eqref{eqn:adjust-gridfns-when-moving-worldtube}
when copying or interpolating data between different refinement levels.
The worldtube boundary is effectively quantized to the coarsest (base)
grid, but we do not find this to be a problem in practice.

%%%%%%%%%%%%%%%%%%%%%%%%%%%%%%%%%%%%%%%%

\subsection{Moving the worldtube}
\label{sect:theory/moving-worldtube}

If the particle's orbit has a sufficiently small eccentricity then
a reasonably-sized time-independent worldtube in $(r_*,\theta)$ can
encompass the particle's entire orbital motion.  However, our main
interest is in the case where the particle's orbit is highly eccentric.
This requires the worldtube to be time-dependent in order to enclose
the particle throughout the particle's entire orbital motion.  In our
computational scheme we move the worldtube in~$(r_*,\theta)$ in
discontinuous jumps so as to always keep the worldtube's coordinate
center within a small distance (typically $\sim 0.5\,M$) of the
particle position.  (More precisely, this is the case after the
startup phase of the computation; we discuss this in detail in
Sec.~\ref{app:details/constraints-on-moving-WT-early-in-evolution}.)

When the worldtube moves, those ``transition'' grid points which
were formerly inside the worldtube and are now outside, or vice versa,
essentially have the computation of $\boxop \varphi_\puncture$
switched between being done analytically versus via finite differencing.
In the continuum limit these two computations agree, but at finite
resolutions they differ slightly.  Therefore, moving the worldtube
introduces numerical noise into the evolved field $\varphi_{\numerical,m}$.

Our actual worldtube-moving algorithm (described in detail in
Appendix~\ref{app:details/moving-worldtube}) incorporates a number of refinements
to help mitigate this numerical noise and achieve the most accurate
numerical evolutions possible:
\begin{itemize}
\item	Basically, the algorithm moves the worldtube any time the
	particle position is ``too far'' from the worldtube center.
\item	When moving the worldtube, the algorithm places the new
	worldtube center somewhat ahead of the particle in the
	direction of the particle's motion.  The algorithm includes
	a small amount of hysteresis so as to avoid unnecessary
	back-and-forth worldtube moves.
\item	The algorithm limits the maximum distance the worldtube
	can be moved at any one time.
\item	The algorithm imposes a minimum time interval between
	worldtube moves.
\end{itemize}

Because $\varphi_{\numerical,m}$ has the jump
discontinuity~\eqref{eqn:varphi-jump-across-WT-bndry} across the
worldtube boundary, each time the worldtube is moved the evolved
fields $\varphi_{\numerical,m}$ and $\Pi_{\numerical,m}$ must be
adjusted at transition grid points:
\begin{subequations}
				\label{eqn:adjust-gridfns-when-moving-worldtube}
\begin{align}
\varphi_{\numerical,m}
	& \leftarrow	\varphi_{\numerical,m}
			  \pm \varphi_{\puncture,m}
									\\*
\Pi_{\numerical,m}
	& \leftarrow	\Pi_{\numerical,m}
			  \pm \partial_t \varphi_{\puncture,m}\,,
									%%%\\*
\end{align}
\end{subequations}
where the ``$+$''~applies to grid points which were formerly inside the
worldtube and are now outside it, and the ``$-$''~applies to grid points
which were formerly outside the worldtube and are now inside it.
%%%%%%%%%%%%%%%%%%%%%%%%%%%%%%%%%%%%%%%%

\subsection{Hyperboloidal slices and compactification}
\label{sect:theory/compactification}

Conceptually,~\eqref{eqn:box-varphi=S-effective-piecewise} should be
solved on the entire spacetime, with outflow boundary conditions on the
event horizon and null infinity ($\Scri^+$).
To accomplish this computationally, we use a
hyperboloidal compactification scheme developed by Zengino\u{g}lu~%%%
\cite{Zenginoglu-2008:hyperboloidal-foliations-and-scri-fixing,%%%
Zenginoglu-2008:hyperboloidal-evolution-with-Einstein-eqns,%%%
Zenginoglu-2011:hyperboloidal-layers-j-comp-phys,%%%
Zenginoglu-Khanna-2011:Kerr-EMRI-waveforms-via-Teukolsky-evolution,%%%
Zenginoglu-Kidder-2010:hyperboloidal-evolution-of-scalar-field-on-Schw,%%%
Zenginoglu-Tiglio-2009:spacelike-matching-to-null-infinity,%%%
Bernuzzi-Nagar-Zenginoglu-2011:Schw-EMRI-waveforms-via-EOB-evolution,%%%
Bernuzzi-Nagar-Zenginoglu-2012:Schw-EMRI-horizon-absorption-effects}.
This scheme has a number of desirable properties, including:
\begin{enumerate}
\item	The hyperboloidal slices reach the event horizon and $\Scri^+$,
	allowing pure-outflow boundary conditions to be posed there.
\item
\label{enum-compactified-no-infinite-blue-shifting}
	The transformed evolution equations do \emph{not} suffer the
	``infinite blue-shifting'' problem (\cf{}~the discussion
	of~\cite{Zenginoglu-2011:hyperboloidal-layers-j-comp-phys})
	in the compactification region -- they have finite and nonzero
	propagation speeds throughout the computational domain,
	and outgoing waves suffer at most $\O(1)$ compression
	(blue-shifting) or expansion (red-shifting) as they
	propagate from the region of the particle to the event horizon
	and to $\Scriplus$.
\item	The transformed evolution equations can be formulated to be
	nonsingular everywhere, with all coefficients having finite
	limiting values near to and on both the event horizon and $\Scri^+$.
\item	The (time-independent) compactification transformation can be
	chosen to be the identity transformation throughout a neighborhood
	of the entire range of the particle's orbital motion.
	This means that the computation of the effective
	source and puncture field, the various adjustments to
	the computations when crossing the worldtube boundary or
	when moving the worldtube, and the computation of the
	self-force from the evolved field $\varphi_m$, are all
	unaffected by the compactification.
\item	The scheme is easy to implement, requiring only relatively
	modest modifications to our previous (non-compactified)
	numerical code.
\end{enumerate}

We primarily follow the version of Zengino\u{g}lu's compactification
scheme described
in~\cite{Zenginoglu-Khanna-2011:Kerr-EMRI-waveforms-via-Teukolsky-evolution},
although with slightly different notation to more conveniently allow
a unified treatment of compactification near the event horizon and
near $\Scri^+$.

For purposes of compactification, it is convenient to rewrite the
evolution equation~\eqref{eqn:box-varphi=S-effective-piecewise}
and~\eqref{eqn:box-m-coordinate-form} in the generic form
\begin{align}
  \Coeff{\partial_{tt}           \varphi} \partial_{tt}           \varphi
+ \Coeff{\partial_{tr_*}         \varphi} \partial_{tr_*}         \varphi
+ \Coeff{\partial_{r_*r_*}       \varphi} \partial_{r_*r_*}       \varphi
	&
								\nonumber\\*
+ \Coeff{\partial_t              \varphi} \partial_t              \varphi
+ \Coeff{\partial_{r_*}          \varphi} \partial_{r_*}          \varphi
	&
								\nonumber\\*
+ \Coeff{\partial_{\theta\theta} \varphi} \partial_{\theta\theta} \varphi
+ \Coeff{\partial_{\theta}       \varphi} \partial_{\theta}       \varphi
	&
								\nonumber\\*
+ \Coeff{                        \varphi}                         \varphi
+ \Coeff{                              1}
	& = 0,
					  \label{eqn:evolution-generic-physical}
									%%%\\*
\end{align}
where we have dropped the subscript on $\varphi_m$, and where
the $\Coeff{\cdot}$~coefficients can be read off from the evolution equations.
($\Coeff{\partial_{tr_*} \varphi} = 0$ for our evolution equations,
but is included for generality.)

To make the equations nonsingular near to and on the event horizon, we
multiply \eqref{eqn:box-varphi=S-effective-piecewise} through by a factor
of $r\Sigma\Delta$.  It is also useful for the coefficients to be finite
near to and at $\Scri^+$, so we further multiply through by a factor of
$(r^2+M^2 \tilde{a}^2)^{-2}$.  The resulting coefficients are
\begin{subequations}
\begin{align}
\Coeff{\partial_{tt} \varphi}
	& =	 \frac{M^2 \tilde{a}^2 \Delta \sin^2 \theta}{(r^2+M^2 \tilde{a}^2)^2} - 1,	\\
\Coeff{\partial_t \varphi}
	& =	- i \frac{4mM^2 \tilde{a}r}{(r^2+M^2 \tilde{a}^2)^2},				\\
\Coeff{\partial_{r_*r_*} \varphi}
	& =	1	,						\\
\Coeff{\partial_{r_*} \varphi}
	& =	- \frac{2M^2 \tilde{a}^2 \Delta}{r (r^2+M^2 \tilde{a}^2)^2} + i\frac{2mM\tilde{a}}{r^2 + M^2 \tilde{a}^2},
									\\
\Coeff{\partial_{\theta\theta} \varphi}
	& =	\frac{\Delta}{(r^2+M^2 \tilde{a}^2)^2},				\\
\Coeff{\partial_\theta \varphi}
	& =	\frac{\Delta \cot\theta}{(r^2+M^2 \tilde{a}^2)^2},			\\
\Coeff{\varphi}
	& =	- \frac{2\Delta}{r (r^2+M^2 \tilde{a}^2)^2} \left( M - \frac{M^2 \tilde{a}^2}{r} \right)
								\nonumber\\
	& \quad	- \frac{m^2 \Delta}{(r^2+M^2 \tilde{a}^2)^2 \sin^2 \theta}
		- i\frac{2mM\tilde{a}\Delta}{r (r^2+M^2 \tilde{a}^2)^2},
									\\
\Coeff{1}
	& =	\begin{cases}\begin{aligned}
		&- \dfrac{r \Sigma \Delta}{(r^2+M^2 \tilde{a}^2)^2} S_{\effective,m}
		&& \text{inside}	\\
		&0
		&& \text{outside}.	%%%\\
		\end{aligned}\end{cases}
\end{align}
\end{subequations}

We define the compactified radial coordinate~$R_*$ by
\begin{equation}
r_* = \frac{R_*}{\Omega(R_*)},
						      \label{eqn:r_*(R_*,Omega)}
\end{equation}
where we choose the (time-independent) conformal factor $\Omega$ so
that the event horizon and $\Scri^+$ are at the (finite) $R_*$ coordinates
$R_*^h$ and $R_*^\Scriplus$ respectively.  More precisely, we introduce
the four parameters $R_*^h < R_*^- < 0 < R_*^+ < R_*^\Scriplus$,
chosen such that the particle and worldtube always lie within the
region $R_*^- < R_* < R_*^+$ (where we will choose the compactification
transformation to be the identity transformation).
We define
\begin{equation}
\Omega(R_*)
	= \begin{cases}
	  \displaystyle
	  1 - \left( \frac{R_*^- - R_*}{R_*^-         - R_*^h} \right)^4
					& \text{if $R_* < R_*^-$}
									\\*
	  1				& \text{if $R_*^- \le R_* \le R_*^+$}
									\\*
	  \displaystyle
	  1 - \left( \frac{R_* - R_*^+}{R_*^\Scriplus - R_*^+} \right)^4
					& \text{if $R_* > R_*^+$}
									%%%\\*
	  \end{cases}
							  \label{eqn:Omega(R_*)}
\end{equation}
so that the compactification transformation is indeed the identity
transformation ($\Omega = 1$ and $r_* = R_*$) throughout the region
$R_*^- < R_* < R_*^+$.  We refer to $R_*^-$ and $R_*^+$ as the inner
and outer compactification radii, respectively.  Our numerical grid
spans the full range $R_*^h \le R_* \le R_*^\Scriplus$.

To ensure the absence of infinite blue-shifting
(``desirable property''~\ref{enum-compactified-no-infinite-blue-shifting}),
the time coordinate must also be transformed.
We define the transformed time coordinate~$T$ by
\begin{equation}
T = t - h(R_*),
								\label{eqn:T(t)}
\end{equation}
where the ``height'' function $h$ is given by
\begin{align}
h(R_*)
	& =
		\begin{cases}
		R_* - r_*		& \text{if $R_* < R_*^-$}
								\\*
		0			& \text{if $R_*^- \le R_* \le R_*^+$}
								\\*
		r_* - R_*		& \text{if $R_* > R_*^+$}
								%%%\\*
		\end{cases}
								\nonumber\\*
	& =
		\begin{cases}
		\displaystyle
		R_* \left( 1 - \frac{1}{\Omega} \right)
					& \text{if $R_* < R_*^-$}
								\\*
		0			& \text{if $R_*^- \le R_* \le R_*^+$}
								\\*
		\displaystyle
		R_* \left( \frac{1}{\Omega} - 1 \right)
					& \text{if $R_* > R_*^+$}
								%%%\\*
		\end{cases}.
\end{align}

\begin{figure}[bp]
\begin{center}
\begin{picture}(86,120)
%%% alignment marks for debugging figure placement
%%\put(86,120){\circle{2}}
%%\put(0,120){\circle{2}}
%%\put(86,0){\circle{2}}
%%\put(0,0){\circle{2}}
%
\put(0,65){\includegraphics[scale=1.0]{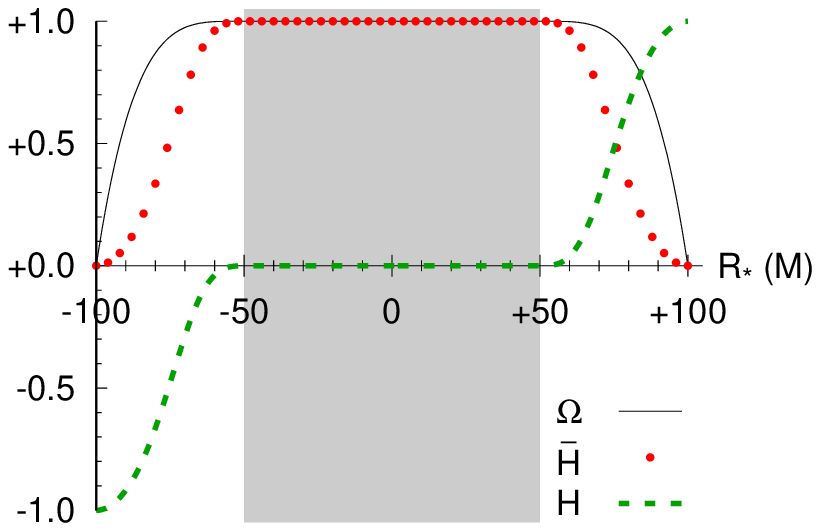}}
\put(0,-4){\includegraphics[scale=1.0]{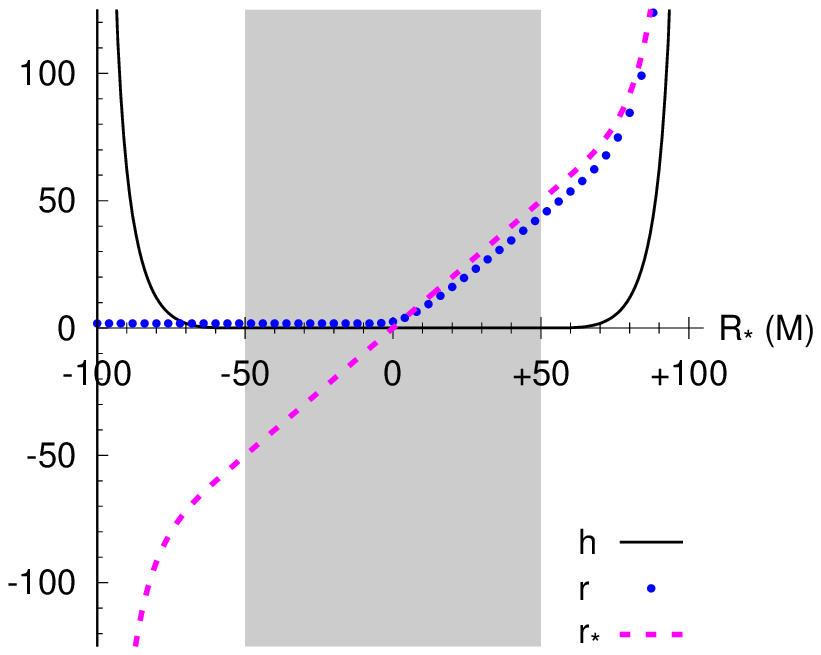}}
\end{picture}
\end{center}
\vspace{-1ex}
\caption{%%%
\label{fig:compactify-example}
	This figure shows an example of the compactification
	for a Kerr spacetime with dimensionless spin~$\tilde{a} = 0.6$.
	The compactification parameters
	(here chosen for visual clarity rather than
	optimum computational efficiency/accuracy) are
	$(R_*^h,R_*^-, R_*^+,R_*^\Scriplus) = (-100,-50,+50,+100)M$.
	The upper subfigure shows $\Omega$, $\Hbar$, and $H$,
	and the lower subfigure shows $h$, $r$, and $r_*$,
	all as functions of $R_*$.  The compactification
	transformation is only nontrivial outside the shaded
	region $R_*^- \le R_* \le R_*^+$; the transformation
	is the identity transformation
	($\Omega = \Hbar = 1$, $H = h = 0$, and $r_* = R_*$)
	in the shaded region.
	For $R_* \ll 0$, $r \to r_h$ ($ = 1.8M$).
	}%%%
\end{figure}

In order to express the equations in a simple form, it is
convenient to define the ``generalized boost'' function
\begin{equation}
\Hbar	= \frac{dR_*}{dr_*}
	= \frac{\Omega^2}{\Omega - R_* \Omega'},
\end{equation}
where $X' := dX/dR_*$ for any quantity $X$, so that
\begin{equation}
\Hbar'	= \frac{2\Omega \Omega'}{\Omega - R_*\Omega'}
	  + \frac{R_*\Omega^2\Omega''}{(\Omega - R_*\Omega')^2}
\end{equation}
We define the ``boost'' function $H$ by
\begin{equation}
H	= \frac{dh}{dr_*}
	=	\begin{cases}
		\Hbar - 1		& \text{if $R_* < R_*^-$}
								\\*
		0			& \text{if $R_*^- \le R_* \le R_*^+$}
								\\*
		1 - \Hbar		& \text{if $R_* > R_*^+$}
								%%%\\*
		\end{cases}
\end{equation}
so that
\begin{equation}
H' =	\begin{cases}
	\Hbar'			& \text{if $R_* < R_*^-$}
								\\*
	0			& \text{if $R_*^- \le R_* \le R_*^+$}
								\\*
	- \Hbar'		& \text{if $R_* > R_*^+$}
								%%%\\*
	\end{cases}.
\end{equation}
Figure~\ref{fig:compactify-example} shows an example of these
quantities and the resultant compactification.

%%%%%%%%%%%%%%%%%%%%
%%%%%%%%%%%%%%%%%%%%

Transforming the generic evolution
equations~\eqref{eqn:evolution-generic-physical} from
$(t,r_*,\theta,\phi)$~coordinates to $(T,R_*,\theta,\phi)$ coordinates,
we see immediately that the coefficients
$\Coeff{\partial_{\theta\theta} \varphi}$,
$\Coeff{\partial_\theta \varphi}$, $\Coeff{\varphi}$, and $\Coeff{1}$
are all unchanged by the transformation.

At all points other than the event horizon or $\Scriplus$, the
nontrivially-transformed coefficients are
\begin{subequations}
\begin{align}
\Coeff{\partial_{TT} \varphi}
	& =	  \frac{1  }{\Hbar} \Coeff{\partial_{tt} \varphi}
  		- \frac{H  }{\Hbar} \Coeff{\partial_{tr_*} \varphi}
		+ \frac{H^2}{\Hbar} \Coeff{\partial_{r_*r_*} \varphi},
									\\*
\Coeff{\partial_{TR_*} \varphi}
	& =		            \Coeff{\partial_{tr_*} \varphi}
		- 2H                \Coeff{\partial_{r_*r_*} \varphi},
									\\*
\Coeff{\partial_{R_*R_*} \varphi}
	& =	  \Hbar		    \Coeff{\partial_{r_*r_*} \varphi},
									\\*
\Coeff{\partial_T \varphi}
	& =	- H'		    \Coeff{\partial_{r_*r_*} \varphi}
		+ \frac{1}{\Hbar}   \Coeff{\partial_t \varphi}
		- \frac{H}{\Hbar}   \Coeff{\partial_{r_*} \varphi},
									\\*
\Coeff{\partial_{R_*} \varphi}
	& =	  \Hbar'	    \Coeff{\partial_{r_*r_*} \varphi}
		+		    \Coeff{\partial_{r_*} \varphi}.
									%%%\\
\end{align}
					\label{eqn:coeffs-compactified/interior}
\end{subequations}
On the event horizon the limiting values of these
(transformed) coefficients are
\begin{subequations}
\begin{align}
\Coeff{\partial_{TT} \varphi}
	& =	-2,
									\\*
\Coeff{\partial_{TR_*} \varphi}
	& =	+2
									\\*
\Coeff{\partial_{R_*R_*} \varphi}
	& =	 0,
									\\*
\Coeff{\partial_T \varphi}
	& =	- i \frac{m\tilde{a}}{r_h},
									\\*
\Coeff{\partial_{R_*} \varphi}
	& =	i \frac{2mM\tilde{a}}{r_h^2 + M^2 \tilde{a}^2},
									\\*
\Coeff{\partial_{\theta\theta} \varphi}
= \Coeff{\partial_\theta \varphi}
	& =	0,
									\\*
\Coeff{\varphi}
= \Coeff{1}
	& =	0,
									%%%\\*
\end{align}
					 \label{eqn:coeffs-compactified/horizon}
\end{subequations}
while at $\Scriplus$ the limiting values are
\begin{subequations}
\begin{align}
\Coeff{\partial_{TT} \varphi}
	& =	2 \frac{M^2 \tilde{a}^2}{\Hbar''^\Scriplus}
		  \left( \frac{\Omega'^\Scriplus}{R_*^\Scriplus} \right)^2
		  \sin^2\theta
		- 2,
									\\*
\Coeff{\partial_{TR_*} \varphi}
	& =	-2,
									\\*
\Coeff{\partial_{R_*R_*} \varphi}
	& =	 0,
									\\*
\Coeff{\partial_T \varphi}
	& =	- i
		  \frac{4mM\tilde{a}}{\Hbar''^\Scriplus}
		  \left( \frac{\Omega'^\Scriplus}{R_*^\Scriplus} \right)^2,
									\\*
\Coeff{\partial_{R_*} \varphi}
	& =	0,
									\\*
\Coeff{\partial_{\theta\theta} \varphi}
	& =	\frac{2}{\Hbar''^\Scriplus}
		\left( \frac{\Omega'^\Scriplus}{R_*^\Scriplus} \right)^2,
									\\*
\Coeff{\partial_\theta \varphi}
	& =	\frac{2}{\Hbar''^\Scriplus}
		\left( \frac{\Omega'^\Scriplus}{R_*^\Scriplus} \right)^2
		\frac{1}{\tan\theta},
									\\*
\Coeff{\varphi}
	& =	- \frac{2 m^2}{\Hbar''^\Scriplus}
		  \left( \frac{\Omega'^\Scriplus}{R_*^\Scriplus} \right)^2
		  \frac{1}{\sin^2\theta},
									\\*
\Coeff{1}
	& =	0,
									%%%\\*
\end{align}
				       \label{eqn:coeffs-compactified/Scri-plus}
\end{subequations}
where
\begin{subequations}
\begin{align}
\Omega'^\Scriplus := \lim_{R_* \to R_*^\Scriplus} \Omega'
	& =	\frac{4}{R_*^\Scriplus - R_*^+},
									\\*
\Hbar''^\Scriplus := \lim_{R_* \to R_*^\Scriplus} \Hbar''
	& =	\frac{2 \Omega'^\Scriplus}{R_*^\Scriplus}.
									%%%\\*
\end{align}
\end{subequations}
While conceptually straightforward, the calculation of
$\Hbar''^\Scriplus$ is somewhat lengthy; we used the
Maple symbolic algebra system
(Version~18 for x86-64~Linux, \cite{Char-etal-1983:Maple-design})
to obtain the result given here.

%%%%%%%%%%%%%%%%%%%%%%%%%%%%%%%%%%%%%%%%

\subsection{1st-order-in-time equations}
\label{sect:theory/1st-order-in-time}

To numerically solve the evolution
equation~\eqref{eqn:box-varphi=S-effective-piecewise}
it is convenient to introduce the auxiliary variable
\begin{equation}
\Pi_{\numerical,m} = \partial_t \varphi_{\numerical,m}
							   \label{eqn:Pi-m-defn}
\end{equation}
so as to obtain a 1st-order-in-time evolution system.
The compactified evolution equation then becomes
\begin{align}
  \Coeff{\partial_{TT}           \varphi} \partial_T              \Pi
+ \Coeff{\partial_{TR_*}         \varphi} \partial_{R_*}          \Pi
+ \Coeff{\partial_{R_*R_*}       \varphi} \partial_{R_*R_*}       \varphi
	&
								\nonumber\\*
+ \Coeff{\partial_T              \varphi}                         \Pi
+ \Coeff{\partial_{R_*}          \varphi} \partial_{R_*}          \varphi
	&
								\nonumber\\*
+ \Coeff{\partial_{\theta\theta} \varphi} \partial_{\theta\theta} \varphi
+ \Coeff{\partial_{\theta}       \varphi} \partial_{\theta}       \varphi
	&
								\nonumber\\*
+ \Coeff{                        \varphi}                         \varphi
+ \Coeff{                              1}
	& =0,
						      \label{eqn:evolution-main}
									%%%\\*
\end{align}
where we have dropped the subscripts on $\varphi_{\numerical,m}$
and $\Pi_{\numerical,m}$.

Our final evolution system
comprises~\eqref{eqn:Pi-m-defn} and~\eqref{eqn:evolution-main}
using the coefficients~\eqref{eqn:coeffs-compactified/interior},
\eqref{eqn:coeffs-compactified/horizon}, and
\eqref{eqn:coeffs-compactified/Scri-plus},
modified by applying L'Hopital's rule on the $z$~axis,
applying boundary conditions (Sec.~\ref{sect:theory/BCs}),
the gradual turnon of the effective source
(Appendix~\ref{app:details/gradual-turnon-of-effective-src}),
the adjustment of $\varphi$ and $\Pi$ when the worldtube is moved
(Sec.~\ref{sect:theory/moving-worldtube}),
and the addition of numerical dissipation
(Appendix~\ref{app:details/numerical-time-evolution}).

%%%%%%%%%%%%%%%%%%%%%%%%%%%%%%%%%%%%%%%%

\subsection{Computing the puncture field and effective source}
\label{sect:theory/computing-puncture-fn-and-effective-src}

There is considerable freedom in the particular choice of puncture
field used to construct an effective source. As mentioned in
Sec.~\ref{sect:theory/effective-src}, we work with a puncture field
which agrees with the Detweiler-Whiting singular field in the first
four orders in its expansion about the worldline.  This ensures
that the computed self-force is finite and uniquely determined, and
that the numerical methods used to compute it converge reasonably
well. Other than that, we shall exploit the freedom to modify the
higher-order terms in the expansion to adapt it to the $m$-mode
scheme.

We begin with a coordinate series approximation for the Detweiler-Whiting
singular field of a scalar charge on an eccentric equatorial geodesic
of the Kerr spacetime, as can be obtained using, e.g., the methods
of \cite{Heffernan:2012su,Heffernan:2012vj}. Our starting point is
thus a coordinate series expansion of the form
\begin{equation}
\label{eqn:Phisum}
\Phi_\singular^{[n]} (x; x_{\particle})
	= \sum_{i=1}^{n} \frac{B^{a(3 i - 3)}}{\rho^{2 i - 1}} \epsilon^{i-2}
	  + \O(\epsilon^{n-1}),
\end{equation}
where
\begin{gather}
B^{a(k)}
	\equiv	b^a_{a_1 a_2 \cdots a_k}(x_{\particle})
		\Delta x^{a_1} \Delta x^{a_2} \cdots \Delta x^{a_k},
									\\[1ex]
\rho^2 = (g_{a b} + u_{a} u_{b}) \Delta x^a \Delta x^b,
									%%%\\
\end{gather}
and $g_{ab}$ and $u^a$ are evaluated on $x_\particle$. Here, we introduce
$\epsilon:=1$ as a formal power-counting parameter used to keep track
of powers of distance from the particle; this amounts to inserting a
factor of $\epsilon$ for each power of
$\Delta x^a
	= [0, r - r_\particle(t), \theta - \pi/2, \phi - \phi_\particle(t)]^a$
appearing either explicitly or implicitly (through powers of $\rho$).
Since we are choosing to include the first four orders in the
expansion of the Detweiler-Whiting singular field, we take $n=4$ and
our approximation neglects terms of order $\epsilon^3$ and higher.

We next make two crucial modifications that make the puncture more
amenable to analytic m-mode decomposition. To motivate these modifications,
consider the general form of the function $\rho$ in the case of
equatorial orbits in Kerr spacetime, which in Boyer-Lindquist coordinates
is given by
\begin{align}
\rho^2 =
	{}& (g_{rr} + u_r u_r) \Delta r^2 + g_{\theta \theta} \Delta \theta^2
								\nonumber \\
	  & + (g_{\phi\phi} + u_\phi u_\phi) \Delta \phi^2
            + u_r u_\phi \Delta r \Delta \phi.
\end{align}
Now, the integration involved in the $m$-mode decomposition of the
$m=0$ mode of the leading-order $1/\rho$ term in the expansion of the
singular field almost has the form of a complete elliptic integral of
the first kind,
$\mathcal{K}(k) \equiv \int_0^{\pi/2} (1-k \sin^2 \phi)^{-1/2} d \phi$,
where the argument $k$ is a function of $x^a_{\particle}$, $u^a$, $\Delta r$
and $\Delta \theta$.  It would be desirable to have it in the exact form
of an elliptic integral, as then it can be efficiently evaluated without
having to resort to numerical quadrature.  Fortunately, the only
modifications required to turn it into elliptic-integral form are to
rewrite $\Delta\phi^2$ in terms of $\sin^2 \Delta \phi$ (or equivalently
$\sin^2\tfrac{\Delta\phi}{2}$ up to an overall factor of $2$ in the
resulting integral), and to eliminate the $\Delta r \Delta \phi$
cross term.  Both of these can be done using methods previously
used in self-force calculations; the former can be achieved using
the ``Q-R'' scheme described in~\cite{Wardell-etal-2012}, and the
latter by combining this with a radially-dependent change of variable,
$\Delta \phi \to \Delta \hat{\phi} - c \Delta r$, where
\begin{equation}
c = \frac{L r_0^3 u^r}
	 {[a^2+r_0 (r_0-2 M)] [a^2 (2 M+r_0)+r_0 \left(L^2+r_0^2\right)]}
\end{equation}
is chosen such that the cross term vanishes. This second trick was
first used by Mino, Nakano and Sasaki \cite{Mino:2001mq} and later
also employed by Haas and Poisson \cite{Haas-Poisson-2006}.

Given these two modifications to $\rho$, we are then left with an
expression for $\hat{\rho} = \rho + \O(\epsilon^2)$ that is of
the form
\begin{equation}
\hat{\rho}^2 = A(r_0, u^a, \Delta r, \Delta \theta)
	       + B(r_0, u^a) \sin^2(\Delta \hat{\phi}/2),
\end{equation}
where $A(r_0, u^a, \Delta r, \Delta \theta)$ is a quadratic polynomial in
$\Delta r$ and $\Delta \theta$. Note that our manipulations introduce an
additional $r$ and $t$ dependence hidden inside the definition of
$\Delta \hat{\phi}$; it is important to take this into account when
computing derivatives of the puncture field, and also when evaluating
it for $\Delta r \ne 0$.  The advantage of working with $\hat{\rho}$
instead of $\rho$ is that the $m=0$ mode of $1/\hat{\rho}$ is
analytically given by a complete elliptic integral of the first kind,
\begin{equation}
\frac{1}{2\pi}\int_{-\pi}^{\pi} \hat{\rho}^{-1}\, d\hat{\phi}
	= \frac{2}{\pi \sqrt{A + B}} \, \mathcal{K}\Big(\tfrac{B}{A+B}\Big),
\end{equation}
and similarly the $m=0$ mode of $\hat{\rho}$ is analytically given by
a complete elliptic integral of the second kind,
\begin{equation}
\frac{1}{2\pi}\int_{-\pi}^{\pi} \hat{\rho}\, d\hat{\phi}
	= \frac{2\, \sqrt{A + B}}{\pi}\, \mathcal{E}\Big(\tfrac{B}{A+B}\Big).
\end{equation}

Returning to the problem of obtaining an $m$-mode decomposed puncture
field, we have to generalize this in three ways:
(i) We need to handle other integer powers of $\hat{\rho}$;
(ii) We need to handle the additional dependence of $\Phi_\singular$
     on $\Delta \hat{\phi}$ other than that appearing in $\hat{\rho}$;
(iii) We need to handle all $m\ge 0$ modes (the fact that the full
      4-dimensional scalar field is real means that the $m<0$ modes
      are trivially related to the $m>0$ modes).
To make things explicit, we use the two previously described modifications
to rewrite our approximation to the singular field,~\eqref{eqn:Phisum},
in the form
\begin{widetext}
\begin{align}
\label{eq:full-phis-m-mode}
\Phi_{\singular} (x; x_{\particle})
	&= \frac{1}{\hat{\rho}^{2n-1}}
	   \Bigg[
	   \sum\limits_{\genfrac{}{}{0pt}{}{i=0}{i~\text{even}}}^{3n-3}
		C_{n,i} \sin^i(\Delta\hat{\phi}/2)
	   +
	   2 \sum\limits_{\genfrac{}{}{0pt}{}{i=0}{i~\text{odd}}}^{3n-3}
		C_{n,i} \sin^{i}(\Delta\hat{\phi}/2) \cos(\Delta\hat{\phi}/2)
	   \Bigg]
	   + \O(\epsilon^{n-1}),
\end{align}
where the coefficients $C_{n,i}$ are functions of $r_0$, $u^a$,
$\Delta r$ and $\Delta \theta$, and where we have replaced
$R=\sin \Delta\hat{\phi}$ with the equivalent expression
$2 \sin (\Delta\hat{\phi}/2) \cos (\Delta\hat{\phi}/2)$.
To define our puncture field, we truncate this expansion at order
$n=4$ and decompose into $m$-modes,
\begin{equation}
\Psi_{\puncture,m}
	= \frac{1}{2\pi}
	  \int_{-\pi}^\pi \Phi_{\singular}^{[4]} e^{-im \hat{\phi}}
	  \, d\hat{\phi}.
\end{equation}
Writing
\begin{align}
e^{-i m \hat{\phi}}
	= &	e^{-i m \hat{\phi}_0} \times
								\nonumber \\
	  &	\sum\limits_{k=0}^{2 m}
		\binom{2 m}{k} (-1)^{k/2} \cos^{2 m-k} (\Delta \hat{\phi}/2)
					  \sin^k       (\Delta \hat{\phi}/2),
\end{align}
and inspecting the form of the integrals, we see that
(apart from a trivial phase factor) the real part of the puncture
is determined purely by the first term in~\eqref{eq:full-phis-m-mode},
while the imaginary part is determined purely by the second term.
Furthermore, in all cases we are left with integrals involving only
even powers of $\sin (\Delta \hat{\phi}/2)$ and $\cos (\Delta \hat{\phi}/2)$.
Then, the three generalisations listed previously can be handled
through the application of two sets of identities,
\begin{align}
\int_{-\pi}^\pi \sin^{2i}(\Delta \hat{\phi}/2)
		\cos^{2j}(\Delta \hat{\phi}/2)
		\hat{\rho}^k
		\, d\hat{\phi}
	& =	\int_{-\pi}^\pi \left[ \frac{\hat{\rho}^2 - A}{B} \right]^i
				\left[ \frac{A+B-\hat{\rho}^2}{B} \right]^j
				\hat{\rho}^k
				\, d\hat{\phi}
\end{align}
and
\begin{subequations}
\begin{align}
\int_{-\pi}^\pi \hat{\rho}^k \, d\hat{\phi}
	& =	\int_{-\pi}^\pi \frac{1}{A(k+2) (A+B)}
				\Bigl[
				(k+3) (2A+B) \hat{\rho}^{k+2}
				-
				(k+4) \hat{\rho}^{k+4}
				\Bigr]
		d\hat{\phi}
					\qquad \text{for $k < -1$},
									\\
\int_{-\pi}^\pi \hat{\rho}^k\, d\hat{\phi}
	& =	\int_{-\pi}^\pi
		\frac{1}{k} \Bigl[
			    A (2-k) (A+B) \hat{\rho}^{k-4}
			    +
			    (k-1) (2A+B) \hat{\rho}^{k-2}
			    \Bigr]
		d\hat{\phi}
					\qquad \text{for $k > 1$}.
\end{align}
\end{subequations}
\end{widetext}
The first of these is a direct consequence of the definition of
$\hat{\rho}$, while the second pair can be obtained from, e.g.,
equation~(1) of \cite[section~1.5.27]{Prudnikov}.
The first identity eliminates all powers of $\sin (\Delta \hat{\phi}/2)$
and $\cos (\Delta \hat{\phi}/2)$ not appearing inside $\hat{\rho}$,
while the second pair of identities may be recursively applied to
rewrite arbitrary (odd integer) powers of $\hat{\rho}$ in terms of
$\hat{\rho}^{-1}$ and $\hat{\rho}$.  Thus we can reduce all cases to
elliptic-integral form and obtain analytic expressions for the puncture
field modes in terms of these easily evaluated elliptic integrals.
In practice, the expressions take the form of an $m$-dependent polynomial
in $\tfrac{A}{B}$ multiplied by $\mathcal{K}$ plus a second polynomial
in $\tfrac{A}{B}$ multiplied by $\mathcal{E}$.

% For an approximation accurate to $\O(\epsilon^n)$, the numerical
% solutions for the field fall off as $m^{-(n+2)}$ for $m$ even and as
% $m^{-(n+3)}$ for $m$ odd. Obviously, only finitely many $m$-modes (typically
% $\sim10$-$20$) can ever be computed numerically; with the error from truncating
% the sum at a finite $m$ putting an upper limit on the accuracy of the self-force
% that can be computed. This may be mitigated, somewhat, by fitting for a large-$m$
% tail, but that fit itself requires more modes and is only ever approximate.
% Here, we propose a much better solution; that is to use the higher order terms in the
% singular field (those that have not been used in computing the effective
% source) to analytically derive expressions for the tail. In many ways, this is
% analogous to the $\ell$-mode regularization scheme, where there is a large-$\ell$ tail
% and one can compute $\ell$-mode regularization parameters.

Given the puncture field~$\Phi_\puncture$, we compute the
effective source~$S_\effective$ via~\eqref{eqn:box-Phi-residual=S-effective}
and then the $m$-mode effective source~$S_{\effective,m}$ via the
Fourier integral~\eqref{eqn:m-mode-decomp-invert}.  Note that the
$\boxop$ operator~\eqref{eqn:4D-box-Phi(phi-tilde)} must be applied
\emph{analytically} to the series expansion for the puncture field
in order to correctly cancel all divergent terms; a numerical
calculation of the $\boxop$~operator would be insufficiently accurate.
The entire computation of $\Phi_{\puncture,m}$ and $S_{\effective,m}$
takes approximately 500~lines of Mathematica code.  The Mathematica
notebook is included in the online supplemental materials accompanying
this paper.

Our final expressions for $\Phi_{\puncture,m}$ and $S_{\effective,m}$
involve
multivariate polynomials in $\Delta r$ and $\Delta\theta$,
the $\mathcal{E}$ and $\mathcal{K}$ elliptic integrals
  (and their derivatives for $S_{\effective,m}$)
and trigonometric polynomials.
The coefficients in these expressions are functions (only) of the
particle position and 4-velocity, so that at each distinct time at
which $\Phi_{\puncture,m}$ and/or $S_{\effective,m}$ need to be
computed, we first precompute these coefficients.
This precomputation is done using C~code
and numerical coefficients which are machine-generated (once) by the
Mathematica program.  The machine-generated C~code is large
(${\sim}\, 10$~megabytes) and involves very lengthy arithmetic expressions
(it contains ${\sim}\, 1.5 \,{\times}\, 10^6$ arithmetic operations);
compiling it is slow and requires large amounts of memory.
Fortunately, the execution of the code (to actually precompute the
coefficients) uses only a small fraction of our code's total CPU
time, so this (machine-generated) code may be compiled without optimization.

The actual evaluation of $\Phi_{\puncture,m}$ and $S_{\effective,m}$
at each grid point is done using hand-written C~code.  In total
(i.e., summed over all grid points and times where the evaluation
is needed) this evaluation uses the majority of our code's total
CPU time; the finite differencing and numerical time-integration
are relatively minor contributors.

%%%%%%%%%%%%%%%%%%%%%%%%%%%%%%%%%%%%%%%%

\subsection{Computing the effective source close to the particle}
\label{sect:theory/esrc-close-to-particle}

As we have noted previously~\cite[section~III.C.3]{Wardell-etal-2012},
our series expressions for the effective source suffer from severe
cancellations when evaluated close to the particle.  Because of the
Fourier integral~\eqref{eqn:m-mode-decomp-invert}, $S_{\effective,m}$
need not -- and typically does not -- vanish at the particle, so the
``interpolate along a ray'' scheme we described
in~\cite[section~III.C.3]{Wardell-etal-2012} is not valid here.

\begin{figure}[bp]
\begin{center}
\includegraphics[scale=1.0]{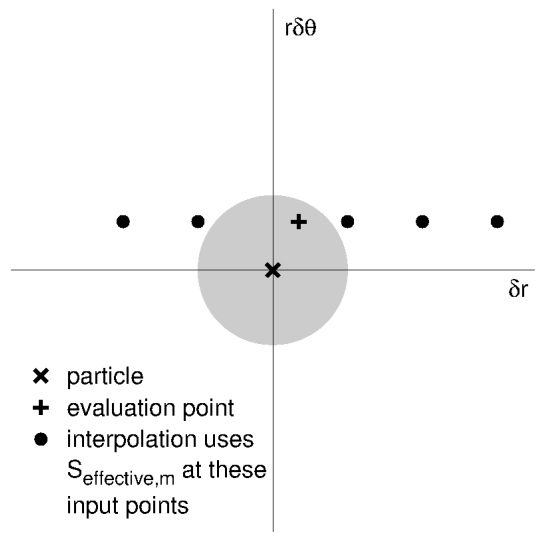}
\end{center}
\vspace{-1ex}
\caption{%%%
\label{fig:interp-esrc2-near-particle}
	This figure shows our interpolation scheme for computing
	the 2-dimensional effective source $S_{\effective,m}$
	near the particle.  We never evaluate $S_{\effective,m}$
	at a Euclidean distance~${} < D_{\min}$ from the particle,
	i.e., within the region shown as the shaded disk.  Instead,
	for an evaluation point within this region we interpolate
	$S_{\effective,m}$ using $S_{\effective,m}$~values
	computed at more distant points.
	}%%%
\end{figure}

Instead, we use the following scheme.  We define a minimum-distance
parameter $D_{\min}$ (typically set to $0.01M$), and if
$(\Delta r)^2 + (r \, \Delta\theta)^2 < D_{\min}^2$, then
we interpolate $S_{\effective,m}$ at $(\Delta r, \Delta\theta)$
using a 4th~order Lagrange interpolating polynomial defined by
the values of $S_{\effective,m}$ at the 5~points
$(-2D_{\min}, \Delta\theta)$, $(-D_{\min}, \Delta\theta)$,
$(+D_{\min}, \Delta\theta)$, $(+2D_{\min}, \Delta\theta)$,
and~$(+3D_{\min}, \Delta\theta)$.
As shown in Fig.~\ref{fig:interp-esrc2-near-particle}, with this
scheme the source is never evaluated closer than a Euclidean
distance~$D_{\min}$ from the particle.  The interpolation is only
needed at at most a few points per slice, so the computational cost
is negligible.

While this scheme has proved adequate for our purposes, it does have
the weakness that if the evaluation point lies in (or very close to)
the equatorial plane $\theta = \pi/2$, then the interpolation molecule
crosses (or almost crosses) the particle position, leading to reduced
accuracy because $\varphi_{\numerical,m}$ is only $C^2$ there.

%%%%%%%%%%%%%%%%%%%%
%%%%%%%%%%%%%%%%%%%%

%%%%%%%%%%%%%%%%%%%%%%%%%%%%%%%%%%%%%%%%

\subsection{Boundary conditions}
\label{sect:theory/BCs}

We implement boundary conditions using finite-differencing ghost zones
which lie immediately adjacent to, but outside, the nominal problem domain.
At each RHS-evaluation time we first use the boundary conditions to
compute $\varphi_{\text{num},m}$ and $\Pi_{\text{num},m}$ at all
ghost-zone grid points.  We then evaluate the RHS (and use this to
time-integrate the evolution equations) at all grid points in the
nominal problem domain.

%%%%%%%%%%%%%%%%%%%%

\subsubsection{Physical boundary conditions}
\label{sect:theory/BCs/physical}

We use pure outflow boundary conditions at the event horizon and $\Scriplus$,
i.e., we apply the interior evolution equations at these grid points,
using (conceptually) 1-sided finite difference molecules for radial
derivatives.%%%
\footnote{%%%
	 For ease of implementation and code organization,
	 we actually implement this by first extrapolating
	 $\varphi$ and $\Pi$ into the radial ghost zones
	 using 5th-order Lagrange polynomial extrapolation,
	 then applying the interior evolution equations
	 using our usual centered finite difference scheme.
	 }%%%

%%%%%%%%%%%%%%%%%%%%

\subsubsection{$z$~axis symmetry boundary conditions}
\label{sect:theory/BCs/z-axis}

As discussed by~\cite[section~IV.C]{Barack-Golbourn-2007},
the $z$~axis symmetry boundary conditions for $\varphi_{\numerical,m}$
(and hence also $\Pi_{\numerical,m}$) depend on $m$.

\begin{description}
\item[\allbf{$m = 0$}]
	In this case $\varphi_{\numerical,m}$ is even across the $z$~axis,
	i.e., $\partial_\theta \varphi_{\numerical,m}
		= \partial_\theta \Pi_{\numerical,m} = 0$.
	The $m^2 / \sin^2 \theta$ term in~\eqref{eqn:evolution-main}
	vanishes identically because $m = 0$, and L'Hopital's rule
	gives the other singular term as
	$\lim_{\theta \to 0}
	 \cot\theta \, \partial_\theta \varphi_{\numerical,m}
		= \partial_{\theta\theta} \varphi_{\numerical,m}$.
\item[\allbf{$m \ne 0$}]
	In this case $\varphi_{\numerical,m}$ is odd across the $z$~axis
	so that $\varphi_{\numerical,m} = \Pi_{\numerical,m} = 0$ there.
	To implement this we specify zero initial data on the $z$~axis
	and replace our evolution equations by
	$\partial_t \varphi_{\numerical,m} = \partial_t \Pi_{\numerical,m} = 0$
	there.
\end{description}

%%%%%%%%%%%%%%%%%%%%

\subsubsection{Equatorial reflection symmetry boundary conditions}
\label{sect:theory/BCs/equator}

If the particle orbit is equatorial (as is the case for all the
numerical computations discussed here), then the entire physical system
has equatorial reflection symmetry, i.e., all fields must be even across
the equator~($\theta = \pi/2$).

%%%%%%%%%%%%%%%%%%%%%%%%%%%%%%%%%%%%%%%%

\subsection{Initial data}
\label{sect:theory/initial-data}

The correct initial data for~\eqref{eqn:box-varphi=S-effective-piecewise}
are unknown (they would represent the equilibrium field configuration
around the particle, which is what we are trying to compute).  Instead,
we follow the usual practice in time-domain self-force computations
(e.g.,~\cite{Dolan-Barack-2011}) and specify arbitrary (zero)
initial data $\varphi_{\numerical,m} = \Pi_{\numerical,m} = 0$
on our initial slice.  This initial data is not a solution of the
sourced evolution equation~\eqref{eqn:box-varphi=S-effective-piecewise},
but we find that the ``junk'' (the deviation of the field configuration
from~\eqref{eqn:box-varphi=S-effective-piecewise}) quickly radiates away
towards the inner and outer boundaries, so that after sufficient time
$\varphi_{\numerical,m}$ relaxes to a solution
of~\eqref{eqn:box-varphi=S-effective-piecewise}
throughout an (expanding) neighborhood of the worldtube.
We see no sign of the persistent (non-radiative) ``Jost junk solutions''
described by~\cite{Field-Hesthaven-Lau-2010,Jaramillo-Sopuerta-Canizares-2011}.
This is to be expected for at least two reasons: (i) the source for our field
equations does not contain the derivative of a Dirac delta function, and (ii)
we are using a second-order-in-space, rather than first-order-in-space
formulation of the field equations.

%%%%%%%%%%%%%%%%%%%%%%%%%%%%%%%%%%%%%%%%

\subsection{Computing the self-force from the evolved fields}
\label{sect:theory/computing-self-force-from-evolved-fields}

Because the physical scalar fields~$\Phi$, $\Phi_\puncture$,
and $\Phi_\residual$ are real,
the Fourier inversion~\eqref{eqn:m-mode-decomp-invert} implies
that $\varphi_{-m} = \conjugate{\varphi_m}$, and similarly for the other
$m$-mode fields.  Hence we only need to (numerically) compute the $m$-modes
$m \ge 0$.

We thus have
\begin{equation}
\Phi_\residual(t,r,\theta,\phi)
	= \sum_{m=0}^\infty \Upsilon^{(\Phi)}_{\residual,m}(t,r,\theta,\phi),
						  \label{eqn:Phi=sum-of-Upsilon}
\end{equation}
where the (real) field $\Upsilon^{(\Phi)}_{\residual,m}$ is given
in a neighborhood of the particle by
\begin{align}
\Upsilon^{(\Phi)}_{\residual,m}&(t,r,\theta,\phi)
	:= \nonumber \\ 
  &\begin{cases}
	   2 \, \realpart{
		  \dfrac{e^{im\tilde{\phi}}}{r}
		  \varphi_{\numerical,m}(t,r,\theta)
			 }
					& \text{if $m \ne 0$}	\\*
	  \dfrac{1}{r} \varphi_{\numerical,m}(t,r,\theta)
					& \text{if $m = 0$}.	%%%\\*
	  \end{cases}
									%%%\\*
\end{align}

We compute the self-force by substituting~\eqref{eqn:Phi=sum-of-Upsilon}
into~\eqref{eqn:F=del-Phi-residual} and differentiating at the particle
position.  A straightforward calculation gives
\begin{equation}
F_a	=	q \sum_{m=0}^\infty
		  \biggl.
		  \left( \Upsilon^{(\partial_a \Phi)}_{\residual,m} \right)
		  \biggr|_\particle,
						    \label{eqn:F=sum-of-Upsilon}
\end{equation}
where the ``self-force modes''
$\Upsilon^{(\partial_a \Phi)}_{\residual,m}
  = \Upsilon^{(\partial_a \Phi)}_{\residual,m}(t,r,\theta)$
are defined in a neighborhood of the particle by
\begin{subequations}
\begin{align}
\Upsilon^{(\partial_t \Phi)}_{\residual,m}
	& =	\begin{cases}
		2 \, \realpart{
		       \dfrac{e^{im\tilde{\phi}}}{r} \Pi_{\numerical,m}
			      }
					& \text{if $m \ne 0$}	\\*
		\dfrac{1}{r} \Pi_{\numerical,m}
					& \text{if $m = 0$},	%%%\\*
		\end{cases}
									\\*
\Upsilon^{(\partial_s \Phi)}_{\residual,m}
	& =	\partial_s \Upsilon^{(\Phi)}_{\residual,m},
									\\*
\Upsilon^{(\partial_\phi \Phi)}_{\residual,m}
	& =	\begin{cases}
		-2m \, \imagpart{
			 \dfrac{e^{im\tilde{\phi}}}{r} \varphi_{\numerical,m}
				}
					& \text{if $m \ne 0$}	\\*
		0
					& \text{if $m = 0$}.	%%%\\*
		\end{cases}
									%%%\\*
\end{align}
\end{subequations}
We compute each self-force mode at the particle by first computing
it in a finite-difference-molecule--sized region about the particle,
then interpolating it to the particle position using the ``C2'' interpolating
function described in Appendix~\ref{app:details/FD-near-particle}.
(For $\Upsilon^{(\partial_s \Phi)}_{\residual,m}$, an alternative
would be to apply a ``differentiating interpolator''%%%
\footnote{%%%
	 An interpolator generally works by (conceptually)
	 locally fitting a fitting function (in our case the
	 C2~interpolant~\protect\eqref{eqn:C2-interpolating-fn})
	 to the data points in a neighbourhood of the
	 interpolation point, then evaluating the fitting
	 function at the interpolation point.  A differentiating
	 interpolator instead evaluates a \emph{derivative}
	 of the fitting function at the interpolation point.
	 This has the effect of interpolating the corresponding
	 derivative of the input data to the interpolation
	 point without ever needing to form a grid function
	 of that derivative.
	 }%%%
{} directly to $\Upsilon^{(\Phi)}_{\residual,m}$.
This would be more elegant and efficient than interpolating a
molecule-sized $\Upsilon^{(\partial_s \Phi)}_{\residual,m}$
grid function.  However, the cost of even the
interpolate-a-molecule-size-grid-function scheme is still only
a minute fraction of the overall self-force computation, so we
did not bother with the additional software complexity of the
differentiating interpolator.)

%%%%%%%%%%%%%%%%%%%%%%%%%%%%%%%%%%%%%%%%

\subsection{The tail series}
\label{sect:theory/tail-series}

In practice we can only numerically compute a finite number of $m$-modes
$0 \le m \le m_{\numerical,\max}$.  We thus partition each of the infinite
sums in~\eqref{eqn:Phi=sum-of-Upsilon} and~\eqref{eqn:F=sum-of-Upsilon}
into a finite ``numerical sum'' plus an infinite ``tail sum'',
\begin{equation}
\sum_{m=0}^\infty
	= \sum_{m=0}^{m_{\numerical,\max}}
	  +
	  \sum_{m=m_{\numerical,\max}{+}1}^\infty,
						  \label{eqn:sum=numerical+tail}
\end{equation}
and account for the tail sum in much the same way as is done in the mode-sum regularization scheme.

To estimate the tail sum for the self-force
computation~\eqref{eqn:F=sum-of-Upsilon},%%%
\footnote{%%%
\label{footnote:how-to-compute-scalar-field-at-particle}
	 The physical scalar field~$\Phi$ at the particle
	 can also be computed by applying similar techniques
	 to the infinite sum~\protect\eqref{eqn:Phi=sum-of-Upsilon}.
	 } %%%
we use the fact that the modes have a known power-law behavior that can
be attributed to the non-smoothness of the residual field. Explicitly, the behavior of the modes
of the residual field is given by
\begin{align}
\biggl.
\left( \Upsilon^{(\Phi)}_{\residual,m} \right)&
\biggr|_\particle = \nonumber \\
  & \sum_{\substack{\alpha \ge n \\ \text{$\alpha$ even}}} k^{(\Phi)}_\alpha f^{(\Phi)}_{\alpha,m} + \left( \Upsilon^{(\Phi)}_{\regular,m} \right)\biggr|_\particle,
\label{eqn:tail-series}
\end{align}
where $\Upsilon^{(\Phi)}_{\regular,m}$ comes from the $C^\infty$ regular field and falls off faster
than any power of $m$; it can therefore be ignored for $m_{\numerical,\max}$ sufficiently large. The
remaining piece of the tail sum is effectively an even power series in $1/m$, starting at an order,
$m^{-n}$, that is determined by the order of the puncture field. In our case $n = 4$, the basis
functions~$f$ for the $m$-dependence are given by
\begin{widetext}
\begin{align}
f^{(\Phi)}_{4,m}
	& =	\frac{1}
		     {
		     (m-\frac{3}{2})
		     (m-\frac{1}{2})
		     (m+\frac{1}{2})
		     (m+\frac{3}{2})
		     }
									\nonumber\\*
f^{(\Phi)}_{6,m}
	& =	\frac{1}
		     {
		     (m-\frac{5}{2})
		     (m-\frac{3}{2})
		     (m-\frac{1}{2})
		     (m+\frac{1}{2})
		     (m+\frac{3}{2})
		     (m+\frac{5}{2})
		     }
									\nonumber\\*
f^{(\Phi)}_{8,m}
	& =	\frac{1}
		     {
		     (m-\frac{7}{2})
		     (m-\frac{5}{2})
		     (m-\frac{3}{2})
		     (m-\frac{1}{2})
		     (m+\frac{1}{2})
		     (m+\frac{3}{2})
		     (m+\frac{5}{2})
		     (m+\frac{7}{2})
		     }
									\nonumber\\*
	 & \vdots
								\nonumber \\*
f^{(\Phi)}_{\alpha,m} &= \frac{1}{\Big(m-\tfrac{\alpha - 1}{2}\Big)_{\alpha}},
\label{eqn:tail-basis-fns}
\end{align}
\end{widetext}
and the coefficient functions, $k$, are given by the $m$-mode decomposition of higher-order terms
(i.e., those that have \emph{not} been included in the definition of the puncture field) in the
series expansion of the Detweiler-Whiting singular field \cite{Heffernan:2012vj}.

Derivatives of the field behave in a similar manner, so that in addition to using this approach for
$\Upsilon^{(\Phi)}_{\residual,m}$, we may also use it for the fields
$\Upsilon^{(X)}_{\residual,m}$, where $X$ is one of $\partial_s \Phi$, $\partial_t \Phi$ or
$\partial_\phi \Phi$. The only caveat is that the $m$-dependence is slightly modified: the
$\phi$ derivative introduces a factor of $m^2$, so $f^{(\partial_\phi\Phi)}_{\alpha,m} = m^2
f^{(\Phi)}_{\alpha,m}$. The $t$ derivative of the Detweiler-Whiting singular field can be written in
terms of $r$ and $\phi$ derivatives, so $f^{(\partial_t\Phi)}_{\alpha,m}$ has both kinds of terms
present.

For any given $X$, $\alpha$ and $m_{\numerical,\max}$, the infinite sum
\begin{equation}
S^{(X)}_{\alpha,m_{\numerical,\max}{+}1}
	:= \sum_{m=m_{\numerical,\max}{+}1}^\infty f^{(X)}_{\alpha,m}
					   \label{eqn:S=sum-of-f-tail-basis-fns}
\end{equation}
can be computed exactly. Using the facts that
\begin{align}
\sum_{m=-\infty}^\infty \Big(m-\tfrac{\alpha - 1}{2}\Big)_{\alpha} = 0
	& \qquad \text{for even $\alpha \ge 2$},
									\\
\sum_{m=-\infty}^\infty m^2 \Big(m-\tfrac{\alpha - 1}{2}\Big)_{\alpha} = 0
	& \qquad \text{for even $\alpha \ge 4$},
									%%%\\
\end{align}
we obtain
\begin{align}
S^{(\Phi)}_{\alpha,m_{\numerical,\max}{+}1}
	& = \frac{1}{(\alpha-1)\Big(m_{\numerical,\max} - \tfrac{\alpha - 3}{2}\Big)_{\alpha - 1}},
									\\
S^{(\partial_\phi \Phi)}_{\alpha,m_{\numerical,\max}{+}1}
	& = \frac{m_{\numerical,\max}(m_{\numerical,\max}+1)}{(\alpha-3)\Big(m_{\numerical,\max} - \tfrac{\alpha - 3}{2}\Big)_{\alpha - 1}}.
									%%%\\
\end{align}

Analytical expressions for the $k^{(X)}_4$~coefficients (in this context
known as ``$m$-mode regularization parameters'') compatible with our
choice of puncture field were given in~\cite{Heffernan:2012vj}.  As they
are extremely lengthy we will not repeat them here; a Mathematica
notebook for computing them is included in the online supplemental
materials which accompany this paper.

While the higher-order coefficients could be analytically determined in a similar manner, we choose
instead an alternative approach. To estimate some finite set~$\alpha \in \alpha_\tailset$ of the remaining
$k^{(X)}_\alpha$~coefficients, we first truncate the series~\eqref{eqn:tail-series} to only the terms
$\alpha=4$ and $\alpha \in \alpha_\tailset$,
\begin{equation}
\biggl.
\left( \Upsilon^{(X)}_{\residual,m} \right)
\biggr|_\particle
	\approx	k^{(X)}_4 f^{(X)}_{4,m}
		+ \sum_{\substack{\text{$\alpha$ even} \\ \alpha \in \alpha_\tailset}}
		  k^{(X)}_\alpha f^{(X)}_{\alpha,m}.
						\label{eqn:tail-series-alpha=4+alpha>=6}
\end{equation}
For a specified particle position, we then estimate the corresponding
set of~$k^{(X)}_\alpha$ by least-squares fitting the numerically-computed
$
\biggl.
\left( \Upsilon^{(X)}_{\residual,m} \right)
\biggr|_\particle
$
with $m_{\fit,\min} \le m \le m_{\fit,\max}$
to the truncated series~\eqref{eqn:tail-series-alpha=4+alpha>=6}.%%%
\footnote{%%%
	 For each~$\alpha$, we normalize $f^{(X)}_{\alpha,m}$ to have
	 unit magnitude at the mean~$m$ in~$\alpha_{\tailset}$.
	 This reduces to a tolerable level what would otherwise
	 be severe numerical ill-conditioning in the least-squares
	 fit~\protect\cite{Thornburg-2010:highly-accurate-self-force}.%%%
	 }
{}  For all analyses reported in this paper we take $\alpha_\tailset$
to be either empty (no tail fit) or $\{6,8\}$.
Table~\ref{tab:configs/low-noise-and-tail-fit-pars} gives $m_{\fit,\min}$ and
$m_{\fit,\max}$ for each of our configurations where a tail fit is done.

Finally, we compute (estimate) each self-force component $F_a$
at each of these times by substituting~\eqref{eqn:sum=numerical+tail},
\eqref{eqn:tail-series}, and~\eqref{eqn:S=sum-of-f-tail-basis-fns}
into~\eqref{eqn:F=sum-of-Upsilon}, giving
\begin{align}
\frac{F_a}{q}
	= {}&	\sum_{m=0}^{m_{\numerical,\max}}
		\biggl.
		\left( \Upsilon^{(X)}_{\residual,m} \right)
		\biggr|_\particle
								\nonumber\\*
	&	{}
		+
		k^{(X)}_4 S^{(X)}_{4,m_{\numerical,\max}{+}1}
								\nonumber\\*
	&	{}
		+
		\sum_{\substack{\text{$\alpha$ even} \\ \alpha \in \alpha_\tailset}}
		k^{(X)}_\alpha S^{(X)}_{\alpha,m_{\numerical,\max}{+}1}.
					    \label{eqn:F=sum-Upsilon+tail-terms}
									%%%\\*
\end{align}

%%%%%%%%%%%%%%%%%%%%%%%%%%%%%%%%%%%%%%%%

\subsection{Selecting the time interval for analysis within an evolution}
\label{sect:theory/selecting-time-interval-for-analysis}

Our discussion in
sections~\ref{sect:theory/computing-self-force-from-evolved-fields}
and~\ref{sect:theory/tail-series} assumed that a time series of the
self-force modes $\Upsilon^{(\partial_a \Phi)}_{\residual,m}$ is
available at a suitable set of points around the orbit for each
$m = 0$, $1$, $2$, \dots, $m_{\numerical,\max}$.  However, as
described in section~\ref{sect:theory/initial-data}, the initial part
of each such time series is contaminated by ``junk'' radiation.
Here we describe how we determine when this junk radiation has decayed
to a negligible level (below our numerical noise level).

The key fact which underlies our algorithm for making this determination
is that since the particle orbit is periodic,%%%
\footnote{%%%
	 More precisely, the particle orbit is periodic
	 modulo an overall rotation in $\phi$, which is
	 ignorable because Kerr spacetime is axisymmetric.%%%
	 }%%%
{} the self-force modes should also be periodic with the orbital period~$T_r$..

Given a time series of some numerically-computed self-force mode
$\Upsilon^{(\partial_a \Phi)}_{\residual,m}$, we define its
``orbit difference'' time series by
\begin{equation}
\Delta \left[ \Upsilon^{(\partial_a \Phi)}_{\residual,m} \right] \! (t)
	:= \left|
	   \Upsilon^{(\partial_a \Phi)}_{\residual,m}(t + T_r)
	   -
	   \Upsilon^{(\partial_a \Phi)}_{\residual,m}(t)
	   \right|.
\end{equation}
The orbit-difference time series is one orbit shorter in duration
than the original time series.

Because of the initial junk radiation, the orbit difference is
initially large.  As the junk radiation radiates away from the
particle and worldtube, the orbit difference decays until it
eventually becomes roughly constant (at a nonzero value due to
finite differencing and other numerical errors) or, in some cases,
varying with the orbital period (since the numerical errors are
similarly periodic).  (This behavior can be seen in
Fig.~\ref{fig:e8-modes-and-orbit-diffs-overview}.)

It is thus quite easy to determine the time when the junk radiation
has decayed to a negligible level by visually inspecting a graph of
the orbit difference as a function of time.  Although this process could
probably be automated by searching backwards in the orbit-difference
time series for a sustained rise (in fact, we implemented such an algorithm),
we find that the visual inspection is valuable for detecting a variety
of other numerical problems which might occur, so we have chosen not
to routinely use an automated algorithm here.

%%%%%%%%%%%%%%%%%%%%%%%%%%%%%%%%%%%%%%%%

\subsection{Selecting a ``low-noise'' subset of times within an evolution}
\label{sect:theory/selecting-low-noise-times}

Because of the interaction between finite differencing and the limited
differentiability of $\varphi_\numerical$ at the particle, as well as
other numerical errors, there is numerical noise in the self-force
modes~$\Upsilon^{(\partial_a \Phi)}_{\residual,m}$.  For highly
eccentric orbits, we find that the higher-$m$ modes may be completely
dominated by numerical noise in the outer parts of the orbit.
(This can be seen in, for example, Figs.~\ref{fig:e8-aligned-modes-3d}
and~\ref{fig:e8-modes-overview/final-orbit}.)

Including these modes in the self-force
sum~\eqref{eqn:F=sum-Upsilon+tail-terms} would add significant numerical
noise to the computed self-force while (in many cases) not adding any
significant ``signal''.  Therefore, it is useful (again, in many although
not all cases) to omit the noisy modes from the self-force
sum~\eqref{eqn:F=sum-Upsilon+tail-terms}, effectively treating
these modes/times as missing data.

To estimate the noise level at any point in an
$\Upsilon^{(\partial_a \Phi)}_{\residual,m}$ time series,
we first define a smoothed time series
$\Smooth{\Upsilon^{(\partial_a \Phi)}_{\residual,m}}$ using
Savitzky-Golay moving-window
smoothing~\cite{Savitzky-Golay-1964:smoothing}, %%%
\cite[section~14.8]{Numerical-Recipes-3rd-edition}.
For all analyses reported in this paper we use a 6th-degree polynomial
over a $\text{current position} \pm 10$-sample moving window
in the time series.

We then define the (absolute) noise time series as
\begin{align}
&\Noise{\Upsilon^{(\partial_a \Phi)}_{\residual,m}} \! (t)
	:= \nonumber \\
& \qquad \RMS_{\text{SG window}(t)}
	   \biggl\{
		   \Upsilon^{(\partial_a \Phi)}_{\residual,m}
	   -
	   \Smooth{\Upsilon^{(\partial_a \Phi)}_{\residual,m}}
	   \biggr\}
\end{align}
and the ``relative noise'' time series as
\begin{align}
&\RelativeNoise{\Upsilon^{(\partial_a \Phi)}_{\residual,m}} \! (t)
	:= \nonumber \\
& \qquad \Noise{\Upsilon^{(\partial_a \Phi)}_{\residual,m}} \! (t)
	   \Bigg/
	   \RMS_{\text{SG window}(t)}
	   \Bigl\{
	   \Upsilon^{(\partial_a \Phi)}_{\residual,m}
	   \Bigr\},
\end{align}
where $\displaystyle \RMS_{\text{SG window}(t)} \bigl\{ \cdots \bigr\}$
is the root-mean-square value over the Savitzky-Golay smoothing window.

Using these definitions we select a ``low-noise'' subset of the time
samples by omitting those samples from the
$\Upsilon^{(\partial_a \Phi)}_{\residual,m}$ time series
which have $m \ge m_{\noise,\min}$ and
$\RelativeNoise{\Upsilon^{(\partial_a \Phi)}_{\residual,m}}
	> \varepsilon_{\relative,\max}$,
where $m_{\noise,\min}$ is a parameter chosen so that time intervals
immediately around zero-crossings in lower-$m$ modes are not falsely
excluded, and where $\varepsilon_{\relative,\max}$ is a parameter chosen to
tune the tolerable level of numerical noise.
Table~\ref{tab:configs/low-noise-and-tail-fit-pars} gives $m_{\noise,\min}$
and $\varepsilon_{\relative,\max}$ for each of our configurations
where smoothing is done.

%%%%%%%%%%%%%%%%%%%%%%%%%%%%%%%%%%%%%%%%

\subsection{Dissipative and conservative parts of the self-force}
\label{sect:theory/diss-cons}

As well as calculating the overall self-force, it is useful to split
the self-force into dissipative and conservative contributions
\cite{Mino-2003,Hinderer-Flanagan-2008,Diaz-Rivera-etal-2004,Barack-Sago-2009,Vega-etal-2013:Schwarzschild-scalar-self-force-via-effective-src}:
the dissipative
part affects the $\O(\mu)$~orbital evolution while the conservative
part only affects the orbital evolution at~$\O(\mu^2)$.  As discussed by
\cite[section~8.1]{Barack-2009:self-force-review}, for equatorial
orbits these can be computed from the even-in-time and odd-in-time
parts of the self-force,%%%
\footnote{%%%
	 It would be possible to similarly compute the
	 dissipative and conservative parts of each individual
	 self-force mode in the sums~\protect\eqref{eqn:F=sum-of-Upsilon}.
	 This would have the advantage that the dissipative
	 part of the self-force could be computed very accurately
	 (its tail sums should converge exponentially fast),
	 with only the conservative
	 part requiring the full tail-sum computation described
	 in section~\protect\ref{sect:theory/tail-series}.  However,
	 for historical reasons we have not taken this approach.
	 }%%%
\begin{subequations}
						   \label{eqn:F-diss-cons-parts}
\begin{align}
F_{\diss,t}	& = F_{\even,t}
				& F_{\cons,t}	& = F_{\odd,t}		\\
F_{\diss,r}	& = F_{\odd,r}
				& F_{\cons,r}	& = F_{\even,r}		\\
F_{\diss,\phi}	& = F_{\even,\phi}
				& F_{\cons,\phi}& = F_{\odd,\phi}	%%%\\
\end{align}
\end{subequations}
where
\begin{subequations}
						      \label{eqn:even-odd-parts}
\begin{align}
F_{\even,a}(t)	& =	\thalf \bigl[ F_a(t) + F_a(T_r-t) \bigr]	\\
F_{\odd,a}(t)	& =	\thalf \bigl[ F_a(t) - F_a(T_r-t) \bigr]	%%%\\
\end{align}
\end{subequations}
with $t$ being the modulo time.

To allow this computation without requiring time interpolation, we
always choose our self-force computation times to be uniformly spaced
in coordinate time~$t$, with a spacing~$\Delta t_\sample$ which
integrally divides the orbital period~$T_r$.

%%%%%%%%%%%%%%%%%%%%%%%%%%%%%%%%%%%%%%%%

\subsection{Summary of computation and data analysis}
\label{sect:theory/summary-of-cmpt-and-data-analysis}

To summarize, our overall computational and data-analysis scheme
involves a sequence of operations:
\begin{itemize}
\item	For each~$m$, we perform a numerical evolution of the
	1st-order-in-time evolution system described in
	section~\ref{sect:theory/1st-order-in-time}.  Our
	evolution code writes out time series of each self-force
	mode~$\Upsilon^{(\partial_a \Phi)}_{\residual,m}$,
	sampled at uniform coordinate-time intervals.
	We always choose the sampling time~$\Delta t_\sample$
	to be the same for each~$m$ and
	(as noted in section~\ref{sect:theory/diss-cons})
	to integrally divide the period~$T_r$ of the particle's radial motion.
\item	For each~$m$, we use the orbit-differences algorithm
	described in
	section~\ref{sect:theory/selecting-time-interval-for-analysis}
	to select a point in each of the self-force modes'
	time series when the initial junk radiation has
	decayed to a level below our numerical noise level.
	For all our further data analysis we use only the modes
	from times $\ge$ this ``self-force computation start time''
	for each~$m$.
\item	For most configurations, for each~$m$ we use the
	noise-estimation and low-noise--selection algorithms described
	in section~\ref{sect:theory/selecting-low-noise-times}
	to select a subset of the self-force
	mode~$\Upsilon^{(\partial_a \Phi)}_{\residual,m}$
	time series which has relatively low numerical noise.
\item	For each modulo time for which we have self-force
	modes (at times $\ge$ the self-force computation
	start time, and with sufficiently low estimated noise),
	we compute the $t$, $r$, and $\phi$ components of the
	self-force using the mode summation and tail-fitting
	algorithms described in
	sections~\ref{sect:theory/computing-self-force-from-evolved-fields}
	and~\ref{sect:theory/tail-series}.
\end{itemize}

%%%%%%%%%%%%%%%%%%%%%%%%%%%%%%%%%%%%%%%%%%%%%%%%%%%%%%%%%%%%%%%%%%%%%%%%%%%%%%%%

\section{Numerical results}
\label{sect:results}

%%%%%%%%%%%%%%%%%%%%%%%%%%%%%%%%%%%%%%%%

\subsection{Configurations and parameters}
\label{sect:results/configs-and-pars}

Tables~\ref{tab:configs/physical-pars}--\ref{tab:configs/compactification-pars}
summarize the main physical and computational parameters for the
configurations presented here.%%%
\footnote{%%%
	 The input parameter files and data analysis scripts
	 for the highest-resolution evolutions for each configuration,
	 as well as for the variant-grid dro6-48 evolutions for
	 the e9~configuration, are included in the online
	 supplemental materials accompanying this paper.
	 }%%%
\footnote{%%%
	 These simulations all used the Karst cluster
	 at Indiana University.%%%
	 }%%%
{}  These configurations fall into four (overlapping) families:
\begin{itemize}
\item	The ns5, n-55, n95, ze4, and e8b configurations are ones which
	have also been calculated by other researchers, allowing us to
	validate our code against their results
	(both published and unpublished).
\item	The e8, e8b, e9, and e95 configurations are (non--zoom-whirl)
	highly eccentric orbits.
\item	The ze4, ze9, zze9, and ze98 configurations are zoom-whirl orbits;
	of these the ze4 configuration is of moderate eccentricity
	while the ze9, zze9, and ze98 configurations are highly eccentric.
\item	The circ-ze4, circ-ze9, circ-zze9, and circ-ze98 configurations are
	circular-orbit configurations with orbital radii matching the
	periastrons of the corresponding zoom-whirl configurations.
\end{itemize}

%%%%%%%%%%%%%%%%%%%%
\begingroup
\squeezetable
\begin{table*}[bp]
\begin{center}
\begin{ruledtabular}
%               name
%                apeEL
%                     min r
%                      max r
%                       period
%                          delta phi per orbit
%               123456789.12
\begin{tabular}{ldddccdddddd}
	&		&		&
	&		&
	&		&
	& \multicolumn{3}{c}{Orbital period}
	& \multicolumn{1}{c}{$\delta\phi$~per}
									\\
\cline{9-11}
	&		&		&
	&		&
	& \multicolumn{1}{c}{min~$r$}
			& \multicolumn{1}{c}{max~$r$}
	& \multicolumn{2}{c}{Radial}
			& \multicolumn{1}{c}{Azimuthal}
	& \multicolumn{1}{c}{orbit}
									\\
\cline{9-10}
Name	& \multicolumn{1}{c}{$\tilde{a}$}
			& \multicolumn{1}{c}{$p$}
					& \multicolumn{1}{c}{$e$}
	& \multicolumn{1}{c}{$\mathcal{E}$ ($M$)}
			& \multicolumn{1}{c}{$\mathcal{L}$ ($M^2$)}
	& \multicolumn{1}{c}{($M$)}
			& \multicolumn{1}{c}{($M$)}
	& \multicolumn{1}{c}{$T_r$ ($M$)}
			& \multicolumn{1}{c}{$\tau_r$ ($M$)}
					& \multicolumn{1}{c}{$T_\phi$ ($M$)}
	& \multicolumn{1}{c}{(orbits)}
									\\
\hline % --------------------------------------------------------------
ns5	& 0.0		& 7.2		& 0.5
	& 0.956\,876	& 3.622\,713
	&  4.8		& 14.4
	& 405.662	& 317.366	& 134.285
	& 3.021
									\\
n-55	& -0.5		& 10.0		& 0.5
	& 0.967\,896	& 4.100\,631
	&  6.667	& 20.0
	& 505.428	& 434.465	& 249.488
	& 2.026
									\\
n95	& 0.9		& 10.0		& 0.5
	& 0.963\,778	& 3.489\,553
	& 6.667		& 20.0
	& 378.408	& 333.027	& 293.070
	& 1.291
									\\
\hline % --------------------------------------------------------------
e8	& 0.6		& 8.0		& 0.8
	& 0.978\,270	& 3.405\,897
	& 4.444		& 40.000
	& 771.968	& 709.796	& 502.435
	& 1.536
									\\
e8b	& 0.8		& 8.0		& 0.8
	& 0.978\,056	& 3.292\,113
	& 4.444		& 40.000
	& 756.641	& 697.570	& 527.812
	& 1.434
									\\
e9	& 0.99		& 7.0		& 0.9
	& 0.986\,565	& 3.052\,860
	&  3.684	& 70.000
	& 1513.855	& 1442.724	& 1060.526
	& 1.427
									\\
e95	& 0.99		& 5.0		& 0.95
	& 0.990\,315	& 2.699\,644
	&   2.564	& 100.000
	& 2436.050	& 2349.870	& 1445.400
	& 1.685
									\\
\hline % --------------------------------------------------------------
ze4	& 0.2		& 6.15		& 0.4
	& 0.945\,536	& 3.366\,468
	&  4.393	& 10.25
	& 354.628	& 255.966	& 95.799
	& 3.702
									\\
ze9	& 0.0		& 7.800\,1	& 0.9
	& 0.988\,333	& 3.904\,885
	&  4.105	& 78.001
	& 2112.079	& 1913.402	& 339.855
	& 6.215
									\\
zze9	& 0.0		& 7.800\,001	& 0.9
	& 0.988\,332	& 3.904\,884
	&  4.105	& 78.000
	& 2224.815	& 1971.883	& 265.734
	& 8.372
									\\
ze98	& 0.99		& 2.4		& 0.98
	& 0.991\,798	& 2.180\,959
	& 1.212		& 120.000
	& 3304.620	& 3021.480	& 215.851
	& 15.310
									\\
\hline % --------------------------------------------------------------
circ-ze4
	& 0.2		& 4.392\,857	& 0.0
	& 0.943\,384	& 3.346\,263
	& 4.392\,857	& 4.392\,857	 
	&		&		& 59.106
	&
									\\
circ-ze9
	& 0.0		& 4.105\,316	& 0.0
	& 0.988\,327	& 3.904\,841
	& 4.105\,316	& 4.105\,316
	&		&		& 52.264
	&
									\\
circ-zze9
	& 0.0		& 4.105\,264	& 0.0
	& 0.988\,332	& 3.904\,884
	& 4.105\,264	& 4.105\,264
	&		&		& 52.263
	&
									\\
circ-ze98
	& 0.99		& 1.212\,121	& 0.0
	& 0.984\,732	& 2.164\,538
	& 1.212\,121	& 1.212\,121
	&		&		& 14.605
									%%%\\
\end{tabular}
\end{ruledtabular}
\end{center}
\vspace{-1ex}
\caption{%%%
\label{tab:configs/physical-pars}
	This table summarizes the main physical parameters
	for the configurations presented in this paper.
	$(\tilde{a},p,e)$ uniquely characterize the spacetime and the
	particle orbit.  $\mathcal{E}$ and $\mathcal{L}$ are
	the particle's specific energy and angular momentum,
	respectively.  ``min~$r$'' and ``max~$r$'' are the particle's
	periastron and apoastron coordinate radii, respectively.
	The orbital period is given in three forms:
	the coordinate time~$T_r$ and proper time along the
	particle orbit~$\tau_r$ of the radial motion, and the long-term mean
	coordinate-time period~$T_\phi$ of the azimuthal~($\phi$) motion
	(i.e., the mean coordinate time~$t$ during which $\phi$
	advances by~$2\pi$).  ``$\delta\phi$~per orbit'' denotes
	the advance in $\phi$ (in units of $2\pi$) during one period
	of the orbit's radial motion (i.e., during a coordinate time~$T_r$);
	this is given by $T_r/T_\phi$ and is $\gg 1$~orbit for a
	zoom-whirl orbit.
	}%%%
\end{table*}
\endgroup
%%%%%%%%%%%%%%%%%%%%

%%%%%%%%%%%%%%%%%%%%
\begingroup
\squeezetable
\begin{table*}[bp]
\begin{center}
\begin{ruledtabular}
%               x Name
%                x N_{sample}
%                |x Delta t_{sample}
%                | x m_{num,max}
%                | |x t_initial
%                | | 01234 Self-force computation start time (m=0,1,2,3,.ge.4)
%                | | |||||01234 Evolution ending time (m=0,1,2,3,4,.ge.5)
%               123456789.12345
\begin{tabular}{llccrrrrrrrrrrr}
	&		&		&	&
	& \multicolumn{5}{c}{Self-force computation start time}
	& \multicolumn{5}{c}{Evolution end time}
									\\
%%%%%%%%%%
\cline{6-10} \cline{11-15}
	&		& \multicolumn{1}{c}{$\Delta t_\sample$}
					&		& \multicolumn{1}{c}{$t_\initial$}
	& \multicolumn{1}{c}{$m=0$}
		& \multicolumn{1}{c}{$m=1$}
			& \multicolumn{1}{c}{$m=2$}
				& \multicolumn{1}{c}{$m=3$}
					& \multicolumn{1}{c}{$m \ge 4$}
	& \multicolumn{1}{c}{$m=0$}
		& \multicolumn{1}{c}{$m=1$}
			& \multicolumn{1}{c}{$m=2$}
				& \multicolumn{1}{c}{$m=3$}
					& \multicolumn{1}{c}{$m \ge 4$}
									\\
%%%%%%%%%%
Name	& $N_\sample$	& \multicolumn{1}{c}{($M$)}
			& $m_{\numerical,\max}$
					& \multicolumn{1}{c}{($M$)}
	& \multicolumn{1}{c}{($M$)}
			& \multicolumn{1}{c}{($M$)}
					& \multicolumn{1}{c}{($M$)}
							& \multicolumn{1}{c}{($M$)}
							& \multicolumn{1}{c}{($M$)}
	& \multicolumn{1}{c}{($M$)}
			& \multicolumn{1}{c}{($M$)}
					& \multicolumn{1}{c}{($M$)}
							& \multicolumn{1}{c}{($M$)}
							& \multicolumn{1}{c}{($M$)}
									\\
\hline % --------------------------------------------------------------
ns5	& \Z\Z\,406	& 0.999		& 20		& 143
	& 8\,000	& 2\,000	& 700		& 365
							& 310
	& 11\,990	& 6\,994	& 3\,997	& 1\,349
							& 1\,349%%%
\rlap{\footnotemark[1]}
									\\
n-55	& \Z\Z\,506	& 0.999		& 20		& 183
	& 7\,200	& 2\,700	& 900		& 480
							& 420
	& 10\,291	& 5\,237	& 3\,721	& 1\,348
							& 1\,348
									\\
n95	& \Z\Z\,378	& 1.001		& 20		& 116
	& 5\,115	& 2\,850	& 890		& 400
							& 350
	& 12\,013	& 8\,009	& 4\,004	& 1\,251
							& 1\,251
									\\
\hline % --------------------------------------------------------------
e8	& \Z\Z\,770	& 1.003		& 20		& 324
	& 7\,000	& 2\,600	& 1\,300	& 800
							& 650
	& 11\,903	& 6\,500	& 4\,184	& 2\,506
							& 2\,506
									\\
e8b	& \Z\Z\,760	& 0.996		& 20		& 289
	& 8\,700	& 4\,200	& 1\,900	& 1\,025
							& 820
	& 11\,728	& 7\,945	& 5\,675	& 3\,405
							& 3\,405
									\\
e9	& \Z{}1\,514	& 1.000		& 20		& 680
	& 10\,000	& 5\,300	& 2\,450	& 1\,300
							& 1\,180%%%
\rlap{\footnotemark[2]}
	& 15\,818	& 8\,249	& 5\,221	& 3\,300
							& 3\,300
									\\
e95	& \Z{}2\,436	& 1.000		& 20		& 1\,090
	& 7\,000	& 4\,700	& 2\,550	& 2\,200
							& 1\,950
	& 13\,270	& 8\,398	& 5\,962	& 5\,400
							& 5\,400
									\\
\hline % --------------------------------------------------------------
ze4	& \Z\Z\,360	& 0.985		& 20		& 111
	& 8\,000	& 2\,000	& 550		& 400
							& 360
	& 11\,821	& 6\,896	& 3\,940	& 1\,182
							& 1\,182
									\\
ze9	& \Z{}2\,112	& 1.000		& 20		& 920
	& 12\,888	& 7\,888	& 2\,500	& 1\,600
							& 1\,600
	& 15\,001	& 10\,000	& 8\,000	& 4\,800
							& 4\,800%%%
\rlap{\footnotemark[1]}
									\\
zze9	& \Z{}2\,224	& 1.000		& 20		& 1\,005
	& 10\,000	& 5\,550	& 3\,350	& 1\,700
							& 1\,575
	& 22\,361	& 14\,461	& 12\,236	& 7\,787
							& 7\,787%%%
\rlap{\footnotemark[1]}
									\\
ze98	& 13\,216	& 0.250		& 12		& 1\,448
	& 11\,600	& 8\,250	& 4\,950	& 4\,950
							& 4\,950%%%
\rlap{\footnotemark[2]}
	& 18\,093	& 14\,871	& 14\,871	& 13\,218
							& 10\,623%%%
\rlap{\footnotemark[1]}
									\\
\hline % --------------------------------------------------------------
circ-ze4
	& \Z\Z\,\Z{}60	& 0.982		& 20		& 0
	& 6\,000	& 2\,000	& 450		& 315
							& 300%%%
\rlap{\footnotemark[2]}
	& 6\,154	& 4\,000	& 940		& 700
							& 600%%%
\rlap{\footnotemark[1]}
									\\
circ-ze9
	& \Z\Z\,\Z{}52	& 1.005		& 20		& 0
	& 10\,000	& 1\,630	& 410		& 325
							& 335%%%
\rlap{\footnotemark[2]}
	& 10\,453	& 1\,980	& 950		& 685
							& 425%%%
\rlap{\footnotemark[1]}
									\\
circ-zze9
	& \Z\Z\,\Z{}52	& 1.005		& 20		& 0
	& 10\,000	& 1\,630	& 410		& 325
							& 335%%%
\rlap{\footnotemark[2]}
	& 10\,453	& 1\,980	& 950		& 685
							& 425%%%
\rlap{\footnotemark[1]}
									%%%\\
\end{tabular}
\end{ruledtabular}
\footnotetext[1]{Some large-$m$ evolutions end earlier.}
\footnotetext[2]{Some large-$m$ evolutions
		 start the self-force computation earlier.}
\end{center}
\vspace{-1ex}
\caption{%%%
\label{tab:configs/evolution-times}
	For each configuration,
	this table gives the number of self-force output samples
	per orbit (more precisely, per radial orbital period for
	the eccentric-orbit configurations, and per azimuthal orbital
	period for the circular-orbit configurations),
	the sampling interval,
	the maximum~$m$ of the numerically-computed modes
	in the self-force sum~\protect\eqref{eqn:F=sum-Upsilon+tail-terms},
	the time at which the self-force computation begins
	(after the initial transients have decayed),
	and the time at which the numerical evolution ends
	(or at which the self-force computation ends, if this is earlier).
	All times are coordinate times in units of $M$ and
	(except for $\Delta t_\sample$) are rounded to the nearest integer.
	For some configurations (footnoted), some large-$m$ evolutions
	use earlier starting and/or ending times (chosen to select
	low--numerical-noise sections of data and/or limited by
	machine failures or queue-time limits).  For the eccentric-orbit
	configurations there is always at least one orbital period
	between the starting and ending times; for the circular-orbit
	configurations the self-force is time-independent so there is
	no need for an extended self-force computation interval.
	The circ-ze98 configuration is omitted because we were
	unable to obtain stable evolutions for it for $m \ge 6$.
	}%%%
\end{table*}
\endgroup
%%%%%%%%%%%%%%%%%%%%

%%%%%%%%%%%%%%%%%%%%
\begin{table}[bp]
\begin{center}
\begin{ruledtabular}
%               12345
%               name
%                --
%                m_{noise,min}
%                 varepsilon_{relative,max}
%                  --
%                  m_{fit,min}
%                   m_{fit,max}
\begin{tabular}{lcccc}
%%%%%%%%%%
		& \multicolumn{2}{c}{low-noise selection}
		& \multicolumn{2}{c}{}
									\\
%%%%%%%%%%
		& \multicolumn{2}{c}{parameters}
		& \multicolumn{2}{c}{tail-fit parameters}
									\\
%%%%%%%%%%
\cline{2-3} \cline{4-5}
%%%%%%%%%%
Name		& $m_{\noise,\min}$
				& $\varepsilon_{\relative,\max}$
		& $m_{\fit,\min}$
				& $m_{\fit,\max}$
									\\
\hline % --------------------------------------------------------------
ns5		& 10		& 0.05
		& \Z{}9		& 18					\\
n-55		& 10		& 0.05
		& \Z{}9		& 18					\\
n95		& 10		& 0.05
		& \Z{}9		& 18					\\
\hline % --------------------------------------------------------------
e8		& \Z{}4		& 0.3\Z
		& 12		& 20					\\
e8b		& \Z{}4		& 0.3\Z
		& 12		& 20					\\
e9		& \Z{}3		& 0.3\Z
		& 12		& 20					\\
e95		& \Z{}2		& 0.3\Z
		& 12		& 20					\\
\hline % --------------------------------------------------------------
ze4		& 10		& 0.05
		& \Z{}8		& 18					\\
ze9		& \Z{}2		& 0.3\Z
		& 12		& 20					\\
zze9		& \Z{}2		& 0.3\Z
		& 12		& 20					\\
ze98		& \multicolumn{2}{c}{no low-noise selection}
		& \multicolumn{2}{c}{--- no tail fit ---}		\\
\hline % --------------------------------------------------------------
circ-ze4	& \multicolumn{2}{c}{no low-noise selection}
		& 12		& 20					\\
circ-ze9	& \multicolumn{2}{c}{no low-noise selection}
		& 12		& 20					\\
circ-zze9	& \multicolumn{2}{c}{no low-noise selection}
		& 12		& 20					%%%\\
\end{tabular}
\end{ruledtabular}
\end{center}
\vspace{-1ex}
\caption{%%%
\label{tab:configs/low-noise-and-tail-fit-pars}
	This table shows the low-noise--selection and tail-fit
	parameters used for computing the self-force for each
	configuration presented in this paper.
	The circ-ze98 configuration is omitted
	because we were unable to obtain stable evolutions
	for it for $m \ge 6$.
	}%%%
\end{table}
%%%%%%%%%%%%%%%%%%%%

%%%%%%%%%%%%%%%%%%%%
\begin{table}[bp]
\begin{center}
\begin{ruledtabular}
%%              123456
\begin{tabular}{lccccc}
	& \multicolumn{5}{c}{Numerical grid}
									\\
%%%%%%%%%%
\cline{2-6}
	& dro4-32	& dro6-48	& dro6-48	& dro8-64
							& dro10-80
									\\
%%%%%%%%%%
Name	& normal	& normal	& variant	& normal
							& normal
									\\
\hline % --------------------------------------------------------------
ns5	& \checkmark	& \checkmark	&		& \checkmark
							&
									\\
n-55	& \checkmark	& \checkmark	&		&
							&
									\\
n95	& \checkmark	& \checkmark	&		& \checkmark
							&
									\\
\hline % --------------------------------------------------------------
e8	& \checkmark	& \checkmark	&		& \checkmark
							&
									\\
e8b	& \checkmark	&\checkmark	&		&
							&
									\\
e9	& \checkmark	& \checkmark	& \checkmark	& \checkmark
							& \checkmark%%%
\rlap{\footnote{$m \le 15$ only}}
									\\
e95	& \checkmark	& \checkmark	&		& \checkmark%%%
\rlap{\footnote{$m=0$, $1$, and $2$ only}}
							&
									\\
\hline % --------------------------------------------------------------
ze4	& \checkmark	& \checkmark	&		&
							&
									\\
ze9	& \checkmark	& \checkmark	&		&
							&
									\\
zze9	& \checkmark	& \checkmark	&		& \checkmark
							&
									\\
ze98	& \checkmark	& \checkmark	&		& \checkmark
							& \checkmark
									\\
\hline % --------------------------------------------------------------
circ-ze4
	& \checkmark	&		&		& \checkmark
							&
									\\
circ-ze9
	& \checkmark	&		&		& \checkmark
							&
									\\
circ-zze9
	& \checkmark	&		&		& \checkmark
							&
									\\
circ-ze98
	& \checkmark	&		&		& \checkmark
							&
									\\
									%%%\\
\end{tabular}
\end{ruledtabular}
\end{center}
\vspace{-1ex}
\caption{%%%
\label{tab:configs/which-grids-used}
	This table shows which numerical grids were used in
	simulating the configurations presented in this paper.
	See tables~\protect\ref{tab:grids/resolutions}
	and~\protect\ref{tab:grids/reflevel-sizes-and-shapes}
	for details of these grids.
	}%%%
\end{table}
%%%%%%%%%%%%%%%%%%%%

%%%%%%%%%%%%%%%%%%%%
\begin{table}[bp]
\begin{center}
\begin{ruledtabular}
%               grid name
%                normal|variant
%                 base (rl=0)
%                   finest (rl=3)
%                 RtRt
%               123456
\begin{tabular}{llcccc}
	&
	& \multicolumn{2}{c}{base grid}
	& \multicolumn{2}{c}{finest grid}
									\\
\cline{3-4} \cline{5-6}
	&
	& $R_*$		& $\theta$ & $R_*$		& $\theta$
									\\
	&
	& ($M$)		& (radians)	& ($M$)		& (radians)
									\\
\hline %---------------------------------------------------------------
dro4-32	& normal
	& $1/4$		& $\pi/72\Z$	& $1/32$	& $\pi/576\Z$	\\
dro6-48	& normal
	& $1/6$		& $\pi/108$	& $1/48$	& $\pi/864\Z$	\\
dro6-48	& variant
	& $1/6$		& $\pi/96\Z$	& $1/48$	& $\pi/768\Z$	\\
dro8-64	& normal
	& $1/8$		& $\pi/144$	& $1/64$	& $\pi/1152$	\\
dro10-80
	& normal
	& $1/1\rlap{0}$	& $\pi/180$	& $1/80$	& $\pi/1440$	\\
									%%%\\
%%%%%%%%%%
\end{tabular}
\end{ruledtabular}
\end{center}
\vspace{-1ex}
\caption{%%%
\label{tab:grids/resolutions}
	This table shows the range of grid resolutions used
	for each of our standard grid structures.  Each grid
	structure has a base grid and 3~refined grids, with
	a 2:1~refinement ratio between adjacent refinement levels.
	See table~\protect\ref{tab:grids/reflevel-sizes-and-shapes}
	for the sizes and shapes of each refinement level.
	}%%%
\end{table}
%%%%%%%%%%%%%%%%%%%%

%%%%%%%%%%%%%%%%%%%%
\begin{table*}[bp]
\begin{center}
\begin{ruledtabular}
%               grid type
%                rl
%                 moves with worldtube?
%                  xxxx grid extent (normal)
%               1234567
\begin{tabular}{lcccccc}
	&
	&
	& \multicolumn{2}{c}{$R_*$}	& \multicolumn{2}{c}{$\theta$}
									\\
\cline{4-5} \cline{6-7}
	& refinement
		& moves with
	& min				& max
	& min				& max
									\\
grid type
	& level	& worldtube?
	&		&		& (radians)	& (radians)
									\\
\hline %---------------------------------------------------------------
normal	& 0	& no
	& $R_*^h$			& $R_*^\Scriplus$
	& $0$				& $\pi/2$
									\\
	& 1	& yes
	& $\WTcenter - 30\,M$		& $\WTcenter + 30\,M$
	& $0$				& $\pi/2$
									\\
	& 2	& yes
	& $\WTcenter - 15\,M$		& $\WTcenter + 15\,M$
	& $\pi/4$			& $\pi/2$
									\\
	& 3	& yes
	& $\WTcenter - \Z{}8\,M$	& $\WTcenter + \Z{}8\,M$
	& $\pi/3$			& $\pi/2$
									\\
\hline %---------------------------------------------------------------
variant	& 0	& no
	& $R_*^h$			& $R_*^\Scriplus$
	& $0$				& $\pi/2$
									\\
	& 1	& yes
	& $\WTcenter - 35\P{.0}\,M$	& $\WTcenter + 40\P{.0}\,M$
	& $0$				& $\pi/2$
									\\
	& 2	& yes
	& $\WTcenter - 18\P{.0}\,M$	& $\WTcenter + 18\P{.0}\,M$
	& $5\pi/24$			& $\pi/2$
									\\
	& 3	& yes
	& $\WTcenter - \Z{}6.5\,M$	& $\WTcenter + \Z{}7.5\,M$
	& $5\pi/16$			& $\pi/2$
									%%%\\
%%%%%%%%%%
\end{tabular}
\end{ruledtabular}
\end{center}
\vspace{-1ex}
\caption{%%%
\label{tab:grids/reflevel-sizes-and-shapes}
	This table shows the size and shape of each refinement level
	in our numerical grids.
	\WTcenter{} is the $R_*$ coordinate of the worldtube center.
	See table~\protect\ref{tab:grids/resolutions} for the grid resolutions.
	}%%%
\end{table*}
%%%%%%%%%%%%%%%%%%%%

%%%%%%%%%%%%%%%%%%%%
\begin{table*}[bp]
\begin{center}
\begin{ruledtabular}
\begin{tabular}{@{\quad}ld@{\,}l}
%%%%%%%%%%
\multicolumn{3}{l}{\underline{Initial startup}}				\\
initial time ($t_\initial$)			& 323.825	& $M$	\\
particle $R_*$ at initial time			&  45.016	& $M$	\\
particle apoastron time				& 385.984	& $M$	\\
particle $R_*$ at apoastron			&  45.889	& $M$	\\
time of first worldtube move ($m=2$)		& 448.706	& $M$	\\
particle $R_*$ at time of first worldtube move	&  45.000	& $M$	\\
time interval from initial time to first worldtube move ($m=2$)
						& 124.881	& $M$	\\[1ex]
%%%%%%%%%%
\multicolumn{3}{l}{\underline{Worldtube}}				\\
$R_*$ (radial) radius (\variablename{WT\_radius})
						&   5.0		& $M$	\\
$\theta$ (angular) radius			& \multicolumn{1}{c}{$\pi/8$}
								& radians
									\\
initial value of worldtube center $R_*$ (\WTcenter)
						&  45.5		& $M$	\\
worldtube center $\theta$			& \multicolumn{1}{c}{$\pi/2$}
								& radians
									\\
move worldtube if
$\bigl|\text{particle $R_*$} - \WTcenter\bigr|
	> f_\move \times \variablename{WT\_radius}$,
where $f_\move = \dots$				&		&	\\
\quad{}initial startup				&   0.10	&	\\
\quad{}main evolution				&   0.05	&	\\
when moving worldtube, place new worldtube center ahead of particle $R_*$
						&		&	\\
\quad{}(where ``ahead'' is defined based on sign of particle $R_*$ 3-velocity)
						&		&	\\
\quad{}by $f_\text{ahead} \times f_\move
			  \times \variablename{WT\_radius}$,
       where $f_\text{ahead} = \dots$
						&   0.9		&	\\
$\text{maximum $R_*$ distance to move worldtube at any one time}
	= f_\maxmove \times \variablename{WT\_radius}$,
	where $f_\maxmove = \dots$
						& 0.1		&	\\
minimum time interval between worldtube moves	& 1.0		& $M$	\\[1ex]
%%%%%%%%%%
\multicolumn{3}{l}{\underline{Overall evolution}}			\\
number of worldtube moves per orbit		& 164		&	%%%\\[1ex]
%%%%%%%%%%
\end{tabular}
\end{ruledtabular}
\end{center}
\vspace{-1ex}
\caption{%%%
\label{tab:configs/computational-pars-misc}
	This table summarizes miscellaneous computational parameters
	for the e8 runs.
	}%%%
\end{table*}
%%%%%%%%%%%%%%%%%%%%

%%%%%%%%%%%%%%%%%%%%
\begin{table}[bp]
\begin{center}
\begin{ruledtabular}
\begin{tabular}{lddrrrr}
%%%%%%%%%%
	& \multicolumn{2}{c}{particle motion}
	& \multicolumn{4}{c}{compactification}
									\\
%%%%%%%%%%
\cline{2-3} \cline{4-7}
Name	& \multicolumn{1}{c}{$\min$ $R_*$}
			& \multicolumn{1}{c}{$\max$ $R_*$}
	& \multicolumn{1}{c}{$R_*^h$}
			& \multicolumn{1}{c}{$R_*^-$}
					& \multicolumn{1}{c}{$R_*^+$}
							& \multicolumn{1}{c}{$R_*^\Scriplus$}
									\\
%%%%%%%%%%
	& \multicolumn{1}{c}{($M$)}
			& \multicolumn{1}{c}{($M$)}
	& \multicolumn{1}{c}{($M$)}
			& \multicolumn{1}{c}{($M$)}
					& \multicolumn{1}{c}{($M$)}
							& \multicolumn{1}{c}{($M$)}
									\\
\hline % --------------------------------------------------------------
ns5	& 4.8		& 14.4
	& -70		& -45		& +70		& +95
									\\
n-55	& 8.370		& 24.395
	& -70		& -45		& +75		& +100
									\\
n95	& 8.390		& 24.397
	& -70		& -45		& +75		& +100
									\\
\hline % --------------------------------------------------------------
e8	& 4.884		& 45.889
	& -75		& -50		& +125		& +150
									\\
e8b	& 4.884		& 45.889
	& -75		& -50		& +125		& +150
									\\
e9	& 3.524		& 77.053
	& -75		& -50		& +160		& +185
									\\
e95	& 0.782		& 107.784
	& -75		& -50		& +190		& +215
									\\
\hline % --------------------------------------------------------------
ze4	& 4.756		& 13.085
	& -70		& -45		& +65		& +90
									\\
ze9	& 4.208		& 85.276
	& -75		& -50		& +135		& +160
									\\
zze9	& 4.208		& 85.275
	& -75		& -50		& +135		& +160
									\\
ze98	& -15.227	& 120.000
	& -90		& -65		& +180		& +205
									\\
\hline % --------------------------------------------------------------
circ-ze4
	& 4.756		& 4.756
	& -70		& -45		& +55		& +80
									\\
circ-ze9
	& 4.208		& 4.208
	& -70		& -45		& +55		& +80
									\\
circ-zze9
	& 4.208		& 4.208
	& -70		& -45		& +55		& +80
									\\
circ-ze98
	& -15.227	& -15.227
	& -90		& -65		& +50		& +75
									%%%\\
\end{tabular}
\end{ruledtabular}
\end{center}
\vspace{-1ex}
\caption{%%%
\label{tab:configs/compactification-pars}
	This table summarizes the compactification
	parameters for the configurations presented
	in this paper.
	}%%%
\end{table}
%%%%%%%%%%%%%%%%%%%%

%%%%%%%%%%%%%%%%%%%%%%%%%%%%%%%%%%%%%%%%

\subsection{Example of data analysis}
\label{sect:results/data-analysis-eg}

Here we give an example of the data analysis ``pipeline'' described in
section~\ref{sect:theory/summary-of-cmpt-and-data-analysis}, for the
e8~configuration, which has $(\tilde{a},p,e) = (0.6,8,0.8)$.

\begin{figure*}[p]
\begin{center}
\includegraphics[scale=1.0]{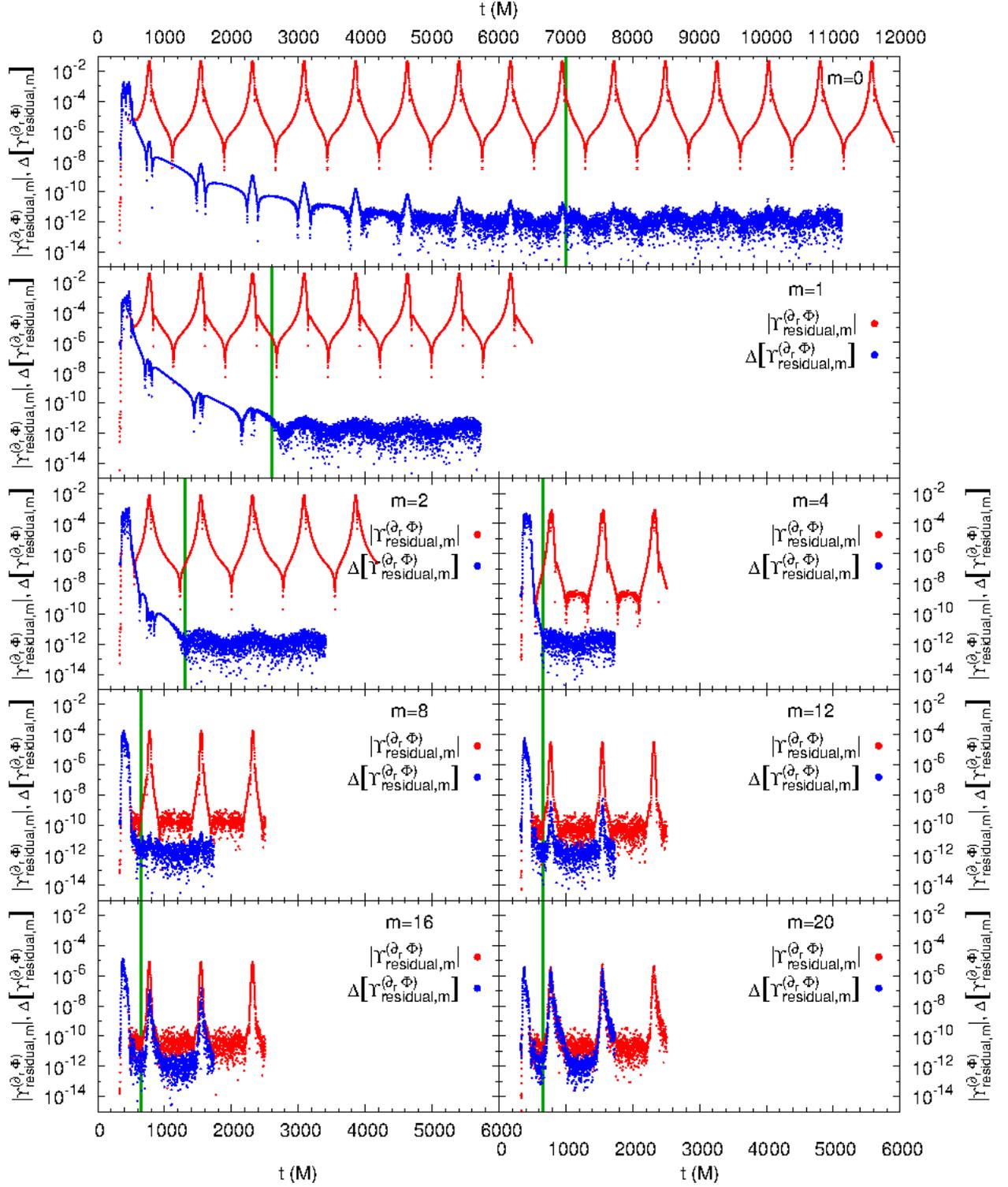}
\end{center}
\vspace{-1ex}
\caption{%%%
\label{fig:e8-modes-and-orbit-diffs-overview}
	This figure shows some of the self-force modes
	$\Upsilon^{(\partial_r \Phi)}_{\residual,m}$ and their
	orbit differences~$\Delta \left[ \Upsilon^{(\partial_r \Phi)}_{\residual,m} \right]$
	for the e8~configuration, which has $(\tilde{a},p,e) = (0.6,8,0.8)$.
	In each subplot the green vertical line marks the self-force
	computation starting time (when the orbit differences
	have decayed to the numerical noise level).  For each~$m$
	the orbit differences are only defined for a time interval
	that is one orbital period shorter than the self-force mode.
	As $m$ increases the initial junk decays faster, so the
	self-force computation starting time can be earlier in the
	evolution.  Correspondingly, we choose shorter numerical
	evolutions for larger~$m$.
	}%%%
\end{figure*}

\begin{figure*}[bp]
\begin{center}
\includegraphics[scale=1.0]{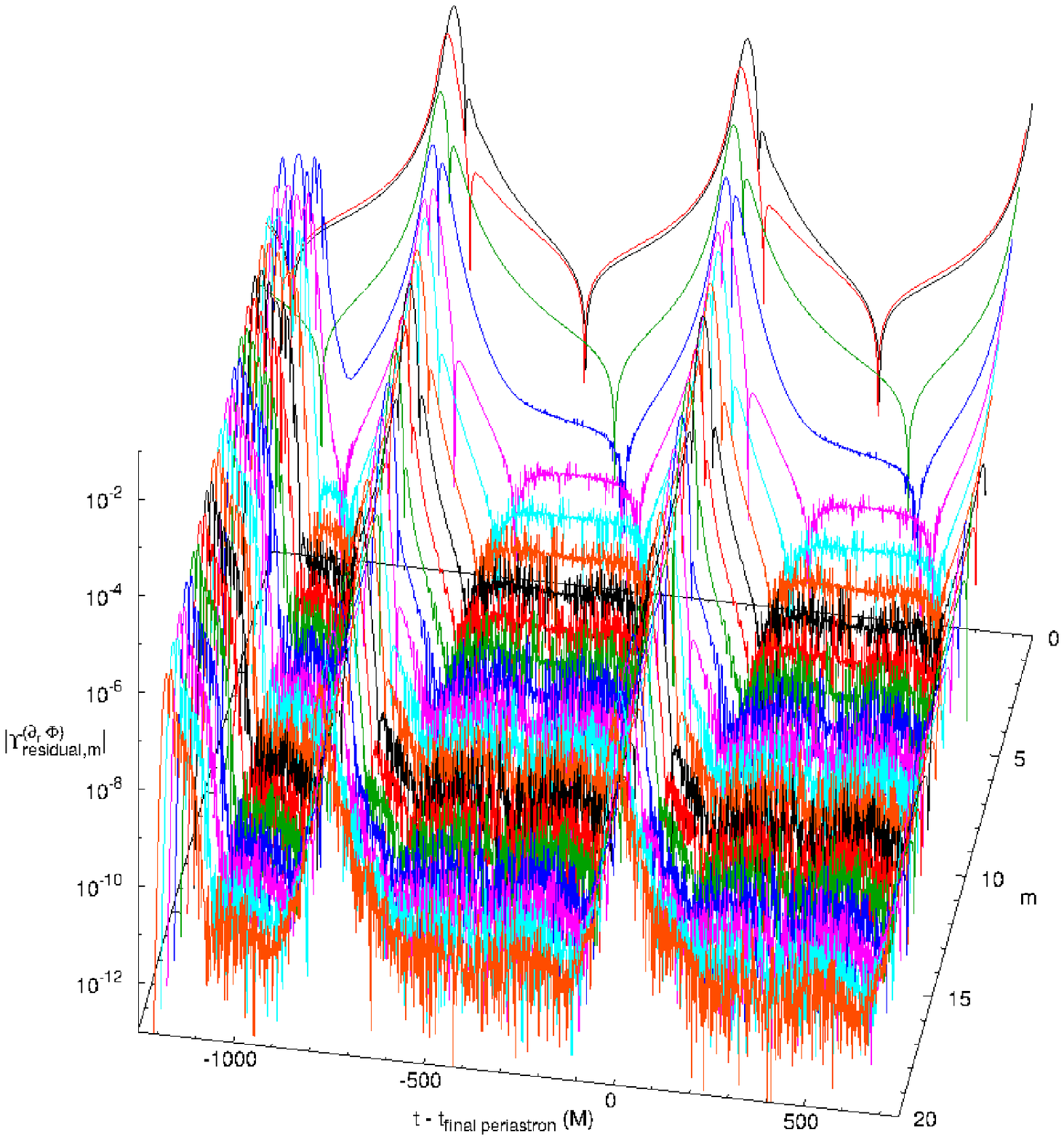}
\end{center}
\vspace{-1ex}
\caption{%%%
\label{fig:e8-aligned-modes-3d}
	This figure shows all of our numerically-computed self-force
	modes~$\Upsilon^{(\partial_r \Phi)}_{\residual,m}$
	for the last $2.6$~orbital periods for each $m \in [0,20]$
	for the e8~configuration, which has $(\tilde{a},p,e) = (0.6,8,0.8)$.
	Compare these with the ``low-noise'' subset of modes
	shown in Fig.~\protect\ref{fig:e8-aligned-smooth-modes-3d}.
	}%%%
\end{figure*}

\begin{figure*}[p]
\begin{center}
\includegraphics[scale=1.0]{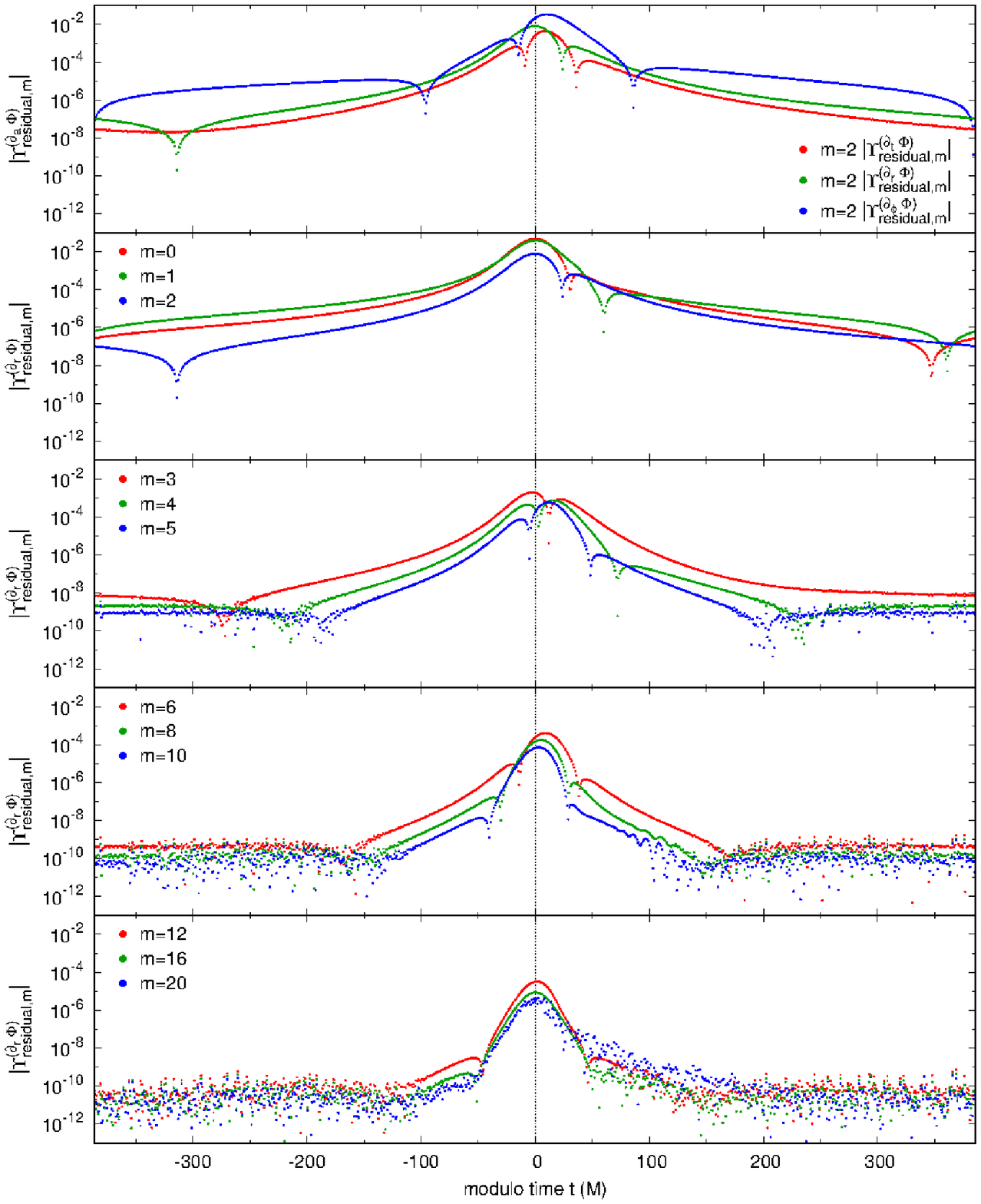}
\end{center}
\vspace{-1ex}
\caption{%%%
\label{fig:e8-modes-overview/final-orbit}
	This figure shows some of the self-force modes
	$\Upsilon^{(\partial_a \Phi)}_{\residual,m}$ for the
	e8~configuration, which has $(\tilde{a},p,e) = (0.6,8,0.8)$.
	The figure shows only data for the final orbit simulated
	for each~$m$.
	Compare these modes with the ``low-noise'' subset of modes
	shown in Fig.~\protect\ref{fig:e8-smooth-modes-overview/final-orbit}.
	}%%%
\end{figure*}

Figure~\ref{fig:e8-modes-and-orbit-diffs-overview} shows a selection
of the modes~$\Upsilon^{(\partial_r \Phi)}_{\residual,m}$ and their orbit
differences~$\Delta \left[ \Upsilon^{(\partial_r \Phi)}_{\residual,m} \right]$
for the entire time span of each $m$'s~evolution.
Figure~\ref{fig:e8-aligned-modes-3d} shows all of the
$\Upsilon^{(\partial_r \Phi)}_{\residual,m}$
for the last $2.85$~orbital periods for each $m \in [0,20]$
for the e8~configuration.
Figure~\ref{fig:e8-modes-overview/final-orbit} shows a selection
of the modes~$\Upsilon^{(\partial_a \Phi)}_{\residual,m}$
in more detail as a function of modulo time.

%%%%%%%%%%%%%%%%%%%%
%%%%%%%%%%%%%%%%%%%%

%%%%%%%%%%%%%%%%%%%%
%%%%%%%%%%%%%%%%%%%%

%%%%%%%%%%%%%%%%%%%%
%%%%%%%%%%%%%%%%%%%%

\begin{figure*}[bp]
\begin{center}
\includegraphics[scale=1.0]{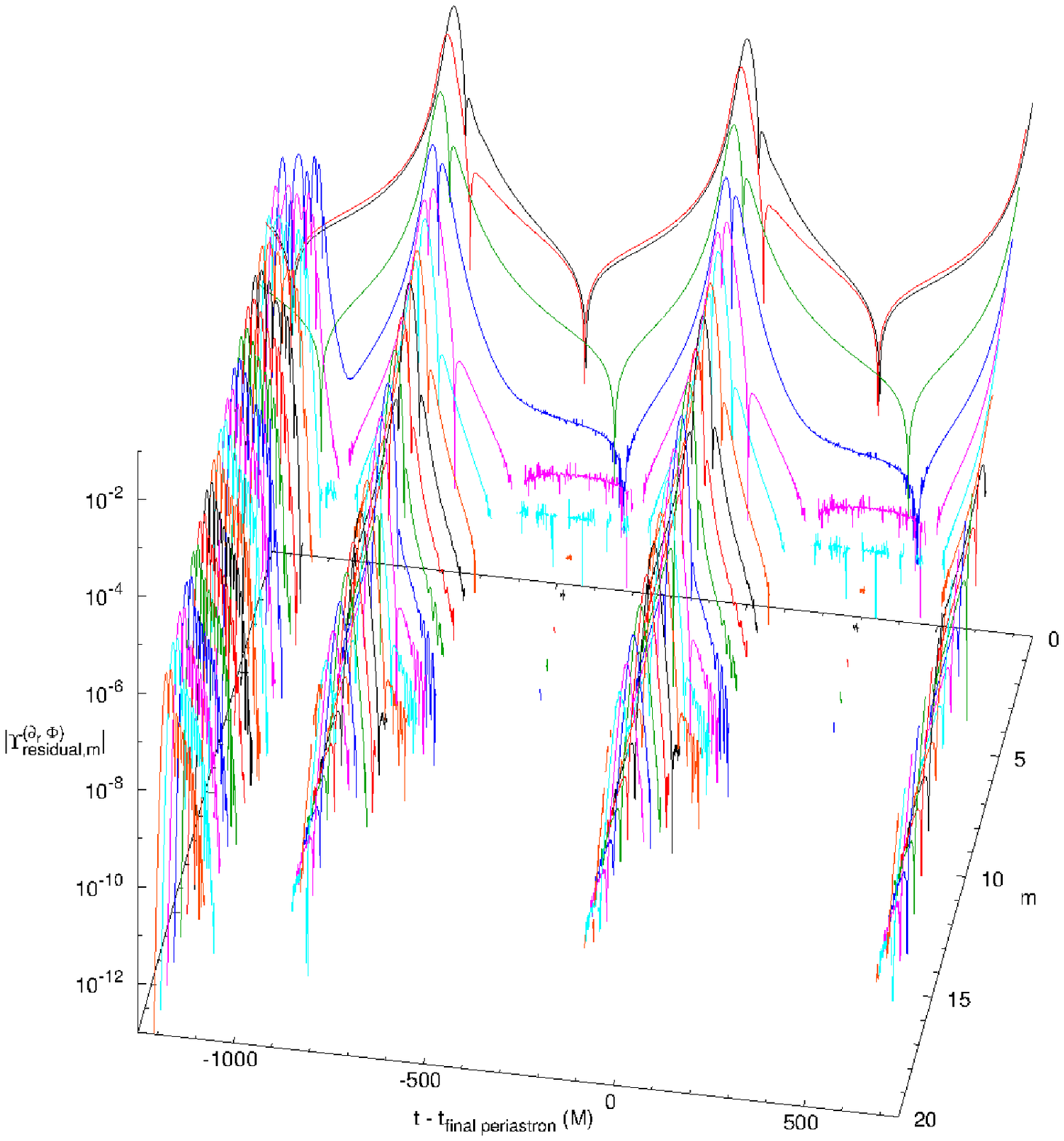}
\end{center}
\vspace{-1ex}
\caption{%%%
\label{fig:e8-aligned-smooth-modes-3d}
	This figure shows the ``low-noise'' numerically-computed self-force
	modes~$\Upsilon^{(\partial_r \Phi)}_{\residual,m}$
	for the last $2.6$~orbital periods for each $m \in [0,20]$
	for the e8~configuration, which has $(\tilde{a},p,e) = (0.6,8,0.8)$.
	Compare these with the full set of modes shown in
	Fig.~\protect\ref{fig:e8-aligned-modes-3d}.
	}%%%
\end{figure*}

\begin{figure*}[p]
\begin{center}
\includegraphics[scale=1.0]{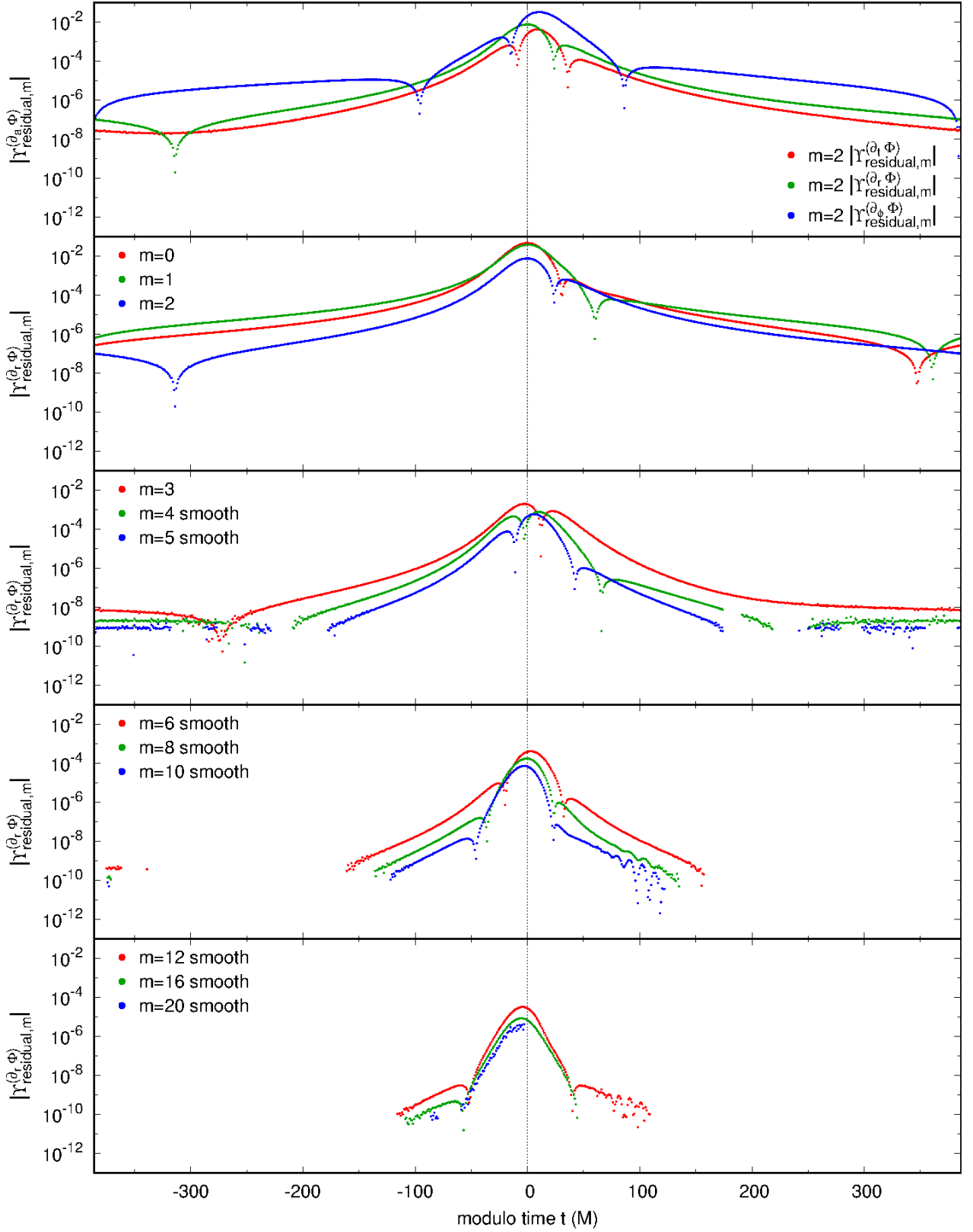}
\end{center}
\vspace{-1ex}
\caption{%%%
\label{fig:e8-smooth-modes-overview/final-orbit}
	This figure shows some of the ``low-noise'' self-force modes
	$\Upsilon^{(\partial_a \Phi)}_{\residual,m}$ for the final orbit
	for each $m$ for the e8~configuration,
	which has $(\tilde{a},p,e) = (0.6,8,0.8)$.
	Compare these modes with the full set of modes shown in
	Fig.~\protect\ref{fig:e8-modes-overview/final-orbit}.
	}%%%
\end{figure*}

After applying the ``low-noise'' selection criteria described in
section~\ref{sect:theory/selecting-low-noise-times},
Fig.~\ref{fig:e8-aligned-smooth-modes-3d} shows the resulting
``low-noise'' subset of the $\Upsilon^{(\partial_r \Phi)}_{\residual,m}$
for the last $2.6$~orbital periods for each $m \in [0,20]$
for the e8~configuration, and
Fig.~\ref{fig:e8-smooth-modes-overview/final-orbit} shows a selection
of the low-noise modes~$\Upsilon^{(\partial_a \Phi)}_{\residual,m}$
in more detail as a function of modulo time.  We use these modes to
compute the self-force using the mode summation and tail-fitting
algorithms described in
sections~\ref{sect:theory/computing-self-force-from-evolved-fields}
and~\ref{sect:theory/tail-series}.

%%%%%%%%%%%%%%%%%%%%
%%%%%%%%%%%%%%%%%%%%

%%%%%%%%%%%%%%%%%%%%
%%%%%%%%%%%%%%%%%%%%

\begin{figure*}[p]
\begin{center}
\includegraphics[scale=1.0]{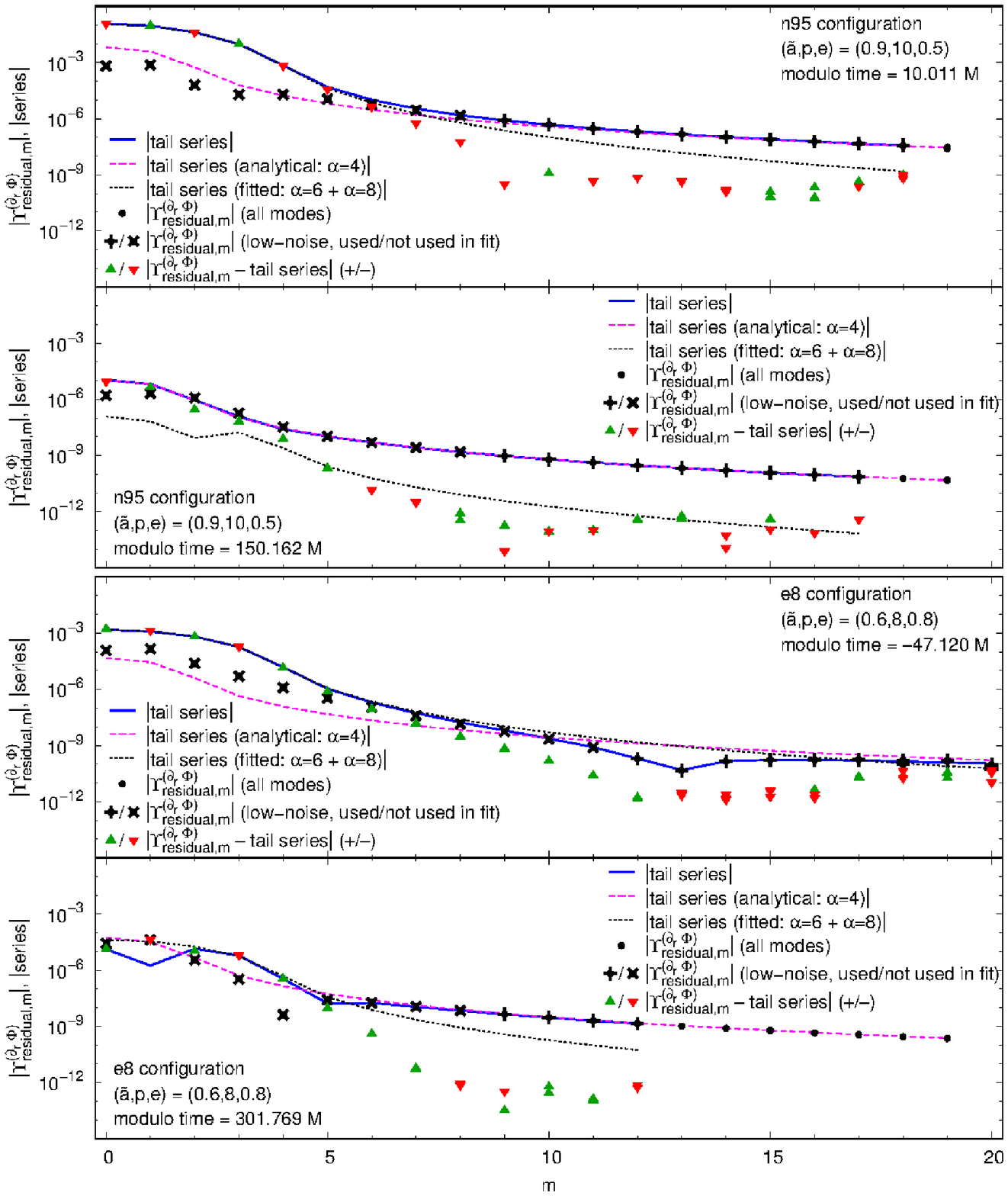}
\end{center}
\vspace{-1ex}
\caption{%%%
\label{fig:n95-e8-tail-fits}
	This figure shows sample fits of the numerically computed
	$\Upsilon^{(\partial_r \Phi)}_{\residual,m}$ to the
	tail series~\protect\eqref{eqn:F=sum-Upsilon+tail-terms}
	for selected times in the n95 and e8~configurations.
	The tail series, fitted tail series, and tail series difference
	$\bigl| \Upsilon^{(\partial_a \Phi)}_{\residual,m}
	 - \text{tail series}\bigr|$
	are only shown for the low-noise modes.
	For this difference, \textcolor{green}{$\blacktriangle$}
	or \textcolor{red}{$\blacktriangledown$} means
	$\Upsilon^{(\partial_a \Phi)}_{\residual,m} - \text{tail series} > 0$
	or~$< 0$, respectively.
	}%%%
\end{figure*}

Figure~\ref{fig:n95-e8-tail-fits} shows some example tail fits
of the low-noise modes~$\Upsilon^{(\partial_r \Phi)}_{\residual,m}$
to the tail series~\eqref{eqn:F=sum-Upsilon+tail-terms} for the
n95 and e8 configurations.

%%%%%%%%%%%%%%%%%%%%
%%%%%%%%%%%%%%%%%%%%

%%%%%%%%%%%%%%%%%%%%%%%%%%%%%%%%%%%%%%%%

\subsection{Convergence of results with numerical resolution}
\label{sect:results/convergence}

When numerically solving partial differential equations, the results
should (must!) converge to a continuum limit.  More precisely (for
finite-difference computations), as the grid is refined, at each event
the results should in general be convergent with the correct convergence
order for the finite differencing scheme~\cite{Choptuik-1991:FD-consistency}.
However, our numerical scheme is an exception: as the particle moves
through the grid, the limited differentiability of our numerical fields
at the particle position introduces finite differencing errors which
fluctuate in a ``bump function'' manner~\cite[appendix~F]{Thornburg-1998}
from one particle position to another.  Moreover, these fluctuations
are typically not coherent between different-resolution evolutions.
Correspondingly, we expect the convergence of our numerical results
to fluctuate from one modulo-time (orbital position) sample to the next.

\begin{figure*}[bp]
\begin{center}
\includegraphics[scale=1.0]{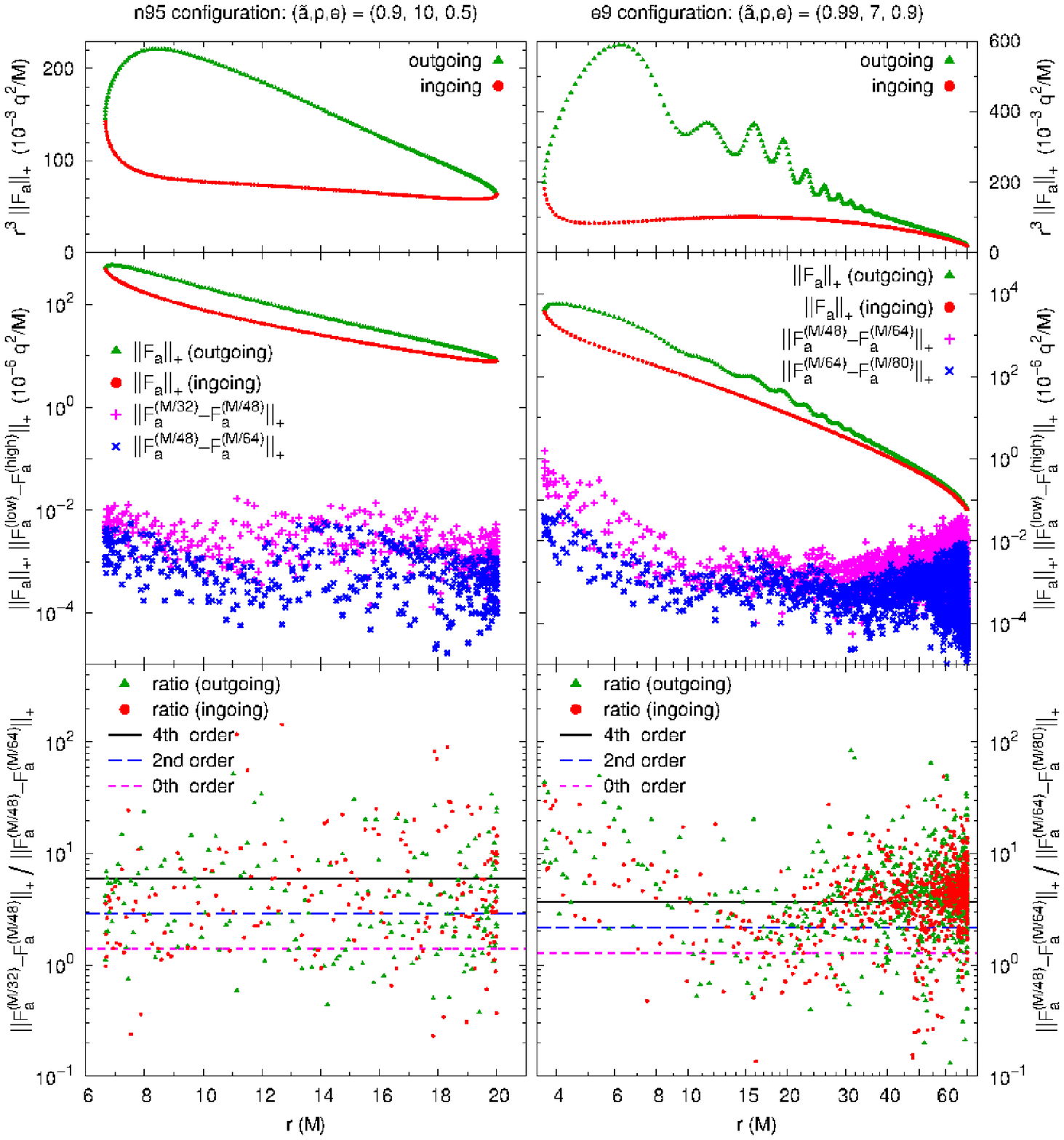}
\end{center}
\vspace{-1ex}
\caption{%%%
\label{fig:n95-e9-convergence}
	This figure shows the convergence of the self-force with
	numerical resolution for the n95 and e9 configurations.
	For the n95 configuration the convergence is calculated
	using the dro4-32, dro6-48, and dro8-64 numerical grids
	(labelled by their finest $\Delta R_*$ of $M/32$, $M/48$,
	and $M/64$ respectively), while
	for the e9 configuration the convergence is calculated
	using the dro6-48, dro8-64, and dro10-80 numerical grids
	(labelled by their finest $\Delta R_*$ of $M/48$, $M/64$,
	and $M/80$ respectively).
	The top subplots in each column shows the configuration's
	self-force loop for the positive-definite pointwise norm
	of the self-force, $\|F_a\|_+$.  The middle subplots
	show the difference norms
	$\bigl\|F_a^{(\text{low})} \,{-}\, F_a^{(\text{medium})}\bigr\|_+$
	and
	$\bigl\|F_a^{(\text{medium})} \,{-}\, F_a^{(\text{high})}\bigr\|_+$.
	The bottom subplots show the convergence ratios
	$\bigl\|F_a^{(\text{low})} \,{-}\, F_a^{(\text{medium})}\bigr\|_+ \Big/
	 \bigl\|F_a^{(\text{medium})} \,{-}\, F_a^{(\text{high})}\bigr\|_+$
	(plotted separately for the ingoing and outgoing legs
	of the orbit), along with the theoretical values
	of this ratio for 0th, 2nd, and 4th~order convergence.
	Notice that there is no systematic difference
	in the convergence ratios between the ingoing (red symbols)
	and outgoing (green symbols) legs of the orbit.
	}%%%
\end{figure*}

Figure~\ref{fig:n95-e9-convergence} illustrates this fluctuating
convergence for the n95 and e9 configurations. 
As expected, the self-force difference norms
$\bigl\|F_a^{(\text{low})} \,{-}\, F_a^{(\text{high})}\bigr\|_+$
and the convergence ratio
$\bigl\|F_a^{(\text{low})} \,{-}\, F_a^{(\text{medium})}\bigr\|_+ \Big/
 \bigl\|F_a^{(\text{medium})} \,{-}\, F_a^{(\text{high})}\bigr\|_+$
fluctuate strongly (typically by an order of magnitude or more)
from one sample to the next.  This makes it difficult to accurately
estimate an overall order of convergence.  However, several conclusions
can be drawn:
\begin{itemize}
\item	For both configurations there is no systematic difference
	in the convergence ratio between the ingoing and outgoing legs
	of the orbit at any given radius~$r$.
\item	For the n95 configuration the convergence order is roughly
	similar everywhere in the orbit, averaging somewhat better
	than 2nd~order.
\item	For the e9 configuration the convergence averages
	much better than 4th~order for $r \ltsim 10\,M$,
	somewhat worse than 2nd~order for $10\,M \ltsim r \ltsim 20\,M$,
	and roughly~4th order for $r \gtsim 25\,M$.

	We have not yet been able to determine the reason for this
	somewhat peculiar convergence behavior.  However, since our
	overall finite differencing scheme is 4th~order accurate
	(in both space and time) in the bulk, achieving an average
	convergence \emph{higher} than this implies that one or more
	of the (e9) evolutions must have insufficient resolution to
	be in the asymptotic-convergence regime.

	Our grid structure for these evolutions
	(Tab.~\ref{tab:grids/reflevel-sizes-and-shapes})
	moves the finest 3~refinement levels with the worldtube,
	which in turn moves so that its center is always very close
	to the particle.  Thus, if the particle is at a sufficiently
	large radius the strong-field region close to the black hole
	will \emph{not} be covered by the finest grid.
	For example, if the particle is at
	$r \,{=}\, R \,{=}\, 10\,M$ ($R_* = 12.8\,M$)
	then the finest grid extends inward only as far as
	$r \,{=}\, R \,{=}\, 4.4\,M$ ($R_* \,{=}\, 4.8\,M$).
	For phenomena nearer the black hole than this, the local
	grid resolution is lower.  As we discuss in
section~\ref{sect-discussion/possible-improvements/accuracy-and-efficiency},
	an adaptive mesh-refinement scheme might well provide
	improved accuracy -- and convergence -- in this situation.
\end{itemize}

%%%%%%%%%%%%%%%%%%%%
%%%%%%%%%%%%%%%%%%%%

On a more qualitative level, figure~\ref{fig:zoom-whirl-ze4-ze9-zze9}
shows visually that the difference between our highest and 2nd-highest
resolution results is very small for the near-periastron parts of the
ze4, ze9, and zze9 orbits.

%%%%%%%%%%%%%%%%%%%%%%%%%%%%%%%%%%%%%%%%

\subsection{Verification that results are independent
	    of the choice of worldtube and other numerical parameters}
\label{sect:results/verify-results-ne-fn(worldtube-etal)}

\begin{figure*}[bp]
\begin{center}
\includegraphics[scale=1.0]{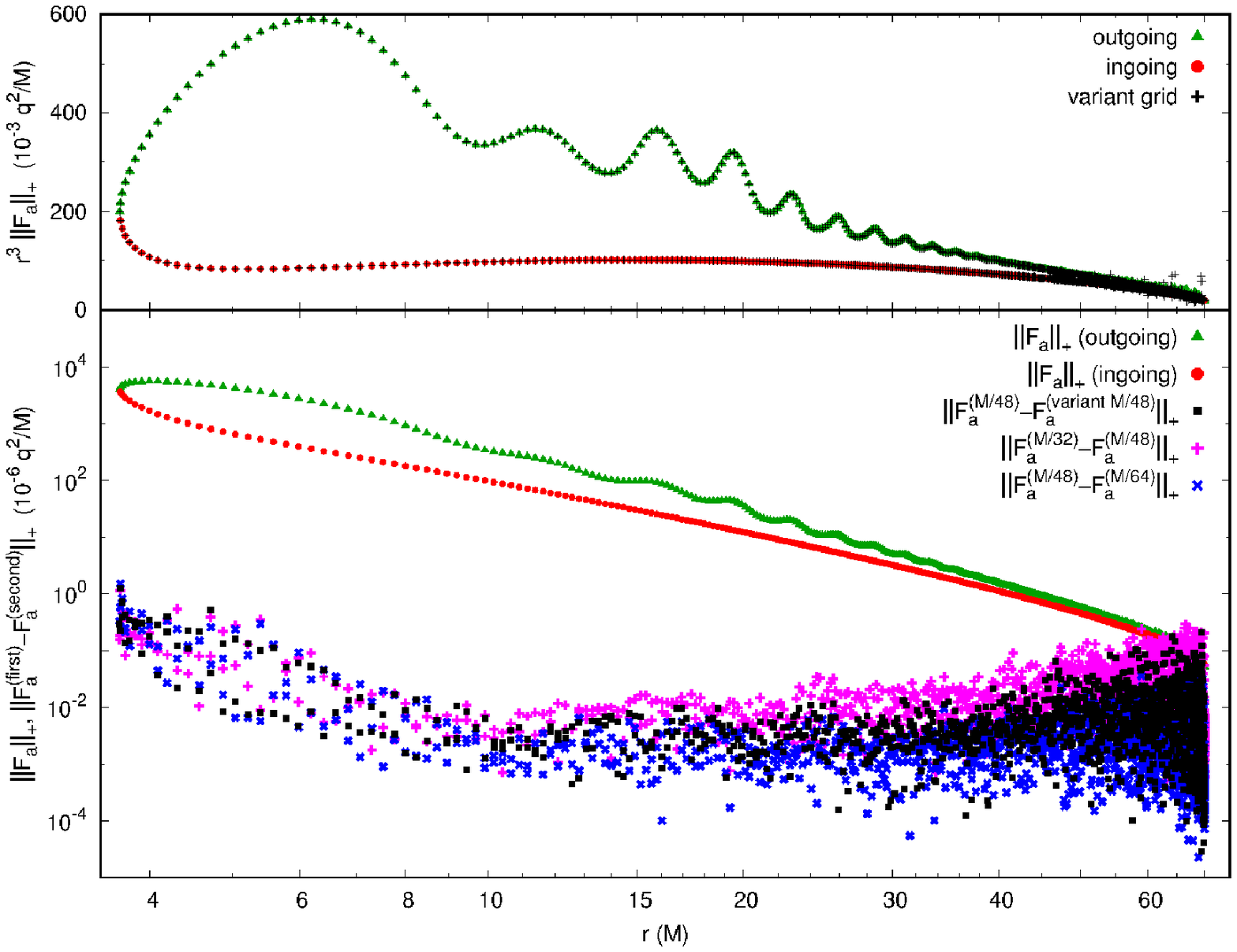}
\end{center}
\vspace{-1ex}
\caption{%%%
\label{fig:e9-variant-vs-normal-grids}
	This figure shows a numerical verification that our
	computed self-force is (approximately) independent of
	the choice of worldtube and other numerical parameters,
	for the e9 configuration
	\big(which has $(\tilde{a},p,e) = (0.99, 7, 0.9)$\big).
  	The top subplot shows the self-force loop for the
	positive-definite pointwise norm of the self-force,
	$\|F_a\|_+$, with computations using the dro6-48 and
	variant dro6-48 worldtube/grids overplotted.
	The points all coincide visually to high accuracy.
	The bottom subplot shows a quantitative assessment,
	the norm of the change in the self-force between the
	dro6-48 and variant dro6-48 worldtube/grid computations,
	$\bigl\|F_a^{(M/48)} \,{-}\, F_a^{(\text{variant $M/48$})}\bigr\|_+$,
	together with the change-in-resolution difference norms
	$\bigl\|F_a^{(M/32)} \,{-}\, F_a^{(M/48)}\bigr\|_+$ and
	$\bigl\|F_a^{(M/48)} \,{-}\, F_a^{(M/64)}\bigr\|_+$ for
	comparison.
	Notice that the variant-grid change in the self-force
	is very small, similar in magnitude to the change-in-resolutions
	change in the self-force.
	}%%%
\end{figure*}

\begingroup
% macros for typesetting algorithms
\newcommand{\assign}{\leftarrow}	% assignment operator
\newcommand{\var}[1]{\texttt{#1}}	% variable name
\newcommand{\kw}[1]{\textbf{#1}}	% keyword

As discussed in section~\ref{sect:theory/worldtube}, our numerically
computed self-force should be independent of the choice of the worldtube.
To test this independence numerically, we compare results for the e9
configuration computed using the normal and variant dro6-48 numerical
grids (these are described in detail in tables~\ref{tab:grids/resolutions}
and~\ref{tab:grids/reflevel-sizes-and-shapes}).  As well as varying
the sizes and positions of each refined grid, these computations also
use different $\Delta R_* \big/ \Delta\theta$ grid aspect ratios
(table~\ref{tab:grids/resolutions}), different worldtube sizes, and
different worldtube-moving parameters $f_\move$ and \var{max\_move\_distance}
(these parameters are defined in figure~\ref{fig:worldtube-move-algorithm}).
Figure~\ref{fig:e9-variant-vs-normal-grids} shows a numerical comparison
of the self-force between these computations.  It is apparent that changing
these parameters changes the computed self-force by only a very small amount
(similar in size to the change induced by a factor-of-$1.5$ change in numerical
resolution).
\endgroup

%%%%%%%%%%%%%%%%%%%%
%%%%%%%%%%%%%%%%%%%%

%%%%%%%%%%%%%%%%%%%%%%%%%%%%%%%%%%%%%%%%

\subsection{Comparison with other researchers' results}
\label{sect:results/cmp-with-other-researchers}

\begin{figure*}[p]
\begin{center}
\includegraphics[scale=1.0]{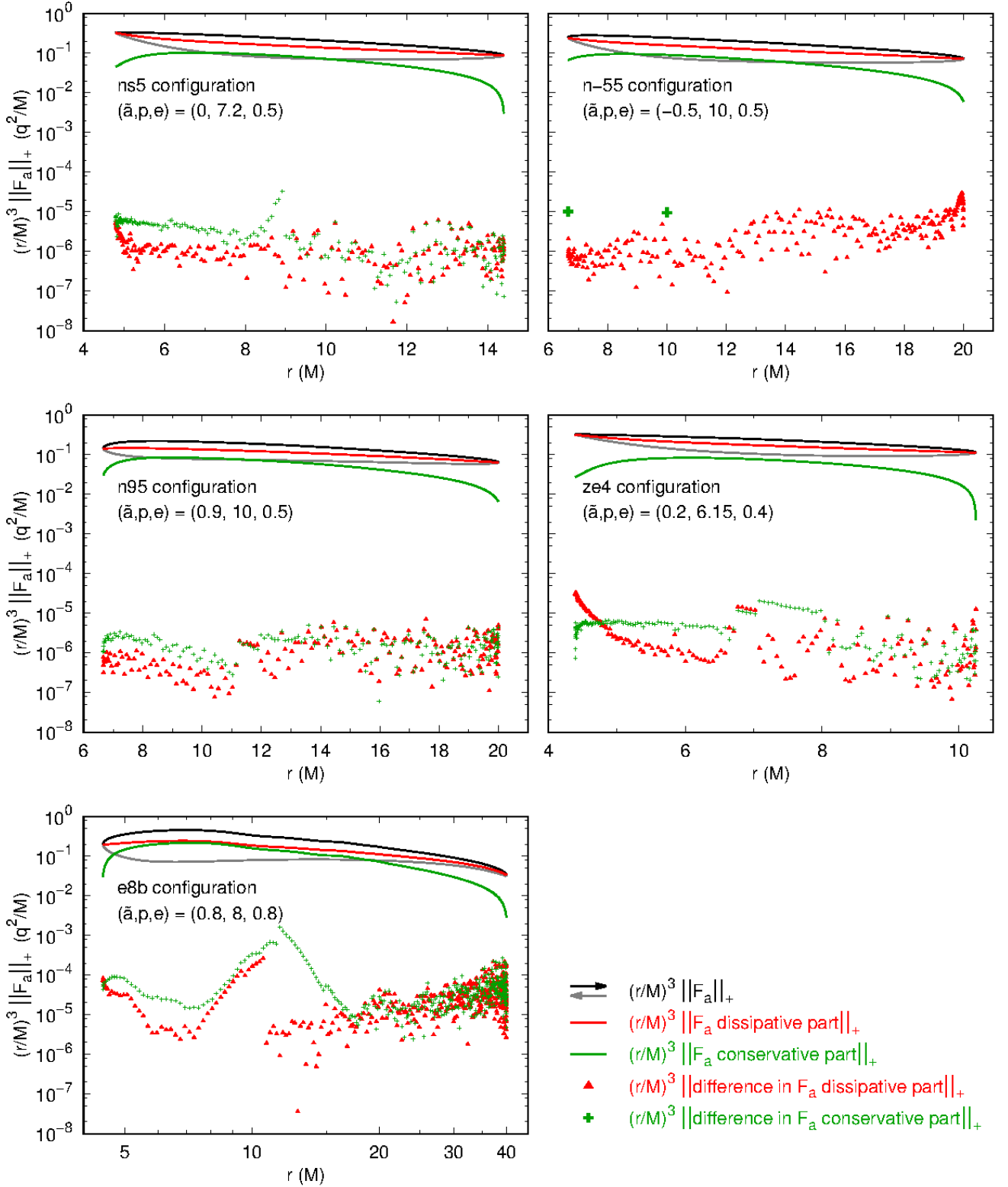}
\end{center}
\vspace{-1ex}
\caption{%%%
\label{fig:self-force-cmp-niels}
	This figure shows the pointwise norms $\| \cdot \|_+$
	of the differences between our computed self-force
	dissipative and conservative parts
	and values computed using Warburton and Barack's
	frequency-domain code~\protect\cite{Warburton-Barack-2011}.
	For the n-55 configuration the conservative part is
	compared with data for $\chi=0$ and $\chi = \pi/2$
	from~\protect\cite[table~II]{Warburton-Barack-2011};
	all other comparisons are with
	unpublished results kindly provided by Warburton.
	}%%%
\end{figure*}

As an external check on the accuracy of our results, we compare these
against results computed using Warburton and Barack's frequency-domain
code~\cite{Warburton-Barack-2011}.
Figure~\ref{fig:self-force-cmp-niels} shows this comparison for the
ns5, n-55, n95, ze4, and e8b configurations.  These span
a considerable range of of black hole spins and particle orbits,
including both prograde and retrograde orbits,
eccentricities ranging up to $e=0.8$ (the e8b configuration),
a zoom-whirl orbit (the ze4 configuration),
and an occurrence of ``wiggles'' (the e8b configuration).

For all but the e8b configuration, the two codes agree everywhere around
the orbit to within approximately one part in $10^5$ (dissipative part)
or one part in $10^4$ (conservative part).  The e8b configuration
has a highly eccentric orbit ($e = 0.8$) that is difficult for the
frequency-domain code to compute accurately, so the somewhat lower
accuracy is expected.  The strong peaks in the e8b difference norms
in the region $8M \ltsim r \ltsim 15M$, and also the similar but less
prominent peaks in the ns5 and ze4 configurations near $r = 9M$ and
$7M \ltsim r \ltsim 8M$ respectively, are probably due to the
frequency-domain code switching between ``inner'' and ``outer''
approximants~\cite{Warburton-pers-comm-2016:FD-inner-outer-approx}.

Overall, the agreement between the two codes is excellent,
particularly given that that they use
different regularizations (effective-source versus mode-sum),
different evolution formulations (time-domain versus frequency-domain),
and were/are independently programmed by disjoint sets of researchers.
This agreement gives quite high confidence that both codes are in
fact computing correct solutions to the $\O(\mu)$-perturbed
scalar-field equations.

%%%%%%%%%%%%%%%%%%%%
%%%%%%%%%%%%%%%%%%%%

%%%%%%%%%%%%%%%%%%%%%%%%%%%%%%%%%%%%%%%%

% *** KLUDGE ***
% prevent latex from crashing with a "too many unprocessed floats" error
%%\JTFIXME{bad page breaks here -- see kludge in .tex file}
%%\clearpage

\subsection{Overview of self-forces}
\label{sect:results/overview-of-self-forces}

\begin{figure*}[p]
\begin{center}
\includegraphics[scale=1.0]{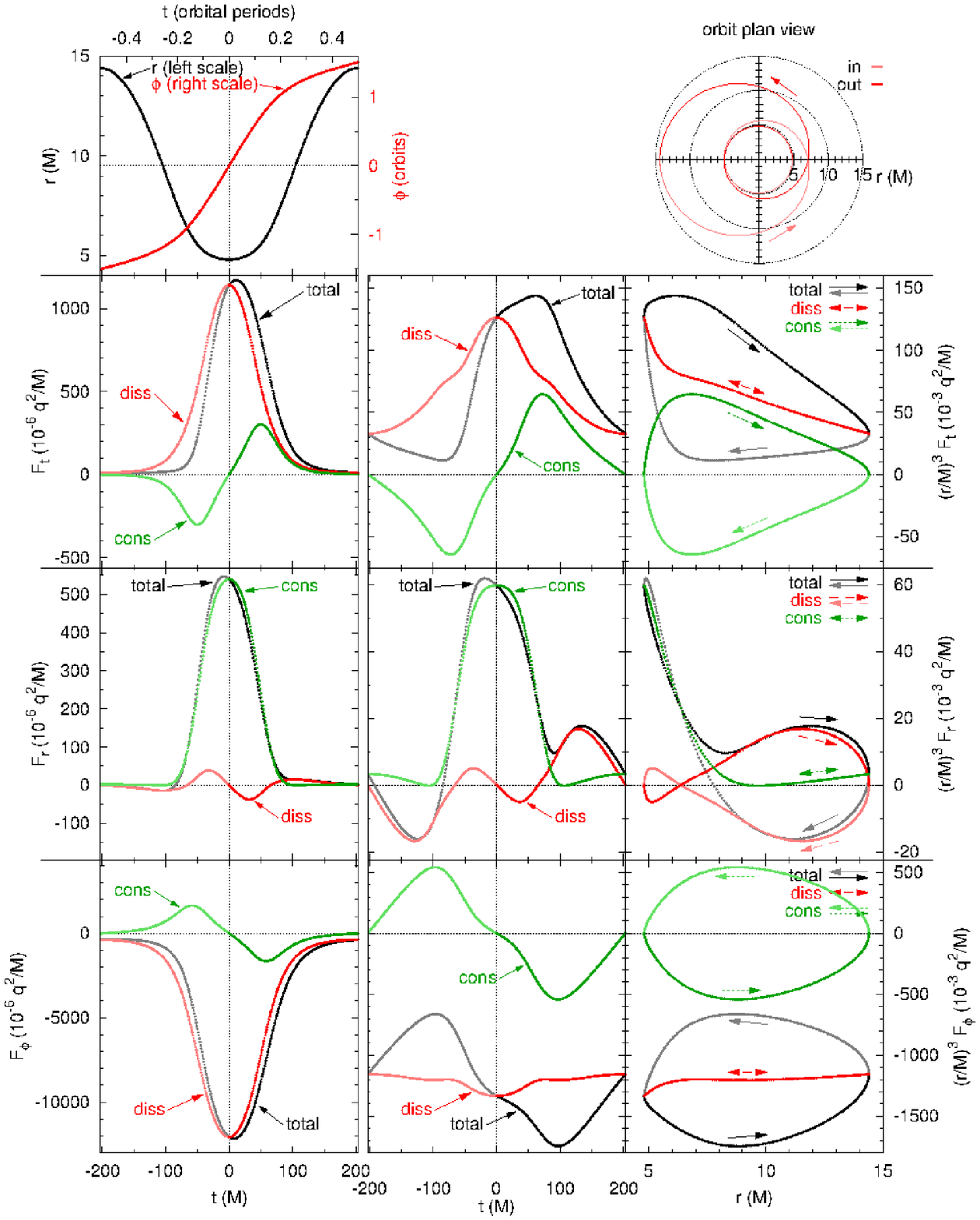}
\end{center}
\vspace{-1ex}
\caption{%%%
\label{fig:ns5-self-force-mosaic}
	This figure shows the self-force for the ns5~configuration,
	which has $(\tilde{a},p,e) = (0,7.2,0.5)$.
	}%%%
\end{figure*}

\begin{figure*}[p]
\begin{center}
\includegraphics[scale=1.0]{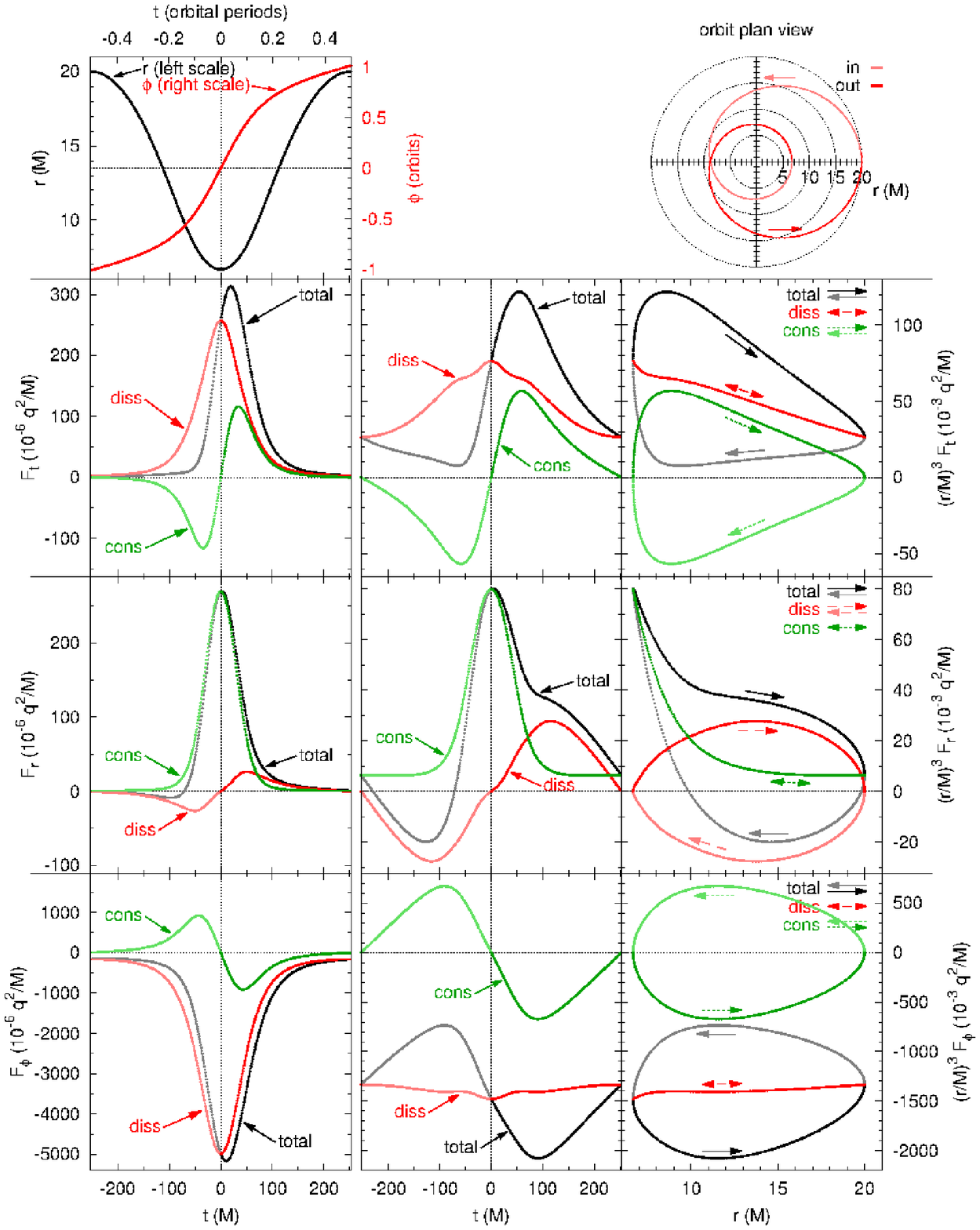}
\end{center}
\vspace{-1ex}
\caption{%%%
\label{fig:n-55-self-force-mosaic}
	This figure shows the self-force for the n-55~configuration,
	which has $(\tilde{a},p,e) = (-0.5,10,0.5)$.
	}%%%
\end{figure*}

\begin{figure*}[p]
\begin{center}
\includegraphics[scale=1.0]{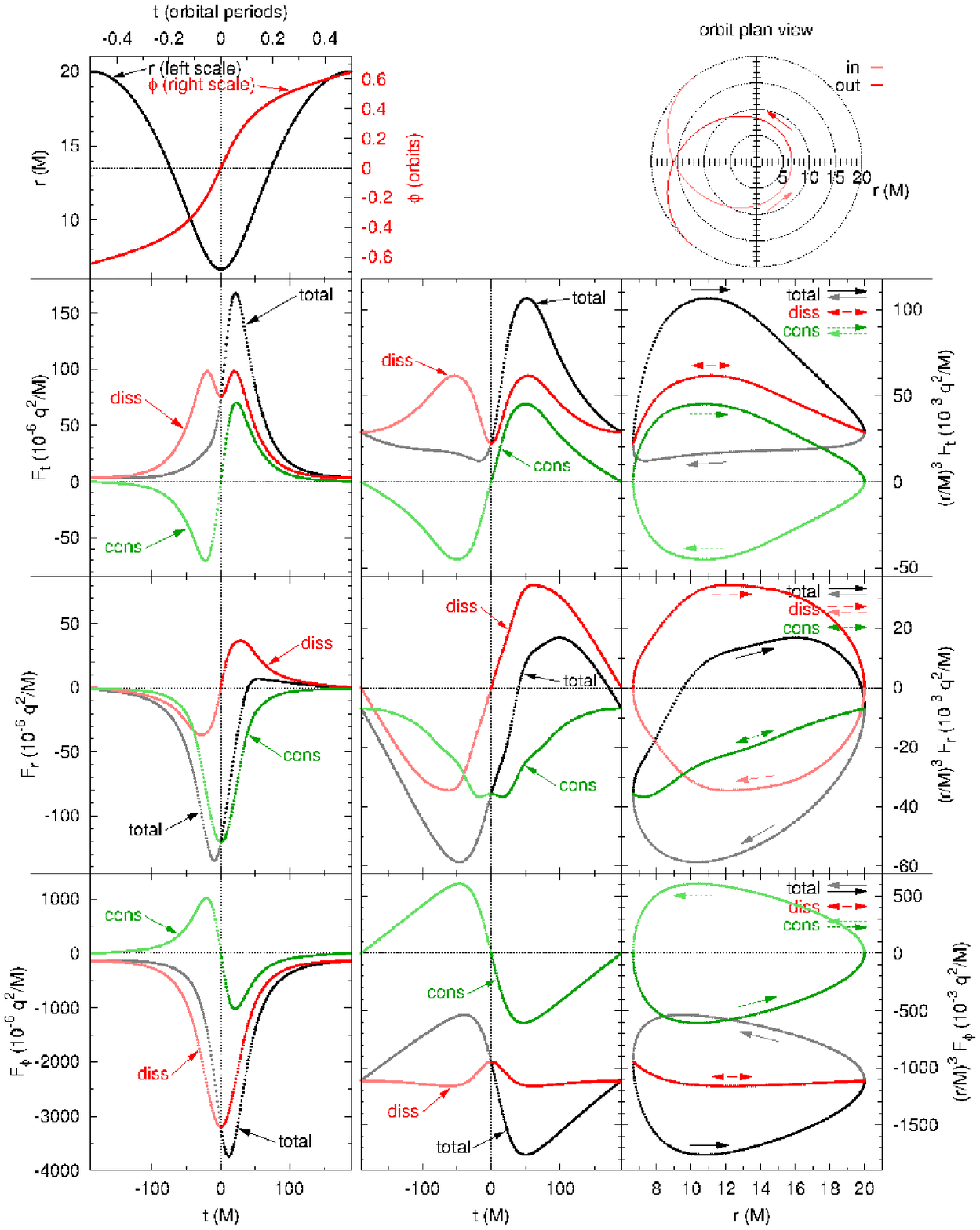}
\end{center}
\vspace{-1ex}
\caption{%%%
\label{fig:n95-self-force-mosaic}
	This figure shows the self-force for the n95~configuration,
	which has $(\tilde{a},p,e) = (0.9,10,0.5)$.
	}%%%
\end{figure*}

\begin{figure*}[p]
\begin{center}
\includegraphics[scale=1.0]{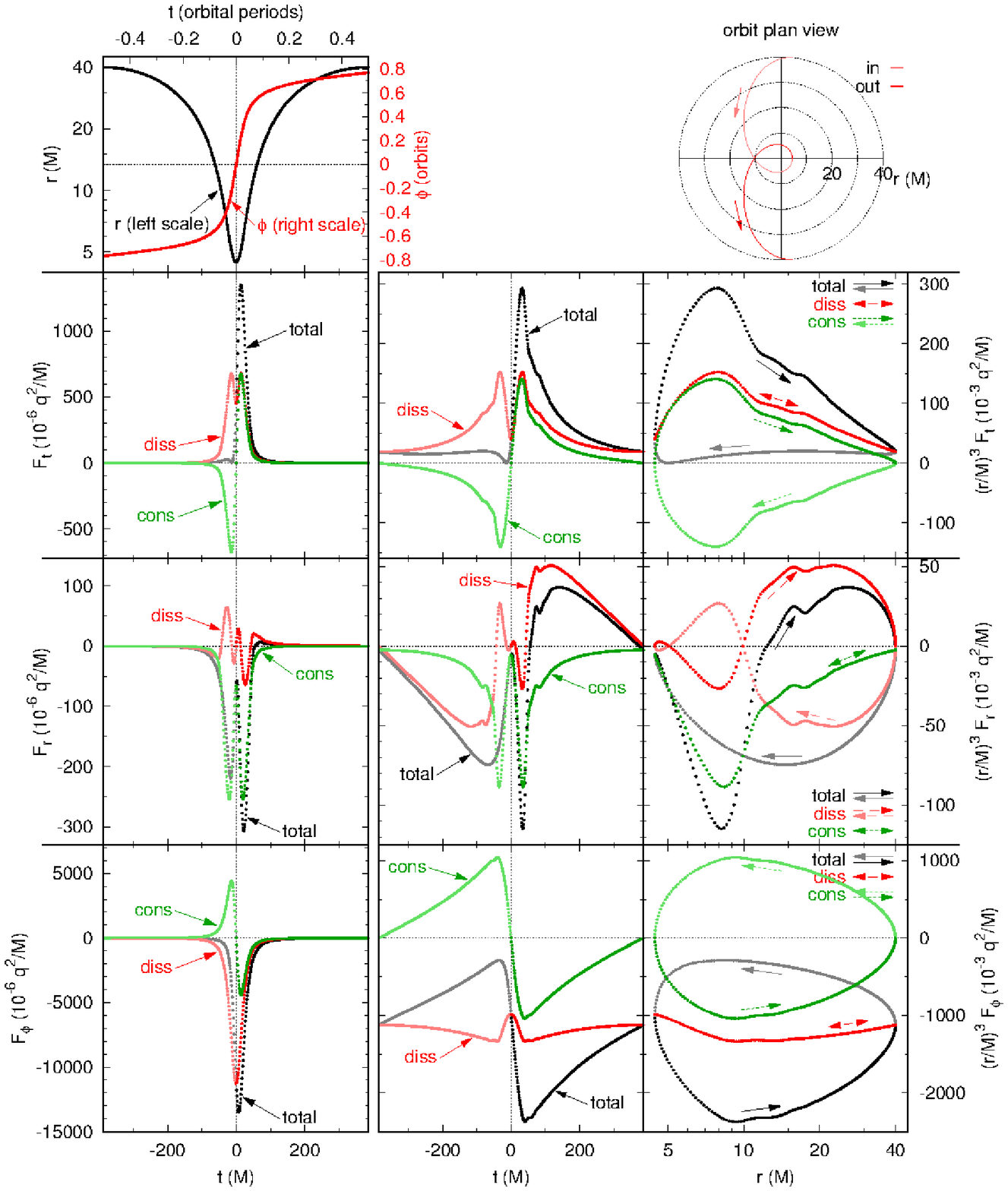}
\end{center}
\vspace{-1ex}
\caption{%%%
\label{fig:e8-self-force-mosaic}
	This figure shows the self-force for the e8~configuration,
	which has $(\tilde{a},p,e) = (0.6,8,0.8)$.
	The self-force loops (right column) are plotted using a
	logarithmic radial scale.
	Notice the wiggle in the self-force on the outgoing leg
	of the orbit, near $t=100\,M$ past periastron,
	at $r \approx 16\,M$; we discuss this in
	section~\protect\ref{sect:results/wiggles}.
	Because the dissipative-conservative
	decomposition~\protect\eqref{eqn:F-diss-cons-parts}
	and~\protect\eqref{eqn:even-odd-parts} is non-local, the
	dissipative and conservative parts of the self-force
	also show wiggles before periastron.
	}%%%
\end{figure*}

\begin{figure*}[p]
\begin{center}
\includegraphics[scale=1.0]{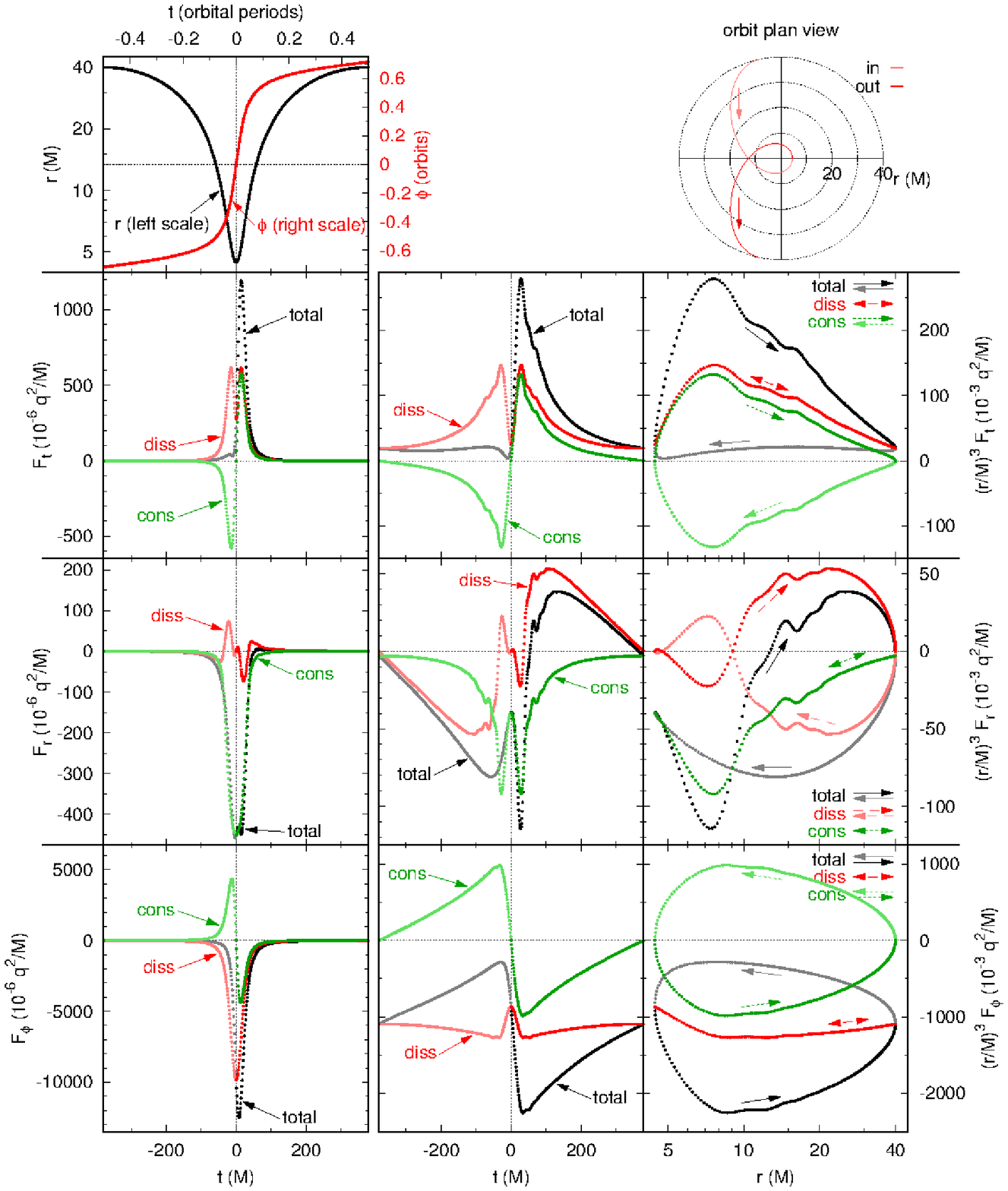}
\end{center}
\vspace{-1ex}
\caption{%%%
\label{fig:e8b-self-force-mosaic}
	This figure shows the self-force for the e8b~configuration,
	which has $(\tilde{a},p,e) = (0.8,8,0.8)$.
	The self-force loops (right column) are plotted using a
	logarithmic radial scale.
	Notice the wiggles in the self-force on the outgoing leg
	of the orbit, between $t \approx 50\,M$ and $100\,M$~past
	periastron, at $r \approx 15\,M$; we discuss this in
	section~\protect\ref{sect:results/wiggles}.
	Because the dissipative-conservative
	decomposition~\protect\eqref{eqn:F-diss-cons-parts}
	and~\protect\eqref{eqn:even-odd-parts} is non-local, the
	dissipative and conservative parts of the self-force
	also show wiggles before periastron.
	}%%%
\end{figure*}

\begin{figure*}[p]
\begin{center}
\includegraphics[scale=1.0]{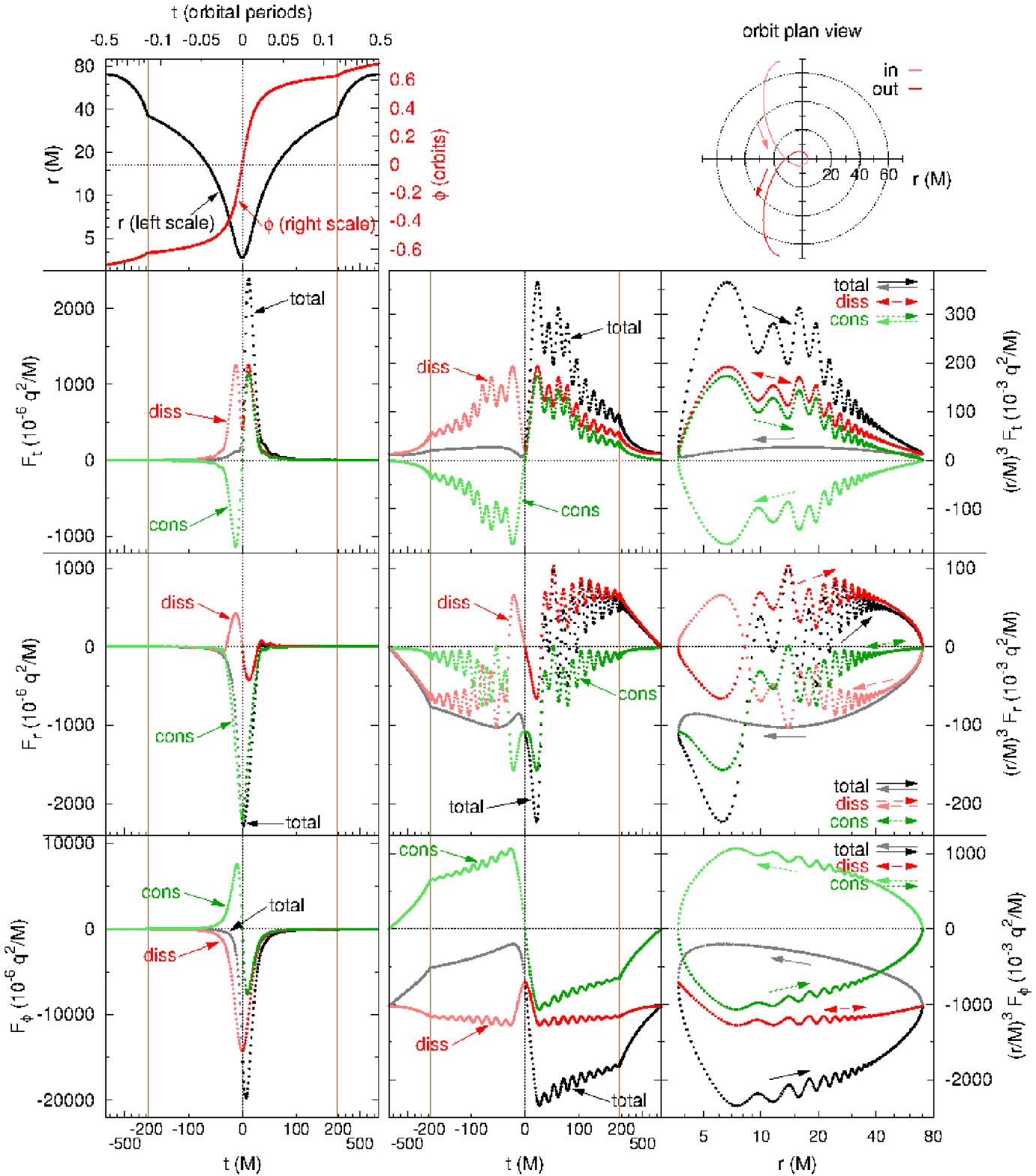}
\end{center}
\vspace{-1ex}
\caption{%%%
\label{fig:e9-self-force-mosaic}
	This figure shows the self-force for the e9~configuration,
	which has $(\tilde{a},p,e) = (0.99,7,0.9)$.
	In the time-domain plots (left and center columns)
	the central $|t| \le 175\,M$ around periastron
	(marked by the vertical lines) is plotted at an expanded
	horizontal scale.
	The self-force loops (right column) are plotted using
	a logarithmic radial scale.
	Notice the many wiggles in the self-force on the
	outgoing leg of the orbit; we discuss these in
	section~\protect\ref{sect:results/wiggles}.
	Because the dissipative-conservative
	decomposition~\protect\eqref{eqn:F-diss-cons-parts}
	and~\protect\eqref{eqn:even-odd-parts} is non-local, the
	dissipative and conservative parts of the self-force
	also show wiggles before periastron.
	}%%%
\end{figure*}

\begin{figure*}[p]
\begin{center}
\includegraphics[scale=1.0]{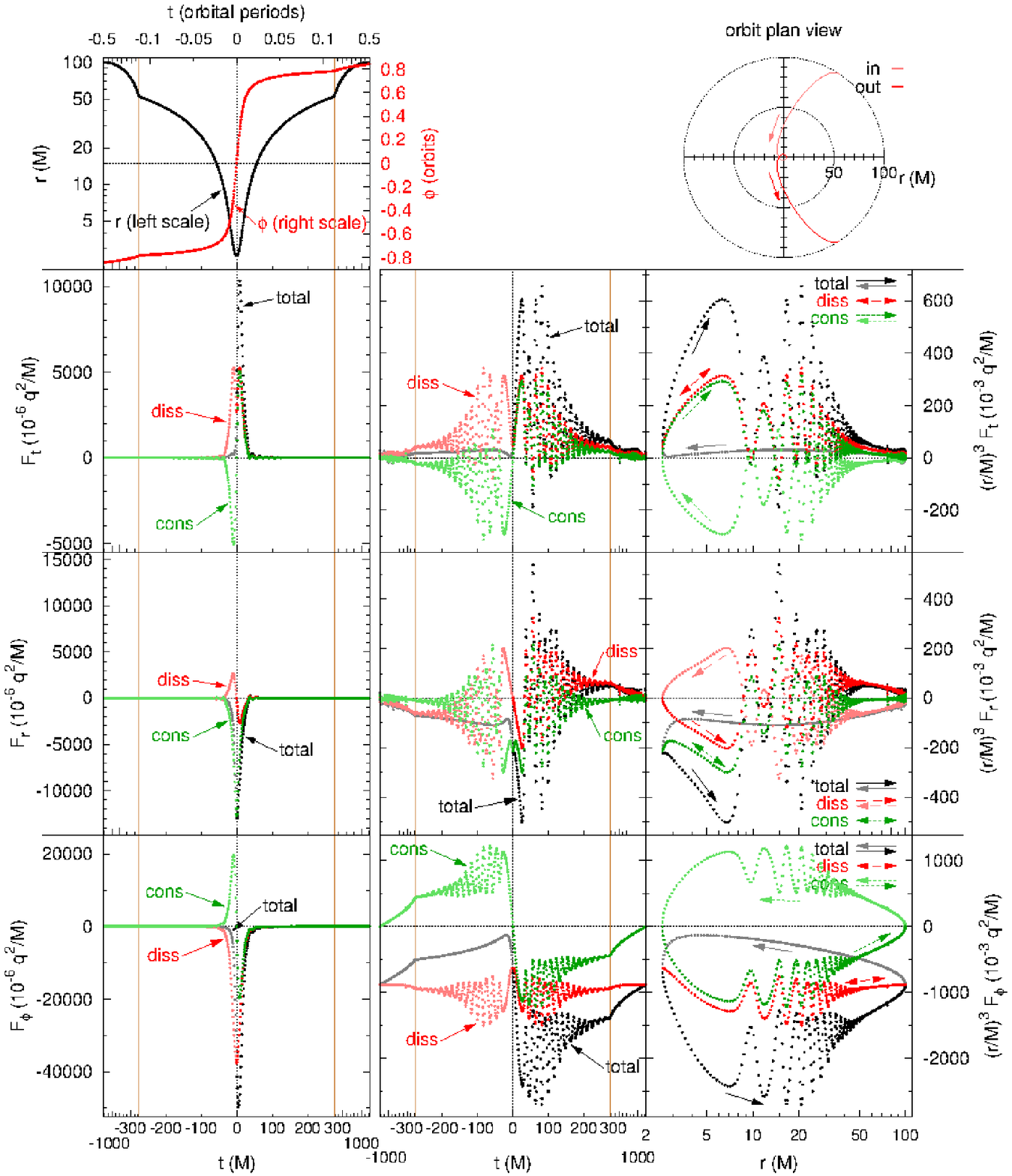}
\end{center}
\vspace{-1ex}
\caption{%%%
\label{fig:e95-self-force-mosaic}
	This figure shows the self-force for the e95~configuration,
	which has $(\tilde{a},p,e) = (0.99,5,0.95)$.
	In the time-domain plots (left and center columns)
	the central $|t| \le 275\,M$
	(marked by the vertical lines) is plotted at an expanded
	horizontal scale.
	The self-force loops (right column) are plotted using
	a logarithmic radial scale.
	Notice the many wiggles in the self-force on the
	outgoing leg of the orbit; we discuss these in
	section~\protect\ref{sect:results/wiggles}.
	Because the dissipative-conservative
	decomposition~\protect\eqref{eqn:F-diss-cons-parts}
	and~\protect\eqref{eqn:even-odd-parts} is non-local, the
	dissipative and conservative parts of the self-force
	also show wiggles before periastron.
	}%%%
\end{figure*}

\begin{figure*}[p]
\begin{center}
\includegraphics[scale=1.0]{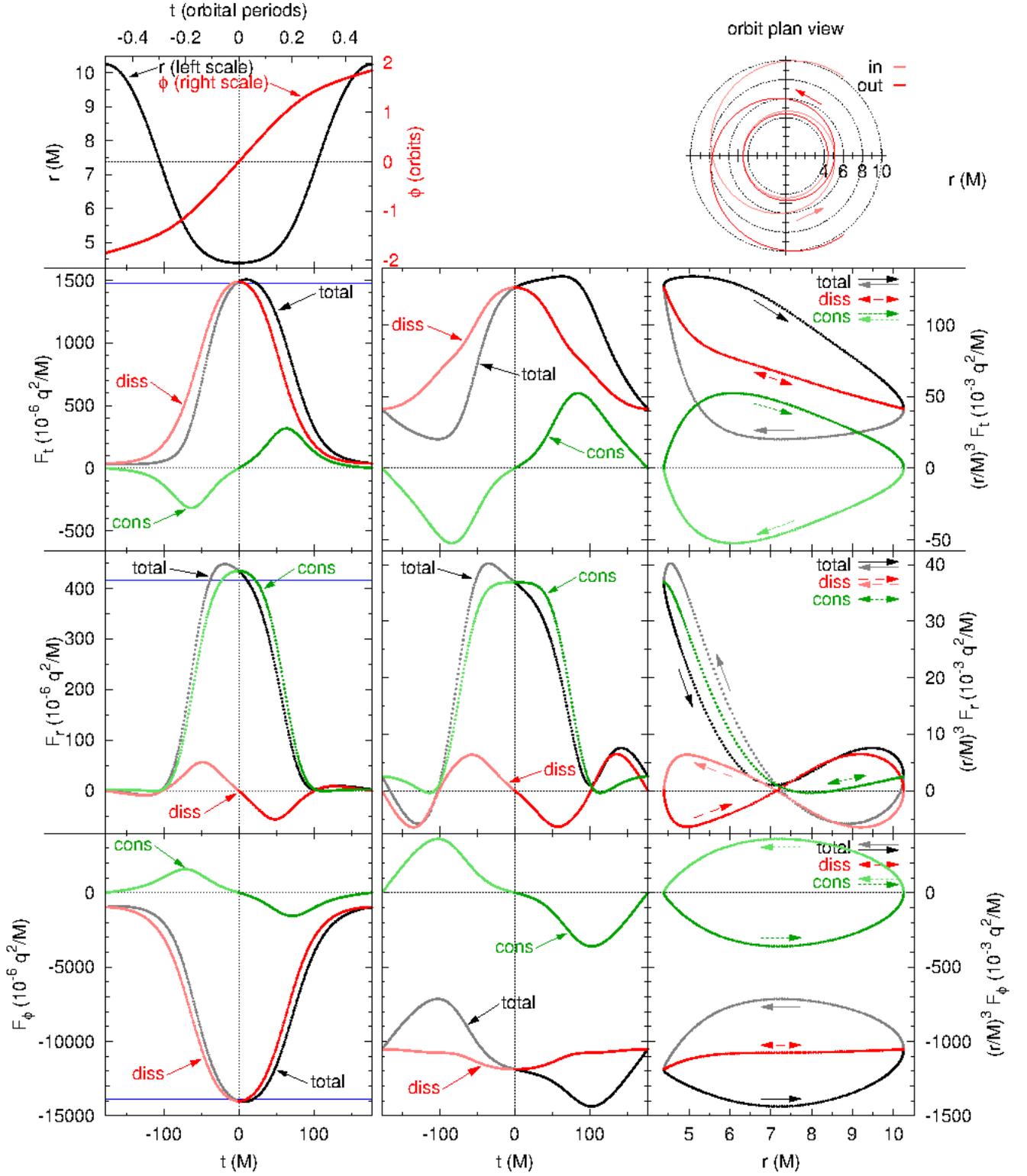}
\end{center}
\vspace{-1ex}
\caption{%%%
\label{fig:ze4-self-force-mosaic}
	This figure shows the self-force for the ze4~configuration,
	which has $(\tilde{a},p,e) = (0.2,6.15,0.4)$.
	This is a mild zoom-whirl orbit; the particle completes
	about~$2$ orbits at $r \approx 4.5\,M$ during the
	approximately~$125\,M$ of the whirl phase.
	In the left column, the horizontal blue line
	in each self-force subplot shows the self-force
	for the circ-ze4 circular-orbit configuration;
	this configuration has the same orbital radius
	as the ze4 configuration's periastron radius.
	The self-force near to and during the whirl phase
	is shown at an expanded scale in
	Fig~\protect\ref{fig:zoom-whirl-ze4-ze9-zze9}.
	}%%%
\end{figure*}

\begin{figure*}[p]
\begin{center}
\includegraphics[scale=1.0]{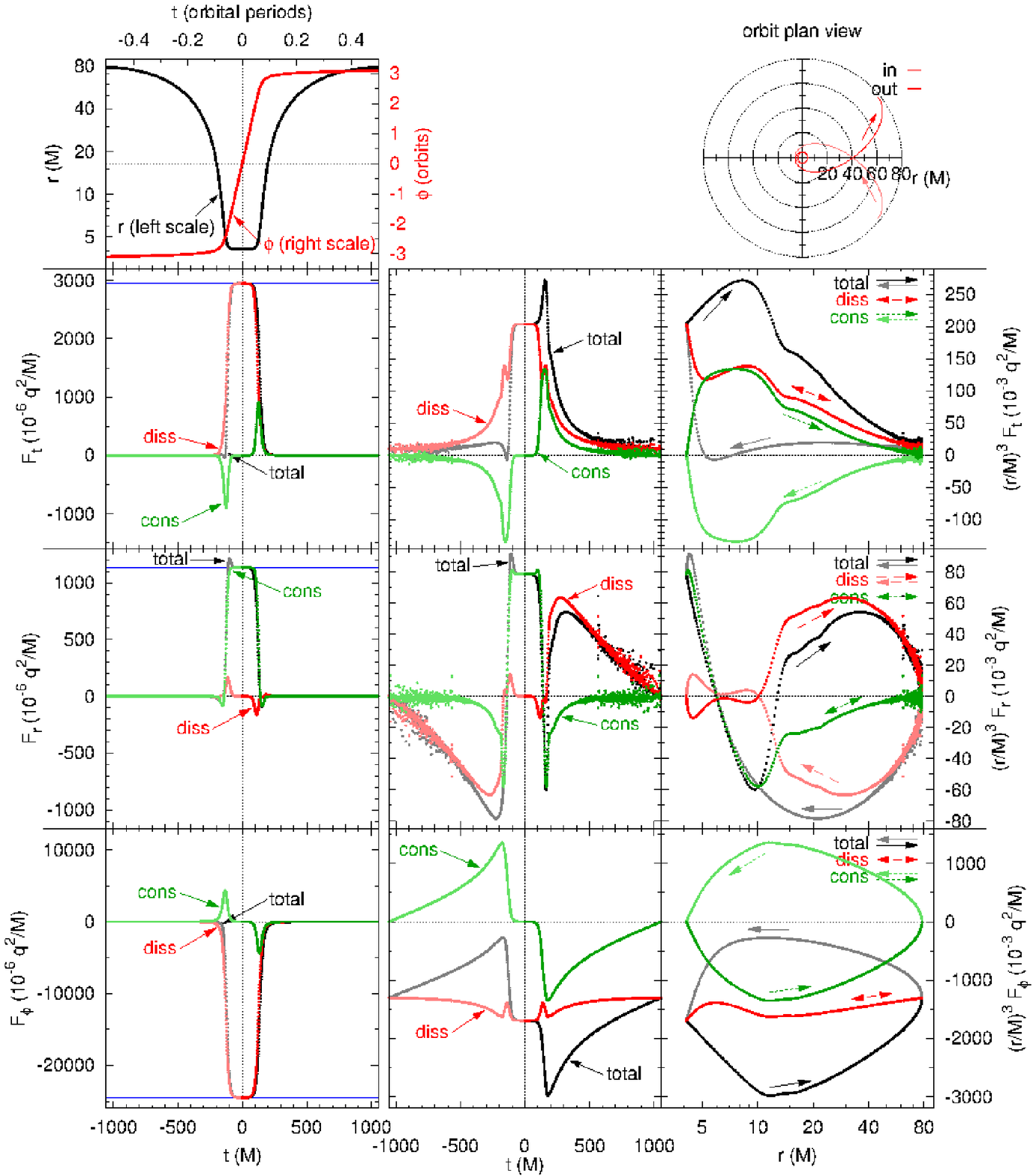}
\end{center}
\vspace{-1ex}
\caption{%%%
\label{fig:ze9-self-force-mosaic}
	This figure shows the self-force for the ze9~configuration,
	which has $(\tilde{a},p,e) = (0.0,7.8001,0.9)$.
	This is a strong zoom-whirl orbit; the particle completes
	about~$5 \tthird$ orbits at $r \approx 4.1\,M$ during the
	approximately~$300\,M$ of the whirl phase.
	During the whirl phase the self-force is large and nearly constant;
	there are also ``spikes'' in $F_r$ at this phase's entry and exit.
	In the left column, the horizontal blue line
	in each self-force subplot shows the self-force
	for the circ-ze9 circular-orbit configuration;
	this configuration has the same orbital radius
	as the ze9 configuration's periastron radius.
	The self-force loops (right column) are plotted using
	a logarithmic radial scale.
	The self-force near to and during the whirl phase
	is shown at an expanded scale in
	Fig.~\protect\ref{fig:zoom-whirl-ze4-ze9-zze9}.
	}%%%
\end{figure*}

\begin{figure*}[p]
\begin{center}
\includegraphics[scale=1.0]{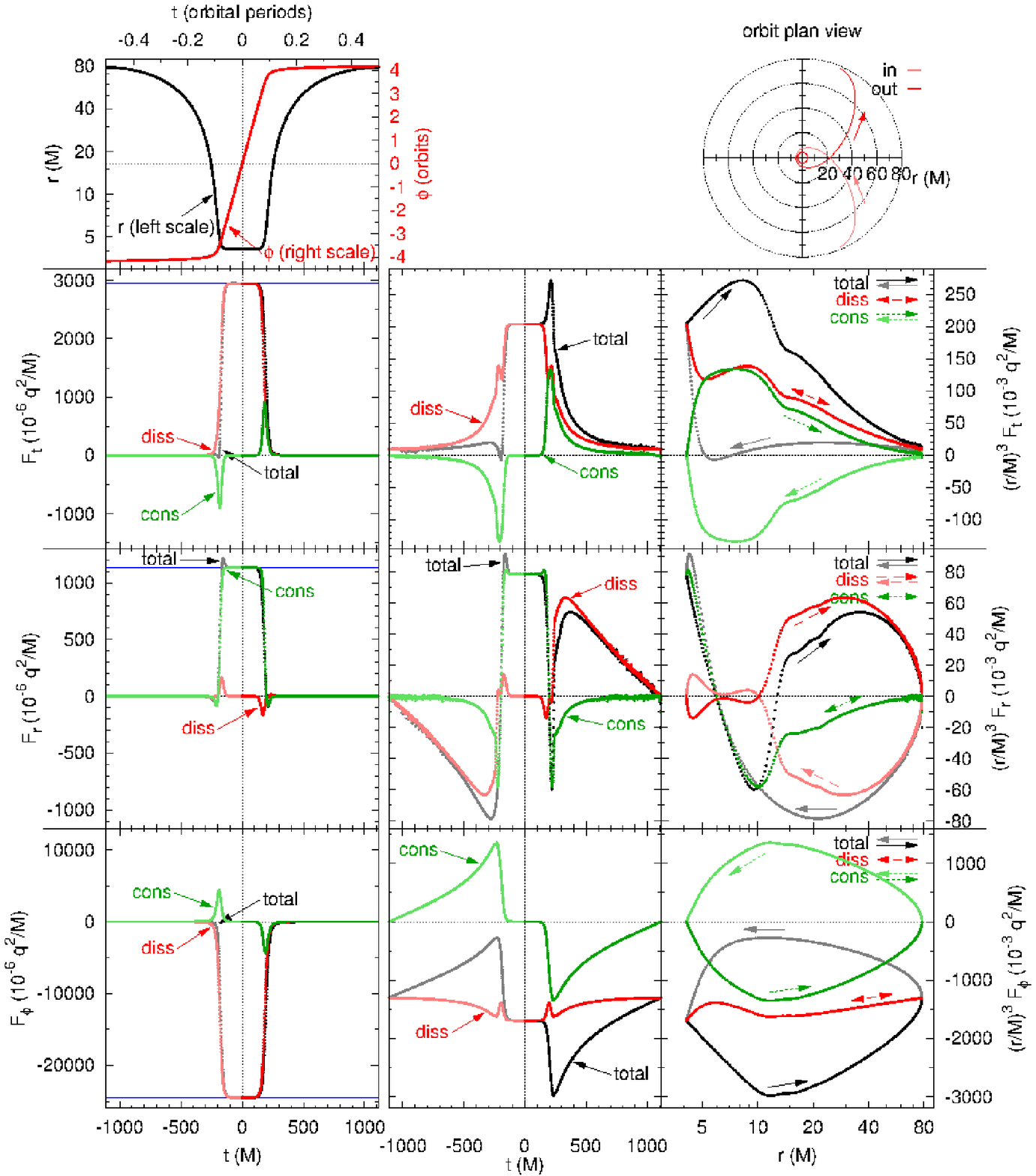}
\end{center}
\vspace{-1ex}
\caption{%%%
\label{fig:zze9-self-force-mosaic}
	This figure shows the self-force for the zze9~configuration,
	which has $(\tilde{a},p,e) = (0.0,7.800\,001,0.9)$.
	This is a very strong zoom-whirl orbit; the particle completes
	about~$7 \tfrac{3}{4}$ orbits at $r \approx 4.1\,M$ during the
	approximately~$450\,M$ of the whirl phase.  
	During the whirl phase the self-force is large and nearly constant;
	there are also ``spikes'' in $F_r$ at this phase's entry and exit.
	In the left column, the horizontal blue line
	in each self-force subplot shows the self-force
	for the circ-zze9 circular-orbit configuration;
	this configuration has the same orbital radius
	as the zze9 configuration's periastron radius.
	The self-force loops (right column) are plotted using
	a logarithmic radial scale.
	The self-force near to and during the whirl phase
	is shown at an expanded scale in
	Fig.~\protect\ref{fig:zoom-whirl-ze4-ze9-zze9}.
	}%%%
\end{figure*}

\begin{figure*}[p]
\begin{center}
\includegraphics[scale=1.0]{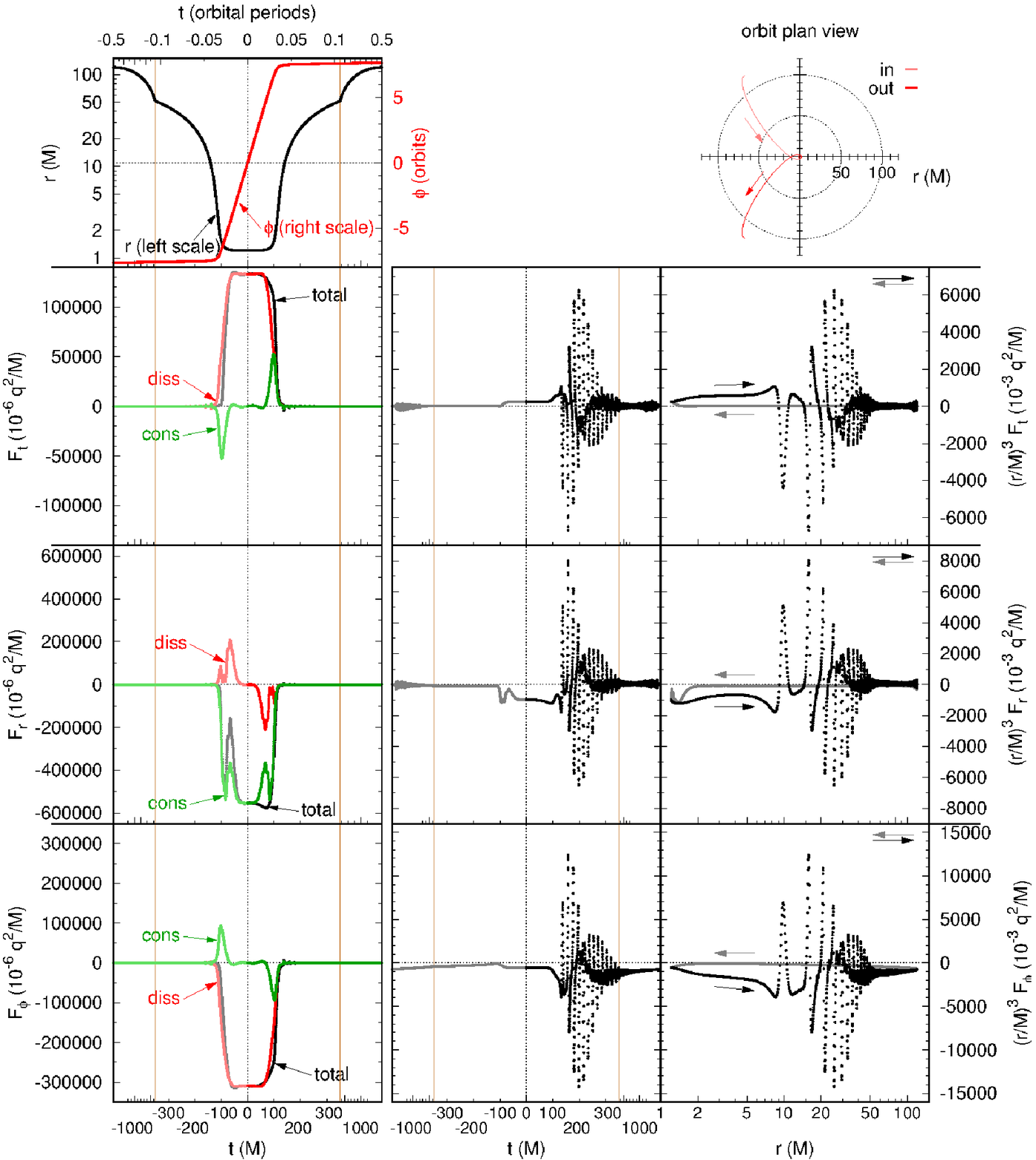}
\end{center}
\vspace{-1ex}
\caption{%%%
\label{fig:ze98-self-force-mosaic}
	This figure shows the self-force for the ze98~configuration,
	which has $(\tilde{a},p,e) = (0.99,2.4,0.98)$.
	This is an extreme zoom-whirl orbit; the particle completes
	about~$15$ orbits at $r \approx 1.2\,M$ during the
	approximately~$220\,M$ of the whirl phase.  
	The self-force loops (right column) are plotted using
	a logarithmic radial scale.
	During the whirl phase the self-force is very large
	(more than $40$~times the peak self-force of any other
	configuration in this study) and shows a variety of
	complicated phenomenology; we discuss this in
	section~\ref{sect:results/zoom-whirl-orbits}.
	Notice the many wiggles in the self-force on the
	outgoing leg of the orbit; we discuss these in
	section~\ref{sect:results/wiggles}.
	The self-force near to and during the whirl phase
	is shown at an expanded scale in
	Fig.~\protect\ref{fig:zoom-whirl-ze98}.
	}%%%
\end{figure*}

Figures~\ref{fig:ns5-self-force-mosaic}--\ref{fig:ze98-self-force-mosaic}
give an overview of the computed self-forces for all our configurations.
To facilitate comparison between the different configurations, these
figures all use a common format (with one exception noted below):
\begin{itemize}
\item	The top row of each figure shows auxiliary information;
	the lower three rows show (respectively) $F_t$, $F_r$, and $F_\phi$.
\item	In the top row, the left plot shows $r$ and $\phi$
	as functions of the coordinate time~$t$,
	while the right plot shows a plan view of the orbit,
	i.e., a parametric plot with
	$x = r(t) \cos\bigl(\phi(t)\bigr)$ and
	$y = r(t) \sin\bigl(\phi(t)\bigr)$.
\item	The coordinate-time scale always runs from $-\thalf T_r$ to
	$+\thalf T_r$, with $t=0$ corresponding to periastron.
	(That is, this ``coordinate time'' is in fact identical
	to the modulo time.)
\item	In the lower three rows of each figure, the left column
	of plots shows each $F_i$ (in units of $10^{-6} q^2/M$)
	as a function of coordinate time~$t$.  For the ze4, ze9, and zze9
	zoom-whirl configurations, these plots also show the self-force
	for the circular-orbit configurations (circ-ze4, circ-ze9, and
	circ-zze9, respectively) with orbital radius equal to the
	zoom-whirl configurations' periastron radius.
\item	In the lower three rows of each figure,
	the center and right columns of plots each show
	the scaled self-force $(r/M)^3 F_i$ (in units of $10^{-3} q^2/M$).
	The center column of plots show $(r/M)^3 F_i$
	as a a function of coordinate time~$t$.
	The right column of plots show $(r/M)^3 F_i$
	as a function of~$r$, forming self-force ``loops'' plots
	of the type introduced
by~\cite{Vega-etal-2013:Schwarzschild-scalar-self-force-via-effective-src}.
\item	In each self-force plot (except the ze98 $r^3 F_i$ plots)
	the total self-force is shown in black and labeled ``total'',
	the dissipative part of the self-force is shown in red
	and labeled ``diss'',
	and the conservative part of the self-force is shown in green
	and labeled ``cons''.  The dissipative and conservative
	parts are omitted in the ze98 $r^3 F_i$ plots to reduce
	clutter.
\item	In each self-force plot the outgoing half of the orbit ($t \ge 0$)
	is shown in fully-saturated color (black, red, or green),
	while the ingoing half of the orbit ($t \le 0$) is shown
	in partially-saturated color (grey, red, or green).
\item	In the self-force loop plots (the right column) the loops
	are labelled with arrows to show the particle's direction
	of motion.  The dissipative part of $F_t$, the conservative
	part of $F_r$, and the dissipative part of $F_\phi$ are
	each independent of the direction of motion.  The conservative
	part of $F_t$, the dissipative part of $F_r$, and the
	conservative part of $F_\phi$ typically differ between
	ingoing (pre-periastron, $t < 0$)
	and outgoing (post-periastron, $t > 0$) motion, forming
	visible loops.
\end{itemize}

%%%%%%%%%%%%%%%%%%%%
%%%%%%%%%%%%%%%%%%%%

%%%%%%%%%%%%%%%%%%%%
%%%%%%%%%%%%%%%%%%%%

%%%%%%%%%%%%%%%%%%%%
%%%%%%%%%%%%%%%%%%%%

%%%%%%%%%%%%%%%%%%%%
%%%%%%%%%%%%%%%%%%%%

%%%%%%%%%%%%%%%%%%%%
%%%%%%%%%%%%%%%%%%%%

%%%%%%%%%%%%%%%%%%%%
%%%%%%%%%%%%%%%%%%%%

%%%%%%%%%%%%%%%%%%%%
%%%%%%%%%%%%%%%%%%%%

%%%%%%%%%%%%%%%%%%%%
%%%%%%%%%%%%%%%%%%%%

%%%%%%%%%%%%%%%%%%%%
%%%%%%%%%%%%%%%%%%%%

%%%%%%%%%%%%%%%%%%%%
%%%%%%%%%%%%%%%%%%%%

%%%%%%%%%%%%%%%%%%%%
%%%%%%%%%%%%%%%%%%%%

%%%%%%%%%%%%%%%%%%%%%%%%%%%%%%%%%%%%%%%%

\subsection{High-eccentricity orbits}
\label{sect:results/highly-eccentric-orbits}

Figures~\ref{fig:e8-self-force-mosaic}--\ref{fig:e95-self-force-mosaic}
show our computed self-force for the e8, e8b, e9, and e95 high-eccentricity
configurations, respectively.

For these configurations the self-force is strongly localized around
the periastron passage.  Even though the particle spends most of its
time at large radii, the $\sim r^{-3}$~far-field scaling of the
self-force with radius implies that the orbital evolution will also
be dominated by the periastron passage.

These configurations also show strong oscillations (``wiggles'')
in the self-force shortly after the periastron passage; we discuss
these in section~\ref{sect:results/wiggles}.

%%%%%%%%%%%%%%%%%%%%%%%%%%%%%%%%%%%%%%%%

\subsection{Zoom-whirl orbits}
\label{sect:results/zoom-whirl-orbits}

\begin{figure*}[p]
\begin{center}
\includegraphics[scale=1.0]{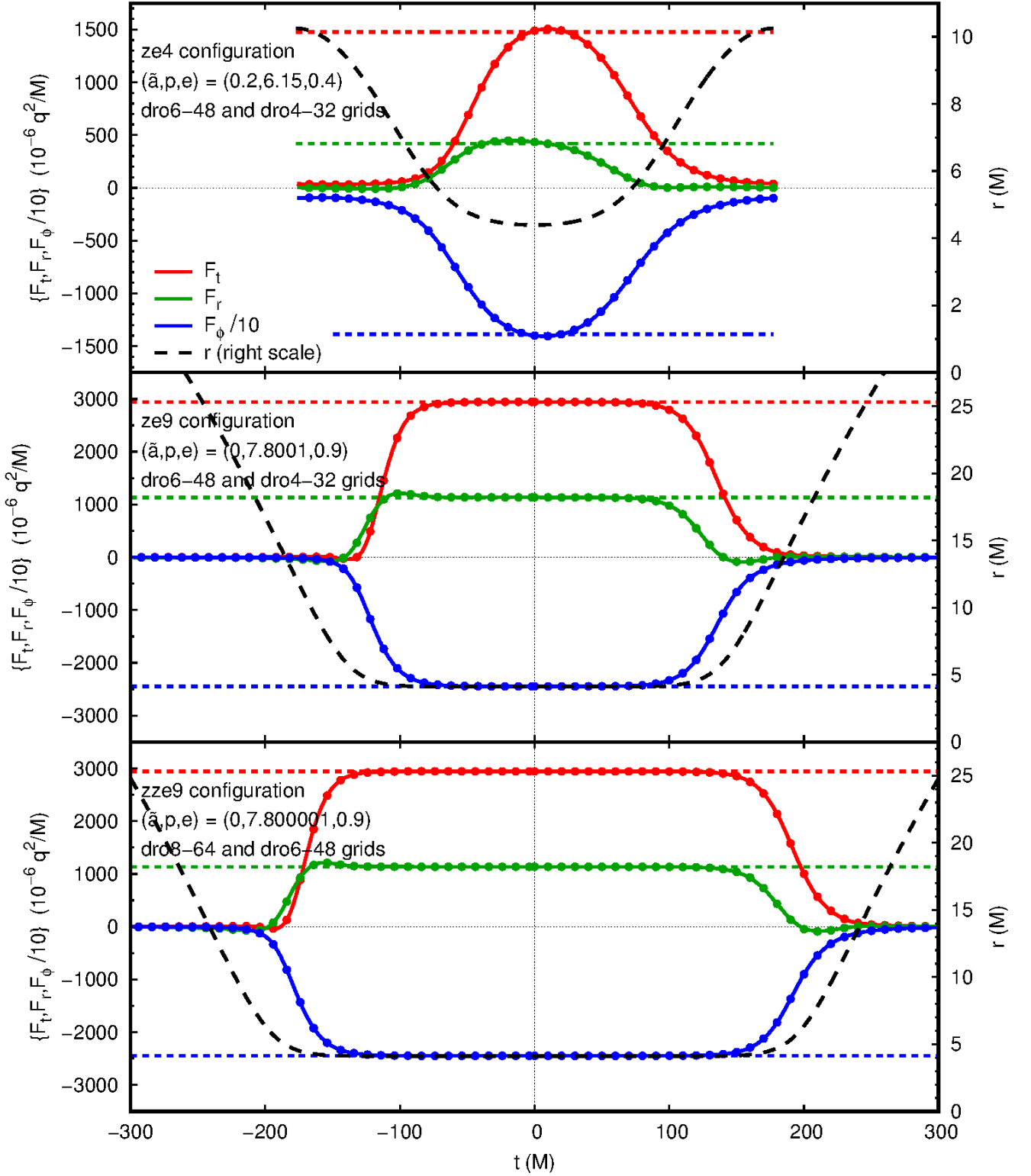}
\end{center}
\vspace{-1ex}
\caption{%%%
\label{fig:zoom-whirl-ze4-ze9-zze9}
	This figure shows the self-force during the whirl phase
	for the ze4, ze9, and zze9 zoom-whirl configurations.
	For each configuration the solid lines show the highest-resolution
	data, while the dots show the lower-resolution data (sampled
	approximately every $10\,M$); these are visually identical.
	The horizontal short-dashed lines show the self-force
	for the corresponding circular-orbit configurations
	(circ-ze4, circ-ze9, and circ-zze9, respectively);
	these have the same orbit radii as the
	zoom-whirl configurations' periastrons.
	}%%%
\end{figure*}

\begin{figure*}[p]
\begin{center}
\includegraphics[scale=1.0]{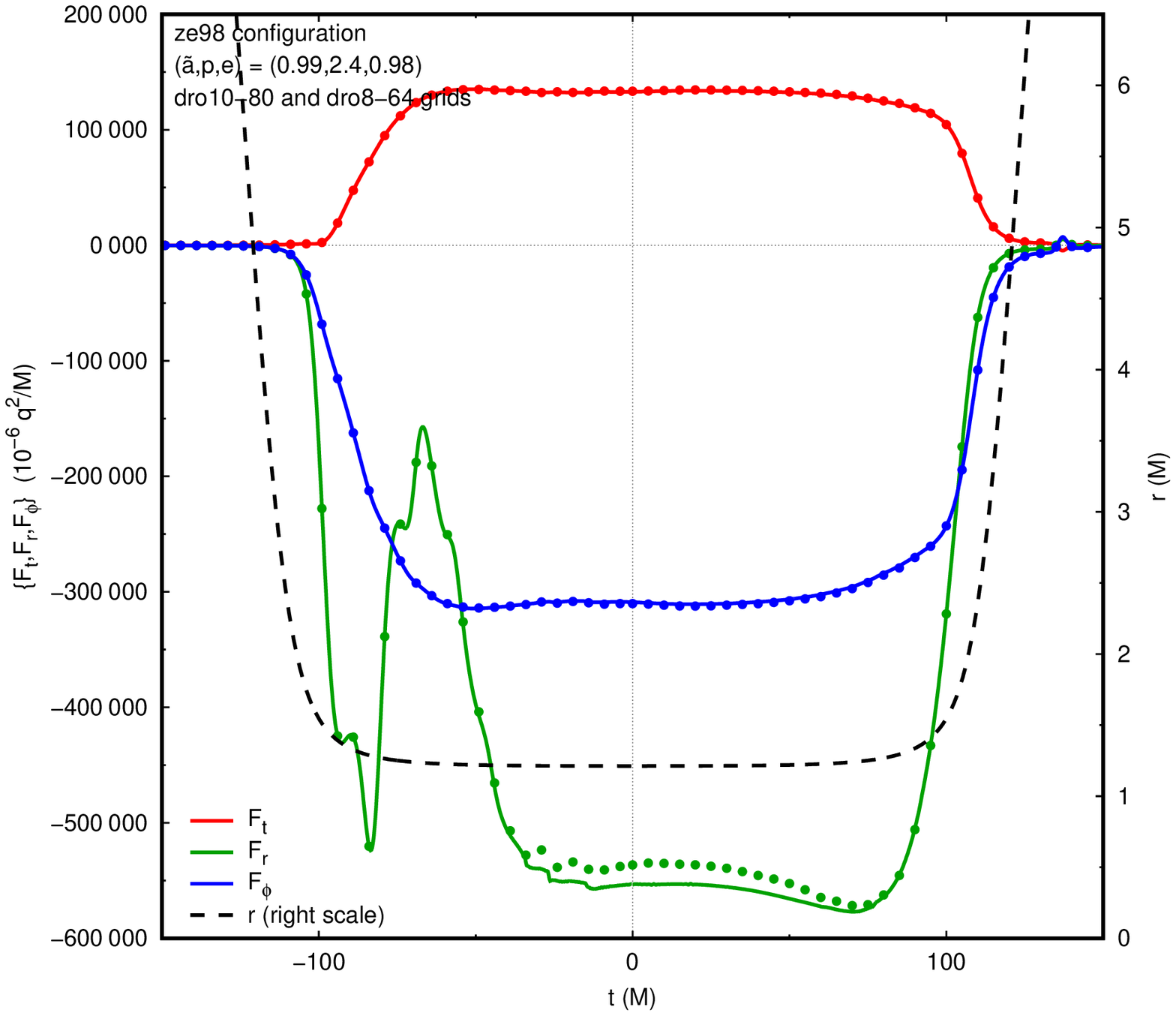}
\end{center}
\vspace{-1ex}
\caption{%%%
\label{fig:zoom-whirl-ze98}
	This figure shows the self-force during the whirl phase
	for the ze98 zoom-whirl configuration.
	The solid lines show the highest-resolution data
	(dro10-80 grids),
	while the dots show the lower-resolution data
	(dro8-64 grids, sampled approximately every $5\,M$);
	these are visually almost identical except for
	a $\ltsim 3.5\%$~difference in $F_r$ at times
	$-35\,M \ltsim t \ltsim 80\,M$.
	}%%%
\end{figure*}

Figures~\ref{fig:ze4-self-force-mosaic}--\ref{fig:ze98-self-force-mosaic}
give an overview of our computed self-force for the ze4, ze9, zze9, and
ze98 zoom-whirl configurations, respectively.
Figures~\ref{fig:zoom-whirl-ze4-ze9-zze9} and~\ref{fig:zoom-whirl-ze98}
show the self-force during the whirl phase in more detail for these
configurations.

Although the self-force is strictly speaking non-local, influenced by
the particle's entire past trajectory, in practice the influence of
distant times is usually small, i.e., the self-force is usually dominated
by the effects of the particle's immediate past.  We thus expect that
if the whirl phase of a zoom-whirl orbit is sufficiently long, the
self-force should be very close to that of a circular orbit at the same
radius.  Figure~\ref{fig:zoom-whirl-ze4-ze9-zze9} shows a numerical test
of this hypothesis for the ze4, ze9, and zze9 configurations, comparing
their whirl-phase self-forces to those of the corresponding circ-ze4,
circ-ze9, and circ-zze9 circular-orbit configurations, respectively.%%%
\footnote{%%%
	 We were unable to calculate the self-force
	 for the circ-ze98 configuration due to numerical
	 instabilities in our evolution code for $m \ge 6$.%%%
	 }%%%
{}  For the ze4 configuration the agreement is only modest, presumably
because of the relatively short whirl phase.  For the ze9 and zze9
configurations the agreement is excellent.

A close examination of Figs.~\ref{fig:ze9-self-force-mosaic}
and~\ref{fig:zze9-self-force-mosaic} shows small ``spikes'' in $F_r$
at the entry/exit to the ze9 and zze9 configurations' whirl phases.
These can be seen at an expanded scale in
Fig.~\ref{fig:zoom-whirl-ze4-ze9-zze9}.
At the whirl-phase entry these configurations' $F_r$ first becomes
slightly negative, then rises to slightly overshoot its whirl-phase
value (this is the ``spike'' visible in
Figs.~\ref{fig:ze9-self-force-mosaic} and~\ref{fig:zze9-self-force-mosaic}),
then decreases slightly to reach the whirl-phase value.  At the
whirl-phase exit $F_r$ decreases smoothly to a slightly negative value,
then rises slightly to its post-whirl (near-zero) value.%%%
\footnote{%%%
	 The visual appearance of these $F_r$ curves in
	 Fig.~\protect\ref{fig:zoom-whirl-ze4-ze9-zze9} somewhat
	 resembles a step function passed through a low-pass filter,
	 although we make no claim that this is in any way
	 the actual mechanism involved.%%%
	 }%%%
{}  Haas~\cite[figure~17]{Haas-2007} has calculated the self-force for
our ze9~configuration and finds similar overshooting behavior.
Barack~\cite{Barack-pers-comm-2016:zoom-whirl-spikes-due-to-radial-accel}
suggests that the underlying cause of this behavior is the particle's
strong radial acceleration when entering/leaving the whirl phase,
but so far as we know no quantitative explanation is known.

For the ze98 configuration (an extreme zoom-whirl orbit),
Fig.~\ref{fig:zoom-whirl-ze98} shows quite complicated phenomenology.
\begin{itemize}
\item	At the entrance to the whirl phase
	(times $-110\,M \ltsim t \ltsim -40\,M$),
	$F_r$ shows small high-frequency oscillations superimposed on
	a larger lower-frequency oscillation; these oscillations last for
	approximately $60\,M$ (about $1/4$~of the entire whirl phase's
	duration).  $F_t$ and $F_\phi$ show small overshoots of their
	whirl-phase values, but no visible high-frequency oscillations.
\item	Well before the exit from the whirl phase
	(times $30\,M \ltsim t \ltsim 75\,M$), while the particle is still
	very close to a circular orbit, $F_r$ increases in amplitude
	by ${\sim}\, 5\%$ (becoming more negative).  Unfortunately,
	while our highest and 2nd-highest-resolution results agree
	on the overall sign of this change, they differ by roughly
	a factor of~$2$ in its magnitude.  (This is the only time at
	which these results differ significantly.)  This suggests that
	higher-resolution data is needed to reliably quantify this feature.
\item	In this same time period (times $30\,M \ltsim t \ltsim 75\,M$)
	$F_t$ and $F_\phi$ both decrease in amplitude.
\item	Shortly before the exit from the whirl phase
	(times $75\,M \ltsim t \ltsim 110\,M$), when the particle
	is significantly departing from a near-circular orbit,
	all components of $F_a$ decrease in magnitude towards
	their post-whirl (small) values.  None of the components
	shows any visible overshoot.
\item	All components of $F_a$ are significantly time-asymmetric
	about the periastron passage.
\end{itemize}

This phenomenology is generally consistent between the dro10-80 and
dro8-64 numerical resolutions.  However, this configuration is a very
difficult one for our numerical evolution scheme%%%
\footnote{%%%
	 At lower resolutions we see numerical instabilities
	 in the ze98 evolutions at times close to periastron.
	 Our numerical evolutions are unstable for $m \ge 6$
	 for the circ-ze98 configuration (a circular orbit at
	 the ze98 configuration's periastron radius).
	 }%%%
{} and it remains possible that some of these features are numerical
artifacts.  We will need to obtain higher-resolution data to resolve
this question.

%%%%%%%%%%%%%%%%%%%%
%%%%%%%%%%%%%%%%%%%%

%%%%%%%%%%%%%%%%%%%%
%%%%%%%%%%%%%%%%%%%%

%%%%%%%%%%%%%%%%%%%%%%%%%%%%%%%%%%%%%%%%

\subsection{Wiggles}
\label{sect:results/wiggles}

\begin{figure*}[bp]
\begin{center}
\includegraphics[scale=1.0]{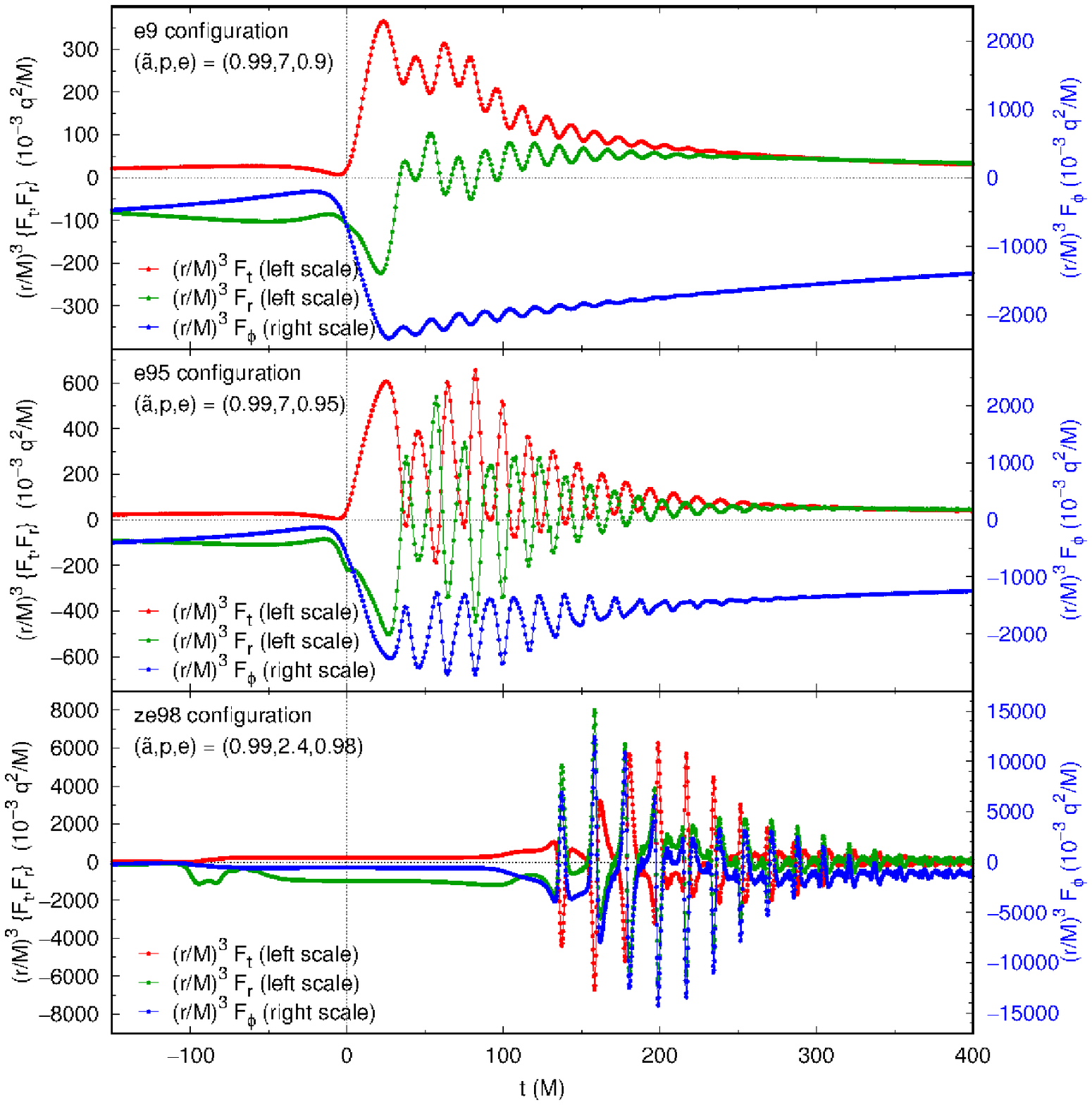}
\end{center}
\vspace{-1ex}
\caption{%%%
\label{fig:e9-e95-ze98-wiggles}
	This figure shows the ``wiggle'' oscillations in
	$(r/M)^3 F_a$ for the e9, e95, and ze98 configurations.
	Notice that wiggles are present only \emph{after} the
	particle's periastron passage ($t=0$).  Notice also that
	for the e95 and ze98 configurations the individual wiggles
	are often non-sinusoidal, with shapes differing between
	different wiggles (this is particularly evident in the
	e95 $(r/M)^3 F_\phi$ wiggles).  The wiggles' shapes also
	differ between different $F_a$ components.
	}%%%
\end{figure*}

In the configurations which combine a highly-spinning black hole and
a prograde high-eccentricity orbit (the e9, e95, and ze98 configurations,
shown in Figs.~\ref{fig:e9-self-force-mosaic},
\ref{fig:e95-self-force-mosaic}, and~\ref{fig:ze98-self-force-mosaic}
respectively), there are prominent and rapid oscillations (``wiggles'')
in $r^3 F_a$ shortly after periastron.  These oscillations are also
visible to a lesser extent in the configurations with moderate
black hole spins and prograde moderate-eccentricity orbits, the
e8 and e8b configurations (shown in Figs.~\ref{fig:e8-self-force-mosaic}
and~\ref{fig:e8b-self-force-mosaic} respectively).
Figure~\ref{fig:e9-e95-ze98-wiggles} shows the wiggles for the
e9, e95, and ze98 configurations at an expanded scale.

%%%%%%%%%%%%%%%%%%%%
%%%%%%%%%%%%%%%%%%%%

Notice that (except for the ze98 configuration, discussed in
section~\ref{sect:results/zoom-whirl-orbits}) the self-force varies
relatively smoothly prior to periastron -- wiggles occur only \emph{after}
the particle's periastron passage ($t=0$).  This suggests that the
wiggles are in some way \emph{caused} by the particle's close passage
by the large black hole.  We will discuss wiggles' phenomenology and
causal mechanisms in a following publication.

%%%%%%%%%%%%%%%%%%%%%%%%%%%%%%%%%%%%%%%%%%%%%%%%%%%%%%%%%%%%%%%%%%%%%%%%%%%%%%%%

\section{Discussion}
\label{sect:discussion}

%%%%%%%%%%%%%%%%%%%%%%%%%%%%%%%%%%%%%%%%

\subsection{Overall assessment}
\label{sect:discussion/overall-assessment}

Our computational scheme combines a number of ingredients:
\begin{itemize}
\item	the initial formulation of the scalar-field toy model for the
	$\O(\mu)$-perturbed scalar-field equations, using a point-particle
	source,
\item	the Barack-Golbourn-Vega-Detweiler effective-source regularization,
\item	our specific choice for the puncture field,
\item	the $m$-mode Fourier decomposition, and the corresponding
	formulation of the puncture field and effective source in
	terms of elliptic integrals,
\item	the introduction of a worldtube, which moves in $(r,\theta)$
	to follow the particle's motion around the orbit,
\item	the Zengino\u{g}lu compactification and hyperboloidal slices,
	and
\item	a finite-difference numerical evolution using Berger-Oliger
	mesh refinement and OpenMP-based parallelization.
\end{itemize}

The initial $\O(\mu)$~perturbation formulation with a point-particle
source is clearly a reasonable starting point for the scalar--self-force
problem.  We discuss possible extensions to this in
section~\ref{sect-discussion/possible-improvements/extensions}.

The Barack-Golbourn-Vega-Detweiler effective-source regularization
scheme works well.  It involves no approximations (a solution of the
regularized equation~\eqref{eqn:box-Phi-residual=S-effective} is an
exact solution of the $\O(\mu)$~field equations), the analytical
computation of the singular field and effective source can be done
with symbolic algebra software, and the resulting regularized equation
is computationally tractable.

In this work, we use a 4th~order puncture for equatorial orbits
in Kerr spacetime. While higher-order, smoother punctures are
available~\cite{Heffernan:2012su,Heffernan:2012vj},
we (like other researchers~\cite{Vega-etal-2009:self-force-3+1-primer,%%%
Dolan-Barack-Wardell-2011,Wardell-etal-2012,%%%
Vega-etal-2013:Schwarzschild-scalar-self-force-via-effective-src})
find that 4th~order represents a good ``sweet spot'' compromise
between a high-order puncture --- which enables high numerical accuracy
and fast convergence at the cost of a having a complicated and
expensive-to-evaluate source --- and a low-order puncture,
which is simple and fast to evaluate, but yields poor convergence
and numerical accuracy.  However, the computation of the effective
source is still computationally expensive.  Further optimization of
this computation would be very useful.

The $m$-mode Fourier decomposition works very well: it provides some
parallelism ``for free'' (each $m$-mode evolution can be performed
independently), it reduces the dimensionality and hence the maximum
CPU and memory usage of each individual evolution, and -- perhaps most
importantly -- it allows different numerical techniques and/or parameters
to be used for different modes' evolutions.  This last advantage may
be of great importance in extending our work to the gravitational case,
where Dolan and Barack~\cite{Dolan-Barack-2013} found that the $m=0$
and $m=1$ modes suffer from gauge instabilities (they were able to
control the $m=0$ gauge modes, but not the $m=1$ modes), while the
$m \ge 2$ modes are stable.

The moving-worldtube scheme works well, allowing highly eccentric
orbits to be simulated while only requiring the (expensive) effective
source computation in a relatively small region of spacetime.  We found
the implementation of the worldtube at a finite-differencing level to be
straightforward (\cf{}~Appendix~\ref{app:details/FD-across-worldtube-boundary})
once the Boolean predicates for where to use adjusted finite differencing 
and where to (pre)compute the puncture field were defined correctly
(\cf{}~appendices~\ref{app:details/computing-where-adjusted-FD-needed}
and~\ref{app:details/computing-where-pfn-needed}).
For orbits of low to moderate eccentricity,
the alternative of using a smooth blending ``window''
function~\cite{Vega-Wardell-Diener-2011:effective-source-for-self-force,%%%
Vega-etal-2013:Schwarzschild-scalar-self-force-via-effective-src}
is also known to work well.  However, extending this to highly
eccentric orbits may require making the window function time-dependent,
which would introduce additional terms into the evolution equations.

Like other researchers
(e.g.,~\cite{Vega-Wardell-Diener-2011:effective-source-for-self-force}),
we find the Zengino\u{g}lu compactification and hyperboloidal slices
to work very well.  They are easy to implement and provide slices
which span the entire spacetime outside the event horizon, allowing
stable and highly accurate horizon and $\Scri^+$ outgoing boundary
conditions.  Slices which reach $\Scri^+$ also allow a direct
computation of the emitted radiation reaching $\Scri^+$, although
for simplicity we have not done so here.

Our numerical evolution uses finite-differencing and Berger-Oliger mesh
refinement techniques which are now standard in numerical relativity.
However, there are three main complications which combine to make the
use of standard adaptive-mesh-refinement frameworks such as
\SubroutineLibrary{Cactus}~\cite{Allen99a,Goodale02a,Cactuscode:web}%%%
\footnote{%%%
	 See~\protect\cite{Dubey-etal-2014:survey-of-AMR-frameworks}
	 for a survey of other such frameworks.%%%
	 }%%%
{} more difficult and less advantageous than would be the case in many
other numerical-relativity calculations:
\begin{itemize}
\item	Our use of a worldtube, and the associated (time-dependent)
	jump discontinuity in the evolved field $\varphi_m$, means that
	interpolation and restriction operators must ``adjust'' the
	field variables when crossing the worldtube boundary
	(\cf{}~Appendix~\ref{app:details/FD-across-worldtube-boundary}).
	This means that standard mesh-refinement software
	requires modification to accommodate the worldtube scheme.
\item	The effective source is expensive to compute, but is only
	needed inside the worldtube, so the overall cost of integrating
	our equations at a single grid point is much larger inside
	the worldtube than outside.  The default domain-decomposition
	parallelization heuristics used by \SubroutineLibrary{Cactus}
	and many other adaptive-mesh-refinement toolkits assume a
	roughly uniform level of computational cost per grid point
	across the problem domain, and thus would give relatively poor
	parallel performance on our computation.
\item	The non-smoothness of the evolved field $\varphi_m$ at the
	puncture (particle) position limits the finite-differencing
	order of accuracy attainable there.  For our 4th~order puncture,
	the accuracy is limited to at best
	$\O\bigl((\Delta R_*)^2, (\Delta\theta)^2\bigr)$ because our
	evolution equation~\eqref{eqn:box-varphi=S-effective-piecewise}
	is 2nd~order in space.  This reduces the benefits gained
	from high-order finite differencing schemes (which are now
	provided by many mesh-refinement software libraries).
\end{itemize}

%%%%%%%%%%%%%%%%%%%%%%%%%%%%%%%%%%%%%%%%

\subsection{Possible improvements}
\label{sect-discussion/possible-improvements}

There are a number of ways in which our results might plausibly be
improved.  While there is an accuracy/performance tradeoff in almost
any finite difference computation, computational improvements can still
usefully be categorized into those which would improve the accuracy of
the self-force computation for a given finest-grid-resolution,
versus those which would improve the efficiency of computing results
using essentially the same numerical scheme, versus those which would
improve both accuracy and efficiency.

%%%%%%%%%%%%%%%%%%%%

\subsubsection{Computational improvements: accuracy}
\label{sect-discussion/possible-improvements/accuracy}

There are several ways in which our computational scheme might be
improved so as to provide more accurate results for the same
finest-grid-resolution.

As noted in section~\ref{sect:theory/esrc-close-to-particle}, our
interpolation scheme for computing the effective source close to the
particle uses an interpolation molecule which crosses the particle
position in some cases, reducing the interpolation accuracy.
An improved interpolation scheme might improve the overall accuracy
of the computation.

As the particle moves through the grid, the limited differentiability
of $\varphi_m$ at the puncture effectively introduces noise into the
evolution and prevents us from obtaining proper (in our case 4th~order)
finite differencing convergence of our results with grid
resolution~\cite{Choptuik-1991:FD-consistency}.
One way to eliminate this noise and obtain proper finite differencing
convergence would be to use finite difference operators which specifically
``know'' the actual functional form of $\varphi_m$ near the puncture.
We have experimented with several finite differencing schemes of this
type, but so far with only limited success.  At present our code uses
the ``C2'' scheme described in Appendix~\ref{app:details/FD-near-particle}.
We find that this lowers the noise level in the computed self-force
by roughly a factor of~$3$, but our results remain quite noisy
and their overall convergence order with respect to grid resolution
is still much lower than we would like.
Further research on finite difference operators which incorporate
more of the puncture's actual singularity structure would be useful.
(We mention one possible finite differencing scheme of this type in
Appendix~\ref{app:details/FD-near-particle}, but we were not able
to obtain stable evolutions with this scheme.)

As noted in section~\ref{sect:results/convergence}, for the
e9~configuration we find poor convergence at small radii
($r \ltsim 10M$).  We do not yet know the cause of this poor
convergence, but fixing it would obviously be highly desirable.

Another possible route to more accurate finite differencing near the
puncture might be to use many mesh-refinement levels of small grids
in the puncture's immediate neighborhood, so as to obtain very high
resolutions at the puncture.  Given a Berger-Oliger--style mesh-refinement
infrastructure, this is not difficult.  However, the interpolations
of the fine-grid boundary values from the coarser grids might limit
the accuracy improvement, even if buffer
zones~\cite{Schnetter-Hawley-Hawke-2004:Carpet-paper} are used.
Further experimentation with this type of grid structure would be
useful.

Raising the order of the puncture would improve the smoothness of
$\varphi_m$ at the puncture, improving the finite-differencing accuracy
there.  However, a higher-order puncture would also yield a much more
complicated and expensive-to-compute effective source.  Our current
choice of a 4th~order puncture seems to be a good compromise between
smoothness and computational expense.

%%%%%%%%%%%%%%%%%%%%

\subsubsection{Computational improvements: efficiency}
\label{sect-discussion/possible-improvements/efficiency}

There are a number of ways in which our computational scheme might be
made more efficient.

At present our code computes the puncture field and effective source
anew at each right-hand-side evaluation whose time coordinate differs
from that of the previous evaluation (this happens 50\% of the time
for the classical 4th-order Runge-Kutta time integration scheme we
currently use).  For periodic orbits (including all equatorial orbits)
the puncture field and effective source are the same (at a given
time-past-periastron) from one orbit to the next, so a much more
efficient choice would be to cache the effective source in memory,
reusing cached values for all of the evolution after the first orbit.
However, such a cache would use a very large amount of memory and
would give no benefit for non-periodic
orbits (including almost all non-equatorial orbits).

Simulations of this type are computationally expensive.  Our code
is currently only partially parallelized, using OpenMP to spread the
computation of the singular field, effective source, and evolution-equation
right-hand-side across multiple cores of a single processor.  This is
easy to implement and typically gives a wall-clock speedup of a factor
of $12$ to $13$ using $16$~cores.  Grid-based parallelism (ultimately
based on message-passing) is an obvious and widely used way of achieving
higher parallelism, and is now well-supported by numerical-relativity
adaptive-mesh-refinement toolkits such as
\SubroutineLibrary{Cactus}~\cite{Allen99a,Goodale02a,Cactuscode:web}
and the Einstein toolkit~\cite{Loffler:2011ay,EinsteinToolkit:web}.
However, Cactus and many other adaptive-mesh-refinement toolkits
generally assume that the cost of computing a grid point is roughly
constant across the problem domain.  Our worldtube scheme strongly
violates that assumption: points inside the worldtube require computing
the effective source and thus cost much more than points outside the
worldtube.  This means that without significant changes to the
domain-decomposition heuristics, standard toolkits would give only
limited parallel speedup for our worldtube scheme. One possible way
to sidestep this issue is to use a domain decomposition for the calculation
of the effective source that is independent of the normal domain
decomposition of the full computational grid; such a method was used
(without mesh refinement), for example, in
\cite{Vega-etal-2013:Schwarzschild-scalar-self-force-via-effective-src}.
Similarly, it might be that
other parallelization techniques such as the task-based model used
by the \SubroutineLibrary{SpECTRE}
code~\cite{Kidder-etal-2016:SpECTRE-discontinuous-Galerkin-code}
would yield better parallel speedup.

%%%%%%%%%%%%%%%%%%%%

\subsubsection{Computational improvements: accuracy and efficiency}
\label{sect-discussion/possible-improvements/accuracy-and-efficiency}

At present our computational scheme uses finite differencing with
Berger-Oliger mesh refinement.  A discontinuous Galerkin
method~\cite{Hesthaven-Warburton-book-2008} might give spectral
(i.e., much better) accuracy/efficiency even with the limited
differentiability of $\varphi_\numerical$ at the particle.  These
methods have been used successfully in other numerical-relativity
and self-force computations by a number of researchers,
e.g.,~\cite{Field-Hesthaven-Lau-2009:discontinuous-Galerkin-1+1-Schw-EMRI,%%%
Field-etal-2010:discontinuous-Galerkin-2nd-order-BSSN,%%%
Brown-etal-2012:1st-order-BSSN-and-discontinuous-Galerkin},
as well as in other areas of computational physics involving
non-smooth solutions,
e.g.,~\cite{Fan-Cai-Ji-2008:discontinuous-Galerkin-Schrodinger-eqn-nonsmooth-solns,%%%
Kidder-etal-2016:SpECTRE-discontinuous-Galerkin-code}.

Within the general framework of finite differencing and Berger-Oliger
mesh refinement, there are a number of ways in which our computational
scheme might be enhanced to better adjust the computations to the
solution dynamics, yielding both improved accuracy (higher effective
grid resolution) and efficiency (fewer high-resolution grid points
``wasted'' on regions of spacetime where $\varphi_m$ is relatively
slowly-varying):
\begin{itemize}
\item	At present our mesh-refinement scheme moves the finer grids with
	the worldtube but does not otherwise adapt to the solution's
	dynamics.  For an orbit with substantial eccentricity, the field
	dynamics near the particle are quite different between the
	particle's periastron and apoastron.
	It seems likely that an adaptive mesh-refinement scheme (of the
	type now widely used in fully-nonlinear binary-black-hole
	simulations) for varying the grid structure around the orbit
	would substantially improve the computation's overall
	accuracy/efficiency.
\item	At present our computational scheme keeps the worldtube size
	and shape fixed throughout the evolution.  An adaptive scheme
	to adjust (optimize) these around the orbit could significantly
	improve the code's accuracy and efficiency.  However, unlike
	the case for adaptive mesh-refinement, there are no existing
	algorithms for making this adjustment.  Further research in
	this area would be valuable.
\item	For a highly eccentric orbit, many of the higher-$m$ self-force
	modes are below our code's noise level during much of the orbit.
	(This can be seen, for example, in Figs.~\ref{fig:e8-aligned-modes-3d}
	and~\ref{fig:e8-modes-overview/final-orbit}.)
	The overall efficiency of the computation could be greatly
	improved by not computing these modes at times when they are
	essentially purely noise.  This would require some means of
	estimating the time intervals in question, and changes to our
	initial-worldtube-setup scheme (described in
	section~\ref{app:details/constraints-on-moving-WT-early-in-evolution})
	to accommodate \hbox{(re)starting} the computation of these modes at a
	time when the particle is moving much faster than near apoastron.
\end{itemize}

%%%%%%%%%%%%%%%%%%%%

\subsubsection{Extensions to more general physical systems}
\label{sect-discussion/possible-improvements/extensions}

In this work we focus on computing the instantaneous scalar
self-force acting on the small body.  One straightforward extension
to this is to also compute the scalar field at the particle, using
the method suggested in
footnote~\ref{footnote:how-to-compute-scalar-field-at-particle}.
Another straightforward extension would be to also compute the
scalar field radiated to infinity ($\Scri^+$).  Given our use of
asymptotically hyperboloidal slices which reach $\Scri^+$, this
information is readily available.  We have preliminary implementations
of both of these extensions; we will discuss their results in a
following publication.

Our present results are limited to (bound, geodesic) equatorial
particle orbits.  Apart from the computational complexity of computing
the effective source (which is probably manageable with some reorganization
of the Mathematica-generated C code),%%%
\footnote{%%%
	 Our preliminary experiments with generalizing our
	 current singular field and effective source to
	 non-equatorial orbits suggest that the complexity
	 of the effective-source coefficients increases
	 by a factor of~${\sim}\, 40$, with a corresponding
	 increase in the size of the machine-generated C~code
	 for computing the coefficients.  This computation
	 would need to be reorganized in order for it to
	 be practical to compile the Mathematica-generated
	 C~code.  We believe this is possible, but haven't
	 yet done so.
	 }%%%
{} there appears to be no fundamental obstacle to allowing non-equatorial
orbits, and this would be a very useful extension.  In particular,
this would allow direct exploration of transient $\theta$-$\phi$
resonances~\cite{Flanagan-Hinderer-2012,Brink-Geyer-Hinderer-2015}.

Our present results are limited to the ``toy model'' of a scalar-field
particle.  Extending these results to a point mass and its gravitational
field perturbations would be very interesting but also challenging.
While the basic effective-source regularization scheme is already known
to be valid for the gravitational case,
Dolan and Barack~\cite{Dolan-Barack-2013} found that the $m=1$
evolutions suffered from linearly-growing-in-time Lorenz gauge modes
which they were not able to control.  Stabilizing these modes, and
more generally achieving long-time-stable evolutions for all~$m$,
is an important area for further research.

Our present results are also limited to $\O(\mu)$~perturbations of
the (Kerr) background spacetime.  LISA could benefit from EMRI waveform
templates with $\sim 10^{-8}$ or better fractional orbital-phase
accuracy~\cite[section~4]{Thornburg-2011:Capra-survey}, which would
require the inclusion of both $\O(\mu^2)$~terms and ``extended-body''
effects caused by the finite size and (in general) nonzero spin of the
small body (see, for example, \cite{Vines-etal-2016:extended-body-effects}
and references therein).

In the longer term, it will also be essential to extend self-force
calculations to include orbital evolution.  This is conceptually
straightforward (though computationally demanding) if the osculating-geodesic
approximation is retained (as was done by
Warburton~\etal{}~\cite{Warburton-etal-2012:Schw-inspiral} in their
pioneering calculation of gravitational inspiral in Schwarzschild
spacetime over a time span of more than $75\,000$~orbits).  However,
going beyond the osculating-geodesic approximation is more difficult.
Diener~\etal{}~\cite{Diener-etal-2012:self-consistent-Schw-orbital-evolution}
have demonstrated that this can be done for a scalar-field particle
in Schwarzschild spacetime, but they were only able to attain relatively
modest accuracies and integration time spans ($\sim 20$~orbits).
Extending their work to higher accuracies and longer integrations
is an important area for further research.

%%%%%%%%%%%%%%%%%%%%%%%%%%%%%%%%%%%%%%%%%%%%%%%%%%%%%%%%%%%%%%%%%%%%%%%%%%%%%%%%

\section*{Author Contributions}

JT developed the numerical evolution and tail-fitting/mode-sum codes,
performed the numerical evolutions, and did the main data analysis.
BW developed the algorithms and symbolic-algebra code for computing
the puncture field and effective source, and for machine-generating
the C code for this computation.   Both authors contributed to the
preparation of this manuscript.

%%%%%%%%%%%%%%%%%%%%%%%%%%%%%%%%%%%%%%%%%%%%%%%%%%%%%%%%%%%%%%%%%%%%%%%%%%%%%%%%

\section*{Acknowledgements}

We are grateful to Niels Warburton for providing unpublished orbit and
self-force results from the code described in~\cite{Warburton-Barack-2011}.
We thank Ian Hinder for performing an eigenvalue analysis of our evolution
system, and for helpful discussions on the origin of our evolution scheme's
very restrictive stability limit for large~$m$.
We thank
Leor Barack, Sam Dolan,
and the other participants of the Capra meetings on Radiation Reaction
for many illuminating conversations.
JT thanks 
Eric Ost for valuable assistance with a computer cluster
used for numerical calculations with early versions of our code,
and Virginia J.~Vitzthum for comments on this manuscript.

This material is based upon work supported by the National Science Foundation
under Grant Number 1417132. B.W. was supported by Science Foundation Ireland
under Grant No.~10/RFP/PHY2847, by the John Templeton Foundation New Frontiers
Program under Grant No.~37426 (University of Chicago) - FP050136-B (Cornell
University), and by the Irish Research Council, which is funded under the
National Development Plan for Ireland.

% acknowledgement for JT using Data Capacitor II filesystem at IU
This material is based upon work supported by the U.S.~National Science
Foundation (NSF) under Grant No.~CNS-0521433.
Any opinions, findings and conclusions, or recommendations expressed
in this material are those of the author(s), and do not necessarily
reflect the views of the NSF.
% acknowledgement for JT using Karst supercomputer at IU
This research was supported in part by Lilly Endowment, Inc., through
its support for the Indiana University Pervasive Technology Institute,
and the Indiana Metabolomics and Cytomics (METACyt) Initiative.

%%%%%%%%%%%%%%%%%%%%%%%%%%%%%%%%%%%%%%%%%%%%%%%%%%%%%%%%%%%%%%%%%%%%%%%%%%%%%%%%
%%%%%%%%%%%%%%%%%%%%%%%%%%%%%%%%%%%%%%%%%%%%%%%%%%%%%%%%%%%%%%%%%%%%%%%%%%%%%%%%
\appendix
%%%%%%%%%%%%%%%%%%%%%%%%%%%%%%%%%%%%%%%%%%%%%%%%%%%%%%%%%%%%%%%%%%%%%%%%%%%%%%%%
%%%%%%%%%%%%%%%%%%%%%%%%%%%%%%%%%%%%%%%%%%%%%%%%%%%%%%%%%%%%%%%%%%%%%%%%%%%%%%%%

\section{$\tilde{\phi}$~derivatives}
\label{app:phi-tilde-derivs}

Clearly $d\tilde{\phi} = d\phi$ if $dr=0$, so for any scalar quantity~$Q$
we have
\begin{equation}
\left. \frac{\partial Q}{\partial \phi} \right|_r
=
\left. \frac{\partial Q}{\partial \tilde{\phi}^{}} \right|_r,
\end{equation}
i.e., (since~$Q$ is arbitrary),
\begin{equation}
\left. \frac{\partial}{\partial \phi} \right|_r
=
\left. \frac{\partial}{\partial \tilde{\phi}} \right|_r
				      \label{eqn:xform-d/dphi|r(d/dphi-tilde|r)}.
\end{equation}

To relate $\bigl. \partial \big/ \partial_r \bigr|_\phi$ and
$\bigl. \partial \big/ \partial_r \bigr|_{\tilde{\phi}}$, consider
two infinitesimally-separated events $X$ and $Y$, with coordinates
\begin{equation}
\begin{array}{ll@{\qquad}l@{\qquad}l}
X:	& r = r_X,
	& \phi = \phi_X,
	& \tilde{\phi} = \tilde{\phi}_X;
									\\*
Y:	& r = r_X + dr,
	& \phi = \phi_X.
	&
									%%%\\*
\end{array}
\end{equation}
Since $\phi$ is the same for events~$X$ and~$Y$,
the definition~\eqref{eqn:dphi-tilde-defn} of $\tilde{\phi}$ implies
that $\tilde{\phi}_Y = \tilde{\phi}_X + d\tilde{\phi}$ with
\begin{equation}
d\tilde{\phi} = \frac{M \tilde{a}}{\Delta} dr.
							  \label{eqn:dphi-tilde}
\end{equation}
Thus for any scalar quantity~$Q$ we have (using the chain rule in
$(r,\phi)$~coordinates)
\begin{equation}
Q_Y - Q_X = dr \cdot \left.
		     \frac{\partial Q}{\partial r}
		     \right|_\phi
	    \qquad
	    \text{since $d\phi = 0$}.
						     \label{eqn:QY-QX-via-r-phi}
\end{equation}
Using the chain rule in $(r,\tilde{\phi})$~coordinates, we also have
\begin{align}
Q_Y - Q_X
	& =	dr \cdot \left.
			 \frac{\partial Q}{\partial r}
			 \right|_{\tilde{\phi}}
		+ d\tilde{\phi} \cdot \left.
				      \frac{\partial Q}
					   {\partial \tilde{\phi}^{}}
				      \right|_r
									\\*
	& =	dr \cdot \left.
			 \frac{\partial Q}{\partial r}
			 \right|_{\tilde{\phi}}
		+ \frac{M \tilde{a}}{\Delta} dr \cdot \left.
					    \frac{\partial Q}
						 {\partial \tilde{\phi}^{}}
					    \right|_r
		\quad
		\text{via~\eqref{eqn:dphi-tilde}},
					       \label{eqn:QY-QX-via-r-phi-tilde}
									%%%\\*
\end{align}
so that (comparing~\eqref{eqn:QY-QX-via-r-phi}
and~\eqref{eqn:QY-QX-via-r-phi-tilde}) we have
\begin{equation}
\left. \frac{\partial Q}{\partial r} \right|_\phi
	= \left.
	  \frac{\partial Q}{\partial r}
	  \right|_{\tilde{\phi}}
	  + \frac{M \tilde{a}}{\Delta} \left.
			     \frac{\partial Q}{\partial \tilde{\phi}^{}}
			     \right|_r,
\end{equation}
i.e., (since~$Q$ is arbitrary),
\begin{equation}
\left. \frac{\partial}{\partial r} \right|_\phi
	= \left.
	  \frac{\partial}{\partial r}
	  \right|_{\tilde{\phi}}
	  + \frac{M \tilde{a}}{\Delta} \left.
			     \frac{\partial}{\partial \tilde{\phi}^{}}
			     \right|_r.
				      \label{eqn:xform-d/dr|phi(d/dr|phi-tilde)}
\end{equation}

%%%%%%%%%%%%%%%%%%%%%%%%%%%%%%%%%%%%%%%%%%%%%%%%%%%%%%%%%%%%%%%%%%%%%%%%%%%%%%%%

\section{Details of our computational scheme}
\label{app:details}

%%%%%%%%%%%%%%%%%%%%%%%%%%%%%%%%%%%%%%%%%%%%%%%%%%%%%%%%%%%%%%%%%%%%%%%%%%%%%%%%

\subsection{Computing $r(r_*)$}
\label{app:details/computing-r(rstar)}

Our computational scheme uses grids which are locally uniform in
$(R_*,\theta)$.  However, the coefficients in many of our equations
are given as explicit functions of $r$, so the code needs to know the
$r$~coordinate of each grid point.  Since $r_*(R)$ is given explicitly
by the compactification~\eqref{eqn:r_*(R_*,Omega)}
and~\eqref{eqn:Omega(R_*)}, it only remains to compute $r(r_*)$.

Given an input value $r_*^\myinput$, the corresponding $r^\myinput$
could be found by using Newton's method to solve the equation
$r_*(r) = r_*^\myinput$ using the definition~\eqref{eqn:rstar(r)}.
However, for positions just outside the event horizon ($r_* \ll 0$)
the near-cancellation in computing $r - r_+$ would make this algorithm
numerically inaccurate.

Instead, we define a new radial coordinate~$y$ by
\begin{equation}
y = \ln \left( \frac{r - r_+}{2M} \right)
							        \label{eqn:y(r)}
\end{equation}
so that
\begin{equation}
r = r_+ + 2M e^y.
								\label{eqn:r(y)}
\end{equation}
The definition~\eqref{eqn:rstar(r)} can then be rewritten as
\begin{widetext}
\begin{equation}
r_* = r + 2M\frac{r_+}{r_+ - r_-} y
	- 2M\frac{r_-}{r_+ - r_-} \ln\left( \frac{r_+ - r_-}{2M} + e^y \right).
							    \label{eqn:rstar(y)}
\end{equation}
\end{widetext}

Given an input value $r_*^\myinput$, we first find the corresponding
$y^\myinput$ by using Newton's method to solve the equation
\begin{equation}
r_*(y) = r_*^\myinput
						   \label{eqn:rstar(y)=rstar-in}
\end{equation}
for $y = y^\myinput$, then computing $r^\myinput$ via~\eqref{eqn:r(y)}.

Newton's method requires an initial guess $y^{(\initial)}$.
If $r_*^\myinput > r_+$ we guess $r^{(\initial)} = r_*$ and
use~\eqref{eqn:y(r)} to compute $y^{(\initial)}$.  Otherwise, we
approximate the right hand side of~\eqref{eqn:rstar(y)} by its first
two terms only, so that
\begin{equation}
y^{(\initial)}
	= \frac{r_+ - r_-}{2M} \left( \frac{r_*^\myinput}{r_+} - 1 \right).
\end{equation}

The Newton's-method solution is moderately expensive for a computation
which (logically) is needed at each grid point: it typically requires
$3$--$10$ iterations, with each iteration needing an \subroutine{exp()}
and a \subroutine{log()} computation as well as $\sim 10$~floating-point
arithmetic operations.  Our code therefore precomputes and caches $r$
for each radial grid point.

%%%%%%%%%%%%%%%%%%%%%%%%%%%%%%%%%%%%%%%%

\subsection{Integrating {K}err geodesics}
\label{app:details/integrating-Kerr-geodesics}

We use the Glampedakis-Kennefick formulation~\cite{Glampedakis-Kennefick-2002}
to integrate the Kerr geodesic equations.%%%
\footnote{%%%
	 Note that we differ slightly
	 from~\protect\cite{Glampedakis-Kennefick-2002} in
	 that we use a dimensionless definition for $p$.%%%
	 }%%%
{}  This parameterizes the radial motion as
\begin{equation}
r = \frac{p M}{1 + e \cos \chi},
\end{equation}
where $p$ is the dimensionless semi-latus rectum and $e$ the eccentricity.
To solve for the particle position we numerically integrate the ODEs
\begin{subequations}
\begin{align}
\frac{dt}{d\chi}	& = \subroutine{RHS\_t}(\chi)			\\
\frac{d\phi}{d\chi}	& = \subroutine{RHS\_$\phi$}(\chi)		\\
\frac{d\tau}{d\chi}	& = \frac{(dr/d\chi) \big/ (e \sin\chi)}
				   {(dr/d\tau) \big/ (e \sin\chi)}
					     \label{eqn:Kerr-geodesic-dtau-dchi}
									%%%\\
\end{align}
\end{subequations}
using $\chi$ is the independent variable.  The right-hand-side
functions \subroutine{RHS\_t} and \subroutine{RHS\_$\phi$} are given
by Glampedakis and Kennefick's equations~(17) and (16) respectively,
while the right hand side of~\eqref{eqn:Kerr-geodesic-dtau-dchi}
is computed using
\begin{subequations}
\begin{align}
\frac{dr/d\chi}{e \sin\chi}
	& =	\frac{p M}{(1 + e \cos\chi)^2}				\\
\frac{dr/d\tau}{e \sin\chi}
	& =	\frac{1}{p} \sqrt{\tilde{V}_r(\chi)}			%%%\\
\end{align}
\end{subequations}
with $\tilde{V}_r(\chi)$ given by Glampedakis and Kennefick's equation~(18).
With this formalism the equations are non-singular at the radial turning
points, and all square roots have their principal values (i.e., there
are no $\pm$~sign ambiguities).  However, integrating to a specified
coordinate time~$t$ requires either an explicit root-finding loop
around the ODE integration or using an ODE integrator with built-in
root-finding capabilities.

We use the \SubroutineLibrary{ODEPACK}
ODE integrator~\cite{Hindmarsh-1983,Radhakrishnan-Hindmarsh:LSODE-report},
whose \subroutine{DLSODAR} subroutine provides ODE integration with
built-in root-finding.  We typically set both the \subroutine{DLSODAR}
relative and absolute error tolerances to $100 \varepsilon$, where
$\varepsilon$ is the floating-point ``machine epsilon''.%%%
\footnote{%%%
	 $\varepsilon$ is the difference between $1.0$ and
	 the next larger floating-point number, approximately
	 $1.1\,{\times}\,10^{-16}$ for IEEE-standard
	 double-precision floating-point
	 arithmetic~\protect\cite{Goldberg91}.
	 }%%%
{}  We set the \subroutine{DLSODAR} \variablename{MXSTEP} parameter
(the maximum number of internal integration steps per \subroutine{DLSODAR}
call) to $10^5$.  This allows \subroutine{DLSODAR} to integrate a full
orbit (and hence determine the orbital period) of an extreme zoom-whirl
orbit like our ze98~configuration in a single call.

The \SubroutineLibrary{ODEPACK} library is written in Fortran~77,
which makes its use somewhat awkward in our context.  Notably,
\SubroutineLibrary{ODEPACK} keeps internal state in static storage
arrays and Fortran common blocks.  In the context of Berger-Oliger
mesh refinement it is natural to use a separate (concurrent) integration
for each refinement level; in our code this requires explicitly saving
and restoring the integrator state to multiplex the multiple concurrent
integrations onto the single-threaded \SubroutineLibrary{ODEPACK}.

The next-generation version of \SubroutineLibrary{ODEPACK}, now known
as \SubroutineLibrary{SUNDIALS}~\cite{Hindmarsh-etal-2005:SUNDIALS},%%%
\footnote{%%%
	 \SubroutineLibrary{SUNDIALS} is available at no cost
	 from \url{https://computation.llnl.gov/casc/sundials/main.html}.
	 }%%%
{} is written in C and (along with other algorithmic and computational
improvements) directly supports multiple concurrent integrations.
This should make it easier to use than the Fortran version.

%%%%%%%%%%%%%%%%%%%%%%%%%%%%%%%%%%%%%%%%

\subsection{Gradual turnon of the effective source}
\label{app:details/gradual-turnon-of-effective-src}

Because of the jump discontinuity in the right hand side
of~\eqref{eqn:box-varphi=S-effective-piecewise}, the process of
radiating away the initial junk generates high-spatial-frequency
noise in $\varphi_{\numerical,m}$ in and near to the worldtube,
leading to high noise levels in the computed self-force time series.
Therefore, we use a gradual turn-on of the effective source,
replacing~\eqref{eqn:box-varphi=S-effective-piecewise} with
\begin{equation}
\boxop_m \varphi_{\numerical,m}
	= \begin{cases}
	  f(t) \, S_{\effective,m}
				& \text{inside the worldtube}	\\*
	  0			& \text{outside the worldtube}	%%%\\*
	  \end{cases},
				\label{eqn:box-varphi=gto-S-effective-piecewise}
\end{equation}
where $f$ is a smooth function which is very small (ideally~$0$) at
the initial time of an evolution and increases to asymptote to~$1$
at late times.  We use
\begin{subequations}
\begin{equation}
f(x) = \thalf \bigl( 1 + \erf(x) \bigr),
						\label{eqn:gradual-turnon-f-erf}
\end{equation}
where the scaled time coordinate~$x$ is defined by
\begin{equation}
x(t) = A + \frac{t - t_\initial}{B},
\end{equation}
where $t_\initial$ is the initial time of the time evolution, and
$A = -5$, and $B = 10\,M$.%%%
\footnote{%%%
	 Note that the expression~\protect\eqref{eqn:gradual-turnon-f-erf}
	 suffers from severe numerical cancellations for $x \ll 0$
	 (i.e., early in the evolution).  Instead, we use the equivalent
	 expression
	 \begin{equation}
	 f(t) = \begin{cases}
		1 - \half \erfc(x)	& \text{if $x \ge 0$}	\\*
		\half \erfc(-x)		& \text{if $x < 0$}	%%%\\*
		\end{cases}
	 \end{equation}
	 which is almost entirely free of numerical cancellation.
 	 }%%%
\end{subequations}%%%
{}  This gives $f(t_\initial) \approx 8 \,{\times}\, 10^{-13}$
(sufficiently small that the noise due to $f$ being nonzero is
below our code's overall numerical noise level from other sources)
and $f > 0.999999$ for $t > t_\initial + 83.6\,M$ (so that our
evolution equation~\eqref{eqn:box-varphi=gto-S-effective-piecewise}
approximates~\eqref{eqn:box-varphi=S-effective-piecewise} to within
one part per million for all later times).

Using the gradual turnon of the effective source, we find that
$\varphi_{\numerical,m}$ is smooth throughout a neighborhood of the
worldtube (apart from being only $C^2$ at the particle and having the
jump discontinuity~\eqref{eqn:varphi-jump-across-WT-bndry} across the
worldtube boundary) once the gradual turnon is complete and the field
configuration has had time to adjust.  In practice this initial startup
phase has a duration of $\delta t_\startup \sim 100\,M$ to $150\,M$.

%%%%%%%%%%%%%%%%%%%%%%%%%%%%%%%%%%%%%%%%

\subsection{Moving the worldtube}
\label{app:details/moving-worldtube}

\begin{figure*}[bp]
\begin{center}
\includegraphics[scale=0.80]{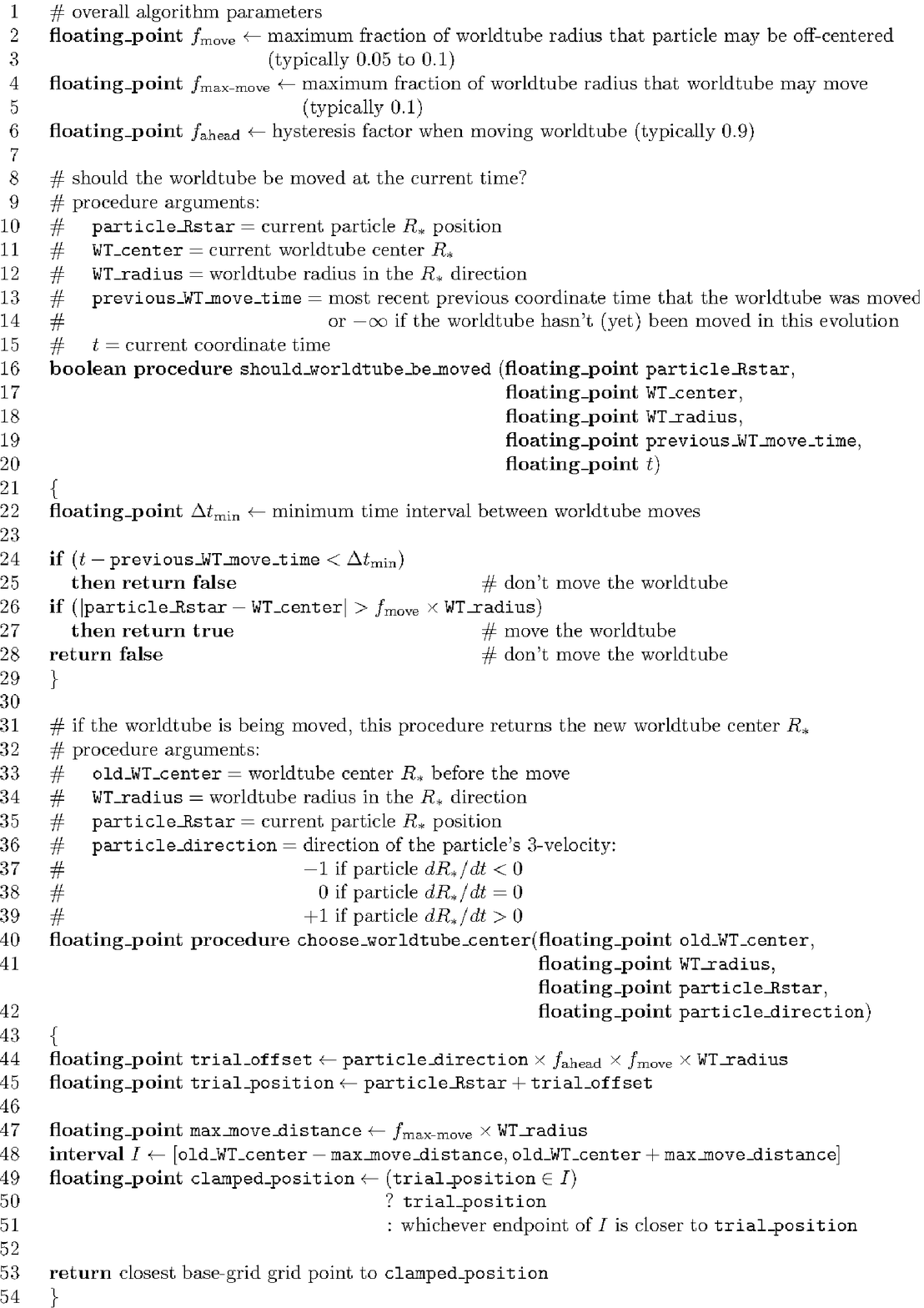}
\end{center}
\caption[Algorithm for moving the worldtube]
        {
        This figure shows our algorithm for moving the worldtube.
	The procedure \variablename{should\_worldtube\_be\_moved()}
	is run at each base-grid time step to determine whether or
	not the worldtube should be moved at the current time, and
	if so, the procedure \variablename{choose\_worldtube\_center()}
	determines the new worldtube position.
        }
\label{fig:worldtube-move-algorithm}
\end{figure*}

Figure~\ref{fig:worldtube-move-algorithm} gives our worldtube-moving
algorithm in detail.  The algorithm is run at each base-grid time step,
and has two parts: determining whether or not the worldtube should be
moved at the current time and, if it should be moved, determining the
new worldtube position.  Table~\ref{tab:configs/computational-pars-misc}
gives the parameters for this algorithm (among others).  In practice,
we find that our computed results are quite insensitive to the precise
values of these parameters
(\cf{}~Sec.~\ref{sect:results/verify-results-ne-fn(worldtube-etal)}).

%%%%%%%%%%%%%%%%%%%%
%%%%%%%%%%%%%%%%%%%%

%%%%%%%%%%%%%%%%%%%%%%%%%%%%%%%%%%%%%%%%

\subsection{Constraints on moving the worldtube early in the evolution}
\label{app:details/constraints-on-moving-WT-early-in-evolution}

When moving the worldtube, the grid-function
adjustments~\eqref{eqn:adjust-gridfns-when-moving-worldtube}
implicitly assume that $\varphi_\numerical$ has the jump
discontinuity~\eqref{eqn:varphi-jump-across-WT-bndry} across the
worldtube boundary.  While this is true once the field is in its
equilibrium configuration, it is \emph{not} true for our initial data
($\varphi_{\numerical,m} = \Pi_{\numerical,m} = 0$,
\cf{}~section~\ref{sect:theory/initial-data}).
When the evolution begins, it takes some time
(in practice~$\sim 100$ to~$150\,M$) for
the gradual turnon of the effective source
(section~\ref{app:details/gradual-turnon-of-effective-src})
to be essentially complete and for the field to
relax to an equilibrium configuration where the worldtube-boundary
jump condition~\eqref{eqn:varphi-jump-across-WT-bndry} is satisfied.

During this initial ``startup'' phase of the computation we do not
know the actual jump conditions satisfied by $\varphi_\numerical$,
so the worldtube can not be moved.  This in turn means that the initial
worldtube must encompass the entire range of motion of the particle
in~$(R_*,\theta)$ during the startup phase.  We use the following
strategy to ensure this (for a equatorial geodesic or near-geodesic
particle orbit) without requiring an excessively-large worldtube:
\begin{itemize}
\item	We first choose a particle apoastron time~$t_\apoastron$.
	Notice that the particle position $R_* = R_*(t)$ is locally
	symmetric about an (any) apoastron time.
\item	We then choose the startup time interval to be symmetric about
	$t_\apoastron$.  That is, given an estimate for the startup
	time interval's duration~$\delta t_\startup$ (typically
	$100\,M$ to $150\,M$), we begin the numerical evolution at
	$t = t_\initial = t_\apoastron - \half \delta t_\startup$,
	so that the startup phase lasts until
	$t = t_{\startup\to\main} = t_\apoastron + \half \delta t_\startup$.
	During the startup time interval the particle first moves outwards,
	then moves back inwards, reaching its initial radius again
	at $t_{\startup\to\main}$.
\item	We initially center the worldtube in~$(R_*,\theta)$ at
	(the base-grid point nearest) the average of
	$x^i_\particle(t_\initial)$ and $x^i_\particle(t_\apoastron)$,
	and choose the worldtube-moving parameters
	so that the worldtube will not be moved during the startup phase.
\item	At the end of the startup phase at $t_{\startup\to\main}$
	(when the particle returns to its initial position, now moving inwards),
	we change the worldtube-moving parameters to values which
	keep the worldtube's coordinate center within approximately
	half a coarse-grid spacing of the particle for the remainder
	of the evolution.  The first worldtube move generally occurs
	immediately after the new parameters take effect.
\end{itemize}

%%%%%%%%%%%%%%%%%%%%%%%%%%%%%%%%%%%%%%%%

\subsection{Finite differencing across the worldtube boundary}
\label{app:details/FD-across-worldtube-boundary}

We numerically implement the
jump condition~\eqref{eqn:varphi-jump-across-WT-bndry}
on the worldtube boundary in the same manner as
Barack and Golbourn~\cite{Barack-Golbourn-2007} and
Dolan and Barack~\cite{Dolan-Barack-2011}.  That is, suppose we are
finite differencing the equations at an ``evaluation'' grid point
which is inside (outside)
the worldtube, using a finite difference molecule which has a
non-empty set $S$ of input grid points which are outside (inside)
the worldtube.  Then instead of applying the finite difference
molecule to the $\varphi_{\numerical,m}$ grid function in the usual
manner, we instead copy $\varphi_{\numerical,m}$ at all the molecule
input points to a (molecule-sized) temporary
grid function~$\varphi_{\numerical,m}^\temp$,
then adjust the values of that temporary grid function to have
the same inside/outside-the-worldtube semantics as the evaluation
point via
\begin{equation}
\varphi_{\numerical,m}^\temp
	\leftarrow \varphi_{\numerical,m}^\temp \mp \varphi_{\puncture,m}
					 \label{eqn:adjusted-FD-across-WT-bndry}
\end{equation}
at each grid point in the set~$S$, then finally apply the usual
finite difference molecule to the adjusted values.

Notice that this ``adjusted finite differencing'' need only be used for
(roughly) those grid points which are within a finite-difference molecule
radius of the worldtube boundary.  (We discuss the precise choice of
those grid points in the following section.)  Because these comprise
only a tiny fraction of all grid points, the adjusted finite differencing
does not itself significantly slow the code.  Rather, its main computational
cost is the test -- at each spatial grid point at each time the
evolution equations are evaluated by the time integrator%%%
\footnote{%%%
	 This evaluation typically happens several times per
	 time step; we discuss our time-evolution algorithms in
	 detail in section~\protect\ref{app:details/numerical-time-evolution}.
	 }%%%
{} -- for whether or not adjusted finite differencing should be used.
As discussed further in the following section, this test costs only
$\sim 10$~arithmetic and logical operations, which is easily tolerable.

%%%%%%%%%%%%%%%%%%%%%%%%%%%%%%%%%%%%%%%%

\subsection{Computing the set of grid points
	    where adjusted finite differencing is needed}
\label{app:details/computing-where-adjusted-FD-needed}

\begin{figure}[bp]
\begin{center}
\includegraphics[scale=1.0]{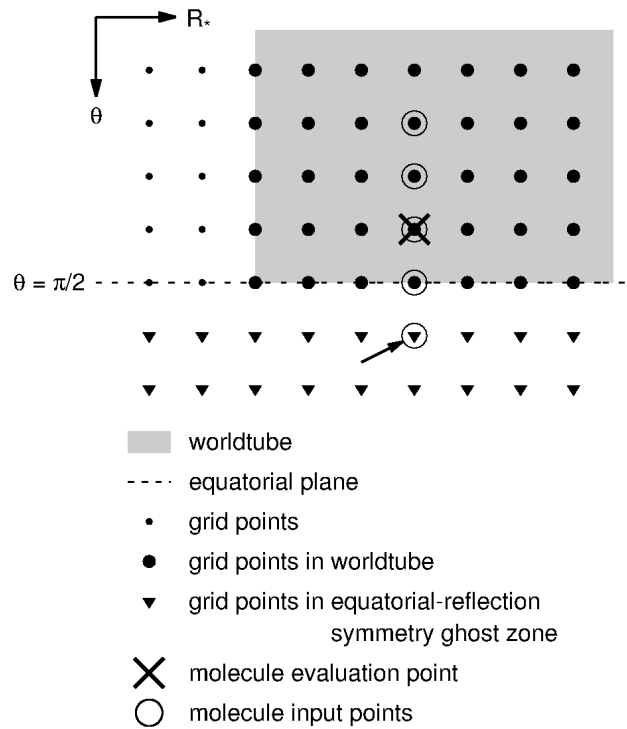}
\end{center}
\vspace{-1ex}
\caption{%%%
\label{fig:WT-equator-bndry-problem}
	This figure shows an example where the naive algorithm
	``use adjusted finite differencing at all grid points
	within a finite-difference molecule radius of the
	worldtube boundary'' would give incorrect results.
	The worldtube is shown by the shaded region; the worldtube
	boundary of interest is the ``equator'' $\theta = \pi/2$
	(shown by the dashed line).
	The molecules being considered are 5-point centered
	molecules in the $\theta$ direction, as would be used
	to approximate $\partial_\theta$ or $\partial_{\theta\theta}$;
	these molecules have radius~$2$ in the $\pm \theta$~directions.
	The molecule evaluation point shown as {\large $\mathbf\times$}
	is only 1~grid point away from the worldtube boundary,
	so the naive algorithm would say that adjusted finite
	differencing should be used for this molecule.  However,
	the arrowed point is actually within the equatorial-reflection
	symmetry ghost zone's ``reflection'' of the worldtube,
	so in terms of the
	adjustment~\protect\eqref{eqn:adjusted-FD-across-WT-bndry}
	this point has \emph{inside}-the-worldtube semantics,
	and hence adjusted finite differencing should \emph{not}
	be used for this molecule.
	}%%%
\end{figure}

In developing our numerical code we found that it was (is) much more
difficult than might be expected to compute the precise set of grid
points where adjusted finite differencing should be done.  As noted
in the previous section, this is approximately the set of all grid points
within a finite-difference molecule radius of the worldtube boundary.
However, in the presence of equatorial reflection symmetry
(cf.~section~\ref{sect:theory/BCs/equator}) this set is not quite correct:
there are certain grid points points near the intersection of the
worldtube boundary with the $\theta = \pi/2$ equatorial-reflection
symmetry plane which \emph{are} within a finite-difference molecule
radius of the worldtube boundary, but where adjusted finite differencing
should (must) \emph{not} be used.  Figure~\ref{fig:WT-equator-bndry-problem}
shows an example of this.

%%%%%%%%%%%%%%%%%%%%
%%%%%%%%%%%%%%%%%%%%

\begin{figure*}[bp]
\begin{center}
\includegraphics[scale=0.80]{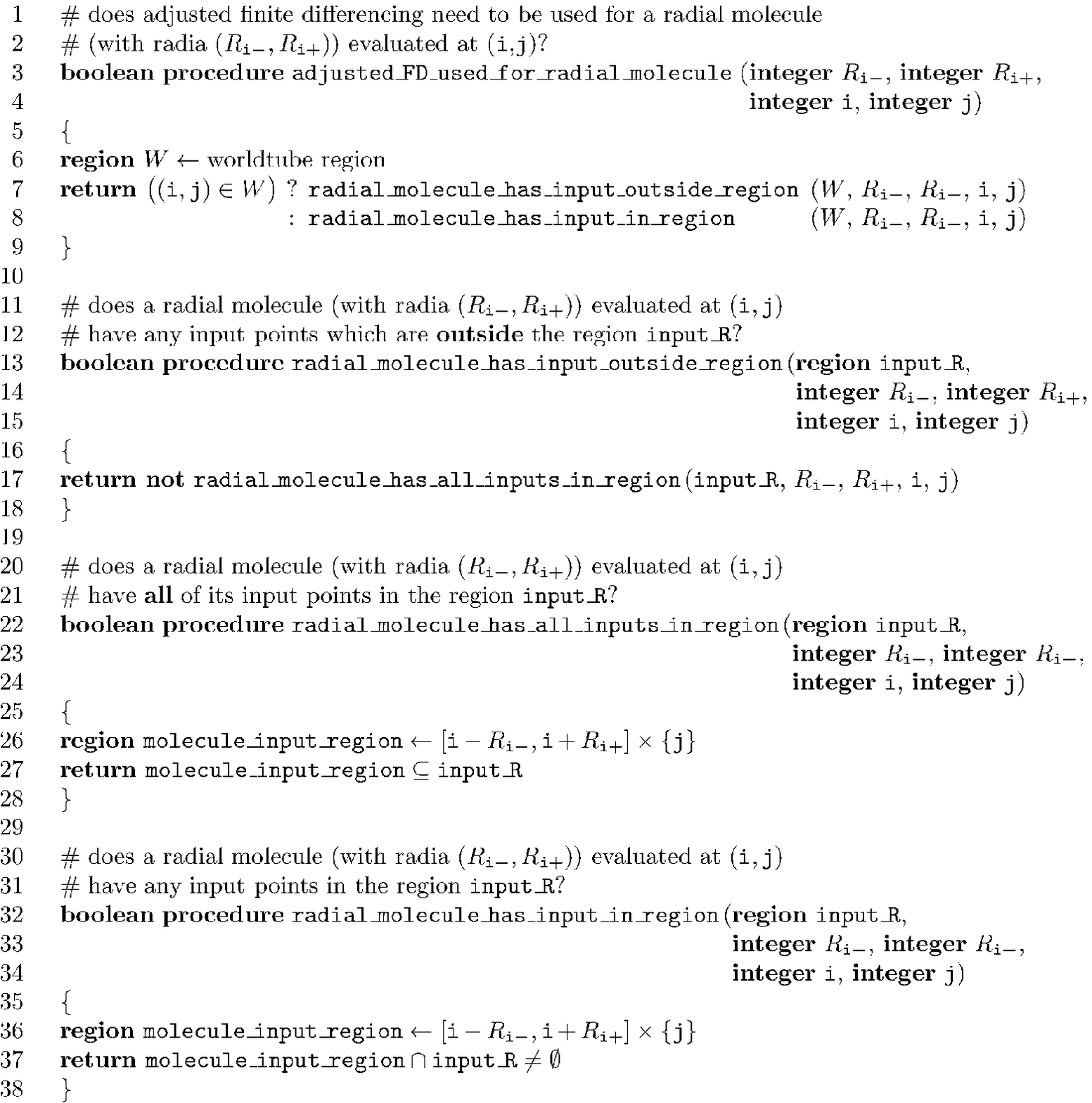}
\end{center}
\caption[Predicates for determining whether
	 the adjustment~\eqref{eqn:adjusted-FD-across-WT-bndry}
	 is needed for a specific finite difference molecule]
        {
        This figure shows our algorithm for computing the
	``should this finite difference operator be adjusted
	  via~\protect\eqref{eqn:adjusted-FD-across-WT-bndry}
	  at this grid point?''
	predicate.  Only the procedures for radial finite difference
	molecules are shown; those for angular molecules are analogous.
        }
\label{fig:adjusted-FD-predicates}
\end{figure*}

The technique we eventually adopted involves two parts:
\begin{enumerate}
\item	We build up the
	``should this finite difference operator be adjusted
	  via~\eqref{eqn:adjusted-FD-across-WT-bndry} at this grid point?''
	predicate in stages via Boolean and set operations on sets of
	grid points.  Figure~\ref{fig:adjusted-FD-predicates} shows the
	resulting algorithm.  With this approach, the semantics of each
	individual function are very clear, which allowed us to develop
	an extensive test suite to help validate the algorithm.
\item	When using equatorial-reflection symmetry, we use a
	numerical grid which spans only the ``northern hemisphere''
	$0 \le \theta \le \pi/2$~radians, \emph{but} we still consider
	the worldtube to be the full region that it would have occupied
	in the absence of equatorial-reflection symmetry, i.e.,
	(assuming an equatorial particle orbit) we take the worldtube
	to be symmetric about the equatorial plane.
\end{enumerate}

%%%%%%%%%%%%%%%%%%%%
%%%%%%%%%%%%%%%%%%%%

%%%%%%%%%%%%%%%%%%%%%%%%%%%%%%%%%%%%%%%%

\subsection{Computing the set of grid points where the puncture field is needed}
\label{app:details/computing-where-pfn-needed}

Given that the adjustment~\eqref{eqn:adjusted-FD-across-WT-bndry}
is to be applied, there remains the problem of computing
$\varphi_{\puncture,m}$ at each finite difference molecule input point.

In our evolution scheme there are (for 5-point centered molecules)
typically 9~molecule input points per evaluation point,%%%
\footnote{%%%
	 There are no $\partial_{R_* \theta}$ terms in our
	 evolution equations; if there were, then (again
	 assuming 5-point centered molecules) there would
	 be 25~molecule input points per evaluation point.%%%
	 }%%%
{} so there is a significant
performance boost from computing $\varphi_{\puncture,m}$ only once
at each grid point where it is needed, rather than 9~times if it were
(re)computed each time it is used at a molecule input point.

[Notice that -- even apart from any performance cost -- we can not
simply compute $\varphi_{\puncture,m}$ at \emph{all} spatial grid points
(at each evaluation time), because (a) $\varphi_{\puncture,m}$ diverges
at the particle, and (b) our series expansions for $\varphi_{\puncture,m}$
may be ill-behaved (e.g., they may involve division by zero) sufficiently
far from the particle (outside the worldtube).]

There are two plausible ways of ensuring that $\varphi_{\puncture,m}$
is computed at the desired set of grid points without trying to compute
it at any point where the computation would blow up:
\begin{itemize}
\item	$\varphi_{\puncture,m}$ could be stored as a ``smart grid function'',
	comprising a standard grid function of complex numbers
	($\varphi_{\puncture,m}$ values) together with an auxiliary
	grid function of Boolean ``valid'' flags recording whether or not
	$\varphi_{\puncture,m}$ has already been computed at the corresponding
	grid point at the current time.  On each access to the grid
	function, the Boolean flag would be checked, and if
	$\varphi_{\puncture,m}$ had not already been computed at that
	grid point at the current time, it would be computed, stored
	(cached) in the corresponding grid function, and the corresponding
	Boolean flag would be set to record that this grid-function
	value was now valid, so that future access could use the
	cached value.
\item	Alternatively, we could use a standard complex grid function
	to store $\varphi_{\puncture,m}$ and \emph{precompute}
	(i.e., compute before starting to compute the
	adjustment~\eqref{eqn:adjusted-FD-across-WT-bndry})
	$\varphi_{\puncture,m}$ at all the grid points where it will
	be needed, storing it in the grid function.
	(At grid points where $\varphi_{\puncture,m}$ will not be needed,
	the grid function can either be left uninitialized or be
	set to dummy values -- these will not affect the result of
	any finite differencing operation.)  The
	adjustment~\eqref{eqn:adjusted-FD-across-WT-bndry} can then
	use the stored $\varphi_{\puncture,m}$ values with no further
	validity checking needed.
\end{itemize}
We have chosen the second option as likely being simpler and more efficient.

The precomputation algorithm does not actually require an exact
computation of the
``is $\varphi_{\puncture,m}$ needed at this grid point?''
predicate: no harm is done if we precompute $\varphi_{\puncture,m}$
at some points where it will not actually be used, so the predicate
need only return true at a (possibly proper) \emph{superset} of the
actual set of grid points where $\varphi_{\puncture,m}$ is needed.
This suggests that the naive algorithm of precomputing
$\varphi_{\puncture,m}$ at every grid point that is within a
finite-difference molecule radius of the worldtube boundary might
well yield correct results.

\begin{figure*}[bp]
\begin{center}
\includegraphics[scale=0.80]{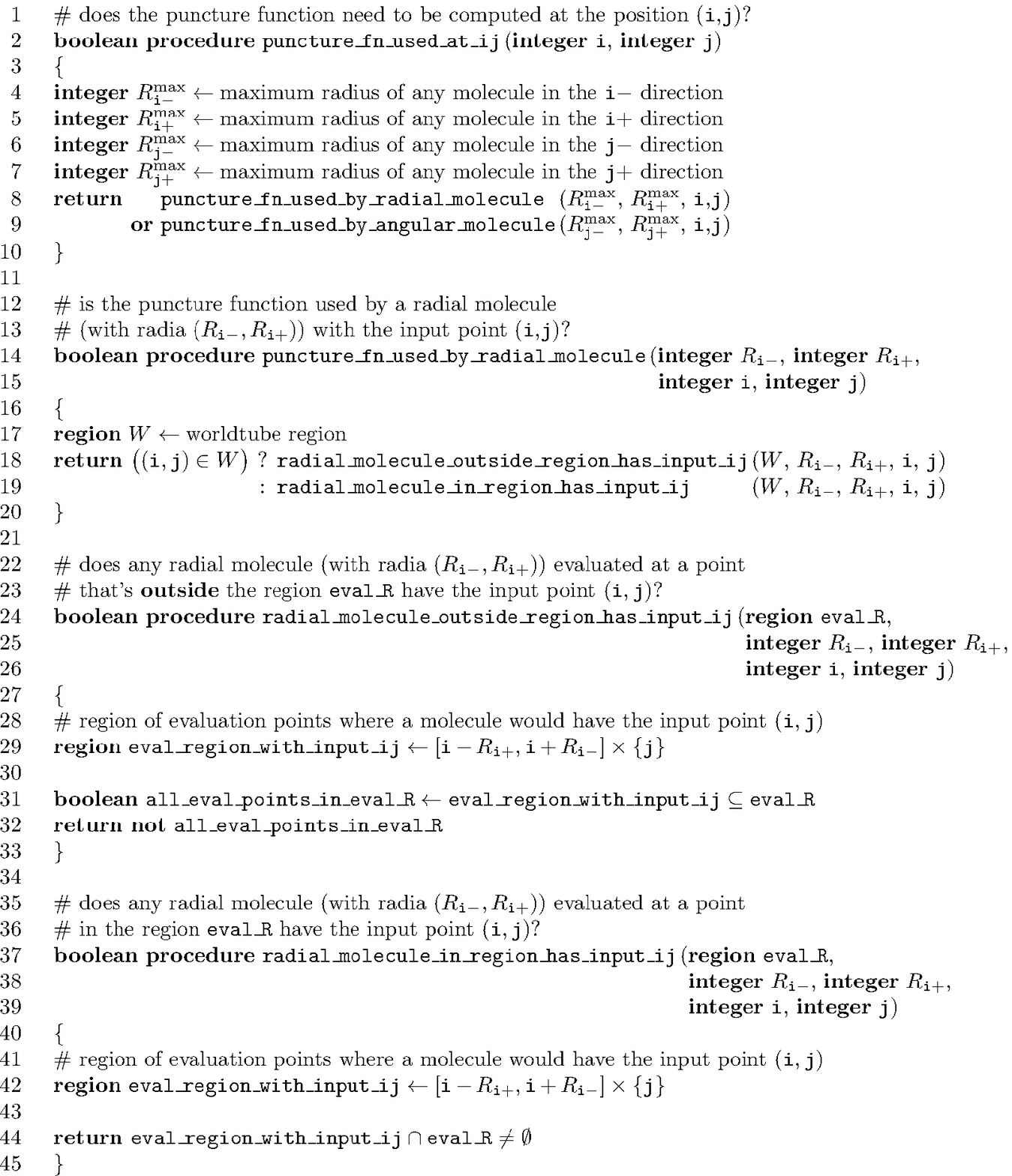}
\end{center}
\caption[Predicates for whether
	 $\varphi_{\puncture,m}$ is needed at a grid point]
        {
        This figure shows our algorithm for computing the
	``is $\varphi_{\puncture,m}$ needed at this grid point?''
	predicate.  Apart from \variablename{puncture\_fn\_used\_at\_ij()},
	only the procedures for radial finite difference molecules
	are shown; those for angular molecules are analogous.
        }
\label{fig:puncture-fn-predicates}
\end{figure*}

However, for consistency and to maximize our confidence that
equatorial-reflection symmetry cases like the one shown in
Fig.~\ref{fig:WT-equator-bndry-problem} are handled correctly, we
choose instead to build up the
``is $\varphi_{\puncture,m}$ needed at this grid point?''
predicate in stages using Boolean and set operations on sets of grid
points, in a manner very similar to our construction of the
``should adjusted finite differencing be used at this grid point?''
algorithm (Fig.~\ref{fig:adjusted-FD-predicates}).
Figure~\ref{fig:puncture-fn-predicates} shows the resulting algorithm
for determining where $\varphi_{\puncture,m}$ is needed.
Like the adjusted--finite-differencing algorithm, this algorithm
has very clear semantics for each individual function, which allowed
us to develop an extensive test suite to help validate the algorithm.

%%%%%%%%%%%%%%%%%%%%
%%%%%%%%%%%%%%%%%%%%

%%%%%%%%%%%%%%%%%%%%%%%%%%%%%%%%%%%%%%%%

\subsection{Numerical time-evolution}
\label{app:details/numerical-time-evolution}

We numerically solve the evolution system~\eqref{eqn:Pi-m-defn}
and~\eqref{eqn:evolution-main} using the method of lines, with
locally-uniform spatial grids in $(R_*,\theta)$.  We discretize all
spatial derivatives using (5-point) 4th-order centered finite
differencing, except that within a few grid points of the particle
we use the ``C2'' finite differencing scheme described in
Appendix~\ref{app:details/FD-near-particle}.
For all results reported here, we use the classical 4th-order
Runge-Kutta method for the time evolution.

We use Berger-Oliger mesh refinement
(\cite{Berger-1982,Berger-1984,Berger86,Berger-1989})
with a 2:1~refinement ratio, full subcycling in time,
and buffer zones~\cite{Schnetter-Hawley-Hawke-2004:Carpet-paper}.
We use 5th-order (6-point) Lagrange polynomial interpolation
in space and time for the coarse-to-fine Berger-Oliger interpolations.
(This requires keeping 6~time levels for all but the finest
refinement level; the latter needs only a single time level.)
For the results reported here we use 4~refinement levels
with the finest 3~refinement levels moved to follow the worldtube
(section~\ref{sect:theory/moving-worldtube} and
table~\ref{tab:grids/reflevel-sizes-and-shapes}).
In the terminology of Berger-Oliger mesh refinement our grid placement
is ``non-adaptive'', in that it does not depend on the values of the
field variables.

While our evolution scheme is stable on moderate time scales, we find
that long-time evolutions can be made much less noisy by adding
6th-order Kreiss-Oliger dissipation in the form
\begin{widetext}
\begin{subequations}
						     \label{eqn:add-dissipation}
\begin{align}
\partial_t \varphi_{\numerical,m}
	& \rightarrow \partial_t \varphi_{\numerical,m}
			  + \varepsilon
			    \bigl(
			    \Diss_{R_*}(\varphi_{\numerical,m})
			    +
			    \Diss_\theta(\varphi_{\numerical,m})
			    \bigr),
									\\*
\partial_t \Pi_{\numerical,m}
	& \rightarrow \partial_t \Pi_{\numerical,m}
			  + \varepsilon
			    \bigl(
			    \Diss_{R_*}(\Pi_{\numerical,m})
			    +
			    \Diss_\theta(\Pi_{\numerical,m})
			    \bigr),
									%%%\\*
\end{align}
\end{subequations}
where
\begin{align}
\bigl(\Diss(g)\bigr)_i
	& =	\frac{1}{64} (\Delta x)^5 \bigl(D_+^3 D_-^3 g\bigr)_i
								\nonumber\\*
	& =	\frac{1}{64 \, \Delta x}
		(g_{i-3} - 6g_{i-2} + 15g_{i-1}
		 - 20g_i + 15g_{i+1} - 6g_{i+2} +g_{i+3}).
						\label{eqn:dissipation-molecule}
									%%%\\*
\end{align}
\end{widetext}
To obtain stable evolutions we found it crucial to add dissipation
\emph{only} at those grid points where the following 3~conditions are
satisfied:
\begin{itemize}
\item	The dissipation molecule does not cross the particle,
	i.e., the the closest grid point to the particle is
	not one of the points $i{-}3$ through $i{+}3$ inclusive
	in the expression~\ref{eqn:dissipation-molecule}.
\item	The dissipation molecule does not cross the worldtube boundary,
	i.e., it does not have input points both inside and outside
	the worldtube.
\item	The dissipation molecule does not have any input points
	outside the union of the nominal grid and any symmetry
	ghost zones.  In practice this prevents dissipation from
	being added close to mesh-refinement boundaries or close
	to the horizon or $\Scri^+$ grid boundaries.
\end{itemize}

We use $\varepsilon = 0.1$ for the evolutions reported here.

Table~\ref{tab:RK4-max-stable-Courant-numbers}
shows the empirically determined Courant-Friedrichs-Lewy (CFL)
stability limit~$\nu_{\max}$~\cite{Courant-Friedrichs-Lewy-1928,%%%
Courant-Friedrichs-Lewy-1967} of our evolution scheme as a function
of~$m$.  Our code chooses the base-grid time step $\Delta t$
by first computing
$\Delta t^{(\preliminary)} = \kappa \nu_{\max} \, \Delta R_*$
(where $\kappa = 0.9$ is a safety-factor parameter), then choosing
$\Delta t$ to be the largest value~$\le \Delta t^{(\preliminary)}$
which integrally divides the output sampling time.  The time steps
for all refined grids are defined by the Berger-Oliger mesh refinement
scheme.

It is clear from table~\ref{tab:RK4-max-stable-Courant-numbers} 
that at large~$m$ our evolution scheme has a very restrictive
CFL stability limit (small~$\nu_{\max}$ and hence small $\Delta t$),
making the evolution quite inefficient.
As discussed in Appendix~\ref{app:details/IMEX-evolution-schemes},
we have experimented with an implicit-explicit (IMEX) time evolution
scheme in an attempt to alleviate the large-$m$ CFL restriction,
but thus far these experiments have not yielded larger stable Courant
numbers.  This remains a topic for further research.

%%%%%%%%%%%%%%%%%%%%
\begin{table}[bp]
\begin{center}
\begin{ruledtabular}
% *** KLUDGE ***
% RevTeX's normally stretches each table to the full column width.
% But this looks really ugly here, so we manually add space to the
% left of the left column and to the right of the right column
% so the inter-column space is more reasonable.
\begin{tabular}{@{\hspace{3.5cm}}r@{\quad}c@{\hspace{3.5cm}}}
$m$	& $\nu_{\max}$	\\
\hline %---------------------------------------------------------------
0	& 0.63\Z	\\
1	& 0.65\Z	\\
2	& 0.59\Z	\\
3	& 0.48\Z	\\
4	& 0.39\Z	\\
5	& 0.33\Z	\\
6	& 0.283		\\
7	& 0.246		\\
8	& 0.217		\\
9	& 0.195		\\
10	& 0.176		\\
11	& 0.160		\\
12	& 0.147		\\
13	& 0.136		\\
14	& 0.127		\\
15	& 0.118		\\
16	& 0.111		\\
17	& 0.105		\\
18	& 0.099		\\
19	& 0.094		\\
20	& 0.089		%%%\\
\end{tabular}
\end{ruledtabular}
\end{center}
\vspace{-1ex}
\caption{%%%
\label{tab:RK4-max-stable-Courant-numbers}
	For each $m \in [0,20]$, this table shows the
	largest Courant number $\nu = \Delta t/\Delta R_*$ for
	which we obtain a stable evolution.  For these
	stability tests we use a dissipation coefficient
	of $\varepsilon = 0.01$ and a 2-refinement-level
	grid with base resolution $\Delta R_* = M/4$;
	the stability limit depends only weakly on these
	parameters.
	}%%%
\end{table}
%%%%%%%%%%%%%%%%%%%%

%%%%%%%%%%%%%%%%%%%%%%%%%%%%%%%%%%%%%%%%

\subsection{Finite differencing near the particle}
\label{app:details/FD-near-particle}

Because $\varphi_m$ is only $C^2$ at the particle, standard finite
difference molecules do not attain their full order of accuracy near
(within roughly a molecule radius of) the particle.  One way to view
this problem is to conceptualize a finite difference molecule as being
derived by fitting a local (sliding-window) Lagrange interpolating
polynomial to the operand grid function, then differentiating that
interpolating polynomial.  This suggests that one way to obtain
more accurate finite differencing near the particle might be to
use a more general interpolating function that better represents
the actual behavior of $\varphi_m$ near the particle.

To this end, without loss of generality, we consider the finite
differencing of a (real or complex) function $g$ which is defined
on a suitable neighborhood of the origin on the real line, using a
1-dimensional numerical grid with grid points at integer coordinates.
Without loss of generality, we assume the particle to be at the
(known) position $p \in [0,\thalf]$.  We consider the
piecewise-polynomial interpolating function
\begin{align}
I(x) = {}
	& a_0 + a_1 (x-p) + a_2 (x-p)^2 			\nonumber\\
	& + \begin{cases}
	    b_3 (x-p)^3 + b_4 (x-p)^4 + b_5 (x-p)^5	\hspace*{-1em}
					&			\\
					& \text{if $x \le 0$}	\\
	    c_3 (x-p)^3 + c_4 (x-p)^4 + c_5 (x-p)^5	\hspace*{-1em}
					&			\\
					& \text{if $x > 0$}.	%%%\\
	    \end{cases}
						 \label{eqn:C2-interpolating-fn}
\end{align}
The 9~coefficients $\{a_0,a_1,a_2, \, b_3,b_4,b_5, \, c_3,c_4,c_5\}$
can be uniquely determined (as functions of the parameter~$p$) by
requiring $I$ to match the specified function $g$ at the 9~adjacent
grid points in the range $-4 \le x \le 4$.
$I$, $dI/dx$, and $d^2I/dx^2$ can then be evaluated at any
desired position to obtain finite difference approximations to $g$,
$dg/dx$, and $d^2g/dx^2$ respectively.  Using a symbolic algebra
system, these finite-difference operators can be written as linear
combinations of the values of $g$ at the grid points, with coefficients
depending only on~$p$ and the evaluation position.

\begin{figure*}[bp]
\begin{center}
\includegraphics[scale=1.0]{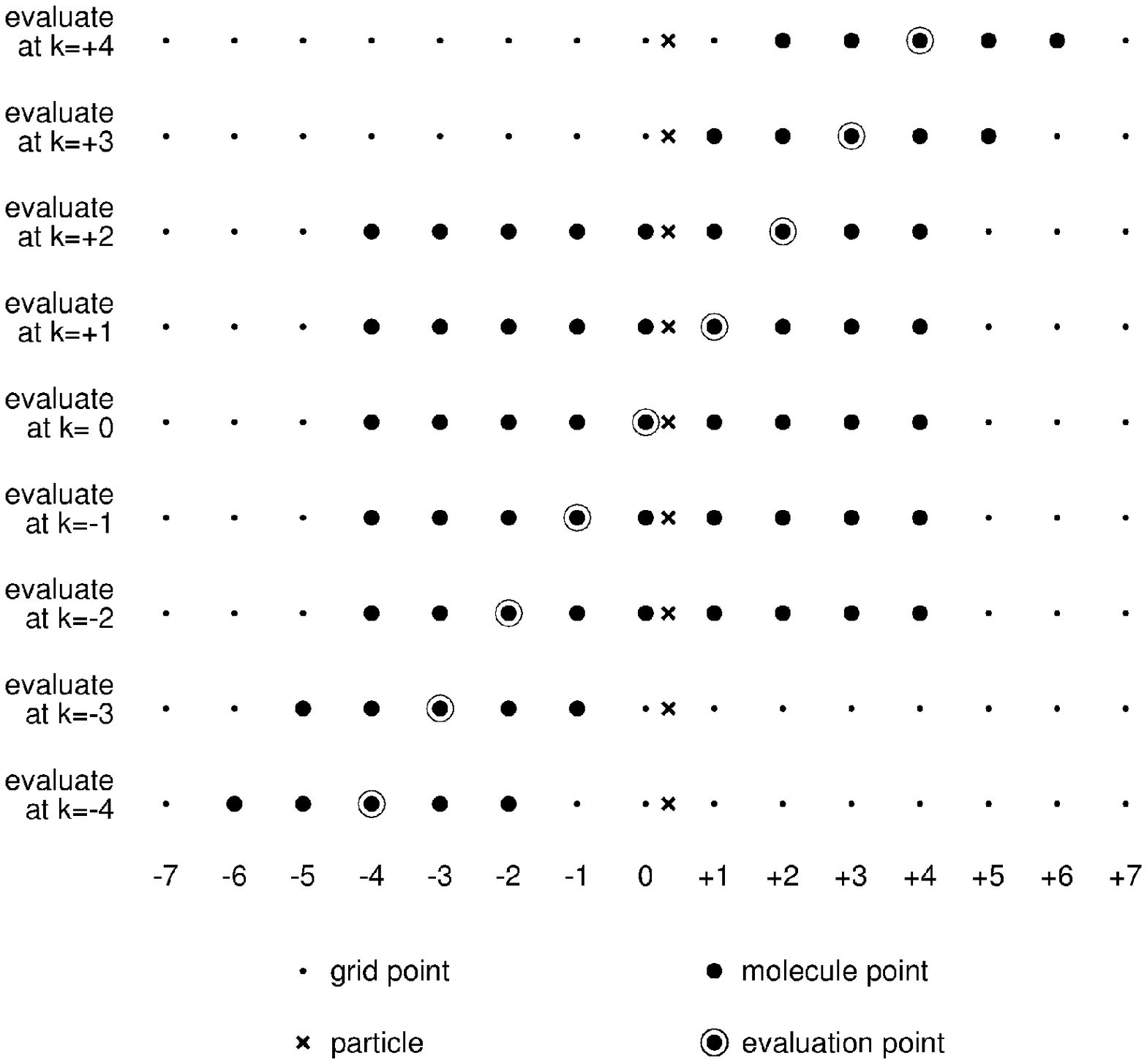}
\end{center}
\vspace{-1ex}
\caption{%%%
\label{fig:C2-FD-near-particle}
	This figure shows our ``C2'' spatial finite differencing scheme
	for use near the particle.  Each row of the diagram shows
	the finite difference molecule used for a different evaluation
	point.  ``\ijcoord{k}''~refers to the integer grid coordinate
	in the $x$~direction, with the origin set so that
	$\ijcoord{k} \,{=}\, 0$ is the grid point closest
	to the particle.  We use molecules based on the piecewise-polynomial
	interpolating function~\protect\eqref{eqn:C2-interpolating-fn}
	for evaluations points $-2 \le \ijcoord{k} \le +2$
	and standard (5-point) centered 4th-order molecules
	at all other evaluation points.
	}%%%
\end{figure*}

Figure~\ref{fig:C2-FD-near-particle} shows how we use these ``C2''
finite-difference operators at various grid points near the particle.
In the present work the particle is always in the background Kerr
spacetime's equatorial plane, and we always place a
$\theta = \text{constant}$ row of grid points on the equatorial plane.
Considering the numerical grid in 2~dimensions $(R_*,\theta)$, with
corresponding integer grid coordinates $(\ijcoord{i},\ijcoord{j})$,
suppose that the closest grid point to the particle is at
$(\ijcoord{i}_\particle,\ijcoord{j}_\equator)$.  Then we
use the C2~scheme for $\partial_{R_*}$ and $\partial_{R_* R_*}$~derivatives
evaluated at grid points on the equator (i.e., for grid points with
$\ijcoord{j} = \ijcoord{j}_\equator$) and $\ijcoord{i}$
near~$\ijcoord{i}_\particle$, in the manner shown in the figure.
We also use this scheme in the $\ijcoord{j}$~direction for $\partial_\theta$
and $\partial_{\theta\theta}$~derivatives evaluated at grid points
with $\ijcoord{i} = \ijcoord{i}_\particle$ and $\ijcoord{j}$~near
the equator.  We use standard (5-point) centered 4th-order molecules
at all other grid points.

%%%%%%%%%%%%%%%%%%%%
%%%%%%%%%%%%%%%%%%%%

As discussed in
section~\ref{sect:theory/computing-self-force-from-evolved-fields},
we also use the interpolating function~$I$ directly in computing
the self-force.

Overall, we find that switching from using 4th-order centered spatial
finite differencing everywhere to using the C2~finite-difference operators
near the particle reduces the noise level in the computed self-force
by about a factor of~$2$ to~$3$.

We also experimented with a more general interpolating function
\begin{widetext}
\begin{equation}
I(x) =	a_0 + a_1 (x-p) + a_2 (x-p)^2
	+ \begin{cases}
	  \phantom{+}
	    b_3 (x-p)^3 + b_{3\ell} (x-p)^3 \log\bigl(K (x-p)^2 \bigr)
					&			\\
	  + b_4 (x-p)^4 + b_{4\ell} (x-p)^4 \log\bigl(K (x-p)^2 \bigr)
					&			\\
	  + b_5 (x-p)^5 + b_{5\ell} (x-p)^5 \log\bigl(K (x-p)^2 \bigr)
					& \text{if $x \le 0$}	\\[1ex]
	  \phantom{+}
	    c_3 (x-p)^3 + c_{3\ell} (x-p)^3 \log\bigl(K (x-p)^2 \bigr)
					&			\\
	  + c_4 (x-p)^4 + c_{4\ell} (x-p)^4 \log\bigl(K (x-p)^2 \bigr)
					&			\\
	  + c_5 (x-p)^5 + c_{5\ell} (x-p)^5 \log\bigl(K (x-p)^2 \bigr)
					& \text{if $x > 0$,}	%%%\\
	  \end{cases}
					       \label{eqn:C2log-interpolatin-fn}
\end{equation}
\end{widetext}
where $K$~is a scaling constant and the 15~coefficients
$\{a_i, b_i, b_{i\ell}, c_i, c_{i\ell}\}$ are determined by solving
a system of 15~linear equations using the values of~$g$ at 15~adjacent
grid points.  (A purely symbolic solution to this linear system proved
impractical, but it is easy to solve numerically.  This needs to be
done once for each choice of the grid spacing and $K$.)  However,
we were not able to obtain stable evolutions with this scheme.

%%%%%%%%%%%%%%%%%%%%%%%%%%%%%%%%%%%%%%%%%%%%%%%%%%%%%%%%%%%%%%%%%%%%%%%%%%%%%%%%

\subsection{Implicit-explicit ({IMEX}) evolution schemes}
\label{app:details/IMEX-evolution-schemes}

As discussed in Appendix~\ref{app:details/numerical-time-evolution},
when using an explicit (Runge-Kutta) time evolution scheme we find that
the CFL stability limit~\cite{Courant-Friedrichs-Lewy-1928,%%%
Courant-Friedrichs-Lewy-1967} is very restrictive for large~$m$,
with the largest stable Courant number being
approximately proportional to~$1/m$.
Examination of the derivative structure of our evolution system, together
with an eigenvalue analysis kindly performed by I.~Hinder, suggests
that $m^2$ coefficient in the $\varphi$~term in the scalar wave
operator~\eqref{eqn:box-m-coordinate-form} may be a major contributor
to the large-$m$ time-step restriction.

We thus consider the use of an implicit time evolution scheme.  More
precisely, we consider the the use of an implicit-explicit (IMEX) time
evolution scheme.  There is a large literature on these schemes; see, for
example,~\cite{Ascher-Ruuth-Wetton-1995:IMEX-methods-for-time-dependent-PDEs,%%%
Ascher-Ruuth-Spiteri-1997:IMEX-methods-for-time-dependent-PDEs,%%%
Pareschi-Russo-2000:IMEX-schemes-for-stiff-PDEs,%%%
Pareschi-Russo-2005:IMEX-RK-schemes-for-hyperbolic-PDEs,%%%
Boscarino-2007,Boscarino-2009,Boscarino-Russo-2009}.
The basic concept of an IMEX scheme is to partition the right-hand-side
function into explicit and implicit parts,
\begin{equation}
\dot{\uu} = \FF(\uu,t) + \GG(\uu,t),
					    \label{eqn:generic-ODE-in-IMEX-form}
\end{equation}
where $\uu$ is the state vector, then treat $\FF$ explicitly and
$\GG$ implicitly.  For our application, we will place all the spatial
derivatives into the explicit term, thus avoiding having to solve an
elliptic system at each time step.

We have chosen the scheme proposed by Boscarino~\cite{Boscarino-2009}
(in particular, his BHR(5,5,3) scheme, variant~2) as being efficient,
relatively easy to implement, and having good accuracy (3rd~order overall)
without the ``order reduction'' problems of many other schemes.%%%
\footnote{%%%
	 We warn the reader of the following typographical
	 errors in~\protect\cite{Boscarino-2009}:
	 \begin{itemize}
	 \item	Equation~(4) should read
		\begin{align}
		U^i = U_n
			& + h \sum_{j=1}^{i-1}
			      \tilde{a}_{ij}
			      F \left( t_n + \tilde{c}_j h, U^j \right)
								\nonumber\\*
			& + h \sum_{j=1}^i
			      a_{ij} \frac{1}{\varepsilon}
			      G \left( t_n + c_j h, U^j \right)
									%%%\\*
		\end{align}
	 \item	In the appendix, in the left (explicit) Butcher tableau,
		the $b$~coefficients should read
		$[b_1\quad{}0\quad{}b_3\quad{}b_4\quad{}\gamma]$.
	 \item	In the appendix, in the right (implicit) Butcher tableau,
		the coefficients for the last stage (the 5th~row of
		the matrix) should be identical to the $b$~coefficients,
		i.e., they should read
		$[b_1\quad{}0\quad{}b_3\quad{}b_4\quad{}\gamma]$.
	 \end{itemize}
	 }%%%

We write a generic implicit-explicit Runge-Kutta scheme for the
ODE~\eqref{eqn:generic-ODE-in-IMEX-form} as%%%
\footnote{%%%
	 Our notation in the remainder of this appendix
	 is somewhat different from Boscarino's; notably,
	 we swap the tilde and non-tilde coefficients.
	 }%%%
\begin{equation}
\uu^{(n+1)} = \uu^{(n)} + h \sum_{i=1}^        s          b _i        \kk _i
			+ h \sum_{i=1}^{\tilde{s}} \tilde{b}_i \tilde{\kk}_i,
\end{equation}
where $h$~is the time step, superscripts ${}^{(n)}$ and ${}^{(n+1)}$
refer to time levels, subscripts refer to Runge-Kutta stages numbered
$1,\dots,s$, and the Runge-Kutta stages are given by
\begin{subequations}
					    \label{eqn:IMEX/RK-implicit-generic}
\begin{align}
\kk_i	& =	F ( \uu_i, t^{(n)} + h c_i ),
									\\
\tilde{\kk}_i
	& =	G ( \uu_i, t^{(n)} + h c_i ),
									\\
\uu_i
	& =	\uu^{(n)}
		+ h \sum_{j <   i}        d _{ij}        \kk _j
		+ h \sum_{j \le i} \tilde{d}_{ij} \tilde{\kk}_j,
									%%%\\
\end{align}
\end{subequations}
with the coefficients $\{b_i\}$, $\{\tilde{b}_i\}$, $\{c_i\}$,
$\{d_{ij}\}$, and~$\{\tilde{d}_{ij}\}$

For example (eliding the evaluation times for clarity), the first
few stages are
\begin{subequations}
\begin{align}
\uu_1	& =	\uu^{(n)}
		+ h \tilde{d}_{11} \GG(\uu_1),
									\\
\uu_2	& =	\uu^{(n)}
		+ h        d _{21} \FF(\uu_1)
		+ h \tilde{d}_{21} \GG(\uu_1) + h \tilde{d}_{22} \GG(\uu_2),
									\\
\uu_3	& =	\uu^{(n)}
		+ h        d _{31} \FF(\uu_1) + h        d _{32} \FF(\uu_2) \nonumber \\ & \qquad
		+ h \tilde{d}_{31} \GG(\uu_1) + h \tilde{d}_{32} \GG(\uu_2)
					      + h \tilde{d}_{33} \GG(\uu_3).
									%%%\\
\end{align}
\end{subequations}

To solve the implicit equations~\eqref{eqn:IMEX/RK-implicit-generic},
we observe that our state vector~$\uu$ is of the form
\begin{equation}
\uu = \left(
    \begin{array}{c}
    \varphi	\\
    \Pi		%%%\\
    \end{array}
    \right),
						  \label{eqn:IMEX/u-2-component}
\end{equation}
so we can write
\begin{equation}
\FF \left(
    \begin{array}{c}
    \varphi	\\
    \Pi		%%%\\
    \end{array}
    \right)
	= \left(
	  \begin{array}{c}
	  y(\varphi,\Pi)	\\
	  z(\varphi,\Pi)	%%%\\
	  \end{array}
	  \right)
						  \label{eqn:IMEX/F-2-component}
\end{equation}
and
\begin{equation}
\GG \left(
    \begin{array}{c}
    \varphi	\\
    \Pi		%%%\\
    \end{array}
    \right)
	= \left(
	  \begin{array}{c}
	  0				\\
	  \tilde{z}(\varphi,\Pi)	%%%\\
	  \end{array}
	  \right)
	= \left(
	  \begin{array}{c}
	  0				\\
	  \alpha \varphi + \beta \Pi	%%%\\
	  \end{array}
	  \right)
						  \label{eqn:IMEX/G-2-component}
\end{equation}
with known coefficients $\alpha$ and $\beta$.
(This use of $\alpha$ is unrelated to its use as a tail-series
exponent in section~\ref{sect:theory/tail-series}.)

The 2-component $\FF$ function~\eqref{eqn:IMEX/F-2-component}
includes all the main evolution equations~\eqref{eqn:Pi-m-defn},
\eqref{eqn:evolution-main}, and~\eqref{eqn:add-dissipation},
as well as all spatial boundary conditions.  Evaluating $F$ requires
computing (or retrieving from a cache) the 2-D puncture field and
effective source.%%%
\footnote{%%%
         Unfortunately, in all IMEX schemes of which we
         are aware it is \emph{not} the case that there
         are repeated evaluations of $\FF$ with different
	 state vectors at the same time coordinate, so
	 there is no reuse possible of the puncture field
	 and effective source from one evaluation to the
	 next.  In contrast (as noted in
section~\protect\ref{sect-discussion/possible-improvements/efficiency}),
         with the classical RK4~scheme 50\% of evaluations
	 are repeated in this way, so -- since the effective
	 source computation dominates the code's overall
	 running time -- there is an easy factor-of-two
	 saving in computational cost by caching and reusing
	 the effective source from one evaluation to the next
	 if the evaluation time is unchanged.%%%
	 }%%%

We have considered a number of possible choices for precisely which
terms from the main evolution equation~\eqref{eqn:evolution-main} should
be treated implicitly (i.e., put into~$G$).
Conceptually, we have $\alpha = \alpha_1 + \alpha_2 + \alpha_3$, where
\begin{subequations}
\begin{align}
\alpha_1
	& =	\text{$0$ or
		      $\displaystyle
		       - \frac{2\Delta}{r (r^2+M^2 \tilde{a}^2)^2}
			 \left( M - \frac{M^2 \tilde{a}^2}{r} \right)$},
									\\
\alpha_2
	& =	\text{$0$ or
		      $\displaystyle
		       - \frac{m^2 \Delta}{(r^2+M^2 \tilde{a}^2)^2 \sin^2 \theta}$},
									\\
\alpha_3
	& =	\text{$0$ or
		      $\displaystyle
		       - i\frac{2mM\tilde{a}\Delta}{r (r^2+M^2 \tilde{a}^2)^2}$},
									\\
\noalign{\hbox{and}}
\beta	& =	\text{$0$ or
		      $\displaystyle
		       - i \frac{4mM^2 \tilde{a}r}{(r^2+M^2 \tilde{a}^2)^2}$}
									%%%\\
\end{align}
\end{subequations}
modified by the compactification
transformation~\eqref{eqn:coeffs-compactified/interior}, together with
the spatial boundary conditions.  This gives 16~possible variant schemes,
depending on which subset of $\{\alpha_1, \alpha_2, \alpha_3, \beta\}$
is nonzero (treated implicitly).  For each of these variants,
\begin{itemize}
\item	$\GG$ is linear in $\varphi$ and $\Pi$ at each grid point,
\item	$\GG$ may be evaluated independently at each grid point, and
\item	this evaluation does \emph{not} require computing
	the 2-D puncture field or effective source.
\end{itemize}
Together, these properties make the scheme efficient and
relatively easy to implement.

Substituting the 2-component $\uu$, $\FF$, and~$\GG$
functions~\eqref{eqn:IMEX/u-2-component}, \eqref{eqn:IMEX/F-2-component},
and~\eqref{eqn:IMEX/G-2-component} into the implicit Runge-Kutta
equations~\eqref{eqn:IMEX/RK-implicit-generic}, we have
\begin{align}
\uu_i
	= \left(
	  \begin{array}{c}
	  \varphi_i	\\
	  \Pi_i		%%%\\
	  \end{array}
	  \right)
		&=       \left(
		  \begin{array}{c}
		  \varphi^{(n)}		\\
		  \Pi    ^{(n)}		%%%\\
		  \end{array}
		  \right)
		  + h \sum_{j < i} d_{ij}
				   \left(
				   \begin{array}{c}
				   y_j		\\
				   z_j		%%%\\
				   \end{array}
				   \right) \nonumber \\ &
		  + h \sum_{j < i} \tilde{d}_{ij}
				   \left(
				   \begin{array}{c}
				   0		\\
				   \tilde{z}_j	%%%\\
				   \end{array}
				   \right)
		  + h \tilde{d}_{ii}
		      \left(
		      \begin{array}{c}
		      0					\\
		      \alpha \varphi_i + \beta \Pi_i	%%%\\
		      \end{array}
		      \right).
\end{align}
We solve this equation at each Runge-Kutta stage by first computing
\begin{subequations}
\begin{equation}
\varphi_i
	= \varphi^{(n)} + h \sum_{j < i} d_{ij} y_j
\end{equation}
and then computing
\begin{equation}
\Pi_i	= \frac{
	       \displaystyle
	       \Pi^{(n)}
	       + h \sum_{j < i}        d _{ij}        z _j
	       + h \sum_{j < i} \tilde{d}_{ij} \tilde{z}_j
	       + h \tilde{d}_{ii} \alpha \varphi_i
	       }
	       {1 - h \tilde{d}_{ii} \beta}.
\end{equation}
\end{subequations}

We have implemented these 16~variant schemes, but unfortunately we
find that all of them have CFL stability limits which are (to within the
$\sim 1\%$~accuracy of our trial-and-error estimation of the stability
limit on test problems) identical to those of the classical RK4~scheme
(table~\ref{tab:RK4-max-stable-Courant-numbers}).
Since the RK4~scheme is simpler and offers a factor-of-two overall speedup
by caching and reusing the effective source at repated evaluation times,
we use it for all the computations presented in this paper.  We hope
to further investigate different partitionings of the right-hand-side
function between $\FF$ and $\GG$ in the future in the hopes of alleviating
the large-$m$ time-step restriction.

%%%%%%%%%%%%%%%%%%%%%%%%%%%%%%%%%%%%%%%%%%%%%%%%%%%%%%%%%%%%%%%%%%%%%%%%%%%%%%%%

%%\bibliography{bw-references,jt-new,aei-references,einsteintoolkit}{}

\begin{thebibliography}{138}%
\makeatletter
\providecommand \@ifxundefined [1]{%
 \@ifx{#1\undefined}
}%
\providecommand \@ifnum [1]{%
 \ifnum #1\expandafter \@firstoftwo
 \else \expandafter \@secondoftwo
 \fi
}%
\providecommand \@ifx [1]{%
 \ifx #1\expandafter \@firstoftwo
 \else \expandafter \@secondoftwo
 \fi
}%
\providecommand \natexlab [1]{#1}%
\providecommand \enquote  [1]{``#1''}%
\providecommand \bibnamefont  [1]{#1}%
\providecommand \bibfnamefont [1]{#1}%
\providecommand \citenamefont [1]{#1}%
\providecommand \href@noop [0]{\@secondoftwo}%
\providecommand \href [0]{\begingroup \@sanitize@url \@href}%
\providecommand \@href[1]{\@@startlink{#1}\@@href}%
\providecommand \@@href[1]{\endgroup#1\@@endlink}%
\providecommand \@sanitize@url [0]{\catcode `\\12\catcode `\$12\catcode
  `\&12\catcode `\#12\catcode `\^12\catcode `\_12\catcode `\%12\relax}%
\providecommand \@@startlink[1]{}%
\providecommand \@@endlink[0]{}%
\providecommand \url  [0]{\begingroup\@sanitize@url \@url }%
\providecommand \@url [1]{\endgroup\@href {#1}{\urlprefix }}%
\providecommand \urlprefix  [0]{URL }%
\providecommand \Eprint [0]{\href }%
\providecommand \doibase [0]{http://dx.doi.org/}%
\providecommand \selectlanguage [0]{\@gobble}%
\providecommand \bibinfo  [0]{\@secondoftwo}%
\providecommand \bibfield  [0]{\@secondoftwo}%
\providecommand \translation [1]{[#1]}%
\providecommand \BibitemOpen [0]{}%
\providecommand \bibitemStop [0]{}%
\providecommand \bibitemNoStop [0]{.\EOS\space}%
\providecommand \EOS [0]{\spacefactor3000\relax}%
\providecommand \BibitemShut  [1]{\csname bibitem#1\endcsname}%
\let\auto@bib@innerbib\@empty
%</preamble>
\bibitem [{\citenamefont {Gair}\ \emph {et~al.}(2004)\citenamefont {Gair},
  \citenamefont {Barack}, \citenamefont {Creighton}, \citenamefont {Cutler},
  \citenamefont {Larson}, \citenamefont {Phinney},\ and\ \citenamefont
  {Vallisneri}}]{Gair-etal-2004:LISA-EMRI-event-rates}%
  \BibitemOpen
  \bibfield  {author} {\bibinfo {author} {\bibfnamefont {J.~R.}\ \bibnamefont
  {Gair}}, \bibinfo {author} {\bibfnamefont {L.}~\bibnamefont {Barack}},
  \bibinfo {author} {\bibfnamefont {T.}~\bibnamefont {Creighton}}, \bibinfo
  {author} {\bibfnamefont {C.}~\bibnamefont {Cutler}}, \bibinfo {author}
  {\bibfnamefont {S.~L.}\ \bibnamefont {Larson}}, \bibinfo {author}
  {\bibfnamefont {E.~S.}\ \bibnamefont {Phinney}}, \ and\ \bibinfo {author}
  {\bibfnamefont {M.}~\bibnamefont {Vallisneri}},\ }\href {\doibase
  10.1088/0264-9381/21/20/003} {\bibfield  {journal} {\bibinfo  {journal}
  {Class. Quant. Grav.}\ }\textbf {\bibinfo {volume} {21}},\ \bibinfo {pages}
  {S1595} (\bibinfo {year} {2004})},\ \Eprint
  {http://arxiv.org/abs/gr-qc/0405137} {gr-qc/0405137} \BibitemShut {NoStop}%
%%CITATION = GR-QC 0405137;%%
\bibitem [{\citenamefont {Barack}\ and\ \citenamefont
  {Cutler}(2004)}]{Barack-Cutler-2004}%
  \BibitemOpen
  \bibfield  {author} {\bibinfo {author} {\bibfnamefont {L.}~\bibnamefont
  {Barack}}\ and\ \bibinfo {author} {\bibfnamefont {C.}~\bibnamefont
  {Cutler}},\ }\href {\doibase 10.1103/PhysRevD.69.082005} {\bibfield
  {journal} {\bibinfo  {journal} {Phys. Rev. D}\ }\textbf {\bibinfo {volume}
  {69}},\ \bibinfo {pages} {082005} (\bibinfo {year} {2004})},\ \Eprint
  {http://arxiv.org/abs/gr-qc/0310125} {gr-qc/0310125} \BibitemShut {NoStop}%
\bibitem [{\citenamefont {Amaro-Seoane}\ \emph {et~al.}(2007)\citenamefont
  {Amaro-Seoane}, \citenamefont {Gair}, \citenamefont {Freitag}, \citenamefont
  {Miller}, \citenamefont {Mandel}, \citenamefont {Cutler},\ and\ \citenamefont
  {Babak}}]{Amaro-Seoane-etal-2007:LISA-IMRI-and-EMRI-review}%
  \BibitemOpen
  \bibfield  {author} {\bibinfo {author} {\bibfnamefont {P.}~\bibnamefont
  {Amaro-Seoane}}, \bibinfo {author} {\bibfnamefont {J.~R.}\ \bibnamefont
  {Gair}}, \bibinfo {author} {\bibfnamefont {M.}~\bibnamefont {Freitag}},
  \bibinfo {author} {\bibfnamefont {M.~C.}\ \bibnamefont {Miller}}, \bibinfo
  {author} {\bibfnamefont {I.}~\bibnamefont {Mandel}}, \bibinfo {author}
  {\bibfnamefont {C.~J.}\ \bibnamefont {Cutler}}, \ and\ \bibinfo {author}
  {\bibfnamefont {S.}~\bibnamefont {Babak}},\ }\href {\doibase
  10.1088/0264-9381/24/17/R01} {\bibfield  {journal} {\bibinfo  {journal}
  {Class. Quant. Grav.}\ }\textbf {\bibinfo {volume} {24}},\ \bibinfo {pages}
  {R113} (\bibinfo {year} {2007})},\ \Eprint
  {http://arxiv.org/abs/astro-ph/0703495} {astro-ph/0703495} \BibitemShut
  {NoStop}%
\bibitem [{\citenamefont {Gair}(2009)}]{Gair-2009:LISA-EMRI-event-rates}%
  \BibitemOpen
  \bibfield  {author} {\bibinfo {author} {\bibfnamefont {J.~R.}\ \bibnamefont
  {Gair}},\ }\href {\doibase 10.1088/0264-9381/26/9/094034} {\bibfield
  {journal} {\bibinfo  {journal} {Class. Quant. Grav.}\ }\textbf {\bibinfo
  {volume} {26}},\ \bibinfo {pages} {094034} (\bibinfo {year} {2009})},\
  \Eprint {http://arxiv.org/abs/arXiv:0811.0188} {arXiv:0811.0188} \BibitemShut
  {NoStop}%
\bibitem [{\citenamefont {Damour}(1987)}]{Damour-in-Hawking-Israel-1987}%
  \BibitemOpen
  \bibfield  {author} {\bibinfo {author} {\bibfnamefont {T.}~\bibnamefont
  {Damour}},\ }in\ \href@noop {} {\emph {\bibinfo {booktitle} {Three Hundred
  Years of Gravitation}}},\ \bibinfo {editor} {edited by\ \bibinfo {editor}
  {\bibfnamefont {S.~W.}\ \bibnamefont {Hawking}}\ and\ \bibinfo {editor}
  {\bibfnamefont {W.}~\bibnamefont {Israel}}}\ (\bibinfo  {publisher}
  {Cambridge University Press},\ \bibinfo {address} {Cambridge, England},\
  \bibinfo {year} {1987})\ Chap.~\bibinfo {chapter} {6}, pp.\ \bibinfo {pages}
  {128--198}\BibitemShut {NoStop}%
\bibitem [{\citenamefont {Blanchet}(2014)}]{Blanchet-2014-living-review}%
  \BibitemOpen
  \bibfield  {author} {\bibinfo {author} {\bibfnamefont {L.}~\bibnamefont
  {Blanchet}},\ }\href {\doibase 10.12942/lrr-2014-2} {\bibfield  {journal}
  {\bibinfo  {journal} {Living Reviews in Relativity}\ }\textbf {\bibinfo
  {volume} {17}},\ \bibinfo {pages} {2} (\bibinfo {year} {2014})}\BibitemShut
  {NoStop}%
\bibitem [{\citenamefont {Futamase}\ and\ \citenamefont
  {Itoh}(2007)}]{Futamase-Itoh-2007:PN-review}%
  \BibitemOpen
  \bibfield  {author} {\bibinfo {author} {\bibfnamefont {T.}~\bibnamefont
  {Futamase}}\ and\ \bibinfo {author} {\bibfnamefont {Y.}~\bibnamefont
  {Itoh}},\ }\href {\doibase 10.12942/lrr-2007-2} {\bibfield  {journal}
  {\bibinfo  {journal} {Living Reviews in Relativity}\ }\textbf {\bibinfo
  {volume} {10}},\ \bibinfo {pages} {2} (\bibinfo {year} {2007})}\BibitemShut
  {NoStop}%
\bibitem [{\citenamefont {Blanchet}(2011)}]{Blanchet-2009:PN-review}%
  \BibitemOpen
  \bibfield  {author} {\bibinfo {author} {\bibfnamefont {L.}~\bibnamefont
  {Blanchet}},\ }in\ \href@noop {} {\emph {\bibinfo {booktitle} {Mass and
  Motion in General Relativity}}},\ \bibinfo {series} {Fundamental Theories of
  Physics}, Vol.\ \bibinfo {volume} {162},\ \bibinfo {editor} {edited by\
  \bibinfo {editor} {\bibfnamefont {L.}~\bibnamefont {Blanchet}}, \bibinfo
  {editor} {\bibfnamefont {A.}~\bibnamefont {Spallicci}}, \ and\ \bibinfo
  {editor} {\bibfnamefont {B.~F.}\ \bibnamefont {Whiting}}}\ (\bibinfo
  {publisher} {Springer-Verlag},\ \bibinfo {address} {Berlin},\ \bibinfo {year}
  {2011})\ pp.\ \bibinfo {pages} {125--166},\ \Eprint
  {http://arxiv.org/abs/arXiv:0907.3596} {arXiv:0907.3596} \BibitemShut
  {NoStop}%
\bibitem [{\citenamefont {Sch\"{a}fer}(2011)}]{Schaefer-2009:PN-review}%
  \BibitemOpen
  \bibfield  {author} {\bibinfo {author} {\bibfnamefont {G.}~\bibnamefont
  {Sch\"{a}fer}},\ }in\ \href@noop {} {\emph {\bibinfo {booktitle} {Mass and
  Motion in General Relativity}}},\ \bibinfo {series} {Fundamental Theories of
  Physics}, Vol.\ \bibinfo {volume} {162},\ \bibinfo {editor} {edited by\
  \bibinfo {editor} {\bibfnamefont {L.}~\bibnamefont {Blanchet}}, \bibinfo
  {editor} {\bibfnamefont {A.}~\bibnamefont {Spallicci}}, \ and\ \bibinfo
  {editor} {\bibfnamefont {B.~F.}\ \bibnamefont {Whiting}}}\ (\bibinfo
  {publisher} {Springer-Verlag},\ \bibinfo {address} {Berlin},\ \bibinfo {year}
  {2011})\ pp.\ \bibinfo {pages} {167--210},\ \Eprint
  {http://arxiv.org/abs/arXiv:0910.2857} {arXiv:0910.2857} \BibitemShut
  {NoStop}%
\bibitem [{\citenamefont {Pretorius}(2007)}]{Pretorius-2007:2BH-review}%
  \BibitemOpen
  \bibfield  {author} {\bibinfo {author} {\bibfnamefont {F.}~\bibnamefont
  {Pretorius}},\ }in\ \href@noop {} {\emph {\bibinfo {booktitle} {Relativistic
  Objects in Compact Binaries: From Birth to Coalescence}}},\ \bibinfo {editor}
  {edited by\ \bibinfo {editor} {\bibfnamefont {M.}~\bibnamefont {Colpi}}}\
  (\bibinfo  {publisher} {Springer-Verlag},\ \bibinfo {year} {2007})\ \Eprint
  {http://arxiv.org/abs/arXiv:0710.1338} {arXiv:0710.1338} \BibitemShut
  {NoStop}%
\bibitem [{\citenamefont {Hannam}\ \emph {et~al.}(2009)\citenamefont {Hannam},
  \citenamefont {Husa}, \citenamefont {Baker}, \citenamefont {Boyle},
  \citenamefont {Br\"{u}gmann}, \citenamefont {Chu}, \citenamefont {Dorband},
  \citenamefont {Herrmann}, \citenamefont {Hinder}, \citenamefont {Kelly},
  \citenamefont {Kidder}, \citenamefont {Laguna}, \citenamefont {Matthews},
  \citenamefont {van Meter}, \citenamefont {Pfeiffer}, \citenamefont {Pollney},
  \citenamefont {Reisswig}, \citenamefont {Scheel},\ and\ \citenamefont
  {Shoemaker}}]{Hannam-etal-2009:Samurai-project}%
  \BibitemOpen
  \bibfield  {author} {\bibinfo {author} {\bibfnamefont {M.}~\bibnamefont
  {Hannam}}, \bibinfo {author} {\bibfnamefont {S.}~\bibnamefont {Husa}},
  \bibinfo {author} {\bibfnamefont {J.~G.}\ \bibnamefont {Baker}}, \bibinfo
  {author} {\bibfnamefont {M.}~\bibnamefont {Boyle}}, \bibinfo {author}
  {\bibfnamefont {B.}~\bibnamefont {Br\"{u}gmann}}, \bibinfo {author}
  {\bibfnamefont {T.}~\bibnamefont {Chu}}, \bibinfo {author} {\bibfnamefont
  {N.}~\bibnamefont {Dorband}}, \bibinfo {author} {\bibfnamefont
  {F.}~\bibnamefont {Herrmann}}, \bibinfo {author} {\bibfnamefont
  {I.}~\bibnamefont {Hinder}}, \bibinfo {author} {\bibfnamefont {B.~J.}\
  \bibnamefont {Kelly}}, \bibinfo {author} {\bibfnamefont {L.~E.}\ \bibnamefont
  {Kidder}}, \bibinfo {author} {\bibfnamefont {P.}~\bibnamefont {Laguna}},
  \bibinfo {author} {\bibfnamefont {K.~D.}\ \bibnamefont {Matthews}}, \bibinfo
  {author} {\bibfnamefont {J.~R.}\ \bibnamefont {van Meter}}, \bibinfo {author}
  {\bibfnamefont {H.~P.}\ \bibnamefont {Pfeiffer}}, \bibinfo {author}
  {\bibfnamefont {D.}~\bibnamefont {Pollney}}, \bibinfo {author} {\bibfnamefont
  {C.}~\bibnamefont {Reisswig}}, \bibinfo {author} {\bibfnamefont {M.~A.}\
  \bibnamefont {Scheel}}, \ and\ \bibinfo {author} {\bibfnamefont
  {D.}~\bibnamefont {Shoemaker}},\ }\href {\doibase 10.1103/PhysRevD.79.084025}
  {\bibfield  {journal} {\bibinfo  {journal} {Phys. Rev. D}\ }\textbf {\bibinfo
  {volume} {79}},\ \bibinfo {pages} {084025} (\bibinfo {year} {2009})},\
  \Eprint {http://arxiv.org/abs/arXiv:0901.2437} {arXiv:0901.2437} \BibitemShut
  {NoStop}%
\bibitem [{\citenamefont {Hannam}(2009)}]{Hannam-2009:2BH-review}%
  \BibitemOpen
  \bibfield  {author} {\bibinfo {author} {\bibfnamefont {M.}~\bibnamefont
  {Hannam}},\ }\href {\doibase 10.1088/0264-9381/26/11/114001} {\bibfield
  {journal} {\bibinfo  {journal} {Class. Quant. Grav.}\ }\textbf {\bibinfo
  {volume} {26}},\ \bibinfo {pages} {114001} (\bibinfo {year} {2009})},\
  \Eprint {http://arxiv.org/abs/arXiv:0901.2931} {arXiv:0901.2931} \BibitemShut
  {NoStop}%
\bibitem [{\citenamefont {Hannam}\ and\ \citenamefont
  {Hawke}(2011)}]{Hannam-Hawke-2010:2BH-in-era-of-Einstein-telescope-review}%
  \BibitemOpen
  \bibfield  {author} {\bibinfo {author} {\bibfnamefont {M.}~\bibnamefont
  {Hannam}}\ and\ \bibinfo {author} {\bibfnamefont {I.}~\bibnamefont {Hawke}},\
  }\href {\doibase 10.1007/s10714-010-1008-2} {\bibfield  {journal} {\bibinfo
  {journal} {General Relativity and Gravitation}\ }\textbf {\bibinfo {volume}
  {43}},\ \bibinfo {pages} {465} (\bibinfo {year} {2011})},\ \Eprint
  {http://arxiv.org/abs/arXiv:0908.3139} {arXiv:0908.3139} \BibitemShut
  {NoStop}%
\bibitem [{\citenamefont {Campanelli}\ \emph {et~al.}(2010)\citenamefont
  {Campanelli}, \citenamefont {Lousto}, \citenamefont {Mundim}, \citenamefont
  {Nakano}, \citenamefont {Zlochower},\ and\ \citenamefont
  {Bischof}}]{Campanelli-etal-2010:2BH-numrel-review}%
  \BibitemOpen
  \bibfield  {author} {\bibinfo {author} {\bibfnamefont {M.}~\bibnamefont
  {Campanelli}}, \bibinfo {author} {\bibfnamefont {C.~O.}\ \bibnamefont
  {Lousto}}, \bibinfo {author} {\bibfnamefont {B.~C.}\ \bibnamefont {Mundim}},
  \bibinfo {author} {\bibfnamefont {H.}~\bibnamefont {Nakano}}, \bibinfo
  {author} {\bibfnamefont {Y.}~\bibnamefont {Zlochower}}, \ and\ \bibinfo
  {author} {\bibfnamefont {H.-P.}\ \bibnamefont {Bischof}},\ }\href {\doibase
  10.1088/0264-9381/27/8/084034} {\bibfield  {journal} {\bibinfo  {journal}
  {Class. Quant. Grav.}\ }\textbf {\bibinfo {volume} {27}},\ \bibinfo {pages}
  {084034} (\bibinfo {year} {2010})},\ \Eprint
  {http://arxiv.org/abs/arXiv:1001.3834} {arXiv:1001.3834} \BibitemShut
  {NoStop}%
\bibitem [{\citenamefont {Bishop}\ \emph {et~al.}(2003)\citenamefont {Bishop},
  \citenamefont {G\'omez}, \citenamefont {Husa}, \citenamefont {Lehner},\ and\
  \citenamefont {Winicour}}]{Bishop-etal-2003}%
  \BibitemOpen
  \bibfield  {author} {\bibinfo {author} {\bibfnamefont {N.~T.}\ \bibnamefont
  {Bishop}}, \bibinfo {author} {\bibfnamefont {R.}~\bibnamefont {G\'omez}},
  \bibinfo {author} {\bibfnamefont {S.}~\bibnamefont {Husa}}, \bibinfo {author}
  {\bibfnamefont {L.}~\bibnamefont {Lehner}}, \ and\ \bibinfo {author}
  {\bibfnamefont {J.}~\bibnamefont {Winicour}},\ }\href {\doibase
  10.1103/PhysRevD.68.084015} {\bibfield  {journal} {\bibinfo  {journal} {Phys.
  Rev. D}\ }\textbf {\bibinfo {volume} {68}},\ \bibinfo {pages} {084015}
  (\bibinfo {year} {2003})},\ \Eprint {http://arxiv.org/abs/gr-qc/0301060}
  {gr-qc/0301060} \BibitemShut {NoStop}%
\bibitem [{\citenamefont {Bishop}\ \emph {et~al.}(2005)\citenamefont {Bishop},
  \citenamefont {G\'omez}, \citenamefont {Lehner}, \citenamefont {Maharaj},\
  and\ \citenamefont {Winicour}}]{Bishop-etal-2005}%
  \BibitemOpen
  \bibfield  {author} {\bibinfo {author} {\bibfnamefont {N.~T.}\ \bibnamefont
  {Bishop}}, \bibinfo {author} {\bibfnamefont {R.}~\bibnamefont {G\'omez}},
  \bibinfo {author} {\bibfnamefont {L.}~\bibnamefont {Lehner}}, \bibinfo
  {author} {\bibfnamefont {M.}~\bibnamefont {Maharaj}}, \ and\ \bibinfo
  {author} {\bibfnamefont {J.}~\bibnamefont {Winicour}},\ }\href {\doibase
  10.1103/PhysRevD.72.024002} {\bibfield  {journal} {\bibinfo  {journal} {Phys.
  Rev. D}\ }\textbf {\bibinfo {volume} {72}},\ \bibinfo {pages} {024002}
  (\bibinfo {year} {2005})},\ \Eprint {http://arxiv.org/abs/gr-qc/0412080}
  {gr-qc/0412080} \BibitemShut {NoStop}%
\bibitem [{\citenamefont {Sopuerta}\ \emph {et~al.}(2006)\citenamefont
  {Sopuerta}, \citenamefont {Sun}, \citenamefont {Laguna},\ and\ \citenamefont
  {Xu}}]{Sopuerta-etal-2006}%
  \BibitemOpen
  \bibfield  {author} {\bibinfo {author} {\bibfnamefont {C.~F.}\ \bibnamefont
  {Sopuerta}}, \bibinfo {author} {\bibfnamefont {P.}~\bibnamefont {Sun}},
  \bibinfo {author} {\bibfnamefont {P.}~\bibnamefont {Laguna}}, \ and\ \bibinfo
  {author} {\bibfnamefont {J.}~\bibnamefont {Xu}},\ }\href {\doibase
  10.1088/0264-9381/23/1/013} {\bibfield  {journal} {\bibinfo  {journal}
  {Class. Quant. Grav.}\ }\textbf {\bibinfo {volume} {23}},\ \bibinfo {pages}
  {251} (\bibinfo {year} {2006})},\ \Eprint
  {http://arxiv.org/abs/gr-qc/0507112} {gr-qc/0507112} \BibitemShut {NoStop}%
\bibitem [{\citenamefont {Sopuerta}\ and\ \citenamefont
  {Laguna}(2006)}]{Sopuerta-Laguna-2006}%
  \BibitemOpen
  \bibfield  {author} {\bibinfo {author} {\bibfnamefont {C.~F.}\ \bibnamefont
  {Sopuerta}}\ and\ \bibinfo {author} {\bibfnamefont {P.}~\bibnamefont
  {Laguna}},\ }\href {\doibase 10.1103/PhysRevD.73.044028} {\bibfield
  {journal} {\bibinfo  {journal} {Phys. Rev. D}\ }\textbf {\bibinfo {volume}
  {73}},\ \bibinfo {pages} {044028} (\bibinfo {year} {2006})},\ \Eprint
  {http://arxiv.org/abs/gr-qc/0512028} {gr-qc/0512028} \BibitemShut {NoStop}%
\bibitem [{\citenamefont {Lousto}\ \emph {et~al.}(2010)\citenamefont {Lousto},
  \citenamefont {Nakano}, \citenamefont {Zlochower},\ and\ \citenamefont
  {Campanelli}}]{Lousto-etal-2010:intermediate-mass-2BH-numrel-Lazarus}%
  \BibitemOpen
  \bibfield  {author} {\bibinfo {author} {\bibfnamefont {C.~O.}\ \bibnamefont
  {Lousto}}, \bibinfo {author} {\bibfnamefont {H.}~\bibnamefont {Nakano}},
  \bibinfo {author} {\bibfnamefont {Y.}~\bibnamefont {Zlochower}}, \ and\
  \bibinfo {author} {\bibfnamefont {M.}~\bibnamefont {Campanelli}},\ }\href
  {\doibase 10.1103/PhysRevLett.104.211101} {\bibfield  {journal} {\bibinfo
  {journal} {Phys. Rev. Lett.}\ }\textbf {\bibinfo {volume} {104}},\ \bibinfo
  {pages} {211101} (\bibinfo {year} {2010})},\ \Eprint
  {http://arxiv.org/abs/arXiv:1001.2316} {arXiv:1001.2316} \BibitemShut
  {NoStop}%
\bibitem [{\citenamefont {Lousto}\ and\ \citenamefont
  {Zlochower}(2011)}]{Lousto-Zlochower-2011:100-to-1-mass-ratio-2BH}%
  \BibitemOpen
  \bibfield  {author} {\bibinfo {author} {\bibfnamefont {C.~O.}\ \bibnamefont
  {Lousto}}\ and\ \bibinfo {author} {\bibfnamefont {Y.}~\bibnamefont
  {Zlochower}},\ }\href {\doibase 10.1103/PhysRevLett.106.041101} {\bibfield
  {journal} {\bibinfo  {journal} {Phys. Rev. Lett.}\ }\textbf {\bibinfo
  {volume} {106}},\ \bibinfo {pages} {041101} (\bibinfo {year} {2011})},\
  \Eprint {http://arxiv.org/abs/arXiv:1009.0292} {arXiv:1009.0292} \BibitemShut
  {NoStop}%
\bibitem [{\citenamefont {Poisson}\ \emph {et~al.}(2011)\citenamefont
  {Poisson}, \citenamefont {Pound},\ and\ \citenamefont
  {Vega}}]{Poisson-Pound-Vega-2011:living-review}%
  \BibitemOpen
  \bibfield  {author} {\bibinfo {author} {\bibfnamefont {E.}~\bibnamefont
  {Poisson}}, \bibinfo {author} {\bibfnamefont {A.}~\bibnamefont {Pound}}, \
  and\ \bibinfo {author} {\bibfnamefont {I.}~\bibnamefont {Vega}},\ }\href
  {\doibase 10.12942/lrr-2011-7} {\bibfield  {journal} {\bibinfo  {journal}
  {Living Reviews in Relativity}\ }\textbf {\bibinfo {volume} {14}},\ \bibinfo
  {pages} {7} (\bibinfo {year} {2011})}\BibitemShut {NoStop}%
\bibitem [{\citenamefont {Geroch}\ and\ \citenamefont
  {Traschen}(1987)}]{Geroch-Traschen-1987}%
  \BibitemOpen
  \bibfield  {author} {\bibinfo {author} {\bibfnamefont {R.}~\bibnamefont
  {Geroch}}\ and\ \bibinfo {author} {\bibfnamefont {J.}~\bibnamefont
  {Traschen}},\ }\href {\doibase 10.1103/PhysRevD.36.1017} {\bibfield
  {journal} {\bibinfo  {journal} {Phys. Rev. D}\ }\textbf {\bibinfo {volume}
  {36}},\ \bibinfo {pages} {1017} (\bibinfo {year} {1987})}\BibitemShut
  {NoStop}%
\bibitem [{\citenamefont {Steinbauer}\ and\ \citenamefont
  {Vickers}(2006)}]{Steinbauer-Vickers-2006}%
  \BibitemOpen
  \bibfield  {author} {\bibinfo {author} {\bibfnamefont {R.}~\bibnamefont
  {Steinbauer}}\ and\ \bibinfo {author} {\bibfnamefont {J.~A.}\ \bibnamefont
  {Vickers}},\ }\href {\doibase 10.1088/0264-9381/23/10/R01} {\bibfield
  {journal} {\bibinfo  {journal} {Class. Quant. Grav.}\ }\textbf {\bibinfo
  {volume} {23}},\ \bibinfo {pages} {R91} (\bibinfo {year} {2006})},\ \Eprint
  {http://arxiv.org/abs/gr-qc/0603078v2} {gr-qc/0603078v2} \BibitemShut
  {NoStop}%
\bibitem [{\citenamefont {Mino}\ \emph {et~al.}(1997)\citenamefont {Mino},
  \citenamefont {Sasaki},\ and\ \citenamefont
  {Tanaka}}]{Mino-Sasaki-Tanaka-1997}%
  \BibitemOpen
  \bibfield  {author} {\bibinfo {author} {\bibfnamefont {Y.}~\bibnamefont
  {Mino}}, \bibinfo {author} {\bibfnamefont {M.}~\bibnamefont {Sasaki}}, \ and\
  \bibinfo {author} {\bibfnamefont {T.}~\bibnamefont {Tanaka}},\ }\href
  {\doibase 10.1103/PhysRevD.55.3457} {\bibfield  {journal} {\bibinfo
  {journal} {Phys. Rev. D}\ }\textbf {\bibinfo {volume} {55}},\ \bibinfo
  {pages} {3457} (\bibinfo {year} {1997})},\ \Eprint
  {http://arxiv.org/abs/gr-qc/9606018} {gr-qc/9606018} \BibitemShut {NoStop}%
\bibitem [{\citenamefont {Quinn}\ and\ \citenamefont
  {Wald}(1997)}]{Quinn-Wald-1997}%
  \BibitemOpen
  \bibfield  {author} {\bibinfo {author} {\bibfnamefont {T.~C.}\ \bibnamefont
  {Quinn}}\ and\ \bibinfo {author} {\bibfnamefont {R.~M.}\ \bibnamefont
  {Wald}},\ }\href {\doibase 10.1103/PhysRevD.56.3381} {\bibfield  {journal}
  {\bibinfo  {journal} {Phys. Rev. D}\ }\textbf {\bibinfo {volume} {56}},\
  \bibinfo {pages} {3381} (\bibinfo {year} {1997})},\ \Eprint
  {http://arxiv.org/abs/gr-qc/9610053} {gr-qc/9610053} \BibitemShut {NoStop}%
\bibitem [{\citenamefont
  {Detweiler}(2001)}]{Detweiler-2001:radiation-reaction-and-self-force}%
  \BibitemOpen
  \bibfield  {author} {\bibinfo {author} {\bibfnamefont {S.}~\bibnamefont
  {Detweiler}},\ }\href {\doibase 10.1103/PhysRevLett.86.1931} {\bibfield
  {journal} {\bibinfo  {journal} {Phys. Rev. Lett.}\ }\textbf {\bibinfo
  {volume} {86}},\ \bibinfo {pages} {1931} (\bibinfo {year} {2001})},\ \Eprint
  {http://arxiv.org/abs/gr-qc/0011039} {gr-qc/0011039} \BibitemShut {NoStop}%
\bibitem [{\citenamefont {Gralla}\ and\ \citenamefont
  {Wald}(2008)}]{Gralla-Wald-2008}%
  \BibitemOpen
  \bibfield  {author} {\bibinfo {author} {\bibfnamefont {S.~E.}\ \bibnamefont
  {Gralla}}\ and\ \bibinfo {author} {\bibfnamefont {R.~M.}\ \bibnamefont
  {Wald}},\ }\href {\doibase 10.1088/0264-9381/25/20/205009} {\bibfield
  {journal} {\bibinfo  {journal} {Class. Quant. Grav.}\ }\textbf {\bibinfo
  {volume} {25}},\ \bibinfo {pages} {205009} (\bibinfo {year} {2008})},\
  \Eprint {http://arxiv.org/abs/arXiv:0806.3293} {arXiv:0806.3293} \BibitemShut
  {NoStop}%
\bibitem [{\citenamefont {Gralla}\ \emph {et~al.}(2009)\citenamefont {Gralla},
  \citenamefont {Harte},\ and\ \citenamefont {Wald}}]{Gralla-Harte-Wald-2009}%
  \BibitemOpen
  \bibfield  {author} {\bibinfo {author} {\bibfnamefont {S.~E.}\ \bibnamefont
  {Gralla}}, \bibinfo {author} {\bibfnamefont {A.~I.}\ \bibnamefont {Harte}}, \
  and\ \bibinfo {author} {\bibfnamefont {R.~M.}\ \bibnamefont {Wald}},\ }\href
  {\doibase 10.1103/PhysRevD.80.024031} {\bibfield  {journal} {\bibinfo
  {journal} {Phys. Rev. D}\ }\textbf {\bibinfo {volume} {80}},\ \bibinfo
  {pages} {024031} (\bibinfo {year} {2009})},\ \Eprint
  {http://arxiv.org/abs/arXiv:0905.2391} {arXiv:0905.2391} \BibitemShut
  {NoStop}%
\bibitem [{\citenamefont {Detweiler}(2005)}]{Detweiler-2005}%
  \BibitemOpen
  \bibfield  {author} {\bibinfo {author} {\bibfnamefont {S.}~\bibnamefont
  {Detweiler}},\ }\href {\doibase 10.1088/0264-9381/22/15/006} {\bibfield
  {journal} {\bibinfo  {journal} {Class. Quant. Grav.}\ }\textbf {\bibinfo
  {volume} {22}},\ \bibinfo {pages} {S681} (\bibinfo {year} {2005})},\ \Eprint
  {http://arxiv.org/abs/gr-qc/0501004} {gr-qc/0501004} \BibitemShut {NoStop}%
\bibitem [{\citenamefont {Barack}(2009)}]{Barack-2009:self-force-review}%
  \BibitemOpen
  \bibfield  {author} {\bibinfo {author} {\bibfnamefont {L.}~\bibnamefont
  {Barack}},\ }\href {\doibase 10.1088/0264-9381/26/21/213001} {\bibfield
  {journal} {\bibinfo  {journal} {Class. Quant. Grav.}\ }\textbf {\bibinfo
  {volume} {26}},\ \bibinfo {pages} {213001} (\bibinfo {year} {2009})},\
  \Eprint {http://arxiv.org/abs/arXiv:0908.1664} {arXiv:0908.1664} \BibitemShut
  {NoStop}%
\bibitem [{\citenamefont {Barack}(2011)}]{Barack-2009:self-force-review2}%
  \BibitemOpen
  \bibfield  {author} {\bibinfo {author} {\bibfnamefont {L.}~\bibnamefont
  {Barack}},\ }in\ \href {\doibase 10.1007/978-90-481-3015-3_12} {\emph
  {\bibinfo {booktitle} {Mass and Motion in General Relativity}}},\ \bibinfo
  {series} {Fundamental Theories of Physics}, Vol.\ \bibinfo {volume} {162},\
  \bibinfo {editor} {edited by\ \bibinfo {editor} {\bibfnamefont
  {L.}~\bibnamefont {Blanchet}}, \bibinfo {editor} {\bibfnamefont
  {A.}~\bibnamefont {Spallicci}}, \ and\ \bibinfo {editor} {\bibfnamefont
  {B.~F.}\ \bibnamefont {Whiting}}}\ (\bibinfo  {publisher} {Springer-Verlag},\
  \bibinfo {address} {Berlin},\ \bibinfo {year} {2011})\ pp.\ \bibinfo {pages}
  {327--366}\BibitemShut {NoStop}%
\bibitem [{\citenamefont {Burko}(2011)}]{Burko-2009:self-force-review}%
  \BibitemOpen
  \bibfield  {author} {\bibinfo {author} {\bibfnamefont {L.~M.}\ \bibnamefont
  {Burko}},\ }in\ \href {\doibase 10.1007/978-90-481-3015-3_14} {\emph
  {\bibinfo {booktitle} {Mass and Motion in General Relativity}}},\ \bibinfo
  {series} {Fundamental Theories of Physics}, Vol.\ \bibinfo {volume} {162},\
  \bibinfo {editor} {edited by\ \bibinfo {editor} {\bibfnamefont
  {L.}~\bibnamefont {Blanchet}}, \bibinfo {editor} {\bibfnamefont
  {A.}~\bibnamefont {Spallicci}}, \ and\ \bibinfo {editor} {\bibfnamefont
  {B.~F.}\ \bibnamefont {Whiting}}}\ (\bibinfo  {publisher} {Springer-Verlag},\
  \bibinfo {address} {Berlin},\ \bibinfo {year} {2011})\ pp.\ \bibinfo {pages}
  {395--414}\BibitemShut {NoStop}%
\bibitem [{\citenamefont {Detweiler}(2011)}]{Detweiler-2009:self-force-review}%
  \BibitemOpen
  \bibfield  {author} {\bibinfo {author} {\bibfnamefont {S.}~\bibnamefont
  {Detweiler}},\ }in\ \href@noop {} {\emph {\bibinfo {booktitle} {Mass and
  Motion in General Relativity}}},\ \bibinfo {series} {Fundamental Theories of
  Physics}, Vol.\ \bibinfo {volume} {162},\ \bibinfo {editor} {edited by\
  \bibinfo {editor} {\bibfnamefont {L.}~\bibnamefont {Blanchet}}, \bibinfo
  {editor} {\bibfnamefont {A.}~\bibnamefont {Spallicci}}, \ and\ \bibinfo
  {editor} {\bibfnamefont {B.~F.}\ \bibnamefont {Whiting}}}\ (\bibinfo
  {publisher} {Springer-Verlag},\ \bibinfo {address} {Berlin},\ \bibinfo {year}
  {2011})\ pp.\ \bibinfo {pages} {271--307},\ \Eprint
  {http://arxiv.org/abs/arXiv:0908.4363} {arXiv:0908.4363} \BibitemShut
  {NoStop}%
\bibitem [{\citenamefont {Poisson}(2011)}]{Poisson-2009:self-force-review}%
  \BibitemOpen
  \bibfield  {author} {\bibinfo {author} {\bibfnamefont {E.}~\bibnamefont
  {Poisson}},\ }in\ \href@noop {} {\emph {\bibinfo {booktitle} {Mass and Motion
  in General Relativity}}},\ \bibinfo {series} {Fundamental Theories of
  Physics}, Vol.\ \bibinfo {volume} {162},\ \bibinfo {editor} {edited by\
  \bibinfo {editor} {\bibfnamefont {L.}~\bibnamefont {Blanchet}}, \bibinfo
  {editor} {\bibfnamefont {A.}~\bibnamefont {Spallicci}}, \ and\ \bibinfo
  {editor} {\bibfnamefont {B.~F.}\ \bibnamefont {Whiting}}}\ (\bibinfo
  {publisher} {Springer-Verlag},\ \bibinfo {address} {Berlin},\ \bibinfo {year}
  {2011})\ pp.\ \bibinfo {pages} {309--325},\ \Eprint
  {http://arxiv.org/abs/arXiv:0909.2994} {arXiv:0909.2994} \BibitemShut
  {NoStop}%
\bibitem [{\citenamefont {Wald}(2011)}]{Wald-2009:self-force-review}%
  \BibitemOpen
  \bibfield  {author} {\bibinfo {author} {\bibfnamefont {R.~M.}\ \bibnamefont
  {Wald}},\ }in\ \href@noop {} {\emph {\bibinfo {booktitle} {Mass and Motion in
  General Relativity}}},\ \bibinfo {series} {Fundamental Theories of Physics},
  Vol.\ \bibinfo {volume} {162},\ \bibinfo {editor} {edited by\ \bibinfo
  {editor} {\bibfnamefont {L.}~\bibnamefont {Blanchet}}, \bibinfo {editor}
  {\bibfnamefont {A.}~\bibnamefont {Spallicci}}, \ and\ \bibinfo {editor}
  {\bibfnamefont {B.~F.}\ \bibnamefont {Whiting}}}\ (\bibinfo  {publisher}
  {Springer-Verlag},\ \bibinfo {address} {Berlin},\ \bibinfo {year} {2011})\
  pp.\ \bibinfo {pages} {253--262},\ \Eprint
  {http://arxiv.org/abs/arXiv:0907.0412} {arXiv:0907.0412} \BibitemShut
  {NoStop}%
\bibitem [{\citenamefont {Sago}\ \emph {et~al.}(2008)\citenamefont {Sago},
  \citenamefont {Barack},\ and\ \citenamefont
  {Detweiler}}]{Sago-Barack-Detweiler-2008}%
  \BibitemOpen
  \bibfield  {author} {\bibinfo {author} {\bibfnamefont {N.}~\bibnamefont
  {Sago}}, \bibinfo {author} {\bibfnamefont {L.}~\bibnamefont {Barack}}, \ and\
  \bibinfo {author} {\bibfnamefont {S.}~\bibnamefont {Detweiler}},\ }\href
  {\doibase 10.1103/PhysRevD.78.124024} {\bibfield  {journal} {\bibinfo
  {journal} {Phys. Rev. D}\ }\textbf {\bibinfo {volume} {78}},\ \bibinfo
  {pages} {124024} (\bibinfo {year} {2008})},\ \Eprint
  {http://arxiv.org/abs/arXiv:0810.2530} {arXiv:0810.2530} \BibitemShut
  {NoStop}%
\bibitem [{\citenamefont {Barack}\ and\ \citenamefont
  {Ori}(2000)}]{Barack-Ori-2000}%
  \BibitemOpen
  \bibfield  {author} {\bibinfo {author} {\bibfnamefont {L.}~\bibnamefont
  {Barack}}\ and\ \bibinfo {author} {\bibfnamefont {A.}~\bibnamefont {Ori}},\
  }\href {\doibase 10.1103/PhysRevD.61.061502} {\bibfield  {journal} {\bibinfo
  {journal} {Phys. Rev. D}\ }\textbf {\bibinfo {volume} {61}},\ \bibinfo
  {pages} {061502(R)} (\bibinfo {year} {2000})},\ \Eprint
  {http://arxiv.org/abs/gr-qc/9912010} {gr-qc/9912010} \BibitemShut {NoStop}%
\bibitem [{\citenamefont {Barack}(2000)}]{Barack-2000}%
  \BibitemOpen
  \bibfield  {author} {\bibinfo {author} {\bibfnamefont {L.}~\bibnamefont
  {Barack}},\ }\href {\doibase 10.1103/PhysRevD.62.084027} {\bibfield
  {journal} {\bibinfo  {journal} {Phys. Rev. D}\ }\textbf {\bibinfo {volume}
  {62}},\ \bibinfo {pages} {084027} (\bibinfo {year} {2000})},\ \Eprint
  {http://arxiv.org/abs/gr-qc/0005042} {gr-qc/0005042} \BibitemShut {NoStop}%
\bibitem [{\citenamefont {Barack}\ \emph {et~al.}(2002)\citenamefont {Barack},
  \citenamefont {Mino}, \citenamefont {Nakano}, \citenamefont {Ori},\ and\
  \citenamefont {Sasaki}}]{Barack-etal-2002}%
  \BibitemOpen
  \bibfield  {author} {\bibinfo {author} {\bibfnamefont {L.}~\bibnamefont
  {Barack}}, \bibinfo {author} {\bibfnamefont {Y.}~\bibnamefont {Mino}},
  \bibinfo {author} {\bibfnamefont {H.}~\bibnamefont {Nakano}}, \bibinfo
  {author} {\bibfnamefont {A.}~\bibnamefont {Ori}}, \ and\ \bibinfo {author}
  {\bibfnamefont {M.}~\bibnamefont {Sasaki}},\ }\href {\doibase
  10.1103/PhysRevLett.88.091101} {\bibfield  {journal} {\bibinfo  {journal}
  {Phys. Rev. Lett.}\ }\textbf {\bibinfo {volume} {88}},\ \bibinfo {pages}
  {091101} (\bibinfo {year} {2002})},\ \Eprint
  {http://arxiv.org/abs/gr-qc/0111001} {gr-qc/0111001} \BibitemShut {NoStop}%
\bibitem [{\citenamefont {Barack}\ and\ \citenamefont
  {Ori}(2002)}]{Barack-Ori-2002}%
  \BibitemOpen
  \bibfield  {author} {\bibinfo {author} {\bibfnamefont {L.}~\bibnamefont
  {Barack}}\ and\ \bibinfo {author} {\bibfnamefont {A.}~\bibnamefont {Ori}},\
  }\href {\doibase 10.1103/PhysRevD.66.084022} {\bibfield  {journal} {\bibinfo
  {journal} {Phys. Rev. D}\ }\textbf {\bibinfo {volume} {66}},\ \bibinfo
  {pages} {084022} (\bibinfo {year} {2002})},\ \Eprint
  {http://arxiv.org/abs/gr-qc/0204093} {gr-qc/0204093} \BibitemShut {NoStop}%
\bibitem [{\citenamefont {Barack}\ and\ \citenamefont
  {Ori}(2003)}]{Barack-Ori-2003}%
  \BibitemOpen
  \bibfield  {author} {\bibinfo {author} {\bibfnamefont {L.}~\bibnamefont
  {Barack}}\ and\ \bibinfo {author} {\bibfnamefont {A.}~\bibnamefont {Ori}},\
  }\href {\doibase 10.1103/PhysRevD.67.024029} {\bibfield  {journal} {\bibinfo
  {journal} {Phys. Rev. D}\ }\textbf {\bibinfo {volume} {67}},\ \bibinfo
  {pages} {024029} (\bibinfo {year} {2003})},\ \Eprint
  {http://arxiv.org/abs/gr-qc/0209072} {gr-qc/0209072} \BibitemShut {NoStop}%
\bibitem [{\citenamefont {Detweiler}\ and\ \citenamefont
  {Whiting}(2003)}]{Detweiler-Whiting-2003}%
  \BibitemOpen
  \bibfield  {author} {\bibinfo {author} {\bibfnamefont {S.}~\bibnamefont
  {Detweiler}}\ and\ \bibinfo {author} {\bibfnamefont {B.~F.}\ \bibnamefont
  {Whiting}},\ }\href {\doibase 10.1103/PhysRevD.67.024025} {\bibfield
  {journal} {\bibinfo  {journal} {Phys. Rev. D}\ }\textbf {\bibinfo {volume}
  {67}},\ \bibinfo {pages} {024025} (\bibinfo {year} {2003})},\ \Eprint
  {http://arxiv.org/abs/gr-qc/0202086} {gr-qc/0202086} \BibitemShut {NoStop}%
\bibitem [{\citenamefont {Detweiler}\ \emph {et~al.}(2003)\citenamefont
  {Detweiler}, \citenamefont {Messaritaki},\ and\ \citenamefont
  {Whiting}}]{Detweiler-Messaritaki-Whiting-2003}%
  \BibitemOpen
  \bibfield  {author} {\bibinfo {author} {\bibfnamefont {S.}~\bibnamefont
  {Detweiler}}, \bibinfo {author} {\bibfnamefont {E.}~\bibnamefont
  {Messaritaki}}, \ and\ \bibinfo {author} {\bibfnamefont {B.~F.}\ \bibnamefont
  {Whiting}},\ }\href {\doibase 10.1103/PhysRevD.67.104016} {\bibfield
  {journal} {\bibinfo  {journal} {Phys. Rev. D}\ }\textbf {\bibinfo {volume}
  {67}},\ \bibinfo {pages} {104016} (\bibinfo {year} {2003})},\ \Eprint
  {http://arxiv.org/abs/gr-qc/0205079} {gr-qc/0205079} \BibitemShut {NoStop}%
\bibitem [{\citenamefont {Haas}\ and\ \citenamefont
  {Poisson}(2006)}]{Haas-Poisson-2006}%
  \BibitemOpen
  \bibfield  {author} {\bibinfo {author} {\bibfnamefont {R.}~\bibnamefont
  {Haas}}\ and\ \bibinfo {author} {\bibfnamefont {E.}~\bibnamefont {Poisson}},\
  }\href {\doibase 10.1103/PhysRevD.74.044009} {\bibfield  {journal} {\bibinfo
  {journal} {Phys. Rev. D}\ }\textbf {\bibinfo {volume} {74}},\ \bibinfo
  {pages} {044009} (\bibinfo {year} {2006})},\ \Eprint
  {http://arxiv.org/abs/gr-qc/0605077} {gr-qc/0605077} \BibitemShut {NoStop}%
\bibitem [{\citenamefont {Anderson}\ and\ \citenamefont
  {Wiseman}(2005)}]{Anderson:2005gb}%
  \BibitemOpen
  \bibfield  {author} {\bibinfo {author} {\bibfnamefont {W.~G.}\ \bibnamefont
  {Anderson}}\ and\ \bibinfo {author} {\bibfnamefont {A.~G.}\ \bibnamefont
  {Wiseman}},\ }\href {\doibase 10.1088/0264-9381/22/15/010} {\bibfield
  {journal} {\bibinfo  {journal} {Class. Quant. Grav.}\ }\textbf {\bibinfo
  {volume} {22}},\ \bibinfo {pages} {S783} (\bibinfo {year} {2005})},\ \Eprint
  {http://arxiv.org/abs/gr-qc/0506136} {arXiv:gr-qc/0506136 [gr-qc]}
  \BibitemShut {NoStop}%
%%CITATION = GR-QC/0506136;%%
\bibitem [{\citenamefont {Casals}\ \emph {et~al.}(2009)\citenamefont {Casals},
  \citenamefont {Dolan}, \citenamefont {Ottewill},\ and\ \citenamefont
  {Wardell}}]{Casals:2009zh}%
  \BibitemOpen
  \bibfield  {author} {\bibinfo {author} {\bibfnamefont {M.}~\bibnamefont
  {Casals}}, \bibinfo {author} {\bibfnamefont {S.~R.}\ \bibnamefont {Dolan}},
  \bibinfo {author} {\bibfnamefont {A.~C.}\ \bibnamefont {Ottewill}}, \ and\
  \bibinfo {author} {\bibfnamefont {B.}~\bibnamefont {Wardell}},\ }\href
  {\doibase 10.1103/PhysRevD.79.124043} {\bibfield  {journal} {\bibinfo
  {journal} {Phys. Rev.}\ }\textbf {\bibinfo {volume} {D79}},\ \bibinfo {pages}
  {124043} (\bibinfo {year} {2009})},\ \Eprint {http://arxiv.org/abs/0903.0395}
  {arXiv:0903.0395 [gr-qc]} \BibitemShut {NoStop}%
%%CITATION = ARXIV:0903.0395;%%
\bibitem [{\citenamefont {Casals}\ \emph {et~al.}(2013)\citenamefont {Casals},
  \citenamefont {Dolan}, \citenamefont {Ottewill},\ and\ \citenamefont
  {Wardell}}]{Casals:2013mpa}%
  \BibitemOpen
  \bibfield  {author} {\bibinfo {author} {\bibfnamefont {M.}~\bibnamefont
  {Casals}}, \bibinfo {author} {\bibfnamefont {S.}~\bibnamefont {Dolan}},
  \bibinfo {author} {\bibfnamefont {A.~C.}\ \bibnamefont {Ottewill}}, \ and\
  \bibinfo {author} {\bibfnamefont {B.}~\bibnamefont {Wardell}},\ }\href
  {\doibase 10.1103/PhysRevD.88.044022} {\bibfield  {journal} {\bibinfo
  {journal} {Phys. Rev.}\ }\textbf {\bibinfo {volume} {D88}},\ \bibinfo {pages}
  {044022} (\bibinfo {year} {2013})},\ \Eprint {http://arxiv.org/abs/1306.0884}
  {arXiv:1306.0884 [gr-qc]} \BibitemShut {NoStop}%
%%CITATION = ARXIV:1306.0884;%%
\bibitem [{\citenamefont {Wardell}\ \emph {et~al.}(2014)\citenamefont
  {Wardell}, \citenamefont {Galley}, \citenamefont {Zengino\u{g}lu},
  \citenamefont {Casals}, \citenamefont {Dolan},\ and\ \citenamefont
  {Ottewill}}]{Wardell-etal-2014:self-force-via-Green-fn}%
  \BibitemOpen
  \bibfield  {author} {\bibinfo {author} {\bibfnamefont {B.}~\bibnamefont
  {Wardell}}, \bibinfo {author} {\bibfnamefont {C.~R.}\ \bibnamefont {Galley}},
  \bibinfo {author} {\bibfnamefont {A.}~\bibnamefont {Zengino\u{g}lu}},
  \bibinfo {author} {\bibfnamefont {M.}~\bibnamefont {Casals}}, \bibinfo
  {author} {\bibfnamefont {S.~R.}\ \bibnamefont {Dolan}}, \ and\ \bibinfo
  {author} {\bibfnamefont {A.~C.}\ \bibnamefont {Ottewill}},\ }\href {\doibase
  10.1103/PhysRevD.89.084021} {\bibfield  {journal} {\bibinfo  {journal} {Phys.
  Rev. D}\ }\textbf {\bibinfo {volume} {89}},\ \bibinfo {pages} {084021}
  (\bibinfo {year} {2014})},\ \Eprint
  {http://arxiv.org/abs/http://arxiv.org/abs/1401.1506}
  {http://arxiv.org/abs/1401.1506} \BibitemShut {NoStop}%
\bibitem [{\citenamefont {Barack}\ and\ \citenamefont
  {Golbourn}(2007)}]{Barack-Golbourn-2007}%
  \BibitemOpen
  \bibfield  {author} {\bibinfo {author} {\bibfnamefont {L.}~\bibnamefont
  {Barack}}\ and\ \bibinfo {author} {\bibfnamefont {D.~A.}\ \bibnamefont
  {Golbourn}},\ }\href {\doibase 10.1103/PhysRevD.76.044020} {\bibfield
  {journal} {\bibinfo  {journal} {Phys. Rev. D}\ }\textbf {\bibinfo {volume}
  {76}},\ \bibinfo {pages} {044020} (\bibinfo {year} {2007})},\ \Eprint
  {http://arxiv.org/abs/arXiv:0705.3620} {arXiv:0705.3620} \BibitemShut
  {NoStop}%
\bibitem [{\citenamefont {Vega}\ and\ \citenamefont
  {Detweiler}(2008)}]{Vega-Detweiler-2008:self-force-regularization}%
  \BibitemOpen
  \bibfield  {author} {\bibinfo {author} {\bibfnamefont {I.}~\bibnamefont
  {Vega}}\ and\ \bibinfo {author} {\bibfnamefont {S.}~\bibnamefont
  {Detweiler}},\ }\href {\doibase 10.1103/PhysRevD.77.084008} {\bibfield
  {journal} {\bibinfo  {journal} {Phys. Rev. D}\ }\textbf {\bibinfo {volume}
  {77}},\ \bibinfo {pages} {084008} (\bibinfo {year} {2008})},\ \Eprint
  {http://arxiv.org/abs/arXiv:0712.4405} {arXiv:0712.4405} \BibitemShut
  {NoStop}%
\bibitem [{\citenamefont {Wardell}(2015)}]{Wardell:2015kea}%
  \BibitemOpen
  \bibfield  {author} {\bibinfo {author} {\bibfnamefont {B.}~\bibnamefont
  {Wardell}},\ }in\ \href {\doibase 10.1007/978-3-319-18335-0_14} {\emph
  {\bibinfo {booktitle} {Equations of Motion in Relativistic Gravity}}},\
  \bibinfo {series} {Fundamental Theories of Physics}, Vol.\ \bibinfo {volume}
  {179},\ \bibinfo {editor} {edited by\ \bibinfo {editor} {\bibfnamefont
  {D.}~\bibnamefont {Puetzfeld}}, \bibinfo {editor} {\bibfnamefont
  {C.}~\bibnamefont {Lämmerzahl}}, \ and\ \bibinfo {editor} {\bibfnamefont
  {B.}~\bibnamefont {Schutz}}}\ (\bibinfo  {publisher} {Springer International
  Publishing},\ \bibinfo {year} {2015})\ pp.\ \bibinfo {pages} {487--522},\
  \Eprint {http://arxiv.org/abs/1501.07322} {arXiv:1501.07322} \BibitemShut
  {NoStop}%
\bibitem [{\citenamefont {Warburton}\ and\ \citenamefont
  {Barack}(2011)}]{Warburton-Barack-2011}%
  \BibitemOpen
  \bibfield  {author} {\bibinfo {author} {\bibfnamefont {N.}~\bibnamefont
  {Warburton}}\ and\ \bibinfo {author} {\bibfnamefont {L.}~\bibnamefont
  {Barack}},\ }\href {\doibase 10.1103/PhysRevD.83.124038} {\bibfield
  {journal} {\bibinfo  {journal} {Phys. Rev. D}\ }\textbf {\bibinfo {volume}
  {83}},\ \bibinfo {pages} {124038} (\bibinfo {year} {2011})},\ \Eprint
  {http://arxiv.org/abs/arXiv:1103.0287} {arXiv:1103.0287} \BibitemShut
  {NoStop}%
\bibitem [{\citenamefont {Wardell}\ \emph
  {et~al.}(2012{\natexlab{a}})\citenamefont {Wardell}, \citenamefont {Vega},
  \citenamefont {Thornburg},\ and\ \citenamefont {Diener}}]{Wardell:2011gb}%
  \BibitemOpen
  \bibfield  {author} {\bibinfo {author} {\bibfnamefont {B.}~\bibnamefont
  {Wardell}}, \bibinfo {author} {\bibfnamefont {I.}~\bibnamefont {Vega}},
  \bibinfo {author} {\bibfnamefont {J.}~\bibnamefont {Thornburg}}, \ and\
  \bibinfo {author} {\bibfnamefont {P.}~\bibnamefont {Diener}},\ }\href
  {\doibase 10.1103/PhysRevD.85.104044} {\bibfield  {journal} {\bibinfo
  {journal} {Phys.Rev.}\ }\textbf {\bibinfo {volume} {D85}},\ \bibinfo {pages}
  {104044} (\bibinfo {year} {2012}{\natexlab{a}})},\ \Eprint
  {http://arxiv.org/abs/1112.6355} {arXiv:1112.6355 [gr-qc]} \BibitemShut
  {NoStop}%
%%CITATION = ARXIV:1112.6355;%%
\bibitem [{\citenamefont {Blanchet}\ \emph {et~al.}(2010)\citenamefont
  {Blanchet}, \citenamefont {Detweiler}, \citenamefont {{Le Tiec}},\ and\
  \citenamefont {Whiting}}]{Blanchet-etal-2010:cmp-3PN-with-self-force}%
  \BibitemOpen
  \bibfield  {author} {\bibinfo {author} {\bibfnamefont {L.}~\bibnamefont
  {Blanchet}}, \bibinfo {author} {\bibfnamefont {S.}~\bibnamefont {Detweiler}},
  \bibinfo {author} {\bibfnamefont {A.}~\bibnamefont {{Le Tiec}}}, \ and\
  \bibinfo {author} {\bibfnamefont {B.~F.}\ \bibnamefont {Whiting}},\ }\href
  {\doibase 10.1103/PhysRevD.81.064004} {\bibfield  {journal} {\bibinfo
  {journal} {Phys. Rev. D}\ }\textbf {\bibinfo {volume} {81}},\ \bibinfo
  {pages} {064004} (\bibinfo {year} {2010})},\ \Eprint
  {http://arxiv.org/abs/arXiv:0910.0207} {arXiv:0910.0207} \BibitemShut
  {NoStop}%
\bibitem [{\citenamefont {Shah}\ \emph {et~al.}(2011)\citenamefont {Shah},
  \citenamefont {Keidl}, \citenamefont {Kim},\ and\ \citenamefont
  {Price}}]{Shah-etal-2011}%
  \BibitemOpen
  \bibfield  {author} {\bibinfo {author} {\bibfnamefont {A.~G.}\ \bibnamefont
  {Shah}}, \bibinfo {author} {\bibfnamefont {T.~S.}\ \bibnamefont {Keidl}},
  \bibinfo {author} {\bibfnamefont {J.~L. F. D.-H.}\ \bibnamefont {Kim}}, \
  and\ \bibinfo {author} {\bibfnamefont {L.~R.}\ \bibnamefont {Price}},\ }\href
  {\doibase 10.1103/PhysRevD.83.064018} {\bibfield  {journal} {\bibinfo
  {journal} {Phys. Rev. D}\ }\textbf {\bibinfo {volume} {83}},\ \bibinfo
  {pages} {064018} (\bibinfo {year} {2011})},\ \Eprint
  {http://arxiv.org/abs/arXiv:1009.4876} {arXiv:1009.4876} \BibitemShut
  {NoStop}%
\bibitem [{\citenamefont {Heffernan}\ \emph {et~al.}(2012)\citenamefont
  {Heffernan}, \citenamefont {Ottewill},\ and\ \citenamefont
  {Wardell}}]{Heffernan:2012su}%
  \BibitemOpen
  \bibfield  {author} {\bibinfo {author} {\bibfnamefont {A.}~\bibnamefont
  {Heffernan}}, \bibinfo {author} {\bibfnamefont {A.}~\bibnamefont {Ottewill}},
  \ and\ \bibinfo {author} {\bibfnamefont {B.}~\bibnamefont {Wardell}},\ }\href
  {\doibase 10.1103/PhysRevD.86.104023} {\bibfield  {journal} {\bibinfo
  {journal} {Phys.Rev.}\ }\textbf {\bibinfo {volume} {D86}},\ \bibinfo {pages}
  {104023} (\bibinfo {year} {2012})},\ \Eprint {http://arxiv.org/abs/1204.0794}
  {arXiv:1204.0794 [gr-qc]} \BibitemShut {NoStop}%
%%CITATION = ARXIV:1204.0794;%%
\bibitem [{\citenamefont {Johnson-McDaniel}\ \emph {et~al.}(2015)\citenamefont
  {Johnson-McDaniel}, \citenamefont {Shah},\ and\ \citenamefont
  {Whiting}}]{Johnson-McDaniel-Shah-Whiting-2015}%
  \BibitemOpen
  \bibfield  {author} {\bibinfo {author} {\bibfnamefont {N.~K.}\ \bibnamefont
  {Johnson-McDaniel}}, \bibinfo {author} {\bibfnamefont {A.~G.}\ \bibnamefont
  {Shah}}, \ and\ \bibinfo {author} {\bibfnamefont {B.~F.}\ \bibnamefont
  {Whiting}},\ }\href {\doibase 10.1103/PhysRevD.92.044007} {\bibfield
  {journal} {\bibinfo  {journal} {Phys. Rev. D}\ }\textbf {\bibinfo {volume}
  {92}},\ \bibinfo {pages} {044007} (\bibinfo {year} {2015})},\ \Eprint
  {http://arxiv.org/abs/arXiv:1503.02638} {arXiv:1503.02638} \BibitemShut
  {NoStop}%
\bibitem [{\citenamefont {Glampedakis}\ and\ \citenamefont
  {Kennefick}(2002)}]{Glampedakis-Kennefick-2002}%
  \BibitemOpen
  \bibfield  {author} {\bibinfo {author} {\bibfnamefont {K.}~\bibnamefont
  {Glampedakis}}\ and\ \bibinfo {author} {\bibfnamefont {D.}~\bibnamefont
  {Kennefick}},\ }\href {\doibase 10.1103/PhysRevD.66.044002} {\bibfield
  {journal} {\bibinfo  {journal} {Phys. Rev. D}\ }\textbf {\bibinfo {volume}
  {66}},\ \bibinfo {pages} {044002} (\bibinfo {year} {2002})}\BibitemShut
  {NoStop}%
\bibitem [{\citenamefont {Barack}\ and\ \citenamefont
  {Lousto}(2005)}]{Barack-Lousto-2005}%
  \BibitemOpen
  \bibfield  {author} {\bibinfo {author} {\bibfnamefont {L.}~\bibnamefont
  {Barack}}\ and\ \bibinfo {author} {\bibfnamefont {C.~O.}\ \bibnamefont
  {Lousto}},\ }\href {\doibase 10.1103/PhysRevD.72.104026} {\bibfield
  {journal} {\bibinfo  {journal} {Phys. Rev. D}\ }\textbf {\bibinfo {volume}
  {72}},\ \bibinfo {pages} {104026} (\bibinfo {year} {2005})},\ \Eprint
  {http://arxiv.org/abs/gr-qc/0510019} {gr-qc/0510019} \BibitemShut {NoStop}%
\bibitem [{\citenamefont {Barack}\ \emph {et~al.}(2008)\citenamefont {Barack},
  \citenamefont {Ori},\ and\ \citenamefont {Sago}}]{Barack-Ori-Sago-2008}%
  \BibitemOpen
  \bibfield  {author} {\bibinfo {author} {\bibfnamefont {L.}~\bibnamefont
  {Barack}}, \bibinfo {author} {\bibfnamefont {A.}~\bibnamefont {Ori}}, \ and\
  \bibinfo {author} {\bibfnamefont {N.}~\bibnamefont {Sago}},\ }\href {\doibase
  10.1103/PhysRevD.78.084021} {\bibfield  {journal} {\bibinfo  {journal} {Phys.
  Rev. D}\ }\textbf {\bibinfo {volume} {78}},\ \bibinfo {pages} {084021}
  (\bibinfo {year} {2008})},\ \Eprint {http://arxiv.org/abs/arXiv:0808.2315}
  {arXiv:0808.2315} \BibitemShut {NoStop}%
\bibitem [{\citenamefont {Barton}\ \emph {et~al.}(2008)\citenamefont {Barton},
  \citenamefont {Lazar}, \citenamefont {Kennefick}, \citenamefont {Khanna},\
  and\ \citenamefont
  {Burko}}]{Barton-etal:cmp-EMRI-frequency-vs-time-domain-methods}%
  \BibitemOpen
  \bibfield  {author} {\bibinfo {author} {\bibfnamefont {J.~L.}\ \bibnamefont
  {Barton}}, \bibinfo {author} {\bibfnamefont {D.~J.}\ \bibnamefont {Lazar}},
  \bibinfo {author} {\bibfnamefont {D.~J.}\ \bibnamefont {Kennefick}}, \bibinfo
  {author} {\bibfnamefont {G.}~\bibnamefont {Khanna}}, \ and\ \bibinfo {author}
  {\bibfnamefont {L.~M.}\ \bibnamefont {Burko}},\ }\href {\doibase
  10.1103/PhysRevD.78.064042} {\bibfield  {journal} {\bibinfo  {journal} {Phys.
  Rev. D}\ }\textbf {\bibinfo {volume} {78}},\ \bibinfo {pages} {064042}
  (\bibinfo {year} {2008})},\ \Eprint {http://arxiv.org/abs/arXiv:0804.1075}
  {arXiv:0804.1075} \BibitemShut {NoStop}%
\bibitem [{\citenamefont {Haas}(2007)}]{Haas-2007}%
  \BibitemOpen
  \bibfield  {author} {\bibinfo {author} {\bibfnamefont {R.}~\bibnamefont
  {Haas}},\ }\href {\doibase 10.1103/PhysRevD.75.124011} {\bibfield  {journal}
  {\bibinfo  {journal} {Phys. Rev. D}\ }\textbf {\bibinfo {volume} {75}},\
  \bibinfo {pages} {124011} (\bibinfo {year} {2007})},\ \Eprint
  {http://arxiv.org/abs/arXiv:0704.0797} {arXiv:0704.0797} \BibitemShut
  {NoStop}%
\bibitem [{\citenamefont {Barack}\ and\ \citenamefont
  {Sago}(2010)}]{Barack-Sago-2010}%
  \BibitemOpen
  \bibfield  {author} {\bibinfo {author} {\bibfnamefont {L.}~\bibnamefont
  {Barack}}\ and\ \bibinfo {author} {\bibfnamefont {N.}~\bibnamefont {Sago}},\
  }\href {\doibase 10.1103/PhysRevD.81.084021} {\bibfield  {journal} {\bibinfo
  {journal} {Phys. Rev. D}\ }\textbf {\bibinfo {volume} {81}},\ \bibinfo
  {pages} {084021} (\bibinfo {year} {2010})},\ \Eprint
  {http://arxiv.org/abs/arXiv:1002.2386} {arXiv:1002.2386} \BibitemShut
  {NoStop}%
\bibitem [{\citenamefont {Hopman}\ and\ \citenamefont
  {Alexander}(2005)}]{Hopman-Alexander-2005}%
  \BibitemOpen
  \bibfield  {author} {\bibinfo {author} {\bibfnamefont {C.}~\bibnamefont
  {Hopman}}\ and\ \bibinfo {author} {\bibfnamefont {T.}~\bibnamefont
  {Alexander}},\ }\href {\doibase 10.1086/431475} {\bibfield  {journal}
  {\bibinfo  {journal} {The Astrophysical Journal}\ }\textbf {\bibinfo {volume}
  {629}},\ \bibinfo {pages} {362} (\bibinfo {year} {2005})},\ \Eprint
  {http://arxiv.org/abs/astro-ph/0503672} {astro-ph/0503672} \BibitemShut
  {NoStop}%
\bibitem [{\citenamefont {Wald}(1984)}]{Wald-1984}%
  \BibitemOpen
  \bibfield  {author} {\bibinfo {author} {\bibfnamefont {R.~M.}\ \bibnamefont
  {Wald}},\ }\href@noop {} {\emph {\bibinfo {title} {General relativity}}}\
  (\bibinfo  {publisher} {The University of Chicago Press},\ \bibinfo {address}
  {Chicago},\ \bibinfo {year} {1984})\BibitemShut {NoStop}%
\bibitem [{\citenamefont {Brill}\ \emph {et~al.}(1972)\citenamefont {Brill},
  \citenamefont {Chrzanowski}, \citenamefont {Pereira}, \citenamefont
  {Fackerell},\ and\ \citenamefont {Ipser}}]{Brill-etal-1972}%
  \BibitemOpen
  \bibfield  {author} {\bibinfo {author} {\bibfnamefont {D.~R.}\ \bibnamefont
  {Brill}}, \bibinfo {author} {\bibfnamefont {P.~L.}\ \bibnamefont
  {Chrzanowski}}, \bibinfo {author} {\bibfnamefont {C.~M.}\ \bibnamefont
  {Pereira}}, \bibinfo {author} {\bibfnamefont {E.~D.}\ \bibnamefont
  {Fackerell}}, \ and\ \bibinfo {author} {\bibfnamefont {J.~R.}\ \bibnamefont
  {Ipser}},\ }\href {\doibase 10.1103/PhysRevD.5.1913} {\bibfield  {journal}
  {\bibinfo  {journal} {Phys. Rev. D}\ }\textbf {\bibinfo {volume} {5}},\
  \bibinfo {pages} {1913} (\bibinfo {year} {1972})}\BibitemShut {NoStop}%
\bibitem [{\citenamefont {Teukolsky}(1973)}]{Teukolsky73}%
  \BibitemOpen
  \bibfield  {author} {\bibinfo {author} {\bibfnamefont {S.~A.}\ \bibnamefont
  {Teukolsky}},\ }\href@noop {} {\bibfield  {journal} {\bibinfo  {journal}
  {Astrophys. J.}\ }\textbf {\bibinfo {volume} {185}},\ \bibinfo {pages} {635}
  (\bibinfo {year} {1973})}\BibitemShut {NoStop}%
\bibitem [{\citenamefont {Vega}\ \emph {et~al.}(2011)\citenamefont {Vega},
  \citenamefont {Wardell},\ and\ \citenamefont
  {Diener}}]{Vega-Wardell-Diener-2011:effective-source-for-self-force}%
  \BibitemOpen
  \bibfield  {author} {\bibinfo {author} {\bibfnamefont {I.}~\bibnamefont
  {Vega}}, \bibinfo {author} {\bibfnamefont {B.}~\bibnamefont {Wardell}}, \
  and\ \bibinfo {author} {\bibfnamefont {P.}~\bibnamefont {Diener}},\ }\href
  {\doibase 10.1088/0264-9381/28/13/134010} {\bibfield  {journal} {\bibinfo
  {journal} {Class. Quant. Grav.}\ }\textbf {\bibinfo {volume} {28}},\ \bibinfo
  {pages} {134010} (\bibinfo {year} {2011})},\ \Eprint
  {http://arxiv.org/abs/arXiv:1101.2925} {arXiv:1101.2925} \BibitemShut
  {NoStop}%
\bibitem [{\citenamefont {Vega}\ \emph {et~al.}(2009)\citenamefont {Vega},
  \citenamefont {Diener}, \citenamefont {Tichy},\ and\ \citenamefont
  {Detweiler}}]{Vega-etal-2009:self-force-3+1-primer}%
  \BibitemOpen
  \bibfield  {author} {\bibinfo {author} {\bibfnamefont {I.}~\bibnamefont
  {Vega}}, \bibinfo {author} {\bibfnamefont {P.}~\bibnamefont {Diener}},
  \bibinfo {author} {\bibfnamefont {W.}~\bibnamefont {Tichy}}, \ and\ \bibinfo
  {author} {\bibfnamefont {S.}~\bibnamefont {Detweiler}},\ }\href {\doibase
  10.1103/PhysRevD.80.084021} {\bibfield  {journal} {\bibinfo  {journal} {Phys.
  Rev. D}\ }\textbf {\bibinfo {volume} {80}},\ \bibinfo {pages} {084021}
  (\bibinfo {year} {2009})},\ \Eprint {http://arxiv.org/abs/arXiv:0908.2138}
  {arXiv:0908.2138} \BibitemShut {NoStop}%
\bibitem [{\citenamefont {Diener}\ \emph {et~al.}(2012)\citenamefont {Diener},
  \citenamefont {Vega}, \citenamefont {Wardell},\ and\ \citenamefont
  {Detweiler}}]{Diener-etal-2012:self-consistent-Schw-orbital-evolution}%
  \BibitemOpen
  \bibfield  {author} {\bibinfo {author} {\bibfnamefont {P.}~\bibnamefont
  {Diener}}, \bibinfo {author} {\bibfnamefont {I.}~\bibnamefont {Vega}},
  \bibinfo {author} {\bibfnamefont {B.}~\bibnamefont {Wardell}}, \ and\
  \bibinfo {author} {\bibfnamefont {S.}~\bibnamefont {Detweiler}},\ }\href
  {\doibase 10.1103/PhysRevLett.108.191102} {\bibfield  {journal} {\bibinfo
  {journal} {Phys. Rev. Lett.}\ }\textbf {\bibinfo {volume} {108}},\ \bibinfo
  {pages} {191102} (\bibinfo {year} {2012})},\ \Eprint
  {http://arxiv.org/abs/arXiv:1112.4821} {arXiv:1112.4821} \BibitemShut
  {NoStop}%
\bibitem [{\citenamefont {Vega}\ \emph {et~al.}(2013)\citenamefont {Vega},
  \citenamefont {Wardell}, \citenamefont {Diener}, \citenamefont {Cupp},\ and\
  \citenamefont
  {Haas}}]{Vega-etal-2013:Schwarzschild-scalar-self-force-via-effective-src}%
  \BibitemOpen
  \bibfield  {author} {\bibinfo {author} {\bibfnamefont {I.}~\bibnamefont
  {Vega}}, \bibinfo {author} {\bibfnamefont {B.}~\bibnamefont {Wardell}},
  \bibinfo {author} {\bibfnamefont {P.}~\bibnamefont {Diener}}, \bibinfo
  {author} {\bibfnamefont {S.}~\bibnamefont {Cupp}}, \ and\ \bibinfo {author}
  {\bibfnamefont {R.}~\bibnamefont {Haas}},\ }\href {\doibase
  10.1103/PhysRevD.88.084021} {\bibfield  {journal} {\bibinfo  {journal} {Phys.
  Rev. D}\ }\textbf {\bibinfo {volume} {88}},\ \bibinfo {pages} {084021}
  (\bibinfo {year} {2013})},\ \Eprint {http://arxiv.org/abs/arXiv:1307.3476}
  {arXiv:1307.3476} \BibitemShut {NoStop}%
\bibitem [{\citenamefont {Barack}\ \emph {et~al.}(2007)\citenamefont {Barack},
  \citenamefont {Golbourn},\ and\ \citenamefont
  {Sago}}]{Barack-Golbourn-Sago-2007}%
  \BibitemOpen
  \bibfield  {author} {\bibinfo {author} {\bibfnamefont {L.}~\bibnamefont
  {Barack}}, \bibinfo {author} {\bibfnamefont {D.~A.}\ \bibnamefont
  {Golbourn}}, \ and\ \bibinfo {author} {\bibfnamefont {N.}~\bibnamefont
  {Sago}},\ }\href {\doibase 10.1103/PhysRevD.76.124036} {\bibfield  {journal}
  {\bibinfo  {journal} {Phys. Rev. D}\ }\textbf {\bibinfo {volume} {76}},\
  \bibinfo {pages} {124036} (\bibinfo {year} {2007})},\ \Eprint
  {http://arxiv.org/abs/arXiv:0709.4588} {arXiv:0709.4588} \BibitemShut
  {NoStop}%
\bibitem [{\citenamefont {Dolan}\ and\ \citenamefont
  {Barack}(2011)}]{Dolan-Barack-2011}%
  \BibitemOpen
  \bibfield  {author} {\bibinfo {author} {\bibfnamefont {S.~R.}\ \bibnamefont
  {Dolan}}\ and\ \bibinfo {author} {\bibfnamefont {L.}~\bibnamefont {Barack}},\
  }\href {\doibase 10.1103/PhysRevD.83.024019} {\bibfield  {journal} {\bibinfo
  {journal} {Phys. Rev. D}\ }\textbf {\bibinfo {volume} {83}},\ \bibinfo
  {pages} {024019} (\bibinfo {year} {2011})},\ \Eprint
  {http://arxiv.org/abs/arXiv:1010.5255} {arXiv:1010.5255} \BibitemShut
  {NoStop}%
\bibitem [{\citenamefont {Dolan}\ \emph {et~al.}(2011)\citenamefont {Dolan},
  \citenamefont {Wardell},\ and\ \citenamefont
  {Barack}}]{Dolan-Barack-Wardell-2011}%
  \BibitemOpen
  \bibfield  {author} {\bibinfo {author} {\bibfnamefont {S.~R.}\ \bibnamefont
  {Dolan}}, \bibinfo {author} {\bibfnamefont {B.}~\bibnamefont {Wardell}}, \
  and\ \bibinfo {author} {\bibfnamefont {L.}~\bibnamefont {Barack}},\ }\href
  {\doibase 10.1103/PhysRevD.84.084001} {\bibfield  {journal} {\bibinfo
  {journal} {Phys. Rev. D}\ }\textbf {\bibinfo {volume} {84}},\ \bibinfo
  {pages} {084001} (\bibinfo {year} {2011})},\ \Eprint
  {http://arxiv.org/abs/arXiv:1010.5255} {arXiv:1010.5255} \BibitemShut
  {NoStop}%
\bibitem [{\citenamefont {Dolan}\ and\ \citenamefont
  {Barack}(2013)}]{Dolan-Barack-2013}%
  \BibitemOpen
  \bibfield  {author} {\bibinfo {author} {\bibfnamefont {S.~R.}\ \bibnamefont
  {Dolan}}\ and\ \bibinfo {author} {\bibfnamefont {L.}~\bibnamefont {Barack}},\
  }\href {\doibase 10.1103/PhysRevD.87.084066} {\bibfield  {journal} {\bibinfo
  {journal} {Phys. Rev. D}\ }\textbf {\bibinfo {volume} {87}},\ \bibinfo
  {pages} {084066} (\bibinfo {year} {2013})},\ \Eprint
  {http://arxiv.org/abs/arXiv:1211.4586} {arXiv:1211.4586} \BibitemShut
  {NoStop}%
\bibitem [{\citenamefont {Krivan}\ \emph {et~al.}(1996)\citenamefont {Krivan},
  \citenamefont {Laguna},\ and\ \citenamefont
  {Papadopoulos}}]{Krivan-Laguna-Papadopuloos-1996:scalar-field-on-Kerr-background}%
  \BibitemOpen
  \bibfield  {author} {\bibinfo {author} {\bibfnamefont {W.}~\bibnamefont
  {Krivan}}, \bibinfo {author} {\bibfnamefont {P.}~\bibnamefont {Laguna}}, \
  and\ \bibinfo {author} {\bibfnamefont {P.}~\bibnamefont {Papadopoulos}},\
  }\href {\doibase 10.1103/PhysRevD.54.4728} {\bibfield  {journal} {\bibinfo
  {journal} {Phys. Rev. D}\ }\textbf {\bibinfo {volume} {54}},\ \bibinfo
  {pages} {4728} (\bibinfo {year} {1996})},\ \Eprint
  {http://arxiv.org/abs/gr-qc/9606003} {gr-qc/9606003} \BibitemShut {NoStop}%
\bibitem [{\citenamefont {Sundararajan}\ \emph {et~al.}(2007)\citenamefont
  {Sundararajan}, \citenamefont {Khanna},\ and\ \citenamefont
  {Hughes}}]{Sundararajan-Khanna-Hughes-2007}%
  \BibitemOpen
  \bibfield  {author} {\bibinfo {author} {\bibfnamefont {P.~A.}\ \bibnamefont
  {Sundararajan}}, \bibinfo {author} {\bibfnamefont {G.}~\bibnamefont
  {Khanna}}, \ and\ \bibinfo {author} {\bibfnamefont {S.~A.}\ \bibnamefont
  {Hughes}},\ }\href {\doibase 10.1103/PhysRevD.76.104005} {\bibfield
  {journal} {\bibinfo  {journal} {Phys. Rev. D}\ }\textbf {\bibinfo {volume}
  {76}},\ \bibinfo {pages} {104005} (\bibinfo {year} {2007})},\ \Eprint
  {http://arxiv.org/abs/gr-qc/0703028} {gr-qc/0703028} \BibitemShut {NoStop}%
\bibitem [{\citenamefont
  {Zengino\u{g}lu}(2008{\natexlab{a}})}]{Zenginoglu-2008:hyperboloidal-foliations-and-scri-fixing}%
  \BibitemOpen
  \bibfield  {author} {\bibinfo {author} {\bibfnamefont {A.}~\bibnamefont
  {Zengino\u{g}lu}},\ }\href {\doibase 10.1088/0264-9381/25/14/145002}
  {\bibfield  {journal} {\bibinfo  {journal} {Class. Quant. Grav.}\ }\textbf
  {\bibinfo {volume} {25}},\ \bibinfo {pages} {145002} (\bibinfo {year}
  {2008}{\natexlab{a}})},\ \Eprint {http://arxiv.org/abs/arXiv:0712.4333}
  {arXiv:0712.4333} \BibitemShut {NoStop}%
\bibitem [{\citenamefont
  {Zengino\u{g}lu}(2008{\natexlab{b}})}]{Zenginoglu-2008:hyperboloidal-evolution-with-Einstein-eqns}%
  \BibitemOpen
  \bibfield  {author} {\bibinfo {author} {\bibfnamefont {A.}~\bibnamefont
  {Zengino\u{g}lu}},\ }\href {\doibase 10.1088/0264-9381/25/19/195025}
  {\bibfield  {journal} {\bibinfo  {journal} {Class. Quant. Grav.}\ }\textbf
  {\bibinfo {volume} {25}},\ \bibinfo {pages} {195025} (\bibinfo {year}
  {2008}{\natexlab{b}})},\ \Eprint {http://arxiv.org/abs/arXiv:0808.0810}
  {arXiv:0808.0810} \BibitemShut {NoStop}%
\bibitem [{\citenamefont
  {Zengino\u{g}lu}(2011)}]{Zenginoglu-2011:hyperboloidal-layers-j-comp-phys}%
  \BibitemOpen
  \bibfield  {author} {\bibinfo {author} {\bibfnamefont {A.}~\bibnamefont
  {Zengino\u{g}lu}},\ }\href {\doibase
  http://dx.doi.org/10.1016/j.jcp.2010.12.016} {\bibfield  {journal} {\bibinfo
  {journal} {J. Comp. Phys.}\ }\textbf {\bibinfo {volume} {230}},\ \bibinfo
  {pages} {2286} (\bibinfo {year} {2011})},\ \Eprint
  {http://arxiv.org/abs/arXiv:1008.3809} {arXiv:1008.3809} \BibitemShut
  {NoStop}%
\bibitem [{\citenamefont {Zengino\u{g}lu}\ and\ \citenamefont
  {Khanna}(2011)}]{Zenginoglu-Khanna-2011:Kerr-EMRI-waveforms-via-Teukolsky-evolution}%
  \BibitemOpen
  \bibfield  {author} {\bibinfo {author} {\bibfnamefont {A.}~\bibnamefont
  {Zengino\u{g}lu}}\ and\ \bibinfo {author} {\bibfnamefont {G.}~\bibnamefont
  {Khanna}},\ }\href {\doibase 10.1103/PhysRevX.1.021017} {\bibfield  {journal}
  {\bibinfo  {journal} {Phys. Rev. X}\ }\textbf {\bibinfo {volume} {1}},\
  \bibinfo {pages} {021017} (\bibinfo {year} {2011})},\ \Eprint
  {http://arxiv.org/abs/arXiv:1108.1816} {arXiv:1108.1816} \BibitemShut
  {NoStop}%
\bibitem [{\citenamefont {Zengino\u{g}lu}\ and\ \citenamefont
  {Kidder}(2010)}]{Zenginoglu-Kidder-2010:hyperboloidal-evolution-of-scalar-field-on-Schw}%
  \BibitemOpen
  \bibfield  {author} {\bibinfo {author} {\bibfnamefont {A.}~\bibnamefont
  {Zengino\u{g}lu}}\ and\ \bibinfo {author} {\bibfnamefont {L.~E.}\
  \bibnamefont {Kidder}},\ }\href {\doibase 10.1103/PhysRevD.81.124010}
  {\bibfield  {journal} {\bibinfo  {journal} {Phys. Rev. D}\ }\textbf {\bibinfo
  {volume} {81}},\ \bibinfo {pages} {124010} (\bibinfo {year} {2010})},\
  \Eprint {http://arxiv.org/abs/arXiv:1004.0760} {arXiv:1004.0760} \BibitemShut
  {NoStop}%
\bibitem [{\citenamefont {Zengino\u{g}lu}\ and\ \citenamefont
  {Tiglio}(2009)}]{Zenginoglu-Tiglio-2009:spacelike-matching-to-null-infinity}%
  \BibitemOpen
  \bibfield  {author} {\bibinfo {author} {\bibfnamefont {A.}~\bibnamefont
  {Zengino\u{g}lu}}\ and\ \bibinfo {author} {\bibfnamefont {M.}~\bibnamefont
  {Tiglio}},\ }\href {\doibase 10.1103/PhysRevD.80.024044} {\bibfield
  {journal} {\bibinfo  {journal} {Phys. Rev. D}\ }\textbf {\bibinfo {volume}
  {80}},\ \bibinfo {pages} {024044} (\bibinfo {year} {2009})},\ \Eprint
  {http://arxiv.org/abs/arXiv:0906.3342} {arXiv:0906.3342} \BibitemShut
  {NoStop}%
\bibitem [{\citenamefont {Bernuzzi}\ \emph {et~al.}(2011)\citenamefont
  {Bernuzzi}, \citenamefont {Nagar},\ and\ \citenamefont
  {Zengino\u{g}lu}}]{Bernuzzi-Nagar-Zenginoglu-2011:Schw-EMRI-waveforms-via-EOB-evolution}%
  \BibitemOpen
  \bibfield  {author} {\bibinfo {author} {\bibfnamefont {S.}~\bibnamefont
  {Bernuzzi}}, \bibinfo {author} {\bibfnamefont {A.}~\bibnamefont {Nagar}}, \
  and\ \bibinfo {author} {\bibfnamefont {A.}~\bibnamefont {Zengino\u{g}lu}},\
  }\href {\doibase 10.1103/PhysRevD.84.084026} {\bibfield  {journal} {\bibinfo
  {journal} {Phys. Rev. D}\ }\textbf {\bibinfo {volume} {84}},\ \bibinfo
  {pages} {084026} (\bibinfo {year} {2011})},\ \Eprint
  {http://arxiv.org/abs/arXiv:1107.5402} {arXiv:1107.5402} \BibitemShut
  {NoStop}%
\bibitem [{\citenamefont {Bernuzzi}\ \emph {et~al.}(2012)\citenamefont
  {Bernuzzi}, \citenamefont {Nagar},\ and\ \citenamefont
  {Zengino\u{g}lu}}]{Bernuzzi-Nagar-Zenginoglu-2012:Schw-EMRI-horizon-absorption-effects}%
  \BibitemOpen
  \bibfield  {author} {\bibinfo {author} {\bibfnamefont {S.}~\bibnamefont
  {Bernuzzi}}, \bibinfo {author} {\bibfnamefont {A.}~\bibnamefont {Nagar}}, \
  and\ \bibinfo {author} {\bibfnamefont {A.}~\bibnamefont {Zengino\u{g}lu}},\
  }\href {\doibase 10.1103/PhysRevD.86.104038} {\bibfield  {journal} {\bibinfo
  {journal} {Phys. Rev. D}\ }\textbf {\bibinfo {volume} {86}},\ \bibinfo
  {pages} {104038} (\bibinfo {year} {2012})},\ \Eprint
  {http://arxiv.org/abs/arXiv:1207.0769} {arXiv:1207.0769} \BibitemShut
  {NoStop}%
\bibitem [{\citenamefont {Char}\ \emph {et~al.}(1983)\citenamefont {Char},
  \citenamefont {Geddes}, \citenamefont {Gentleman},\ and\ \citenamefont
  {Gonnet}}]{Char-etal-1983:Maple-design}%
  \BibitemOpen
  \bibfield  {author} {\bibinfo {author} {\bibfnamefont {B.}~\bibnamefont
  {Char}}, \bibinfo {author} {\bibfnamefont {K.~O.}\ \bibnamefont {Geddes}},
  \bibinfo {author} {\bibfnamefont {W.~M.}\ \bibnamefont {Gentleman}}, \ and\
  \bibinfo {author} {\bibfnamefont {G.~H.}\ \bibnamefont {Gonnet}},\ }in\
  \href@noop {} {\emph {\bibinfo {booktitle} {Lecture Notes in Computer Science
  162: Computer Algebra}}},\ \bibinfo {editor} {edited by\ \bibinfo {editor}
  {\bibfnamefont {J.~A.}\ \bibnamefont {van Hulzen}}}\ (\bibinfo  {publisher}
  {Springer-Verlag},\ \bibinfo {year} {1983})\ pp.\ \bibinfo {pages}
  {101--115}\BibitemShut {NoStop}%
\bibitem [{\citenamefont {Heffernan}\ \emph {et~al.}(2014)\citenamefont
  {Heffernan}, \citenamefont {Ottewill},\ and\ \citenamefont
  {Wardell}}]{Heffernan:2012vj}%
  \BibitemOpen
  \bibfield  {author} {\bibinfo {author} {\bibfnamefont {A.}~\bibnamefont
  {Heffernan}}, \bibinfo {author} {\bibfnamefont {A.}~\bibnamefont {Ottewill}},
  \ and\ \bibinfo {author} {\bibfnamefont {B.}~\bibnamefont {Wardell}},\ }\href
  {\doibase 10.1103/PhysRevD.89.024030} {\bibfield  {journal} {\bibinfo
  {journal} {Phys.Rev.}\ }\textbf {\bibinfo {volume} {D89}},\ \bibinfo {pages}
  {024030} (\bibinfo {year} {2014})},\ \Eprint {http://arxiv.org/abs/1211.6446}
  {arXiv:1211.6446 [gr-qc]} \BibitemShut {NoStop}%
%%CITATION = ARXIV:1211.6446;%%
\bibitem [{\citenamefont {Wardell}\ \emph
  {et~al.}(2012{\natexlab{b}})\citenamefont {Wardell}, \citenamefont {Vega},
  \citenamefont {Thornburg},\ and\ \citenamefont {Diener}}]{Wardell-etal-2012}%
  \BibitemOpen
  \bibfield  {author} {\bibinfo {author} {\bibfnamefont {B.}~\bibnamefont
  {Wardell}}, \bibinfo {author} {\bibfnamefont {I.}~\bibnamefont {Vega}},
  \bibinfo {author} {\bibfnamefont {J.}~\bibnamefont {Thornburg}}, \ and\
  \bibinfo {author} {\bibfnamefont {P.}~\bibnamefont {Diener}},\ }\href
  {\doibase 10.1103/PhysRevD.85.104044} {\bibfield  {journal} {\bibinfo
  {journal} {Phys. Rev. D}\ }\textbf {\bibinfo {volume} {85}},\ \bibinfo
  {pages} {104044} (\bibinfo {year} {2012}{\natexlab{b}})},\ \Eprint
  {http://arxiv.org/abs/arXiv:1112.6355} {arXiv:1112.6355} \BibitemShut
  {NoStop}%
\bibitem [{\citenamefont {Mino}\ \emph {et~al.}(2002)\citenamefont {Mino},
  \citenamefont {Nakano},\ and\ \citenamefont {Sasaki}}]{Mino:2001mq}%
  \BibitemOpen
  \bibfield  {author} {\bibinfo {author} {\bibfnamefont {Y.}~\bibnamefont
  {Mino}}, \bibinfo {author} {\bibfnamefont {H.}~\bibnamefont {Nakano}}, \ and\
  \bibinfo {author} {\bibfnamefont {M.}~\bibnamefont {Sasaki}},\ }\href
  {\doibase 10.1143/PTP.108.1039} {\bibfield  {journal} {\bibinfo  {journal}
  {Prog.Theor.Phys.}\ }\textbf {\bibinfo {volume} {108}},\ \bibinfo {pages}
  {1039} (\bibinfo {year} {2002})},\ \Eprint
  {http://arxiv.org/abs/gr-qc/0111074} {arXiv:gr-qc/0111074 [gr-qc]}
  \BibitemShut {NoStop}%
%%CITATION = GR-QC/0111074;%%
\bibitem [{\citenamefont {Prudnikov}\ \emph {et~al.}(1986)\citenamefont
  {Prudnikov}, \citenamefont {Brychkov},\ and\ \citenamefont
  {Marichev}}]{Prudnikov}%
  \BibitemOpen
  \bibfield  {author} {\bibinfo {author} {\bibfnamefont {A.~P.}\ \bibnamefont
  {Prudnikov}}, \bibinfo {author} {\bibfnamefont {Y.~A.}\ \bibnamefont
  {Brychkov}}, \ and\ \bibinfo {author} {\bibfnamefont {O.~I.}\ \bibnamefont
  {Marichev}},\ }\href@noop {} {\emph {\bibinfo {title} {Integrals and
  Series}}},\ Vol.~\bibinfo {volume} {1}\ (\bibinfo  {publisher} {Gordon and
  Breach},\ \bibinfo {year} {1986})\BibitemShut {NoStop}%
\bibitem [{\citenamefont {Field}\ \emph
  {et~al.}(2010{\natexlab{a}})\citenamefont {Field}, \citenamefont
  {Hesthaven},\ and\ \citenamefont {Lau}}]{Field-Hesthaven-Lau-2010}%
  \BibitemOpen
  \bibfield  {author} {\bibinfo {author} {\bibfnamefont {S.~E.}\ \bibnamefont
  {Field}}, \bibinfo {author} {\bibfnamefont {J.~S.}\ \bibnamefont
  {Hesthaven}}, \ and\ \bibinfo {author} {\bibfnamefont {S.~R.}\ \bibnamefont
  {Lau}},\ }\href {\doibase 10.1103/PhysRevD.81.124030} {\bibfield  {journal}
  {\bibinfo  {journal} {Phys. Rev. D}\ }\textbf {\bibinfo {volume} {81}},\
  \bibinfo {pages} {124030} (\bibinfo {year} {2010}{\natexlab{a}})},\ \Eprint
  {http://arxiv.org/abs/arXiv:1001.2578} {arXiv:1001.2578} \BibitemShut
  {NoStop}%
\bibitem [{\citenamefont {Jaramillo}\ \emph {et~al.}(2011)\citenamefont
  {Jaramillo}, \citenamefont {Sopuerta},\ and\ \citenamefont
  {Canizares}}]{Jaramillo-Sopuerta-Canizares-2011}%
  \BibitemOpen
  \bibfield  {author} {\bibinfo {author} {\bibfnamefont {J.~L.}\ \bibnamefont
  {Jaramillo}}, \bibinfo {author} {\bibfnamefont {C.~F.}\ \bibnamefont
  {Sopuerta}}, \ and\ \bibinfo {author} {\bibfnamefont {P.}~\bibnamefont
  {Canizares}},\ }\href {\doibase 10.1103/PhysRevD.83.061503} {\bibfield
  {journal} {\bibinfo  {journal} {Phys. Rev. D}\ }\textbf {\bibinfo {volume}
  {83}},\ \bibinfo {pages} {061503} (\bibinfo {year} {2011})},\ \Eprint
  {http://arxiv.org/abs/arXiv:1101.2324} {arXiv:1101.2324} \BibitemShut
  {NoStop}%
\bibitem [{\citenamefont
  {Thornburg}(2010)}]{Thornburg-2010:highly-accurate-self-force}%
  \BibitemOpen
  \bibfield  {author} {\bibinfo {author} {\bibfnamefont {J.}~\bibnamefont
  {Thornburg}},\ }\href@noop {} {\enquote {\bibinfo {title} {Highly accurate
  and efficient self-force computation using time-domain methods: Error
  estimates, validation, and optimization},}\ } (\bibinfo {year} {2010}),\
  \bibinfo {note} {arXiv:1006.3788},\ \Eprint
  {http://arxiv.org/abs/arXiv:1006.3788} {arXiv:1006.3788} \BibitemShut
  {NoStop}%
\bibitem [{\citenamefont {Savitzky}\ and\ \citenamefont
  {Golay}(1964)}]{Savitzky-Golay-1964:smoothing}%
  \BibitemOpen
  \bibfield  {author} {\bibinfo {author} {\bibfnamefont {A.}~\bibnamefont
  {Savitzky}}\ and\ \bibinfo {author} {\bibfnamefont {M.~J.~E.}\ \bibnamefont
  {Golay}},\ }\href {\doibase doi:10.1021/ac60214a047} {\bibfield  {journal}
  {\bibinfo  {journal} {Analytical Chemistry}\ }\textbf {\bibinfo {volume}
  {36}},\ \bibinfo {pages} {1627} (\bibinfo {year} {1964})}\BibitemShut
  {NoStop}%
\bibitem [{\citenamefont {Press}\ \emph {et~al.}(2007)\citenamefont {Press},
  \citenamefont {Teukolsky}, \citenamefont {Vetterling},\ and\ \citenamefont
  {Flannery}}]{Numerical-Recipes-3rd-edition}%
  \BibitemOpen
  \bibfield  {author} {\bibinfo {author} {\bibfnamefont {W.~H.}\ \bibnamefont
  {Press}}, \bibinfo {author} {\bibfnamefont {S.~A.}\ \bibnamefont
  {Teukolsky}}, \bibinfo {author} {\bibfnamefont {W.~T.}\ \bibnamefont
  {Vetterling}}, \ and\ \bibinfo {author} {\bibfnamefont {B.~P.}\ \bibnamefont
  {Flannery}},\ }\href@noop {} {\emph {\bibinfo {title} {Numerical Recipes}}},\
  \bibinfo {edition} {3rd}\ ed.\ (\bibinfo  {publisher} {Cambridge University
  Press},\ \bibinfo {year} {2007})\BibitemShut {NoStop}%
\bibitem [{\citenamefont {Mino}(2003)}]{Mino-2003}%
  \BibitemOpen
  \bibfield  {author} {\bibinfo {author} {\bibfnamefont {Y.}~\bibnamefont
  {Mino}},\ }\href {\doibase 10.1103/PhysRevD.67.084027} {\bibfield  {journal}
  {\bibinfo  {journal} {Phys. Rev. D}\ }\textbf {\bibinfo {volume} {67}},\
  \bibinfo {pages} {084027} (\bibinfo {year} {2003})},\ \Eprint
  {http://arxiv.org/abs/gr-qc/0302075} {gr-qc/0302075} \BibitemShut {NoStop}%
\bibitem [{\citenamefont {Hinderer}\ and\ \citenamefont
  {Flanagan}(2008)}]{Hinderer-Flanagan-2008}%
  \BibitemOpen
  \bibfield  {author} {\bibinfo {author} {\bibfnamefont {T.}~\bibnamefont
  {Hinderer}}\ and\ \bibinfo {author} {\bibfnamefont {E.~E.}\ \bibnamefont
  {Flanagan}},\ }\href {\doibase 10.1103/PhysRevD.78.064028} {\bibfield
  {journal} {\bibinfo  {journal} {Phys. Rev. D}\ }\textbf {\bibinfo {volume}
  {78}},\ \bibinfo {pages} {064028} (\bibinfo {year} {2008})},\ \Eprint
  {http://arxiv.org/abs/arXiv:0805.3337} {arXiv:0805.3337} \BibitemShut
  {NoStop}%
\bibitem [{\citenamefont {Diaz-Rivera}\ \emph {et~al.}(2004)\citenamefont
  {Diaz-Rivera}, \citenamefont {Messaritaki}, \citenamefont {Whiting},\ and\
  \citenamefont {Detweiler}}]{Diaz-Rivera-etal-2004}%
  \BibitemOpen
  \bibfield  {author} {\bibinfo {author} {\bibfnamefont {L.~M.}\ \bibnamefont
  {Diaz-Rivera}}, \bibinfo {author} {\bibfnamefont {E.}~\bibnamefont
  {Messaritaki}}, \bibinfo {author} {\bibfnamefont {B.~F.}\ \bibnamefont
  {Whiting}}, \ and\ \bibinfo {author} {\bibfnamefont {S.}~\bibnamefont
  {Detweiler}},\ }\href {\doibase 10.1103/PhysRevD.70.124018} {\bibfield
  {journal} {\bibinfo  {journal} {Phys. Rev. D}\ }\textbf {\bibinfo {volume}
  {70}},\ \bibinfo {pages} {124018} (\bibinfo {year} {2004})}\BibitemShut
  {NoStop}%
\bibitem [{\citenamefont {Barack}\ and\ \citenamefont
  {Sago}(2009)}]{Barack-Sago-2009}%
  \BibitemOpen
  \bibfield  {author} {\bibinfo {author} {\bibfnamefont {L.}~\bibnamefont
  {Barack}}\ and\ \bibinfo {author} {\bibfnamefont {N.}~\bibnamefont {Sago}},\
  }\href {\doibase 10.1103/PhysRevLett.102.191101} {\bibfield  {journal}
  {\bibinfo  {journal} {Phys. Rev. Lett.}\ }\textbf {\bibinfo {volume} {102}},\
  \bibinfo {pages} {191101} (\bibinfo {year} {2009})},\ \Eprint
  {http://arxiv.org/abs/arXiv:0902.0573} {arXiv:0902.0573} \BibitemShut
  {NoStop}%
\bibitem [{\citenamefont {Choptuik}(1991)}]{Choptuik-1991:FD-consistency}%
  \BibitemOpen
  \bibfield  {author} {\bibinfo {author} {\bibfnamefont {M.~W.}\ \bibnamefont
  {Choptuik}},\ }\href {\doibase 10.1103/PhysRevD.44.3124} {\bibfield
  {journal} {\bibinfo  {journal} {Phys. Rev. D}\ }\textbf {\bibinfo {volume}
  {44}},\ \bibinfo {pages} {3124} (\bibinfo {year} {1991})}\BibitemShut
  {NoStop}%
\bibitem [{\citenamefont {Thornburg}(1999)}]{Thornburg-1998}%
  \BibitemOpen
  \bibfield  {author} {\bibinfo {author} {\bibfnamefont {J.}~\bibnamefont
  {Thornburg}},\ }\href@noop {} {\bibfield  {journal} {\bibinfo  {journal}
  {Phys. Rev. D}\ }\textbf {\bibinfo {volume} {59}},\ \bibinfo {pages} {104007}
  (\bibinfo {year} {1999})},\ \Eprint {http://arxiv.org/abs/gr-qc/9801087}
  {gr-qc/9801087} \BibitemShut {NoStop}%
\bibitem [{\citenamefont
  {Warburton}(2016)}]{Warburton-pers-comm-2016:FD-inner-outer-approx}%
  \BibitemOpen
  \bibfield  {author} {\bibinfo {author} {\bibfnamefont {N.}~\bibnamefont
  {Warburton}},\ }\href@noop {} {\enquote {\bibinfo {title} {personal
  communication},}\ } (\bibinfo {year} {2016})\BibitemShut {NoStop}%
\bibitem [{\citenamefont
  {Barack}(2016)}]{Barack-pers-comm-2016:zoom-whirl-spikes-due-to-radial-accel}%
  \BibitemOpen
  \bibfield  {author} {\bibinfo {author} {\bibfnamefont {L.}~\bibnamefont
  {Barack}},\ }\href@noop {} {\enquote {\bibinfo {title} {personal
  communication},}\ } (\bibinfo {year} {2016})\BibitemShut {NoStop}%
\bibitem [{\citenamefont {Allen}\ \emph {et~al.}(1999)\citenamefont {Allen},
  \citenamefont {Goodale},\ and\ \citenamefont {Seidel}}]{Allen99a}%
  \BibitemOpen
  \bibfield  {author} {\bibinfo {author} {\bibfnamefont {G.}~\bibnamefont
  {Allen}}, \bibinfo {author} {\bibfnamefont {T.}~\bibnamefont {Goodale}}, \
  and\ \bibinfo {author} {\bibfnamefont {E.}~\bibnamefont {Seidel}},\ }in\
  \href@noop {} {\emph {\bibinfo {booktitle} {7th Symposium on the Frontiers of
  Massively Parallel Computation-Frontiers 99}}}\ (\bibinfo  {publisher}
  {IEEE},\ \bibinfo {address} {New York},\ \bibinfo {year} {1999})\BibitemShut
  {NoStop}%
\bibitem [{\citenamefont {Goodale}\ \emph {et~al.}(2003)\citenamefont
  {Goodale}, \citenamefont {Allen}, \citenamefont {Lanfermann}, \citenamefont
  {Mass{\'o}}, \citenamefont {Radke}, \citenamefont {Seidel},\ and\
  \citenamefont {Shalf}}]{Goodale02a}%
  \BibitemOpen
  \bibfield  {author} {\bibinfo {author} {\bibfnamefont {T.}~\bibnamefont
  {Goodale}}, \bibinfo {author} {\bibfnamefont {G.}~\bibnamefont {Allen}},
  \bibinfo {author} {\bibfnamefont {G.}~\bibnamefont {Lanfermann}}, \bibinfo
  {author} {\bibfnamefont {J.}~\bibnamefont {Mass{\'o}}}, \bibinfo {author}
  {\bibfnamefont {T.}~\bibnamefont {Radke}}, \bibinfo {author} {\bibfnamefont
  {E.}~\bibnamefont {Seidel}}, \ and\ \bibinfo {author} {\bibfnamefont
  {J.}~\bibnamefont {Shalf}},\ }in\ \href@noop {} {\emph {\bibinfo {booktitle}
  {Vector and Parallel Processing -- VECPAR'2002, 5th International Conference,
  Lecture Notes in Computer Science}}}\ (\bibinfo  {publisher} {Springer},\
  \bibinfo {address} {Berlin},\ \bibinfo {year} {2003})\BibitemShut {NoStop}%
\bibitem [{Cactus developers()}]{Cactuscode:web}%
  \BibitemOpen
  Cactus developers,\ \href {http://www.cactuscode.org/} {\enquote {\bibinfo
  {title} {{Cactus Computational Toolkit}},}\ }\BibitemShut {NoStop}%
\bibitem [{\citenamefont {Dubey}\ \emph {et~al.}(2014)\citenamefont {Dubey},
  \citenamefont {Almgren}, \citenamefont {Bell}, \citenamefont {Berzins},
  \citenamefont {Brandt}, \citenamefont {Bryan}, \citenamefont {Colella},
  \citenamefont {Graves}, \citenamefont {Lijewski}, \citenamefont
  {L\"{o}ffler}, \citenamefont {O'Shea}, \citenamefont {Schnetter},
  \citenamefont {Straalen},\ and\ \citenamefont
  {Weide}}]{Dubey-etal-2014:survey-of-AMR-frameworks}%
  \BibitemOpen
  \bibfield  {author} {\bibinfo {author} {\bibfnamefont {A.}~\bibnamefont
  {Dubey}}, \bibinfo {author} {\bibfnamefont {A.}~\bibnamefont {Almgren}},
  \bibinfo {author} {\bibfnamefont {J.}~\bibnamefont {Bell}}, \bibinfo {author}
  {\bibfnamefont {M.}~\bibnamefont {Berzins}}, \bibinfo {author} {\bibfnamefont
  {S.}~\bibnamefont {Brandt}}, \bibinfo {author} {\bibfnamefont
  {G.}~\bibnamefont {Bryan}}, \bibinfo {author} {\bibfnamefont
  {P.}~\bibnamefont {Colella}}, \bibinfo {author} {\bibfnamefont
  {D.}~\bibnamefont {Graves}}, \bibinfo {author} {\bibfnamefont
  {M.}~\bibnamefont {Lijewski}}, \bibinfo {author} {\bibfnamefont
  {F.}~\bibnamefont {L\"{o}ffler}}, \bibinfo {author} {\bibfnamefont
  {B.}~\bibnamefont {O'Shea}}, \bibinfo {author} {\bibfnamefont
  {E.}~\bibnamefont {Schnetter}}, \bibinfo {author} {\bibfnamefont {B.~V.}\
  \bibnamefont {Straalen}}, \ and\ \bibinfo {author} {\bibfnamefont
  {K.}~\bibnamefont {Weide}},\ }\href {\doibase 10.1016/j.jpdc.2014.07.001}
  {\bibfield  {journal} {\bibinfo  {journal} {J. Parallel Distrib. Comput.}\
  }\textbf {\bibinfo {volume} {74}},\ \bibinfo {pages} {3217} (\bibinfo {year}
  {2014})},\ \Eprint
  {http://arxiv.org/abs/http://cdmbuntu.lib.utah.edu/utils/getfile/collection/uspace/id/10677/filename/10677.pdf}
  {http://cdmbuntu.lib.utah.edu/utils/getfile/collection/uspace/id/10677/filename/10677.pdf}
  \BibitemShut {NoStop}%
\bibitem [{\citenamefont {Schnetter}\ \emph {et~al.}(2004)\citenamefont
  {Schnetter}, \citenamefont {Hawley},\ and\ \citenamefont
  {Hawke}}]{Schnetter-Hawley-Hawke-2004:Carpet-paper}%
  \BibitemOpen
  \bibfield  {author} {\bibinfo {author} {\bibfnamefont {E.}~\bibnamefont
  {Schnetter}}, \bibinfo {author} {\bibfnamefont {S.~H.}\ \bibnamefont
  {Hawley}}, \ and\ \bibinfo {author} {\bibfnamefont {I.}~\bibnamefont
  {Hawke}},\ }\href {\doibase 10.1088/0264-9381/21/6/014} {\bibfield  {journal}
  {\bibinfo  {journal} {Class. Quant. Grav.}\ }\textbf {\bibinfo {volume}
  {21}},\ \bibinfo {pages} {1465} (\bibinfo {year} {2004})},\ \Eprint
  {http://arxiv.org/abs/gr-qc/0310042} {gr-qc/0310042} \BibitemShut {NoStop}%
\bibitem [{\citenamefont {L{\"{o}}ffler}\ \emph {et~al.}(2012)\citenamefont
  {L{\"{o}}ffler}, \citenamefont {Faber}, \citenamefont {Bentivegna},
  \citenamefont {Bode}, \citenamefont {Diener}, \citenamefont {Haas},
  \citenamefont {Hinder}, \citenamefont {Mundim}, \citenamefont {Ott},
  \citenamefont {Schnetter}, \citenamefont {Allen}, \citenamefont
  {Campanelli},\ and\ \citenamefont {Laguna}}]{Loffler:2011ay}%
  \BibitemOpen
  \bibfield  {author} {\bibinfo {author} {\bibfnamefont {F.}~\bibnamefont
  {L{\"{o}}ffler}}, \bibinfo {author} {\bibfnamefont {J.}~\bibnamefont
  {Faber}}, \bibinfo {author} {\bibfnamefont {E.}~\bibnamefont {Bentivegna}},
  \bibinfo {author} {\bibfnamefont {T.}~\bibnamefont {Bode}}, \bibinfo {author}
  {\bibfnamefont {P.}~\bibnamefont {Diener}}, \bibinfo {author} {\bibfnamefont
  {R.}~\bibnamefont {Haas}}, \bibinfo {author} {\bibfnamefont {I.}~\bibnamefont
  {Hinder}}, \bibinfo {author} {\bibfnamefont {B.~C.}\ \bibnamefont {Mundim}},
  \bibinfo {author} {\bibfnamefont {C.~D.}\ \bibnamefont {Ott}}, \bibinfo
  {author} {\bibfnamefont {E.}~\bibnamefont {Schnetter}}, \bibinfo {author}
  {\bibfnamefont {G.}~\bibnamefont {Allen}}, \bibinfo {author} {\bibfnamefont
  {M.}~\bibnamefont {Campanelli}}, \ and\ \bibinfo {author} {\bibfnamefont
  {P.}~\bibnamefont {Laguna}},\ }\href {\doibase
  doi:10.1088/0264-9381/29/11/115001} {\bibfield  {journal} {\bibinfo
  {journal} {Class. Quantum Grav.}\ }\textbf {\bibinfo {volume} {29}},\
  \bibinfo {pages} {115001} (\bibinfo {year} {2012})},\ \Eprint
  {http://arxiv.org/abs/arXiv:1111.3344 [gr-qc]} {arXiv:1111.3344 [gr-qc]}
  \BibitemShut {NoStop}%
\bibitem [{EinsteinToolkit()}]{EinsteinToolkit:web}%
  \BibitemOpen
  EinsteinToolkit,\ \href {http://einsteintoolkit.org/} {\enquote {\bibinfo
  {title} {{Einstein Toolkit}: Open software for relativistic astrophysics},}\
  }\BibitemShut {NoStop}%
\bibitem [{\citenamefont {Kidder}\ \emph {et~al.}(2016)\citenamefont {Kidder},
  \citenamefont {Field}, \citenamefont {Foucart}, \citenamefont {Schnetter},
  \citenamefont {Teukolsky}, \citenamefont {Bohn}, \citenamefont {Deppe},
  \citenamefont {Diener}, \citenamefont {H\'{e}bert}, \citenamefont {Lippuner},
  \citenamefont {Miller}, \citenamefont {Ott}, \citenamefont {Scheel},\ and\
  \citenamefont
  {Vincent}}]{Kidder-etal-2016:SpECTRE-discontinuous-Galerkin-code}%
  \BibitemOpen
  \bibfield  {author} {\bibinfo {author} {\bibfnamefont {L.~E.}\ \bibnamefont
  {Kidder}}, \bibinfo {author} {\bibfnamefont {S.~E.}\ \bibnamefont {Field}},
  \bibinfo {author} {\bibfnamefont {F.}~\bibnamefont {Foucart}}, \bibinfo
  {author} {\bibfnamefont {E.}~\bibnamefont {Schnetter}}, \bibinfo {author}
  {\bibfnamefont {S.~A.}\ \bibnamefont {Teukolsky}}, \bibinfo {author}
  {\bibfnamefont {A.}~\bibnamefont {Bohn}}, \bibinfo {author} {\bibfnamefont
  {N.}~\bibnamefont {Deppe}}, \bibinfo {author} {\bibfnamefont
  {P.}~\bibnamefont {Diener}}, \bibinfo {author} {\bibfnamefont
  {F.}~\bibnamefont {H\'{e}bert}}, \bibinfo {author} {\bibfnamefont
  {J.}~\bibnamefont {Lippuner}}, \bibinfo {author} {\bibfnamefont
  {J.}~\bibnamefont {Miller}}, \bibinfo {author} {\bibfnamefont {C.~D.}\
  \bibnamefont {Ott}}, \bibinfo {author} {\bibfnamefont {M.~A.}\ \bibnamefont
  {Scheel}}, \ and\ \bibinfo {author} {\bibfnamefont {T.}~\bibnamefont
  {Vincent}},\ }\href@noop {} {\enquote {\bibinfo {title} {{SpECTRE}: A
  task-based discontinuous {G}alerkin code for relativistic astrophysics},}\ }
  (\bibinfo {year} {2016}),\ \bibinfo {note} {1609.00098},\ \Eprint
  {http://arxiv.org/abs/1609.00098} {1609.00098} \BibitemShut {NoStop}%
\bibitem [{\citenamefont {Hesthaven}\ and\ \citenamefont
  {Warburton}(2008)}]{Hesthaven-Warburton-book-2008}%
  \BibitemOpen
  \bibfield  {author} {\bibinfo {author} {\bibfnamefont {J.~S.}\ \bibnamefont
  {Hesthaven}}\ and\ \bibinfo {author} {\bibfnamefont {T.}~\bibnamefont
  {Warburton}},\ }\href@noop {} {\emph {\bibinfo {title} {Nodal discontinuous
  {G}alerkin methods: algorithms, analysis, and applications}}},\ Texts in
  applied mathematics\ (\bibinfo  {publisher} {Springer},\ \bibinfo {address}
  {New York, London},\ \bibinfo {year} {2008})\BibitemShut {NoStop}%
\bibitem [{\citenamefont {Field}\ \emph {et~al.}(2009)\citenamefont {Field},
  \citenamefont {Hesthaven},\ and\ \citenamefont
  {Lau}}]{Field-Hesthaven-Lau-2009:discontinuous-Galerkin-1+1-Schw-EMRI}%
  \BibitemOpen
  \bibfield  {author} {\bibinfo {author} {\bibfnamefont {S.~E.}\ \bibnamefont
  {Field}}, \bibinfo {author} {\bibfnamefont {J.~S.}\ \bibnamefont
  {Hesthaven}}, \ and\ \bibinfo {author} {\bibfnamefont {S.~R.}\ \bibnamefont
  {Lau}},\ }\href {\doibase http://dx.doi.org/10.1088/0264-9381/26/16/165010}
  {\bibfield  {journal} {\bibinfo  {journal} {Classical and Quantum Gravity}\
  }\textbf {\bibinfo {volume} {26}},\ \bibinfo {pages} {165010} (\bibinfo
  {year} {2009})},\ \Eprint {http://arxiv.org/abs/arXiv:0902.1287}
  {arXiv:0902.1287} \BibitemShut {NoStop}%
\bibitem [{\citenamefont {Field}\ \emph
  {et~al.}(2010{\natexlab{b}})\citenamefont {Field}, \citenamefont {Hesthaven},
  \citenamefont {Lau},\ and\ \citenamefont
  {Mroue}}]{Field-etal-2010:discontinuous-Galerkin-2nd-order-BSSN}%
  \BibitemOpen
  \bibfield  {author} {\bibinfo {author} {\bibfnamefont {S.~E.}\ \bibnamefont
  {Field}}, \bibinfo {author} {\bibfnamefont {J.~S.}\ \bibnamefont
  {Hesthaven}}, \bibinfo {author} {\bibfnamefont {S.~R.}\ \bibnamefont {Lau}},
  \ and\ \bibinfo {author} {\bibfnamefont {A.~H.}\ \bibnamefont {Mroue}},\
  }\href {\doibase 10.1103/PhysRevD.82.104051} {\bibfield  {journal} {\bibinfo
  {journal} {Phys. Rev. D}\ }\textbf {\bibinfo {volume} {82}},\ \bibinfo
  {pages} {104051} (\bibinfo {year} {2010}{\natexlab{b}})},\ \Eprint
  {http://arxiv.org/abs/arXiv:1008.1820} {arXiv:1008.1820} \BibitemShut
  {NoStop}%
\bibitem [{\citenamefont {Brown}\ \emph {et~al.}(2012)\citenamefont {Brown},
  \citenamefont {Diener}, \citenamefont {Field}, \citenamefont {Hesthaven},
  \citenamefont {Herrmann}, \citenamefont {Mrou\'{e}}, \citenamefont {Sarbach},
  \citenamefont {Schnetter}, \citenamefont {Tiglio},\ and\ \citenamefont
  {Wagman}}]{Brown-etal-2012:1st-order-BSSN-and-discontinuous-Galerkin}%
  \BibitemOpen
  \bibfield  {author} {\bibinfo {author} {\bibfnamefont {J.~D.}\ \bibnamefont
  {Brown}}, \bibinfo {author} {\bibfnamefont {P.}~\bibnamefont {Diener}},
  \bibinfo {author} {\bibfnamefont {S.~E.}\ \bibnamefont {Field}}, \bibinfo
  {author} {\bibfnamefont {J.~S.}\ \bibnamefont {Hesthaven}}, \bibinfo {author}
  {\bibfnamefont {F.}~\bibnamefont {Herrmann}}, \bibinfo {author}
  {\bibfnamefont {A.~H.}\ \bibnamefont {Mrou\'{e}}}, \bibinfo {author}
  {\bibfnamefont {O.}~\bibnamefont {Sarbach}}, \bibinfo {author} {\bibfnamefont
  {E.}~\bibnamefont {Schnetter}}, \bibinfo {author} {\bibfnamefont
  {M.}~\bibnamefont {Tiglio}}, \ and\ \bibinfo {author} {\bibfnamefont
  {M.}~\bibnamefont {Wagman}},\ }\href {\doibase 10.1103/PhysRevD.85.084004}
  {\bibfield  {journal} {\bibinfo  {journal} {Phys. Rev. D}\ }\textbf {\bibinfo
  {volume} {85}},\ \bibinfo {pages} {084004} (\bibinfo {year} {2012})},\
  \Eprint {http://arxiv.org/abs/arXiv:1202.1038} {arXiv:1202.1038} \BibitemShut
  {NoStop}%
\bibitem [{\citenamefont {Fan}\ \emph {et~al.}(2008)\citenamefont {Fan},
  \citenamefont {Cai},\ and\ \citenamefont
  {Ji}}]{Fan-Cai-Ji-2008:discontinuous-Galerkin-Schrodinger-eqn-nonsmooth-solns}%
  \BibitemOpen
  \bibfield  {author} {\bibinfo {author} {\bibfnamefont {K.}~\bibnamefont
  {Fan}}, \bibinfo {author} {\bibfnamefont {W.}~\bibnamefont {Cai}}, \ and\
  \bibinfo {author} {\bibfnamefont {X.}~\bibnamefont {Ji}},\ }\href {\doibase
  http://dx.doi.org/10.1016/j.jcp.2007.10.023} {\bibfield  {journal} {\bibinfo
  {journal} {J. Comput. Phys.}\ }\textbf {\bibinfo {volume} {227}},\ \bibinfo
  {pages} {2387} (\bibinfo {year} {2008})}\BibitemShut {NoStop}%
\bibitem [{\citenamefont {\'{E}anna \'{E}.~Flanagan}\ and\ \citenamefont
  {Hinderer}(2012)}]{Flanagan-Hinderer-2012}%
  \BibitemOpen
  \bibfield  {author} {\bibinfo {author} {\bibnamefont {\'{E}anna
  \'{E}.~Flanagan}}\ and\ \bibinfo {author} {\bibfnamefont {T.}~\bibnamefont
  {Hinderer}},\ }\href@noop {} {\bibfield  {journal} {\bibinfo  {journal}
  {Phys. Rev. Lett.}\ }\textbf {\bibinfo {volume} {109}},\ \bibinfo {pages}
  {071102} (\bibinfo {year} {2012})},\ \Eprint
  {http://arxiv.org/abs/arXiv:1009.4923} {arXiv:1009.4923} \BibitemShut
  {NoStop}%
\bibitem [{\citenamefont {Brink}\ \emph {et~al.}(2015)\citenamefont {Brink},
  \citenamefont {Geyer},\ and\ \citenamefont
  {Hinderer}}]{Brink-Geyer-Hinderer-2015}%
  \BibitemOpen
  \bibfield  {author} {\bibinfo {author} {\bibfnamefont {J.}~\bibnamefont
  {Brink}}, \bibinfo {author} {\bibfnamefont {M.}~\bibnamefont {Geyer}}, \ and\
  \bibinfo {author} {\bibfnamefont {T.}~\bibnamefont {Hinderer}},\ }\href
  {\doibase 10.1103/PhysRevD.91.083001} {\bibfield  {journal} {\bibinfo
  {journal} {Phys. Rev. D}\ }\textbf {\bibinfo {volume} {91}},\ \bibinfo
  {pages} {083001} (\bibinfo {year} {2015})},\ \Eprint
  {http://arxiv.org/abs/arXiv:1501.07728} {arXiv:1501.07728} \BibitemShut
  {NoStop}%
\bibitem [{\citenamefont {Thornburg}(2011)}]{Thornburg-2011:Capra-survey}%
  \BibitemOpen
  \bibfield  {author} {\bibinfo {author} {\bibfnamefont {J.}~\bibnamefont
  {Thornburg}},\ }\href@noop {} {\bibfield  {journal} {\bibinfo  {journal} {GW
  Notes}\ }\textbf {\bibinfo {volume} {5}},\ \bibinfo {pages} {3} (\bibinfo
  {year} {2011})},\ \Eprint {http://arxiv.org/abs/arXiv:1102.2857}
  {arXiv:1102.2857} \BibitemShut {NoStop}%
\bibitem [{\citenamefont {Vines}\ \emph {et~al.}(2016)\citenamefont {Vines},
  \citenamefont {Kunst}, \citenamefont {Steinhoff},\ and\ \citenamefont
  {Hinderer}}]{Vines-etal-2016:extended-body-effects}%
  \BibitemOpen
  \bibfield  {author} {\bibinfo {author} {\bibfnamefont {J.}~\bibnamefont
  {Vines}}, \bibinfo {author} {\bibfnamefont {D.}~\bibnamefont {Kunst}},
  \bibinfo {author} {\bibfnamefont {J.}~\bibnamefont {Steinhoff}}, \ and\
  \bibinfo {author} {\bibfnamefont {T.}~\bibnamefont {Hinderer}},\ }\href
  {\doibase 10.1103/PhysRevD.93.103008} {\bibfield  {journal} {\bibinfo
  {journal} {Phys. Rev. D}\ }\textbf {\bibinfo {volume} {93}},\ \bibinfo
  {pages} {103008} (\bibinfo {year} {2016})},\ \Eprint
  {http://arxiv.org/abs/arXiv:1601.07529} {arXiv:1601.07529} \BibitemShut
  {NoStop}%
\bibitem [{\citenamefont {Warburton}\ \emph {et~al.}(2012)\citenamefont
  {Warburton}, \citenamefont {Akcay}, \citenamefont {Barack}, \citenamefont
  {Gair},\ and\ \citenamefont {Sago}}]{Warburton-etal-2012:Schw-inspiral}%
  \BibitemOpen
  \bibfield  {author} {\bibinfo {author} {\bibfnamefont {N.}~\bibnamefont
  {Warburton}}, \bibinfo {author} {\bibfnamefont {S.}~\bibnamefont {Akcay}},
  \bibinfo {author} {\bibfnamefont {L.}~\bibnamefont {Barack}}, \bibinfo
  {author} {\bibfnamefont {J.~R.}\ \bibnamefont {Gair}}, \ and\ \bibinfo
  {author} {\bibfnamefont {N.}~\bibnamefont {Sago}},\ }\href {\doibase
  10.1103/PhysRevD.85.061501} {\bibfield  {journal} {\bibinfo  {journal} {Phys.
  Rev. D}\ }\textbf {\bibinfo {volume} {85}},\ \bibinfo {pages} {061501R}
  (\bibinfo {year} {2012})},\ \Eprint {http://arxiv.org/abs/arXiv:1111.6908}
  {arXiv:1111.6908} \BibitemShut {NoStop}%
\bibitem [{\citenamefont {Hindmarsh}(1983)}]{Hindmarsh-1983}%
  \BibitemOpen
  \bibfield  {author} {\bibinfo {author} {\bibfnamefont {A.~C.}\ \bibnamefont
  {Hindmarsh}},\ }\bibfield  {booktitle} {\emph {\bibinfo {booktitle}
  {Scientific Computing}},\ }\href {http://www.netlib.org/odepack/index.html}
  {\bibfield  {journal} {\bibinfo  {journal} {IMACS Transactions on Scientific
  Computing}\ }\textbf {\bibinfo {volume} {1}},\ \bibinfo {pages} {55}
  (\bibinfo {year} {1983})},\ \bibinfo {note} {article also available at
  http://www.llnl.gov/CASC/nsde/pubs/u88007.pdf}\BibitemShut {NoStop}%
\bibitem [{\citenamefont {Radhakrishnan}\ and\ \citenamefont
  {Hindmarsh}(1993)}]{Radhakrishnan-Hindmarsh:LSODE-report}%
  \BibitemOpen
  \bibfield  {author} {\bibinfo {author} {\bibfnamefont {K.}~\bibnamefont
  {Radhakrishnan}}\ and\ \bibinfo {author} {\bibfnamefont {A.~C.}\ \bibnamefont
  {Hindmarsh}},\ }\href
  {https://computation.llnl.gov/casc/nsde/pubs/u113855.pdf} {\emph {\bibinfo
  {title} {Description and Use of {LSODE}, the {L}ivermore {S}olver for
  {O}rdinary {D}ifferential {E}quations}}},\ \bibinfo {type} {Tech. Rep.}\
  \bibinfo {number} {UCRL-ID-113855}\ (\bibinfo  {institution} {{L}awrence
  {L}ivermore National {L}aboratory},\ \bibinfo {year} {1993})\ \bibinfo {note}
  {{NASA} Reference Publication 1327}\BibitemShut {NoStop}%
\bibitem [{\citenamefont {Goldberg}(1991)}]{Goldberg91}%
  \BibitemOpen
  \bibfield  {author} {\bibinfo {author} {\bibfnamefont {D.}~\bibnamefont
  {Goldberg}},\ }\href {http://citeseer.ist.psu.edu/goldberg91what.html}
  {\bibfield  {journal} {\bibinfo  {journal} {ACM Computing Surveys}\ }\textbf
  {\bibinfo {volume} {23}},\ \bibinfo {pages} {5} (\bibinfo {year}
  {1991})}\BibitemShut {NoStop}%
\bibitem [{\citenamefont {Hindmarsh}\ \emph {et~al.}(2005)\citenamefont
  {Hindmarsh}, \citenamefont {Brown}, \citenamefont {Grant}, \citenamefont
  {Lee}, \citenamefont {Serban}, \citenamefont {Shumaker},\ and\ \citenamefont
  {Woodward}}]{Hindmarsh-etal-2005:SUNDIALS}%
  \BibitemOpen
  \bibfield  {author} {\bibinfo {author} {\bibfnamefont {A.~C.}\ \bibnamefont
  {Hindmarsh}}, \bibinfo {author} {\bibfnamefont {P.~N.}\ \bibnamefont
  {Brown}}, \bibinfo {author} {\bibfnamefont {K.~E.}\ \bibnamefont {Grant}},
  \bibinfo {author} {\bibfnamefont {S.~L.}\ \bibnamefont {Lee}}, \bibinfo
  {author} {\bibfnamefont {R.}~\bibnamefont {Serban}}, \bibinfo {author}
  {\bibfnamefont {D.~E.}\ \bibnamefont {Shumaker}}, \ and\ \bibinfo {author}
  {\bibfnamefont {C.~S.}\ \bibnamefont {Woodward}},\ }\href {\doibase
  10.1145/1089014.1089020} {\bibfield  {journal} {\bibinfo  {journal} {ACM
  Trans. Math. Softw.}\ }\textbf {\bibinfo {volume} {31}},\ \bibinfo {pages}
  {363} (\bibinfo {year} {2005})}\BibitemShut {NoStop}%
\bibitem [{\citenamefont {Berger}(1982)}]{Berger-1982}%
  \BibitemOpen
  \bibfield  {author} {\bibinfo {author} {\bibfnamefont {M.~J.}\ \bibnamefont
  {Berger}},\ }\emph {\bibinfo {title} {Adaptive Mesh Refinement for Hyperbolic
  Partial Differential Equations}},\ \href@noop {} {Ph.D. thesis},\ \bibinfo
  {school} {Stanford University} (\bibinfo {year} {1982}),\ \bibinfo {note}
  {{U}niversity {M}icrofilms \#DA 83-01196}\BibitemShut {NoStop}%
\bibitem [{\citenamefont {Berger}\ and\ \citenamefont
  {Oliger}(1984)}]{Berger-1984}%
  \BibitemOpen
  \bibfield  {author} {\bibinfo {author} {\bibfnamefont {M.~J.}\ \bibnamefont
  {Berger}}\ and\ \bibinfo {author} {\bibfnamefont {J.}~\bibnamefont
  {Oliger}},\ }\href {\doibase 10.1016/0021-9991(84)90073-1} {\bibfield
  {journal} {\bibinfo  {journal} {J. Comput. Phys.}\ }\textbf {\bibinfo
  {volume} {53}},\ \bibinfo {pages} {484} (\bibinfo {year} {1984})}\BibitemShut
  {NoStop}%
\bibitem [{\citenamefont {Berger}(1986)}]{Berger86}%
  \BibitemOpen
  \bibfield  {author} {\bibinfo {author} {\bibfnamefont {M.~J.}\ \bibnamefont
  {Berger}},\ }\href@noop {} {\bibfield  {journal} {\bibinfo  {journal} {SIAM
  Journal of Scientific and Statistical Computing}\ }\textbf {\bibinfo {volume}
  {7}},\ \bibinfo {pages} {904} (\bibinfo {year} {1986})}\BibitemShut {NoStop}%
\bibitem [{\citenamefont {Berger}\ and\ \citenamefont
  {Colella}(1989)}]{Berger-1989}%
  \BibitemOpen
  \bibfield  {author} {\bibinfo {author} {\bibfnamefont {M.~J.}\ \bibnamefont
  {Berger}}\ and\ \bibinfo {author} {\bibfnamefont {P.}~\bibnamefont
  {Colella}},\ }\href {\doibase 10.1016/0021-9991(89)90035-1} {\bibfield
  {journal} {\bibinfo  {journal} {J. Comput. Phys.}\ }\textbf {\bibinfo
  {volume} {82}},\ \bibinfo {pages} {64} (\bibinfo {year} {1989})}\BibitemShut
  {NoStop}%
\bibitem [{\citenamefont {Courant}\ \emph {et~al.}(1928)\citenamefont
  {Courant}, \citenamefont {Friedrichs},\ and\ \citenamefont
  {Lewy}}]{Courant-Friedrichs-Lewy-1928}%
  \BibitemOpen
  \bibfield  {author} {\bibinfo {author} {\bibfnamefont {R.}~\bibnamefont
  {Courant}}, \bibinfo {author} {\bibfnamefont {K.}~\bibnamefont {Friedrichs}},
  \ and\ \bibinfo {author} {\bibfnamefont {H.}~\bibnamefont {Lewy}},\
  }\href@noop {} {\bibfield  {journal} {\bibinfo  {journal} {Mathematische
  Annalen}\ }\textbf {\bibinfo {volume} {100}},\ \bibinfo {pages} {32}
  (\bibinfo {year} {1928})},\ \bibinfo {note} {(English translation in
  \cite{Courant-Friedrichs-Lewy-1967})}\BibitemShut {NoStop}%
\bibitem [{\citenamefont {Courant}\ \emph {et~al.}(1967)\citenamefont
  {Courant}, \citenamefont {Friedrichs},\ and\ \citenamefont
  {Lewy}}]{Courant-Friedrichs-Lewy-1967}%
  \BibitemOpen
  \bibfield  {author} {\bibinfo {author} {\bibfnamefont {R.}~\bibnamefont
  {Courant}}, \bibinfo {author} {\bibfnamefont {K.}~\bibnamefont {Friedrichs}},
  \ and\ \bibinfo {author} {\bibfnamefont {H.}~\bibnamefont {Lewy}},\
  }\href@noop {} {\bibfield  {journal} {\bibinfo  {journal} {IBM Journal of
  Research and Development}\ }\textbf {\bibinfo {volume} {11}},\ \bibinfo
  {pages} {215} (\bibinfo {year} {1967})},\ \bibinfo {note} {(English
  translation of \cite{Courant-Friedrichs-Lewy-1928})}\BibitemShut {NoStop}%
\bibitem [{\citenamefont {Ascher}\ \emph {et~al.}(1995)\citenamefont {Ascher},
  \citenamefont {Ruuth},\ and\ \citenamefont
  {Wetton}}]{Ascher-Ruuth-Wetton-1995:IMEX-methods-for-time-dependent-PDEs}%
  \BibitemOpen
  \bibfield  {author} {\bibinfo {author} {\bibfnamefont {U.~M.}\ \bibnamefont
  {Ascher}}, \bibinfo {author} {\bibfnamefont {S.~J.}\ \bibnamefont {Ruuth}}, \
  and\ \bibinfo {author} {\bibfnamefont {B.~T.~R.}\ \bibnamefont {Wetton}},\
  }\href {\doibase 10.1137/0732037} {\bibfield  {journal} {\bibinfo  {journal}
  {SIAM J. Numerical Analysis}\ }\textbf {\bibinfo {volume} {32}},\ \bibinfo
  {pages} {797} (\bibinfo {year} {1995})}\BibitemShut {NoStop}%
\bibitem [{\citenamefont {Ascher}\ \emph {et~al.}(1997)\citenamefont {Ascher},
  \citenamefont {Ruuth},\ and\ \citenamefont
  {Spiteri}}]{Ascher-Ruuth-Spiteri-1997:IMEX-methods-for-time-dependent-PDEs}%
  \BibitemOpen
  \bibfield  {author} {\bibinfo {author} {\bibfnamefont {U.~M.}\ \bibnamefont
  {Ascher}}, \bibinfo {author} {\bibfnamefont {S.~J.}\ \bibnamefont {Ruuth}}, \
  and\ \bibinfo {author} {\bibfnamefont {R.~J.}\ \bibnamefont {Spiteri}},\
  }\href {\doibase 10.1016/S0168-9274(97)00056-1} {\bibfield  {journal}
  {\bibinfo  {journal} {Applied Numerical Mathematics}\ }\textbf {\bibinfo
  {volume} {25}},\ \bibinfo {pages} {151} (\bibinfo {year} {1997})}\BibitemShut
  {NoStop}%
\bibitem [{\citenamefont {Pareschi}\ and\ \citenamefont
  {Russo}(2000)}]{Pareschi-Russo-2000:IMEX-schemes-for-stiff-PDEs}%
  \BibitemOpen
  \bibfield  {author} {\bibinfo {author} {\bibfnamefont {L.}~\bibnamefont
  {Pareschi}}\ and\ \bibinfo {author} {\bibfnamefont {G.}~\bibnamefont
  {Russo}},\ }\href@noop {} {\bibfield  {journal} {\bibinfo  {journal} {Recent
  Trends in Numerical Analysis}\ }\textbf {\bibinfo {volume} {3}},\ \bibinfo
  {pages} {269} (\bibinfo {year} {2000})}\BibitemShut {NoStop}%
\bibitem [{\citenamefont {Pareschi}\ and\ \citenamefont
  {Russo}(2005)}]{Pareschi-Russo-2005:IMEX-RK-schemes-for-hyperbolic-PDEs}%
  \BibitemOpen
  \bibfield  {author} {\bibinfo {author} {\bibfnamefont {L.}~\bibnamefont
  {Pareschi}}\ and\ \bibinfo {author} {\bibfnamefont {G.}~\bibnamefont
  {Russo}},\ }\href {\doibase 10.1007/s10915-004-4636-4} {\bibfield  {journal}
  {\bibinfo  {journal} {Journal of Scientific Computing}\ }\textbf {\bibinfo
  {volume} {25}},\ \bibinfo {pages} {129} (\bibinfo {year} {2005})},\ \Eprint
  {http://arxiv.org/abs/arXiv:1009.2757} {arXiv:1009.2757} \BibitemShut
  {NoStop}%
\bibitem [{\citenamefont {Boscarino}(2007)}]{Boscarino-2007}%
  \BibitemOpen
  \bibfield  {author} {\bibinfo {author} {\bibfnamefont {S.}~\bibnamefont
  {Boscarino}},\ }\href {\doibase 10.1137/060656929} {\bibfield  {journal}
  {\bibinfo  {journal} {SIAM J. Numerical Analysis}\ }\textbf {\bibinfo
  {volume} {45}},\ \bibinfo {pages} {1600} (\bibinfo {year}
  {2007})}\BibitemShut {NoStop}%
\bibitem [{\citenamefont {Boscarino}(2009)}]{Boscarino-2009}%
  \BibitemOpen
  \bibfield  {author} {\bibinfo {author} {\bibfnamefont {S.}~\bibnamefont
  {Boscarino}},\ }\href {\doibase 10.1016/j.apnum.2008.10.003} {\bibfield
  {journal} {\bibinfo  {journal} {Applied Numerical Mathematics}\ }\textbf
  {\bibinfo {volume} {59}},\ \bibinfo {pages} {1515} (\bibinfo {year}
  {2009})}\BibitemShut {NoStop}%
\bibitem [{\citenamefont {Boscarino}\ and\ \citenamefont
  {Russo}(2009)}]{Boscarino-Russo-2009}%
  \BibitemOpen
  \bibfield  {author} {\bibinfo {author} {\bibfnamefont {S.}~\bibnamefont
  {Boscarino}}\ and\ \bibinfo {author} {\bibfnamefont {G.}~\bibnamefont
  {Russo}},\ }\href {\doibase 10.1137/080713562} {\bibfield  {journal}
  {\bibinfo  {journal} {SIAM J. Scientific Computing}\ }\textbf {\bibinfo
  {volume} {31}},\ \bibinfo {pages} {1926} (\bibinfo {year}
  {2009})}\BibitemShut {NoStop}%
\end{thebibliography}

%%%%%%%%%%%%%%%%%%%%%%%%%%%%%%%%%%%%%%%%%%%%%%%%%%%%%%%%%%%%%%%%%%%%%%%%%%%%%%%%
%%%%%%%%%%%%%%%%%%%%%%%%%%%%%%%%%%%%%%%%%%%%%%%%%%%%%%%%%%%%%%%%%%%%%%%%%%%%%%%%
%%%%%%%%%%%%%%%%%%%%%%%%%%%%%%%%%%%%%%%%%%%%%%%%%%%%%%%%%%%%%%%%%%%%%%%%%%%%%%%%

\end{document}